\documentclass[aps,groupedaddress,floatfix,preprintnumbers,nofootinbib,eqsecnum]{revtex4}
\usepackage{amssymb,amsmath,graphicx,multirow}

\usepackage{amssymb, amsmath}
\usepackage{graphics,graphicx,epsfig,ulem}
\usepackage{float}

\usepackage{multirow}
\usepackage{longtable}
\usepackage{color}
\usepackage{xcolor}
\usepackage{slashed}

\begin{document}

\title{Towards a Non-Supersymmetric String Phenomenology}
\author{ Steven Abel$^{1}$\footnote{E-mail address:
      {\tt s.a.abel@durham.ac.uk}},
       Keith R. Dienes$^{2,3}$\footnote{E-mail address:
      {\tt dienes@email.arizona.edu}},
      Eirini Mavroudi$^{1}$\footnote{E-mail address:
      {\tt irene.mavroudi@durham.ac.uk}}}
\affiliation{
     $^1\,$IPPP, Durham University, Durham, DH1 3LE, UK \\
     $^2\,$Department of Physics, University of Arizona, Tucson, AZ  85721  USA \\
     $^3\,$Department of Physics, University of Maryland, College Park, MD  20742  USA }

\begin{abstract}
  {\rm
Over the past three decades, considerable effort has been devoted to studying
the rich and diverse phenomenologies of heterotic strings exhibiting spacetime supersymmetry.
Unfortunately, during this same period, there has been relatively little work studying
the phenomenologies associated with 
their non-supersymmetric counterparts.
The primary reason for this relative lack of attention is the fact that strings
without spacetime supersymmetry   
are generally unstable, exhibiting large one-loop dilaton tadpoles.
In this paper, we demonstrate that this hurdle can be overcome in a class
of tachyon-free four-dimensional string models realized through coordinate-dependent 
compactifications.  
Moreover, as we shall see, it is possible to construct
models in this class 
whose low-lying states resemble the Standard Model (or even
potential unified extensions thereof) --- all without any light superpartners,
and indeed without supersymmetry at any energy scale.
The existence of such models thus opens the door to general studies of non-supersymmetric
string phenomenology, and in this paper we proceed to discuss a variety of theoretical
and phenomenological issues 
associated with such non-supersymmetric strings.
On the theoretical side,
we discuss the finiteness properties of such strings, 
the general characteristics of their mass spectra,
the magnitude and behavior of their one-loop cosmological constants, 
and their interpolation properties.
By contrast, on the phenomenological side,
the properties we discuss
are more model-specific and 
include their construction techniques, 
their natural energy scales,
their particle and charge
assignments, and the magnitudes of their associated Yukawa couplings and scalar masses.
  }
\end{abstract}

\maketitle

\def\beq{\begin{equation}}
\def\eeq{\end{equation}}
\def\beqn{\begin{eqnarray}}
\def\eeqn{\end{eqnarray}}
\def\half{{\textstyle{1\over 2}}}
\def\quarter{{\textstyle{1\over 4}}}

\def\calO{{\cal O}}
\def\calC{{\cal C}}
\def\calE{{\cal E}}
\def\calT{{\cal T}}
\def\calM{{\cal M}}
\def\calN{{\cal N}}
\def\calF{{\cal F}}
\def\calS{{\cal S}}
\def\calY{{\cal Y}}
\def\calV{{\cal V}}
\def\ibar{{\overline{\imath}}}
\def\chibar{{\overline{\chi}}}
\def\ttwo{{\vartheta_2}}
\def\tthree{{\vartheta_3}}
\def\tfour{{\vartheta_4}}
\def\ttwob{{\overline{\vartheta}_2}}
\def\tthreeb{{\overline{\vartheta}_3}}
\def\tfourb{{\overline{\vartheta}_4}}
\def\Str{{{\rm Str}\,}}

\def\bfell{{\boldsymbol \ell}}
\def\xx{\hspace{0.3cm}}
\def\xxl{\hspace{0.295cm}}
\def\xxh{\hspace{0.242cm}}
\def\yy{\hspace{0.115cm}}
\def\yyr{\hspace{-0.02cm}}

\def\qbar{{\overline{q}}}
\def\mm{{\tilde m}}
\def\nn{{\tilde n}}
\def\rep#1{{\bf {#1}}}
\def\ie{{\it i.e.}\/}
\def\eg{{\it e.g.}\/}

\newcommand{\newc}{\newcommand}
\newc{\gsim}{\lower.7ex\hbox{$\;\stackrel{\textstyle>}{\sim}\;$}}
\newc{\lsim}{\lower.7ex\hbox{$\;\stackrel{\textstyle<}{\sim}\;$}}

\newcommand{\red}[1]{\textcolor{red}{#1}}

\hyphenation{su-per-sym-met-ric non-su-per-sym-met-ric}
\hyphenation{space-time-super-sym-met-ric}
\hyphenation{mod-u-lar mod-u-lar--in-var-i-ant}


\def\inbar{\,\vrule height1.5ex width.4pt depth0pt}

\def\IC{\relax\hbox{$\inbar\kern-.3em{\rm C}$}}
\def\IQ{\relax\hbox{$\inbar\kern-.3em{\rm Q}$}}
\def\IR{\relax{\rm I\kern-.18em R}}
 \font\cmss=cmss10 \font\cmsss=cmss10 at 7pt
\def\IZ{\relax\ifmmode\mathchoice
 {\hbox{\cmss Z\kern-.4em Z}}{\hbox{\cmss Z\kern-.4em Z}}
 {\lower.9pt\hbox{\cmsss Z\kern-.4em Z}} {\lower1.2pt\hbox{\cmsss
 Z\kern-.4em Z}}\else{\cmss Z\kern-.4em Z}\fi}

\long\def\@caption#1[#2]#3{\par\addcontentsline{\csname
  ext@#1\endcsname}{#1}{\protect\numberline{\csname
  the#1\endcsname}{\ignorespaces #2}}\begingroup \small
  \@parboxrestore \@makecaption{\csname
  fnum@#1\endcsname}{\ignorespaces #3}\par \endgroup}
\catcode`@=12

\input epsf


\tableofcontents

\section{Introduction}  
\setcounter{footnote}{0}

It is an undeniable fact of nature
that the world as we know it is non-supersymmetric.
Supersymmetry is nowhere to be found amongst
the elementary particles
or their fundamental interactions --- it has certainly not appeared
at presently accessible energy scales (including those probed
at the first run of the LHC), nor 
are there currently any 
signs of  
its imminent appearance 
at higher energies.
While many 
continue to feel that a discovery of supersymmetry is 
more likely than not, 
there is increasing room for skepticism 
--- especially as current data already imply that supersymmetry would no longer be capable
of serving its primary theoretical function
of providing, in and of itself, a complete solution to the gauge hierarchy problem.
Indeed, the recent discovery of the Higgs 
makes the issue of stabilizing the gauge hierarchy all the 
more pressing.

Given the observational absence of supersymmetry, model-builders
face a stark choice.  One possibility is to 
assume that nature is
fundamentally supersymmetric 
and then to determine how this supersymmetry might be broken in 
order to yield the non-supersymmetric world we observe.
The second 
is to imagine that nature is, by contrast,
non-supersymmetric at {\it all} \ levels, and then to proceed
entirely along non-supersymmetric lines.  
This distinction is especially marked in the case of string theory
because non-supersymmetric and 
supersymmetric theories are radically different, even when they might seem to be closely related.

For most of the modern history of string theory,
it is the first option which has received the greatest attention.
As a result, string phenomenology has largely 
consisted of constructing realistic or semi-realistic string models with 
${\cal N}=1$ supersymmetry --- where
``realistic'' is usually taken to mean that the string spectrum
bears a resemblance to the Minimal Supersymmetric Standard Model (MSSM) ---
and then calling upon some field-theoretic 
SUSY-breaking mechanism (such as gaugino condensation) to break the ${\cal N}=1$ SUSY.~
However
the second option, that of constructing 
entirely non-supersymmetric string  models 
whose low-energy limits correspond directly to the Standard Model,
has received relatively little attention.
It is the purpose of this paper to 
stimulate a 
dedicated and systematic exploration of this second path.

It is important to understand that in string theory, the distinction between these two 
possibilities is {\it not}\/ a question of the {\it energy scale}\/ at which
supersymmetry is broken, with the first path corresponding to a
relatively low SUSY-breaking scale 
and the second corresponding to one at the Planck scale.  
Rather, the question 
is whether {\it the string model itself}\/, 
at its own fundamental energy scale, is supersymmetric or not.
For example, as we shall 
see, it is possible to build string 
models in which the massless states resemble the Standard Model
and in which their 
erstwhile 
superpartners have masses that can be dialed
to literally any value --- even TeV-scale values!  
However these are still {\it non-supersymmetric}\/
string models, and it would be wrong to view such string models
as having been supersymmetric at high energy scales
but subsequently subjected to some sort of SUSY-breaking mechanism
at lower energies.

This last point is critical, and may require 
additional clarification.
It is common to speak
of the various steps taken in the {\it construction}\/ of a given string model
as if they occurred in an actual 
time-ordered or energy-ordered sequence.
Within such heuristic language, one might then refer to a ``SUSY-breaking'' step in the construction.
For example, one 
mechanism 
that we shall discuss extensively in this paper 
is ``Scherk-Schwarz'' compactification, and one often refers to this
as a ``Scherk-Schwarz'' breaking of supersymmetry.
While we shall even use this terminology ourselves throughout this paper, we wish to be completely
clear that this is not a breaking of supersymmetry according to any time-ordered or energy-ordered dynamics.
Indeed, more proper language would simply indicate that there exists a well-defined 
set of procedures that enable us to produce one self-consistent 
string model from another.  Moreover, under certain circumstances, such procedures 
may produce a non-supersymmetric string model from a supersymmetric 
one. However, the resulting non-supersymmetric string model is no less fundamental than the original:
both sit independently within the landscape of possible string models
as potential independent descriptions of physics at all energy scales.

Our interest in this paper 
concerns the methods by which suitable non-supersymmetric 
strings might be constructed, and the low-energy phenomenologies to which they give rise.
We are certainly not the first to focus on such non-supersymmetric strings and their phenomenologies.
Indeed, over the past thirty years, despite the torrent of work focusing
on supersymmetric string model-building,
there has nevertheless also been a steady trickle of work focusing on the 
properties of {\it non-supersymmetric}\/ strings.
For example, ever since the construction of the first known example of a non-supersymmetric
string theory without tree-level tachyons ---
the original ten-dimensional $SO(16)\times SO(16)$ heterotic string~\cite{SOsixteen} --- 
there has also been a steady line of work studying the diverse properties of 
such strings.
This includes studies of their
one-loop cosmological constants~\cite{Rohm,nonSUSYgauge,Itoyama:1986ei,Itoyama:1987rc,Moore,
 Dienes:1990ij,KutasovSeiberg,missusy,supertraces,Kachru:1998hd,KachSilvothers,
  Shiu:1998he,Iengo:1999sm,DhokerPhong,Faraggi:2009xy},
their finiteness properties~\cite{missusy,supertraces,Angelantonj:2010ic},
and their strong/weak coupling duality symmetries~\cite{Bergman:1997rf,julie1,julie2,Faraggi:2007tj}.
There have even been studies of the landscapes of such strings~\cite{Dienes:2006ut,Dienes:2012dc},
and of course all studies of 
strings at finite temperature are also implicitly studies
of such non-supersymmetric strings~(for early work in this area, 
see, \eg, Refs.~\cite{finitetemp,AtickWitten,wasKounnasRostand,Kounnas:1989dk,earlystringpapersfiniteT}).
In general, the non-supersymmetric string models 
which were studied were either non-supersymmetric by construction
or exhibited a form of spontaneous supersymmetry 
breaking~\cite{Rohm, Ferrara:1987es, Ferrara:1987qp, Ferrara:1988jx, Kiritsis:1997ca, Dudas:2000ff,
Scrucca:2001ni, Borunda:2002ra, Angelantonj:2006ut}, achieved through
a stringy version of the Scherk-Schwarz mechanism~\cite{scherkschwarz} ---
indeed, potentially viable models within this class were constructed
in Refs.~\cite{Lust:1986kj, Lerche:1986ae,Lerche:1986cx, nonSUSYgauge, Chamseddine:1988ck, Font:2002pq,
Faraggi:2007tj,
Blaszczyk:2014qoa,
Angelantonj:2014dia}.
Moreover, although our main focus in this paper concerns weakly coupled heterotic strings,
non-supersymmetric string models have also been 
explored in a wide variety of other configurations~\cite{othernonsusy, Sagnotti:1995ga,Sagnotti:1996qj,Angelantonj:1998gj,Blumenhagen:1999ns,Sugimoto:1999tx,Aldazabal:1999tw,Angelantonj:1999xc,Forger:1999ev,Moriyama:2001ge,Angelantonj:2003hr,Angelantonj:2004yt,Dudas:2004vi,GatoRivera:2007yi}.  One
interesting aspect of Scherk-Schwarz breaking in the brane context is 
that the sizes of large extra dimensions can also play the role of order parameters for 
supersymmetry breaking. This leads to interesting 
relations between scales in various 
schemes~\cite{Antoniadis:1988jn, Antoniadis:1990ew, 
Antoniadis:1992fh,Antoniadis:1996hk,Benakli:1998pw,Bachas:1999es,Dudas:2000bn}.

Despite this body of work, however, the possibility of developing a genuinely non-supersymmetric string phenomenology for the weakly coupled heterotic string has not attracted the attention it deserves. 
Undoubtedly the major stumbling block has been the question of stability.
As we shall discuss in Sect.~II, non-supersymmetric strings 
generally give rise to non-zero tadpole diagrams.  
The existence of such tadpole diagrams
is problematic, indicating 
that such strings are generally formulated on unstable vacua.
In principle, one can absorb dilaton tadpoles via the
Fischler-Susskind mechanism~\cite{Fischler:1986ci,Fischler:1986tb}, as in Refs.~\cite{Lust:1986kj,Dudas:2000ff}.
However, if such tadpoles are unsuppressed, the new background produced
is expected to be very different from the initial one, thereby invalidating the original construction. 
Thus, in any complete discussion of non-supersymmetric string phenomenology, the question of overcoming
the instabilities associated with the dilaton tadpoles will be central.

In this paper, we show that it is possible to overcome this hurdle and build non-supersymmetric
heterotic string models which are essentially stable --- \ie, to build heterotic string models for which the degree
of instability associated with the dilaton tadpole
is exponentially suppressed compared with what might otherwise be expected.
Moreover, we demonstrate through explicit construction that 
such models can at the same time also exhibit semi-realistic phenomenological properties,
such as a particle 
content resembling that of the Standard Model 
or a Pati-Salam-type extension thereof.
Indeed, 
no light superpartners are 
predicted for many of the Standard-Model particles.
Thus, we shall see that it is possible to suppress the 
dilaton tadpole (and thereby suppress the resulting non-supersymmetric
instabilities) while simultaneously retaining a promising low-energy phenomenology ---
all without any remnant of supersymmetry in the corresponding string spectrum.

Of course, we are not claiming that such string models are completely satisfactory
as bona-fide models of the universe, or even as phenomenologically acceptable string vacua.
For example, just as with typical {\it supersymmetric}\/ string models, these non-supersymmetric
heterotic string models generally contain many unfixed moduli whose vacuum expectation values (VEV's) 
ultimately remain to be 
determined.
Our point, however, is that these non-supersymmetric models  
are just as viable as the supersymmetric ones, in the sense that the primary {\it extra}\/
instabilities which arise due to the absence of spacetime supersymmetry have been exponentially
suppressed.
This places such non-supersymmetric 
string models on an essentially equal footing with their more traditional supersymmetric 
counterparts, and thereby opens the door to 
a study of the phenomenology 
of {\it non}\/-supersymmetric heterotic strings
which mirrors that developed over the past thirty years for their
supersymmetric cousins.

This paper is organized in three main parts.
The first part, consisting of Sects.~II through IV, 
lays the groundwork
for our study in a way which is 
completely general and independent of any particular
model-construction formalism.
In Sect.~II we begin by discussing 
the finiteness and stability properties of non-supersymmetric heterotic strings,
while in Sect.~III we focus on the so-called {\it interpolating models}\/  
and explain why these strings are particularly relevant for questions of stability.
In Sect.~IV, we then discuss the one-loop cosmological 
constants associated with such strings.  In so doing, 
we focus on their leading and subleading behaviors,
paying particular attention to the role played by off-shell 
string states and the contributions they provide.
These cosmological constants are important 
for such strings 
because they
determine the sizes of the dilaton tadpoles that are ultimately 
responsible for their instability.

The stage having thus been set, 
the second part of this paper focuses
on the actual construction of semi-realistic non-supersymmetric heterotic string models with 
suppressed cosmological constants.
This is done in several steps, for which a roadmap is provided in Sect.~V.A.~
First, in Sect.~V, we introduce a particular six-dimensional free-fermionic string model from 
which our ultimate four-dimensional string models emerge upon compactification.
Then, in Sect.~VI, we consider two different types of compactification for this model --- one
a traditional SUSY-preserving $\IZ_2$ orbifold compactification and the other a SUSY-breaking
coordinate-dependent compactification (CDC) --- and analyze the four-dimensional models that result.
Finally, in Sect.~VII, we show how the latter of these four-dimensional string models
can be altered in different ways in order to achieve our 
desired goal, namely models with exponentially suppressed dilaton-tadpole instabilities.
Indeed, 
we present several models of this type, one whose low-energy
spectrum resembles the Standard Model,
and others resembling either Pati-Salam-like or GUT-like ``unified'' extensions thereof.

The third and final part of this paper then begins an exploration of the properties of
these models.  In Sect.~VIII we study two theoretical properties of these models:
the behavior of
the degeneracies associated with their physical-state spectra as functions of energy,
and the behavior of their cosmological constants as functions of their compactification
radii.  In each case, we find special features which are unique 
and which reflect the fact that these models exhibit enhanced stability properties relative 
to typical non-supersymmetric string models.
Then, in Sect.~IX, we begin a study of their phenomenological properties,
focusing on particle assignments, Yukawa couplings, and scalar masses. 
Finally, in Sect.~X, we discuss the different possibilities for the overall energy scales
that can be associated with such models.

Sect.~XI contains our Conclusions and outlines some avenues for further study.
This paper also contains four Appendices.
The first three respectively describe our notation
and conventions;  the formalism for analyzing the spectra of fermionic strings;
and our proof of a claim concerning the generation of  
tachyons in models with supersymmetry broken by discrete torsion.
The final Appendix then summarizes the explicit definitions of our models in Sect.~VII.

\section{Preliminaries:~  General results concerning the finiteness
and stability of non-supersymmetric heterotic strings}
\setcounter{footnote}{0}

This paper focuses on perturbative non-supersymmetric heterotic strings.
Accordingly, 
in this section, we review those aspects of such strings
which will be important for our work,  particularly those aspects related to their
finiteness and stability properties.
Along the way, we will also clarify some potentially subtle points,
dispel some common misconceptions,
and offer a perhaps unique perspective on these issues  which may not be widely appreciated 
in the string community.

\subsection{Partition functions of the heterotic string}

The stability properties of a given closed string 
can often be analyzed by studying its one-loop partition function $Z(\tau)$, where $\tau$ is
the one-loop modular parameter.  
In general, such partition functions are nothing
but the traces over the left- and right-moving string Fock spaces,
\beq
  Z(\tau) ~=~ {\rm Tr}\, (-1)^F \, \qbar^{H_R} q^{H_L}~,
\eeq
where $q\equiv e^{2\pi i\tau}$, where $(H_R,H_L)$ are the right- and left-moving
world-sheet Hamiltonians whose eigenvalues are the right- and left-moving world-sheet energies $(E_R,E_L)$,
and where $F$ denotes the spacetime fermion number.
If the string in question is formulated in $D$ spacetime dimensions, then this partition function $Z(\tau)$
may be expanded in a double power-series of the form
\beq
  Z(\tau) ~=~ {\tau_2}^{1-D/2} \sum_{m,n} \, a_{mn} \, \qbar^m q^n~
\label{partfunctgen}
\eeq
where $\tau_1\equiv {\rm Re}\,\tau$, $\tau_2\equiv {\rm Im}\,\tau$, and $a_{m,n}$ denotes the net number of spacetime bosonic
minus spacetime fermionic string states with world-sheet energies $(E_R,E_L)=(m,n)$.
In general, the spacetime mass $M$ of a given $(m,n)$  state is given by $\alpha' M^2\sim m+n$ where
$M_{\rm string}\equiv 1/\sqrt{\alpha'}$ is the string scale;  as a result we see that states with $m+n<0$ are tachyonic.  In general, states with $m=n$ are deemed ``physical'' and can survive as in- or out-states;  these are the states that are usually described as being part of the string spectrum. 
By contrast, states with $m\not=n$ are deemed ``unphysical'' or ``off-shell'';  such states cannot serve as in- or out-states, but can nevertheless propagate within (and hence contribute to) loop amplitudes.
For heterotic strings, $m\geq -1/2$ and $n\geq -1$.
The leading factors of $\tau_2$  
in Eq.~(\ref{partfunctgen})
come from the ``traces'' over the continuum of states corresponding to the uncompactified 
transverse spacetime dimensions.

Modular invariance is a fundamental symmetry of all closed perturbative strings, and requires that 
the partition function $Z(\tau)$ be invariant under $T: \, \tau\to\tau+1$ and $S:\, \tau\to -1/\tau$.
Invariance under the $T$ transformation therefore requires that 
any $(m,n)$ string state have $m-n\in\IZ$.
Of course, only if $m-n=0$ will the corresponding state be physical (``level-matched'').

If the heterotic string in question has a spectrum exhibiting spacetime supersymmetry, we then have
$a_{mn}=0$ for all $(m,n)$.
As a result, we find that $Z(\tau)=0$ for supersymmetric theories.
String theories with $Z(\tau)\not=0$ are therefore necessarily 
non-supersymmetric, which is one reason the partition function is a particularly powerful tool 
for explorations of non-supersymmetric theories.

As already implicit in the above discussion, 
there are two pieces of string folklore which are incorrect but which nevertheless survive in some quarters. 
They therefore deserve explicit refutation.
\begin{itemize}
\item{}  The first is that modular invariance somehow requires supersymmetry.
This confusion occurred historically when these two features appeared to be correlated 
in the earliest string models.  However, it is well known that there is no such correlation in general:
there exist many examples of non-supersymmetric string theories which are
nevertheless modular invariant.  Indeed, all of the string theories we shall discuss in this 
paper are of this variety.

\item  Second, it is also a piece of string folklore --- indeed, one which has the same historical roots --- 
that string theories without supersymmetry must be tachyonic.  
However, like its cousin, this assertion is also false, at least at tree level:  there exist many examples of
non-supersymmetric string models whose tree-level spectra are entirely tachyon-free.
Note that freedom from
tachyonic states in this context merely requires that $a_{nn}=0$ for all $n<0$ --- \ie, that the number
of bosonic tachyons match the number of fermionic tachyons at all tachyonic mass levels.  
However, since fermionic tachyonic states with $m=n<0$ are generally forbidden by Lorentz invariance,
the claim that $a_{nn}=0$ for all $n<0$ actually implies that there are no tachyonic states of any spin whatsoever.
However, we must emphasize that it is at present only a tree-level statement that there exist
tachyon-free non-supersymmetric string theories.
Once quantum effects are included, the vacua of such non-supersymmetric theories 
can generally shift, and tachyons might be generated at higher loops.  The status of this 
issue will be discussed below, and is best studied on a case-by-case basis.

\end{itemize}

Finally, it will be relevant for our later work to understand the relation between
partition functions of string theories in different spacetime dimensions.
In general, if a $D$-dimensional string theory with partition function $Z^{(D)}$ is compactified
on a $d$-dimensional volume $V_d$, resulting in a $(D-d)$-dimensional string theory 
with partition function $Z^{(D-d)}$,
we can identify
\beq
        Z^{(D)} ~=~ \lim_{V_d\to \infty} \left[ {1\over \calM^d V_d} \, Z^{(D-d)}\right]~
\label{identifyZs}
\eeq
where ${\cal M}\equiv M_{\rm string}/(2\pi)=1/(2\pi \sqrt{\alpha'})$ is the reduced string scale.
Likewise, if we are dealing with closed strings, then the $V_d\to 0$ limit
will also generally produce a $D$-dimensional string.  
This is the result of T-duality.
In such cases, we continue to have the same relation as in Eq.~(\ref{identifyZs}) but with $V_d$ replaced
with a suitably identified {\it T-dual}\/ volume $\tilde V_d$.

\subsection{Proto-gravitons:~  A general theorem}

In general there can be many different kinds of physical and unphysical states which contribute to $Z(\tau)$.
However, {\it every non-supersymmetric string model
necessarily contains off-shell tachyonic states with $(m,n)=(0,-1)$}\/.
This is a general theorem~\cite{Dienes:1990ij} 
which holds regardless of the specific class of non-supersymmetric string 
model under study, and regardless of the particular GSO projections that might be imposed. 
As a result, every non-supersymmetric string theory must have $a_{0,-1}\not=0$.

It is easy to understand the origins of this result.
We know that every string model contains 
a completely NS/NS sector from which the gravity multiplet arises:
\beq
        \hbox{graviton} ~~\subset ~~
              \tilde \psi_{-1/2}^\mu |0\rangle_R ~\otimes~ ~\alpha_{-1}^\nu |0\rangle_L~.
\label{graviton}
\eeq
Here $|0\rangle_{R,L}$ are the right- and left-moving vacua of the heterotic string, 
$\tilde \psi_{-1/2}^\mu$ represents the excitation
of the right-moving world-sheet Neveu-Schwarz fermion $\tilde \psi^\mu$,
and $\alpha_{-1}^\nu$ represents the excitation
of the left-moving coordinate boson $X^\nu$.
Indeed, no self-consistent GSO projection can possibly 
eliminate this gravity multiplet from the string spectrum.
However, given that the graviton is always in the string spectrum, 
then there must also exist in the string spectrum a corresponding state
for which the left-moving coordinate oscillator is {\it not}\/ excited:
\beq
        \hbox{proto-graviton:}~~~~~~~~~~~~
              \tilde \psi_{-1/2}^\mu |0\rangle_R ~\otimes~ |0\rangle_L~.
\label{protograviton}
\eeq
This ``proto-graviton'' state has world-sheet energies $(E_R,E_L)=(m,n)=(0,-1)$, and is thus off-shell  
and tachyonic.  However, it always exists in the string spectrum so long as the graviton exists.
(In CFT language, they are part of the same Verma module, with the graviton appearing as a descendant
of the proto-graviton.)  We shall see explicit examples of this below.

Of course, in 
a supersymmetric theory, there is likewise a ``proto-gravitino'' state
which also exists in the spectrum:
\beq
        \hbox{proto-gravitino:}~~~~~~~~~~~~
         \lbrace \tilde \psi_{0}\rbrace^\mu  |0\rangle_R ~\otimes~ ~ |0\rangle_L~,
\label{protogravitino}
\eeq
where  $\lbrace \tilde \psi_{0}\rbrace^\alpha$ schematically indicates the Ramond zero-mode
combinations which collectively give rise to the spacetime Lorentz spinor index $\mu$.
Indeed, this state is ultimately related to the gravitino in exactly the same way as the proto-graviton
is related to the graviton.
Thus, in a supersymmetric theory, the contributions from the proto-graviton and proto-gravitino
states cancel in the full
partition function, so that the full partition function lacks 
a contribution $\sim q^{-1}$ (and in fact vanishes entirely).
However, {\it any GSO projection which eliminates the gravitino from the string spectrum 
and thereby produces a non-supersymmetric string will also correspondingly eliminate the proto-gravitino state}\/.  
This is ultimately because GSO projections are completely insensitive to the excitations of the coordinate bosons.
There will therefore be nothing to cancel against the contribution from the proto-graviton state, and the resulting non-supersymmetric partition function will necessarily have
$a_{0,-1}>0$. 

In general, as evident from Eq.~(\ref{protograviton}),
the proto-graviton states transform as vectors under the transverse spacetime Lorentz symmetry $SO(D-2)$.
Thus, any non-supersymmetric string theory in $D$ spacetime dimensions must have a partition function
which begins with the contribution 
\beq
          Z(\tau) ~=~ {D-2 \over q } ~+~ ...
\label{protogravitoncontributions}
\eeq
Checking this term thus provides a convenient 
method of verifying the overall normalization of a given string partition function.

\subsection{One-loop cosmological constants and their dominant contributors}

In string theory, conformal invariance kills the tree-level contribution to the 
vacuum energy density (cosmological
constant).  As a result, the dominant contribution to this quantity
comes at one-loop order.
In general, for any string model in $D$ dimensions with partition function $Z(\tau)$,
the corresponding $D$-dimensional one-loop vacuum energy density
may be evaluated as
\beq
         \Lambda^{(D)} ~\equiv~
     -\half \,{\cal M}^D\, \int_{\cal F} {d^2 \tau\over {\tau_2}^2}
             Z(\tau)~
\label{cosconstdef}
\eeq
where $D$ is the number of uncompactified spacetime dimensions,
where ${\cal M}$ is the reduced string scale defined above,
and where
\beq
   {\cal F}~\equiv ~\lbrace \tau:  ~|{\rm Re}\,\tau|\leq \half,~
 {\rm Im}\,\tau>0, ~|\tau|\geq 1\rbrace
\label{Fdef}
\eeq
is the fundamental domain
of the modular group.
In general, it is convenient to regard the fundamental domain $\calF$ as being composed of
two separate regions, an ``upper'' region with $\tau_2\geq 1$ and a ``lower'' region with
$\tau_2<1$.  The upper region extends across the full width $-1/2\leq \tau_1\leq +1/2$;
in this region, the $\tau_1$-integration then guarantees that only the states with $m=n$ survive
as contributors to $\Lambda$.
However, even the unphysical states with $m-n\in\IZ\not=0$ will make contributions
to $\Lambda$ through integration over the lower
region within $\calF$.  Thus, {\it all}\/ states --- both physical and unphysical --- are
relevant in calculations of $\Lambda$. 

In the following we shall usually disregard the prefactor $\half {\cal M}^D$ in Eq.~(\ref{cosconstdef}) and
regard $\Lambda$ as a pure number, but we note that a proper definition does indeed require it.
We shall, however, retain the minus sign in Eq.~(\ref{Fdef}) in all discussions below.
Furthermore, we observe that if a $D$-dimensional string with partition function $Z^{(D)}$ 
is compactified on a $d$-dimensional volume $V_d$, resulting in a $(D-d)$-dimensional string
with partition function $Z^{(D-d)}$, then $\Lambda^{(D-d)}$ will typically diverge as $V_d\to \infty$.
In such cases, we can alternatively define $\tilde \Lambda^{(D-d)} \equiv  \Lambda^{(D-d)}/V_d$;
note that $\tilde \Lambda^{(D-d)}$ continues to describe the $(D-d)$-dimensional theory
but now has the mass dimensions appropriate for a $D$-dimensional [rather than $(D-1)$-dimensional] 
vacuum energy density.
Substituting the result in Eq.~(\ref{identifyZs}), 
we then find that 
\beq
   \Lambda^{(D)} ~=~ \lim_{V_d\to\infty} \tilde \Lambda^{(D-d)}~.
\eeq
The same relations also hold in the $V_d\to0$ limit, provided we replace $V_d$ with the appropriate T-dual
volume $\tilde V_d$.

As apparent from its definition in Eq.~(\ref{cosconstdef}), the cosmological constant $\Lambda$ is real.
This means that when evaluating $\Lambda$, only the {\it symmetric}\/ part of the 
physical-state degeneracy matrix $a_{mn}$ 
within the partition function $Z$ in Eq.~(\ref{partfunctgen}) is relevant.
More precisely, for any degeneracy matrix $a_{mn}$, we can define a corresponding matrix~\cite{Dienes:1990ij} 
\beq
        a'_{mn} ~\equiv~ \begin{cases}
                                 a_{mn} + a_{nm}  & {\rm for~} m>n\cr
                                 a_{mn} & {\rm for}~ m=n\cr
                                 0      & {\rm otherwise}~. 
                         \end{cases}
\eeq
As long as any two degeneracy matrices have the same $a'_{mn}$, they will
give rise to the same cosmological constant.
Conversely, it follows that any purely imaginary partition function  
with $a_{mn}= -a_{nm}$ for all $(m,n)$ will give rise to $a'_{mn}=0$ and hence $\Lambda=0$.
Of course, no self-consistent non-supersymmetric heterotic-string partition function can possibly
be purely imaginary:
we have already seen that the 
partition function of any non-supersymmetric
heterotic string must have $a_{0,-1}>0$, yet
we cannot also have
the non-zero value for $a_{-1,0}$ that would also be required for a purely imaginary partition function
since this would violate
the heterotic-string bound $m\geq -1/2$.
It {\it is}\/ possible, however, for two string models to have partition functions which {\it differ}\/ by a purely
imaginary function and thereby share the same cosmological constant. 
This feature, along with the existence of additional modular functions 
which also integrate to zero, turns out to be responsible for a large one-loop degeneracy
within the space of non-supersymmetric four-dimensional string vacua~\cite{Dienes:1990ij}. 

It will be important for later purposes to have some sense of the relative sizes
of the contributions to the cosmological constant (\ref{cosconstdef})
that come from individual $(m,n)$ string states.
In general, a given state with $(E_R,E_L)=(m,n)$ contributes a term $\qbar^m q^n$
to the partition function, thus making a contribution 
proportional to
\beq
        I_{m,n}^{(D)} ~\equiv~ \int_{\cal F} {d^2\tau\over \tau_2^2} \, \tau_2^{1-D/2} \, \qbar^m q^n
\label{integral}
\eeq
to the cosmological constant.
Note that modular invariance requires that $m-n\in \IZ$.
Note that for unphysical states (\ie, states with $m\not=n$),
the modular integral in Eq.~(\ref{integral}) vanishes in the 
rectangular upper $(\tau_2\geq 1)$ portion of the fundamental domain $\calF$ but nevertheless receives contributions
from the curved lower $(\tau_2<1)$ portion.

It is a common supposition that massless physical states (\ie, states with $m=n=0$) 
make the dominant contributions to vacuum amplitudes.
Indeed, it is easy to verify that $I_{nn}\sim e^{-4\pi n}$ for large $n$,
confirming the trend that the contributions from
heavy physical states are exponentially suppressed relative to those from lighter states.
[As we shall discuss, the numbers of states at each mass level actually {\it grow}\/
as a function of the mass, like $\exp(c\sqrt{n})$.  Ultimately this is not sufficient to overcome the
mass-suppression factor $\exp(-4\pi n)$, which is why the sum over contributions from increasingly massive
states is ultimately convergent.] 
One can also demonstrate that the contributions from states with $m\not=n$ 
are generally suppressed relative to those with $m=n$, even for fixed total energy/mass $m+n$.

However, for relatively light states, we find:
\beq
\begin{tabular}{||r|r|| c | c||}
\hline
\hline
~~$m$~~ & ~~$n$~~ & ~~~$I^{(10)}_{m,n}$~~~& ~~~$I^{(4)}_{m,n}$~~~\\
\hline
\hline
 $  0  $ & $ -1  $&  $-14.258$ &  $-12.192$\\
\hline
 $  1  $ & $ -1  $&  $\phantom{-}0.014$ & $\phantom{-}0.010$\\
 $  1/2  $ & $ -1/2  $&  $-0.038$  & $-0.032$ \\
 $  0 $& $  0  $&  $\phantom{-}0.257$ & $\phantom{-}0.549$ \\
\hline
 $  2 $&$   -1  $&  ~~~$-2.569\times 10^{-5}$~~~ & ~~~$ -1.803\times 10^{-5}$~~~\\
 $  3/2 $&$   -1/2 $&  $\phantom{-}4.682 \times 10^{-5}$  & $\phantom{-}3.456 \times 10^{-5}$\\
 $  1 $&$   0  $&  $-1.029\times 10^{-4}$ & $-8.463\times 10^{-5}$\\
 $  1/2  $&$ 1/2  $&  $\phantom{-}3.021\times 10^{-4}$ & $\phantom{-}3.304\times 10^{-4}$\\
\hline
\hline
\end{tabular}
\eeq
We thus see that the states which make the largest contributions to the cosmological constant
are actually the {\it off-shell tachyonic states}\/  with $(E_R,E_L)=(m,n)=(0,-1)$!
Indeed, the contributions from these states are actually bigger than those from the physical
massless states by a factor of $\sim 55$ for $D=10$ and $\sim 22$ for $D=4$.

What makes this feature 
particularly critical is the fact, already discussed above, that
every closed string model
contains such off-shell tachyonic states with $(m,n)=(0,-1)$, regardless of the particular
GSO projections that may be imposed.
Indeed, these are nothing but the proto-graviton states discussed above.
Thus, these large contributions to $\Lambda$ are necessarily present for any non-supersymmetric
string model, and any attempt to cancel $\Lambda$ must therefore find a way of cancelling
these contributions as well.

\subsection{Misaligned SUSY, supertraces, and finiteness without SUSY}

We have already seen that supersymmetric string theories 
necessarily have vanishing one-loop partition functions,  
\ie, $a_{nn}=0$ for all $n$.  From this it follows that  $Z(\tau)=0$ and $\Lambda=0$.
This cancellation is at the root of the extraordinary finiteness properties which
unbroken supersymmetry bestows upon the theories that exhibit it, and which make 
it an excellent candidate for
solving the hierarchy problems associated with both the Higgs mass as well
as the cosmological constant.
One way of quantifying these finiteness effects 
is through the calculation of {\it supertraces}\/, which are
essentially statistics-weighted sums over the entire spectrum of the theory:
\beq
              \Str \calM^{2\beta} ~\equiv~ \sum_{{\rm states}~i} (-1)^F (M_i)^{2\beta}~.
\label{supertracedefFT}
\eeq
In theories with unbroken supersymmetry, the direct pairing of degenerate bosonic and
fermionic states then implies the vanishing of all supertraces:
\beq
        \Str \calM^{2\beta}~=~0 ~~~~~~~~~{\rm for~all~~} \beta\geq 0~.
\label{sutracevanishing}
\eeq
These supertraces are important because they relate directly to hierarchy issues by
governing the quantum-mechanical sensitivities
of light energy scales (such as the Higgs mass $m_H$ or  
the cosmological constant $\Lambda$) to heavy mass scales (\eg, a cutoff $\lambda$):
\beqn
     \delta m_H^2 &\sim& (\Str \calM^0) \lambda^2 + (\Str \calM^2) \log \lambda + ... \nonumber\\
     \Lambda &\sim&  (\Str \calM^0) \lambda^4 + (\Str \calM^2) \lambda^2 + (\Str \calM^4) \log \lambda ~ + ...
\label{divergences}
\eeqn
Indeed, these relations 
hold supermultiplet by supermultiplet across the entire spectrum.
(Of course in the case of the 
cosmological constant all states in the theory are included,
while in the case of the Higgs we include only those states 
to which it couples.)
Thus, as a result of Eq.~(\ref{sutracevanishing}),
we see that unbroken supersymmetry 
in principle 
solves the hierarchy problems associated with both the hierarchy problems associated with
the Higgs mass and the cosmological constant.
In fact, unbroken supersymmetry even goes one step further, and causes $\Lambda$ to vanish outright.
Of course, nature does not exhibit unbroken supersymmetry, which greatly complicates this situation.

At first glance, one might not expect any similar properties to hold for {\it non}\/-supersymmetric
strings.  However, string theories still generally possess a measure of finiteness which transcends the possible
appearance of supersymmetry in their spacetime spectra. 
Indeed, this finiteness stems
directly from the nature of strings as extended objects.
For closed strings, this spatial extent implies that quantum loop diagrams
take the form of closed surfaces of various genus which are therefore subject to certain symmetries 
for which there are no field-theoretic analogues.  
For example, a one-loop amplitude in a closed string theory
takes the form of a torus, which then requires that all corresponding amplitudes be modular invariant.
Higher-loop amplitudes are likewise subject to multi-loop versions of modular invariance.
Indeed, modular invariance is nothing but a loop extension of conformal invariance. 
Thus, at least for closed strings, modular invariance sits at the root of the extra finiteness
properties that such string theories enjoy --- even without spacetime supersymmetry.

At a formal level, it is well understood how modular invariance achieves this miracle:
the expected field-theoretic one-loop divergences that would be expected to occur 
without supersymmetry 
all reside in the ultraviolet $\tau\to 0$ region of the modular-group fundamental 
domain, yet the modular symmetries that emerge when calculating quantum effects
enable us to truncate our integrations so that this region is excluded.
However, what is less appreciated is how modular invariance --- which is a symmetry of the 
partition function $Z(\tau)$ itself --- manages to achieve this miracle at the level of the
actual string spectrum whose trace this partition function represents.
What constraints does modular invariance enforce on the string spectrum so as to achieve
finiteness without supersymmetry?

The answer is quite remarkable.
It turns out that modular invariance provides a powerful restriction on the degree to 
which supersymmetry can actually be broken in string theory:
{\it in any tachyon-free closed string theory, 
spacetime supersymmetry may be broken  
but a residual so-called ``misaligned supersymmetry'' must always remain in the string
spectrum}~\cite{missusy}.
Indeed, misaligned supersymmetry is a general feature
of non-supersymmetric string models, and serves as the way in which the spectrum of
a given non-supersymmetric string theory
manages to configure itself at all mass levels so as to maintain finiteness even without spacetime supersymmetry.

\begin{figure*}[h!]
\begin{center}
  \epsfxsize 3.0 truein \epsfbox{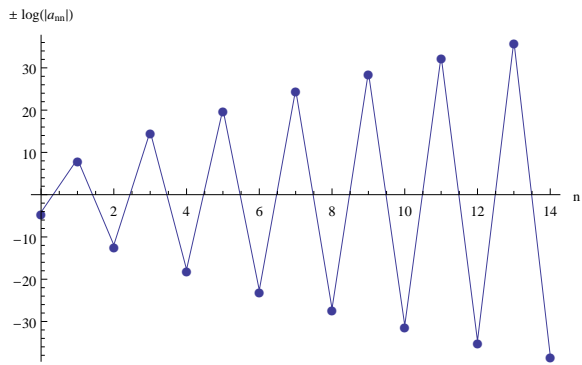}
\end{center}
\caption{A sketch of the boson/fermion oscillations which are the  
hallmark of a hidden ``misaligned supersymmetry'' existing in the spectrum of 
all tachyon-free non-supersymmetric closed strings.
For each mass level $n$ we 
plot $\pm \log(|a_{nn}|)$, where $a_{nn}$ is the number of bosonic minus fermionic
states at that level. The overall sign is chosen according to the sign of $a_{nn}$;  thus
positive values indicate surpluses of bosonic states over fermionic states, and negative
values indicate the reverse.  The points are connected in order of increasing $n$ in order
to stress the alternating, oscillatory behavior of the boson and fermion surpluses throughout 
the string spectrum.  These oscillations represent the manner in which
non-supersymmetric string theories continue to exhibit the finiteness for which
string theory is famous, even without supersymmetry.} 
\label{fig:missusy}
\end{figure*}

The detailed mathematics behind misaligned supersymmetry is beyond the scope of this paper
and is presented in Ref.~\cite{missusy}.
However the most phenomenological imprint of misaligned supersymmetry is easy to characterize.
In supersymmetric theories, we have equal numbers of bosonic and fermionic states at each mass level.
All associated cancellations thus occur level by level across the entire string spectrum.
In non-supersymmetric theories, 
by contrast, such cancellations do not occur level by level, but instead occur
through conspiracies amongst the contributions from all different levels across
the entire string spectrum.
Indeed, at one mass level there might be a surplus of bosonic states.  However
there will then be {\it an even larger}\/ surplus of fermionic states at an even higher level,
followed by an even larger surplus of bosonic states at an even higher energy level,
and so forth.  In other words, as illustrated in Fig.~\ref{fig:missusy},
surpluses of bosonic and/or fermionic states tend to {\it oscillate}\/
and grow exponentially as one progresses upwards through the string spectrum.

The plot shown in Fig.~\ref{fig:missusy} is only an idealized sketch for a simple system;
for actual semi-realistic string models the detailed oscillation patterns can be quite complex.
However, the general imprint of the underlying ``misaligned supersymmetry''
remains the same:  the string spectrum generally exhibits 
repeating patterns of bosonic/fermionic surplus oscillations
within an exponentially growing ``envelope'' function $\Phi(n)\sim |a_{nn}|\sim e^{c\sqrt{n}}$.
A given bosonic surplus may have magnitude $\Phi(n_i)$, but this requires
a corresponding fermionic surplus of magnitude $\Phi(n_i+ \Delta n)$, which in turn
requires a corresponding bosonic surplus of magnitude $\Phi(n_i+2\Delta n)$, and so forth.
Of course, this exponential growth in the numbers of bosonic and fermionic states
is a well-known feature of all string theories, supersymmetric or not, and leads directly
to the string Hagedorn transition~\cite{Hagedorn}.
Indeed, in the case of a supersymmetric theory, these bosonic and fermionic numbers of states
are matched at each level,
so even though each number separately experiences
an exponential growth, their  {\it difference}\/ $a_{nn}$ remains strictly zero.  In such
cases there is no oscillation at all in a plot such as that in Fig.~\ref{fig:missusy}. 
However, in the case of a non-supersymmetric string, bosonic and fermionic states are simply
``misaligned''.  This then results in the wobbling oscillatory behavior\footnote{
    A more precise description of misaligned supersymmetry is as follows~\cite{missusy}.
    In general, the oscillation patterns exhibited by the bosonic and 
    fermionic surpluses in a given string model can be fairly complex, with different
    sectors of the theory corresponding to different bosonic and fermionic envelope functions
    $\Phi^{(i)}_B(n)$ and $\Phi^{(i)}_F(n)$.  However, there exist well-defined methods of analytically
    generating {\it exact}\/ expressions for these envelope functions~\cite{HR},
    and one finds that in general these envelope functions take the form of a leading ``Hagedorn''
    exponential, followed by an infinite series
    of subleading exponential functions, followed ultimately by terms which are
    polynomial in $n$.
    Given this, the precise statement of misaligned supersymmetry is that the sum of the bosonic
    envelope functions necessarily 
    experiences a relative cancellation against the sum of the fermionic envelope functions, 
       {\it i.e.}\/,
    \beq
          {\sum_i \Phi^{(i)}_B(n) - \sum_i\Phi^{(i)}_F(n) 
          \over \sum_i \Phi^{(i)}_B(n) + \sum_i\Phi^{(i)}_F(n)} ~\to~ 0~ ~~~~~~~~ {\rm as}\/ ~~n\to\infty~.
    \eeq
    The precise degree to which $\sum_i\Phi^{(i)}_B (n)$ cancels directly against
    $\sum_i\Phi^{(i)}_F (n)$ 
    as a function of $n$ depends on the off-shell tachyonic structure of the theory 
    and its overall stability properties;  this cancellation is even 
    conjectured to be complete under certain circumstances.  
    Further details can be found in Ref.~\cite{missusy}.} 
sketched in Fig.~\ref{fig:missusy}.

 {\it This, then, is the limit to which supersymmetry can be broken in any closed, tachyon-free string theory:
it can at most be ``misaligned'' in the manner described above.}\/
It is important to realize how profoundly different this is from any usual field-theoretic
breaking of supersymmetry.
In field theory, a breaking of supersymmetry will generally induce a splitting between 
the different states within a given supermultiplet.  Indeed, this splitting might then lead to 
a surplus of bosonic states with one mass and a surplus of fermionic states with a different mass.
However, in such situations it is still possible to identify the original states and their superpartners 
in a pairwise fashion, and mentally reassemble the multiplet.
In string theory, by contrast, there are two profound differences.
First, in string theory the superpartners are often {\it gone}\/ --- literally  projected out of the spectrum ---  
and it is not always clear that one can identify any other heavier 
state in the spectrum as the would-be superpartner.
But even if one can occasionally find a would-be superpartner of a given state 
at a higher mass level, it will still generally
not be possible to achieve a pairwise identification of bosons with fermions across the string spectrum.
The reason for this is already clear from the sketch in Fig.~\ref{fig:missusy}:
the magnitudes of the alternating surpluses are actually {\it growing}\/ because  each  
surplus has a magnitude which samples the common monotonically growing envelope function $\Phi(n)$ 
at a different value of $n$.
These alternating surpluses thus no longer cancel in any pairwise fashion, and it is only through a conspiracy 
between the physics at {\it all}\/ mass levels across the entire string spectrum that finiteness is achieved.

Just as in the case of unbroken supersymmetry,
the finiteness inherent in  {\it misaligned}\/ supersymmetry
is also encoded in supertrace cancellations. 
Unlike the situation in field theory,
however, string theories have infinite towers of states 
with exponentially growing degeneracies.
As a result, we must first define a {\it regulated}\/ string supertrace~\cite{supertraces}:
\beq
    \Str \, M^{2\beta} ~\equiv ~
             \lim_{y\to 0} \, \sum_{\rm states}\,
           (-1)^F  \, M^{2\beta}\, e^{-y\alpha' M^2}~; 
\eeq
note that the regulator $y$ leads to a convergent sum over states
and is then removed once the sum is evaluated.  
This regulator also respects modular invariance, as required for such analyses.
Second, using the standard normalizations appropriate for closed strings, 
we can identify any state with world-sheet energies $(m,n)$ as
having spacetime mass $M^2=2(m+n)M^2_{\rm string}$. 
With these conventions, 
we then find that the spectrum of any non-supersymmetric, tachyon-free
closed string model in $D$ uncompactified spacetime dimensions satisfies~\cite{supertraces} 
\beq
       \Str\, M^0 ~=~ 0~,~~~ \Str\, M^2 ~=~ 0~,~~~ ...~~~ \Str\, M^{D-4} ~=~ 0~,
\label{supertraces1}
\eeq
as well as 
\beq
      \Str\, M^{D-2} ~=~   6\, (-4\pi)^{D/2} \,\left( D/2  -1\right)! \, {\Lambda\over M_{\rm string}^2}~.
\label{supertraces2}
\eeq 
Indeed, these results hold for any such string model regardless of its method of construction,
its compactification manifold, or the details of its low-energy phenomenology.
Unlike the case in field theory, however, these cancellations do not happen level by level,
but rather represent the collective behavior of states at all mass levels simultaneously.
Indeed, they are the direct consequence of the hidden ``misaligned supersymmetry'' which
necessarily remains in the spectra of such theories,
and it is only through the boson/fermion oscillations associated with misaligned supersymmetry
that these supertrace constraints are ultimately satisfied.

One remarkable feature of these supertrace relations is the fact that the supertrace $\Str\,M^{D-2}$ is
actually proportional to the cosmological constant $\Lambda$!
This indicates that in string theory, the issues of finiteness and cancellation of the cosmological
constant are directly tied together in a way that they are not connected in field theory.
Indeed, we see upon comparison
with our expectations from quantum field
theory that modular invariance
and misaligned supersymmetry
are so powerful as symmetries
that they effectively soften
the divergences of a vacuum amplitude
such as the cosmological constant
by four powers of mass.
These issues are discussed more fully in Refs.~\cite{supertraces,heretic}.
Further, we shall see below that in a given string model the value of the cosmological constant 
is directly connected to its ultimate stability. 
Thus, in string theory, we see that issues of hierarchy, finiteness, and stability are all deeply
connected to each other.
As a consequence, any mechanism that leads to a suppression of the cosmological constant
of a given string model will simultaneously help to stabilize this model as well 
as enhance its finiteness properties and resolve its apparent field-theoretic hierarchy problems ---
all without supersymmetry.

\subsection{The myth of ``asymptotic supersymmetry''}

We now turn to another issue concerning the spectrum of non-supersymmetric strings,
namely the notion of ``asymptotic supersymmetry''.

As we have seen, modular invariance and misaligned supersymmetry are powerful tools 
which drive many of the results we have presented thus far.
However, there also exist additional constraints on the spectra of non-supersymmetric
tachyon-free string theories.
For example, given the definition of the cosmological constant $\Lambda$ in Eq.~(\ref{cosconstdef}),
it might seem that one requires knowledge of both the
physical ($m=n$) and unphysical ($m\not= n$) states
in order to calculate the cosmological constant.  In particular,
as discussed above, 
unphysical states make a non-zero contribution to $\Lambda$ 
when their contributions to the partition function are integrated
over the lower ($\tau_2<1$) region of the fundamental domain.
However, the physical and unphysical string states are related to each
other through modular invariance, and it turns out that modular invariance
enables us to determine the total contribution to $\Lambda$ 
from the unphysical states in terms of the total contribution to $\Lambda$ 
from just the physical states.
This implies that it should be possible to express our final
expression for $\Lambda$ in Eq.~(\ref{cosconstdef}) directly in terms
of only the diagonal elements $g_{nn}$,
and indeed it has been shown that~\cite{KutasovSeiberg} 
\beq
      \int_{\cal F} {d^2 \tau\over {\tau_2}^2}  Z(\tau) ~=~ {\pi\over 3} \lim_{\tau_2\to 0} g(\tau_2)
    ~~~~~{\rm where}
    ~~~
           g(\tau_2) ~\equiv~ \int_{-1/2}^{1/2} d \tau_1 \, Z(\tau_1,\tau_2)~.
\label{KS}
\eeq
This result holds for all critical closed string theories 
which are modular invariant and free of physical tachyons.

If $Z(\tau)$ has the general form $Z=\tau_2^{k} \sum_{m,n} a_{mn} \qbar^m q^n$, then
$g(\tau_2) = \tau_2^k \sum_n a_{nn} e^{-4\pi \tau_2 n}$.
In other words, we see that 
$g(\tau_2)$ represents a ``projection'' of the full partition function $Z$ onto those states which are physical (\ie, satisfy left/right level-matching).
The $\tau_2\to 0$ limit can then usually be evaluated by converting the sum within $g(\tau_2)$ into an integral.
In general, this integral will produce precisely the number of factors of $\tau_2$ needed to 
cancel the $\tau_2^k$ prefactor 
and allow a finite $\tau_2\to 0$ limit to emerge.

The result in Eq.~(\ref{KS}) is an important one which can 
be interpreted as constraining the effective numbers of
degrees of freedom in string theory.
However this result  
is extremely subtle, and can lead to many erroneous conclusions if
incorrectly interpreted.  For example,
at first glance 
it might seem that this relation asserts that off-shell states (those with $m\not=n$) make no contributions to $\Lambda$, since the right side of Eq.~(\ref{KS}) manifestly cares only about physical states.
However, such states are clearly required for modular invariance, without which this theorem would not apply.
Or, to phrase things differently, the existence of such off-shell states in the full partition function $Z$ 
on the left side of this relation has a direct relationship, through modular transformations, 
to the densities $g(\tau_2)$ of physical states with $m=n$ which come into the right side of this relation.
Thus, one must be extremely careful when attempting to draw conclusions from this relation.

As another example, let us consider what some have called ``asymptotic supersymmetry''.
Soon after the relation in Eq.~(\ref{KS}) was derived, it was assumed in many quarters 
that the only configuration of states which could possibly be consistent with Eq.~(\ref{KS}) 
is one for which
\beq
               a_{nn}~\stackrel{?}{\to}~ 0~~~~~ {\rm as}~~ n\to\infty~.
\label{asymsusy}
\eeq
Indeed, this relation would imply that bosonic and fermionic states somehow fall into alignment
as $n\to\infty$, giving rise to what has been dubbed ``asymptotic supersymmetry''. 
However, as we have seen, no such thing occurs.
Instead, what occurs for large $n$ is that the values of $a_{nn}$ 
continue to grow exponentially, with signs that continue to oscillate between positive and negative values.
Even for large $n$, the net degeneracies $a_{nn}$ never head towards zero;  the spacings between energy
levels remain fixed, and the oscillations between bosonic and fermionic surpluses continue unabated with
ever-growing amplitudes.
Indeed, as already noted above, there is no sense (not even for large $n$) in which one can establish
a pairwise cancellation between bosonic and fermionic surpluses of the sort that would be  
required in order for Eq.~(\ref{asymsusy}) to hold.
Asymptotic supersymmetry, then, is but a myth.
Indeed, it is only through the more subtle oscillations of {\it misaligned}\/ supersymmetry that Eq.~(\ref{KS}) is
satisfied, and through which the corresponding supertrace relations in Eq.~(\ref{supertraces1}) and (\ref{supertraces2})
emerge.

\subsection{SUSY breaking in string theory:~  Hard, soft, or somewhere in between?}

Closely related to the above issue is 
another subtlety concerning the ultimate nature of supersymmetry breaking in string theory.

We have already shown that there is a fundamental limit on the extent to which
spacetime supersymmetry can be broken in string theory:  the supersymmetry 
can be broken only to the extent that a ``misaligned'' supersymmetry remains in the
spectrum.  As we have seen, this misaligned supersymmetry then implies that the resulting
string spectrum 
will obey the supertrace relations
in Eqs.~(\ref{supertraces1}) and (\ref{supertraces2}).

In field theory, supertrace relations of this form are often the  
tell-tale sign of what would be considered to be 
a ``soft'' supersymmetry breaking.
Indeed, this interpretation has encouraged 
a general belief 
that many of the 
specific construction methods which lead to strings with broken
supersymmetry (for example, so-called ``Scherk-Schwarz'' compactifications)
are tantamount to soft supersymmetry breaking.

However, the main novelty in supersymmetry breaking in string theory
--- as opposed to field theory ---
can be expressed
in terms of the Wilsonian renormalization group (RG), which implies that 
a symmetry is broken
``softly'' only if there is some scale above which the modes that are integrated
out are entirely symmetric.  
Indeed,  
this 
does not mean an ``almost cancellation'' of the ultraviolet (UV) contributions but an
{\it exact}\/ cancellation, provided the RG scale is taken high enough. 
However, as we have seen above,
this is {\it never}\/ the case in string theory ---
there is simply no scale above which 
our theory is exactly (or even asymptotically) supersymmetric,
and indeed the modes only look less supersymmetric the further
into the UV one goes. 
As a result, supersymmetry cannot be said to be softly broken in string theory. 

This highlights one of the profound differences between field theory and string
theory.  In field theory, supertrace relations of the forms in
Eqs.~(\ref{supertraces1}) and (\ref{supertraces2}) are
indicators of the softness of supersymmetry breaking.
However, through misaligned supersymmetry, 
string theory finds a different way 
to guarantee relations such as these --- even without the appearance of supersymmetry
at any mass scale.

\subsection{Stability issues, dilaton tadpoles, and higher-loop tachyons}

Finally we conclude this section with some brief comments concerning the overall question of whether such 
non-supersymmetric strings are truly self-consistent.  Do they truly represent legitimate ground states
of the perturbative string?

First, we should remark that even {\it supersymmetric}\/ strings are not true string ground states ---
they typically have flat directions which correspond to massless moduli.
These directions are flat to all orders in perturbation theory.
As a result, such strings may exhibit a sort of equilibrium, 
but this is not a stable equilibrium.  For perturbative strings, 
most attempts to lift these flat directions result in runaway behavior wherein
moduli fields tend to run off to infinity.
This behavior is highly problematic for string phenomenology
because one of these moduli is the dilaton $\phi$ whose vacuum expectation value (VEV)
sets the value of the string coupling.  Such runaway behavior then sends these couplings
to phenomenologically unacceptable values.

\begin{figure*}[h!]
\begin{center}
  \epsfxsize 5.5 truein \epsfbox {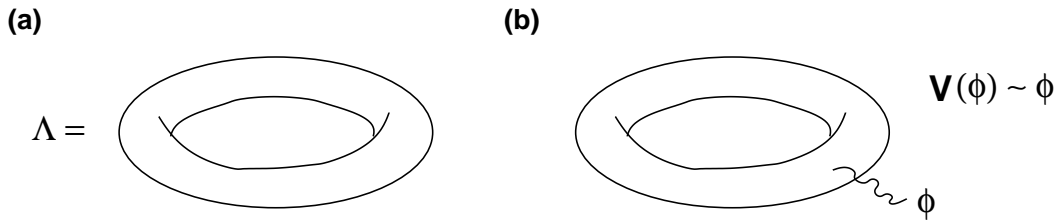}
\end{center}
\caption{(a)  The one-loop zero-point function (cosmological constant) $\Lambda$.
         (b)  The one-loop one-point dilaton ``tadpole'' diagram.   In general, the value of the dilaton tadpole 
   is always proportional to $\Lambda$.  As a result, a non-zero cosmological constant
   implies a non-vanishing one-loop dilaton tadpole diagram, which in turn indicates
   that our effective potential has a linear term in $\phi$.  Thus a non-zero cosmological
      constant indicates that our string ground state is not stable.}
\label{tadpole}
\end{figure*}

For {\it non}\/-supersymmetric strings,
the situation is considerably worse.
As we have seen, closed non-supersymmetric strings generally give rise to non-vanishing one-loop
cosmological constants (or one-loop zero-point functions) $\Lambda$.
However, as illustrated in Fig.~\ref{tadpole}, it turns out that 
for such strings the dilaton tadpole diagram is directly proportional to $\Lambda$.
Consequentially any string with non-zero cosmological constant will also have a non-zero
dilaton tadpole.

The existence of a non-zero dilaton tadpole is extremely dangerous.
A non-zero dilaton tadpole implies that a dilaton can simply be absorbed
into (or emitted from) the vacuum with no other consequences.  
Such a process can repeat itself {\it ad nauseum};
in other words, our ground state is not stable.
Phrased more mathematically, the existence of such a diagram
implies that our dilaton potential $V(\phi)$ contains a non-zero linear term 
proportional to $\phi$ itself.
We are therefore no longer sitting at the minimum of the potential,
and our vacuum is destabilized.

Of course, a similar kind of vacuum destabilization also occurs
when we try to build self-consistent {\it supersymmetric}\/ string models.
In particular, in such cases we often find within the resulting spectrum a pseudo-anomalous $U(1)_X$ gauge
    symmetry.  Although this gauge symmetry is not really anomalous, it leads to
    an effective Fayet-Iliopoulos $D$-term which can break spacetime supersymmetry
    and destabilize the string vacuum.  
     In order to ``fix" this problem,
    the standard procedure is to 
    shift the ground state slightly:  one assigns a VEV to
    certain moduli in the theory in order to break this anomalous $U(1)_X$ gauge symmetry
    and thereby cancel the $D$-term.  This has the effect of making the model stable again.
    Of course, this VEV also has other effects, including the breaking of other gauge symmetries in addition
    to $U(1)_X$ and the generation of intermediate mass scales for various light states in the string model.

The vacuum destabilization that we face within
the context of {\it non}\/-supersymmetric strings
is much more serious, however.
For one thing, we have no assurance
that there exists a ``nearby'' vacuum which continues to be non-supersymmetric
but for which stability is restored.
Indeed, we have no guarantee that any such vacuum exists at all!
For example, one can always attempt to absorb non-zero dilaton tadpoles via the
Fischler-Susskind mechanism~\cite{Fischler:1986ci,Fischler:1986tb}.
However, if such tadpoles are unsuppressed, the new background produced
is expected to be very different from the initial one, thereby invalidating the original construction. 
Moreover, perhaps even more dangerously, such vacuum destabilization 
has the potential to significantly affect our claims about the nature of 
the spectra of such non-supersymmetric strings.
For example, all of our results above concerning the generic spectra of non-supersymmetric strings
have focused on the {\it tree}\/-level spectra of
such theories and require that these spectra be tachyon-free.
For a purely classical theory, such statements have meaning.
However, at a quantum level, the emergence of such tadpoles 
and the corresponding destabilization of our vacuum 
generically imply that tachyonic states will emerge at higher order, even if they
are absent at tree level.
Tachyons thus continue to represent a risk for non-supersymmetric strings.

Over the past thirty years, there has been considerable effort in constructing
stable non-supersymmetric strings --- \ie, non-supersymmetric strings with vanishing
cosmological constants~\cite{Moore,Dienes:1990ij,Kachru:1998hd,KachSilvothers,Shiu:1998he}.
Unfortunately, at the present time, it is not yet known whether such strings truly exist.
If they do, it has been conjectured~\cite{heretic} that 
they could lead to an entirely new approach towards solving the hierarchy problem --- indeed,
an entirely new approach to string phenomenology in general.
However, this remains an unsolved problem.

As a result,
the first and most critical issue we face when attempting to formulate a non-supersymmetric
string theory as our underlying platform for a non-supersymmetric string phenomenology
is to ensure that we are working with a string that is {\it as close to stable as possible}\/.
Of course, we do not need to require complete stability, with $\Lambda=0$.  
Indeed, it is sufficient that $\Lambda$ be sufficiently small that it matches observational limits.  
Or, equivalently, $\Lambda$ must be sufficiently small so that field-theoretic stabilization mechanisms could then be used to finish the job of ensuring absolute stability.  
Finally, it should also be observed that even our notion of ``stability''
only requires stability on the timescale associated with the age of the universe.  
Any ``rolling'' which happens sufficiently slowly can certainly be tolerated within the cosmological
history of the universe.

\section{Interpolating models}
\setcounter{footnote}{0}

The issues discussed in Sect.~II 
are completely generic, and hold
for {\it all}\/ non-supersymmetric closed strings.
However, as we have seen, the critical issue is to construct non-supersymmetric
strings for which the corresponding one-loop vacuum energy (or cosmological constant) $\Lambda$
is as small as possible.  Towards this end, in this paper we shall henceforth
focus on a particular class of models which
naturally lead to such small cosmological constants --- the so-called ``interpolating'' models.
In this section we shall discuss the construction and basic properties of such strings
while continuing to remain as model-independent as possible.

\subsection{Why interpolation?}

In general, string theories (whether open or closed)
are defined by a single mass scale $M_{\rm string}\equiv 1/\sqrt{\alpha'}$, where $\alpha'$ is the
string tension.
However, the background geometry in which they are formulated, and in which they propagate, 
can itself introduce additional mass scales.
For example, strings with world-sheet supersymmetries
have critical dimensions $D=10$, which
requires that the six extra spacetime 
dimensions be compactified.  The process of compactification then introduces
a new scale mass $M_c$ into the problem, namely the mass scale associated
with the six-dimensional compactification volume $M_c \equiv (V_{6})^{-1/6}$.
 
These two mass scales in turn determine the masses associated with our individual
string states.
For example, string {\it oscillator}\/ states
reflect the internal twisting and stretching 
of the string, and thus 
depend solely on the string tension:
$M^{\rm (osc)}_\ell\sim \sqrt{\ell} M_{\rm string}$.
On the other hand, because the string sits within a compactified geometry,
each string oscillator state will also come with an infinite tower of KK excitations.
Relative to the mass of the zero-mode state,
these excited KK modes have masses set by the {\it compactification}\/  
scale: $M^{\rm (KK)}_m \sim m M_{c}$.
Finally, a closed string state
can stretch and 
wrap around compactified geometry.
Such winding-mode states thus have masses which depend on both $M_{\rm string}$ and $M_c$
simultaneously:
$M_n^{\rm (winding)}\sim n M_{\rm string}^2/M_c$.  
All three kinds of states 
typically appear together within the spectra 
of closed strings.

If a given string model exhibits spacetime supersymmetry, the bosonic states will match the fermionic
states and the corresponding cosmological constant $\Lambda$ will vanish.  Otherwise, we generically expect $\Lambda$
to inherit the fundamental scales associated with the states whose one-loop vacuum contributions it represents --- 
namely $M_{\rm string}$ and $M_c$.
Of course, the most natural assumption is that $M_c\sim M_{\rm string}$, for this configuration of scales
is ``minimal'' and does not require a dynamical mechanism
by which a hierarchy of scales might be generated.
Indeed, in such models it is not always clear how to separate oscillator states from KK states
and/or winding states;  there even exist examples of such models which transcend the 
notion of having a compactification geometry altogether and in which no 
compactification geometry can even be identified.
Nevertheless, in such models we typically obtain a cosmological constant of order $\Lambda\sim M_{\rm string}$.
Of course, even within such string models, there remains the possibility that $\Lambda$ might still
vanish through some other mechanism.  For example, the proposals in 
Refs.~\cite{Moore,Dienes:1990ij,Kachru:1998hd} all rely on different kinds of symmetry arguments 
for cancelling $\Lambda$ within closed string models for which $M_c\sim M_{\rm string}$.
Unfortunately, no string models
have ever been constructed exhibiting the symmetries proposed in Refs.~\cite{Moore,Dienes:1990ij},
and the 
mechanism proposed in Ref.~\cite{Kachru:1998hd} may actually fail at higher loops~\cite{Shiu:1998he,Iengo:1999sm}.

An alternate possibility is to consider models in which $M_{\rm string}$ is fixed
but $M_c$ is taken to be a free, adjustable variable.
Indeed, we can go even further and imagine that our compactification volume is characterized
by many different compactification scales $M_c^{(i)}$, each of which we might consider a free parameter;
such a scenario would emerge, for example, if our $d$-dimensional compactification manifold is an $d$-torus
with different radii of compactification $R_i$, $i=1,...,d$.
In general, as the volume of compactification $V_d$ is taken to infinity, we effectively produce a
string model in $d$ additional spacetime dimensions.  We may call this higher-dimensional model $M_1$.  
Moreover, for closed strings, T-duality ensures that we also 
produce a model in $d$ additional spacetime dimensions
if $V_d\to 0$;  we may call this model $M_2$.  Thus, our models with variable compactification volumes
can be said to {\it interpolate}\/ between these two higher-dimensional endpoints, models $M_1$ and $M_2$.

Such interpolating models offer a number of distinct advantages when it comes to suppressing the cosmological
constant.  If the model $M_1$ is supersymmetric, we are assured that $\Lambda=0$ when $V_d\to \infty$.
Moreover, if $M_2$ is non-supersymmetric, then spacetime supersymmetry is likely 
to be broken for all finite $V_d$.
It therefore stands to reason that we can dial $V_d$ to a sufficiently large value in order to obtain
a cosmological constant of whatever small size we wish.
Even more compellingly, there is a widespread belief that spacetime supersymmetry, if it exists at all in
nature, is broken at the TeV scale, with superpartners having masses $\sim {\cal O}({\rm TeV}\/)$.
Indeed, as first suggested in Refs.~\cite{Rohm, Antoniadis:1990ew}, these sorts of scenarios 
with large compactification volumes
are relatively easy to incorporate with the interpolating-model framework
with $M_c\sim {\cal O}({\rm TeV})$.

Within such setups, it might at first glance seem that all we are doing is dialing the
scale of SUSY breaking, just as we might do in field theory:
after all, the closer we come to a supersymmetric limit, the more our lightest bosonic and fermionic states
line up and the smaller our cosmological constant becomes.
However, we stress that this is {\it not}\/ what is happening across the string spectrum.
First, as we have already discussed in Sect.~II, the spectra of these interpolating models
--- just like those of {\it any}\/ non-supersymmetric string model ---
exhibit a {\it misaligned supersymmetry}\/, with boson/fermion oscillations.  As we have discussed,
this is quite different from a field-theoretic breaking of spacetime supersymmetry,
and our supertrace relations are not satisfied supermultiplet by supermultiplet.

But second, and equally importantly, we also find that the {\it scale}\/ of the cosmological constant
need not necessarily be tied to the effective scale of the supersymmetry breaking.
In particular, there is a sense in which we can consider the scale of supersymmetry
breaking in these models to be given by $M_c$, yet we shall see that in these models there are circumstances
under which it is possible
for the cosmological constant to be {\it exponentially suppressed}\/, 
with $\Lambda\sim {\cal O}(e^{-M_c/M_{\rm string}})$.
As a result, such interpolating models offer the intriguing possibility of separating the
effective scale of supersymmetry breaking
from the scale of the cosmological constant, thereby bestowing a certain enhanced stability 
on these models even if the effective scale of supersymmetry breaking is relatively large. 
This can also be important phenomenologically because 
the magnitude of our first non-vanishing supertrace 
in Eq.~(\ref{supertraces2}) is set by $\Lambda$ rather than  by the expected $M_c$;  again, this is only consistent
because our supertrace relations are not satisfied supermultiplet by supermultiplet across the entire string spectrum.
 
Because of these properties,
interpolating models will be the centerpiece of this paper. 
Consequently, in the rest of this section we shall provide
an introduction to the interpolating framework.
First, we shall discuss the structure of these models,
how supersymmetry is broken within these models,
and what features emerge in their resulting spectra.
This part of our discussion will be completely general.
We shall then proceed to provide several concrete examples
of such interpolating models which illustrate these features.

\subsection{The structure of interpolating string models}

We shall now discuss the structure of heterotic interpolating string models.
As might be imagined, there are many different construction techniques that might be followed,
depending on the spacetime dimension of our original model, the number of spacetime dimensions to be compactified,
and so forth.  However, these constructions all share certain common features.
Accordingly, in this section we shall concentrate on the simplest case of heterotic interpolating
models in which our compactification manifold is a circle with a $\IZ_2$ twist.
Such models were originally introduced in Refs.~\cite{Rohm, Itoyama:1986ei, Itoyama:1987rc} and later
in Ref.~\cite{Antoniadis:1990ew};  moreover, through the temperature/radius correspondence,
such models also serve as the finite-temperature extensions of zero-temperature heterotic string models
and thus appear frequently in studies of string thermodynamics~\cite{earlystringpapersfiniteT}.
 
In general, there are three steps in the construction of an interpolating string model
of this type:
\begin{itemize}
\item{}   First, we select a suitable higher-dimensional heterotic string model which will ultimately serve
          as one of the endpoints of our interpolation.
\item{}   Second, we compactify this model on a circle of arbitrary radius $R$.  Because of the
T-dual nature of this compactification, this process results in a model which trivially interpolates
between our original higher-dimensional string model as $R\to\infty$ and (the T-dual of) itself as $R\to 0$.
\item{}  Finally, we introduce a {\it twist}\/ into this compactified model.  As we shall see,
the choice of twist is constrained by a number of factors which relate to the self-consistency
of the resulting interpolation.  However, this twist is ultimately what allows our interpolating
model to interpolate between two {\it different}\/ endpoints.  Thus, it is this twist which 
allows spacetime supersymmetry to be broken within the interpolation.
\end{itemize}

We shall now consider these steps in order.

First, in general,
we begin with a chosen $D$-dimensional heterotic string theory in 
$D$ spacetime dimensions, with partition function $Z(\tau)$.
Note that the form in which $Z(\tau)$ is most naturally expressed
will depend on the precise kind of model we have chosen,
but in all cases it will have a double power-series expansion of the form
in Eq.~(\ref{partfunctgen}).

Next, we now consider
what happens when we compactify this theory on a circle of radius $R$.
For any compactification radius $R$, we define the corresponding dimensionless
inverse radius $a\equiv \sqrt{\alpha'}/R$.
Any field compactified on a circle with this radius then accrues integer
momentum and winding modes around this circle,
resulting in left- and right-moving spacetime momenta of the forms
\beq
   p_R~=~ {1\over \sqrt{2\alpha'}} (ma-n/a)~,~~~~~~~~
   p_L~=~ {1\over \sqrt{2\alpha'}} (ma+n/a)~.
\eeq
The quantities $m$ and $n$ respectively represent the momentum
and winding quantum numbers of the field in question.
The contribution to the partition function
from such modes then takes the form of the double summation
\beq
     Z_{\rm circ}(\tau,R)~=~
        \sqrt{\tau_2}\, \sum_{m,n\in\IZ} \,
         \overline{q}^{\alpha' p_R^2/2} q^{\alpha' p_L^2/2}
        ~=~
     \sqrt{ \tau_2}\,
    \sum_{m,n\in\IZ} \,
      \overline{q}^{(ma-n/a)^2/4}  \,q^{(ma+n/a)^2/4}~.
\label{Zcircdef}
\eeq
Note that $Z_{\rm circ}\to 1/a$ as $a\to 0$,
while $Z_{\rm circ}\to a$ as $a\to \infty$.

The trace $Z_{\rm circ}$ is sufficient for
compactifications on a circle.
Indeed, for such an untwisted compactification, each field within our original $D$-dimensional string theory
accrues the same set of momentum and winding modes.  Thus, the total partition function of our resulting
$(D-1)$-dimensional string theory representing the untwisted
compactification is simply given by 
\beq
         Z(R) ~=~ Z(\tau) \, Z_{\rm circ}(R)~.
\eeq
Note that we reproduce our original ten-dimensional theory as $R\to\infty$ or $R\to 0$. 
At the level of the partition function, this follows from the observation
that 
$Z_{\rm circ}\to 1/a$ or $Z_{\rm circ}\to a$ as $a\to 0$ or $a\to\infty$,
where the extra $a$-dependent prefactor in each case represents the diverging volume of compactification
that generally relates $D$ and $(D-1)$-dimensional partition functions to each other.

For our purposes, however, we are actually interested in
the $(D-1)$-dimensional string theories which represent the
compactifications of our original $D$-dimensional string theory
on {\it twisted}\/ circles, \ie, on $\IZ_2$ orbifolds of the circle.
Towards this end, we
introduce (as in the original papers~\cite{Rohm})
four new functions $\calE_{0,1/2}$ and $\calO_{0,1/2}$
which are the same as the summation in
$Z_{\rm circ}$ in Eq.~(\ref{Zcircdef}) except that their summation
variables are restricted as follows:
\beqn
       \calE_0 &=& \lbrace  m\in\IZ,~n~{\rm even}\rbrace\nonumber\\
       \calE_{1/2} &=& \lbrace  m\in\IZ+\half ,~n~{\rm even}\rbrace\nonumber\\
       \calO_0 &=& \lbrace  m\in\IZ,~n~{\rm odd}\rbrace\nonumber\\
       \calO_{1/2} &=& \lbrace  m\in\IZ+\half ,~n~{\rm odd}\rbrace~.
\label{EOfunctions}
\eeqn
These functions
are to be distinguished from a related 
set of functions with the same names in which the roles of $m$ and $n$ are exchanged.
This therefore establishes our conventions for these functions.

Note that under the modular transformation $T:\tau\to\tau+1$,
the first three functions are invariant    while
$\calO_{1/2}$ picks up an overall minus sign;
likewise, under $S:\tau\to-1/\tau$,
these functions mix amongst themselves according to
\beq
  \begin{pmatrix}
       \calE_0 \cr \calE_{1/2} \cr \calO_0  \cr  \calO_{1/2}  
  \end{pmatrix} (-1/\tau) ~=~
   \half 
   \begin{pmatrix}   1 & 1 & 1 & 1 \cr
                   1 & 1 & -1 & -1 \cr
                   1 & -1 & 1 & -1 \cr
                   1 & -1 & -1 & 1 \cr
   \end{pmatrix}
   \begin{pmatrix}  \calE_0 \cr \calE_{1/2} \cr \calO_0  \cr  \calO_{1/2}  
   \end{pmatrix}  (\tau) ~.
\label{Smixing}
\eeq
In the $a\to 0$ limit,
$\calO_0$ and $\calO_{1/2}$ each vanish while $\calE_0,\calE_{1/2} \to 1/a$;
by contrast, as $a\to\infty$,
$\calE_{1/2}$ and $\calO_{1/2}$ each vanish while $\calE_0,\calO_0 \to a/2$.
Clearly, $\calE_{0}+\calO_{0}= Z_{\rm circ}$.

Given these conventions, we now discuss the structure of the $(D-1)$-dimensional string models 
that result when a given $D$-dimensional string model 
is compactified on a $\IZ_2$ twisted circle.

In general, we must choose a particular twist before we can specify what happens to the 
theory under such a compactification.  Physically, this choice of twist
is equivalent to a choice of the Wilson line,
or equivalently a choice of gauge-field flux threaded through the compactification circle.
In a nutshell, as we shall further discuss below, the only twists that are  possible are
those producing $(D-1)$-dimensional interpolating models which 
yield valid $D$-dimensional string models in {\it both}\/ their
$R\to \infty$ and $R\to 0$ limits.
In other words, the resulting $(D-1)$-dimensional twisted string models must always {\it interpolate}\/
between a given $D$-dimensional string model (call it $M_1$) as $R\to\infty$ and (the T-dual of) another
$D$-dimensional string model (call it $M_2$) as $R\to0$.
Such a $(D-1)$-dimensional model can thus be considered to be a twisted compactification
of the $D$-dimensional model $M_1$, and the possible twists correspond to the possible 
choices for self-consistent $D$-dimensional models $M_2$
at the other end of the interpolation.  

Since we are only considering $\IZ_2$ twists,
the only restriction on our choice of $M_2$ is that $M_2$ must be a $\IZ_2$ twist of $M_1$ directly
in $D$ dimensions --- or equivalently that $M_1$ and $M_2$ be related to each other through $\IZ_2$ twists.
Thus, the set  
of possible twists --- and thus the set of possible $(D-1)$-dimensional interpolating compactifications 
of a given $D$-dimensional 
supersymmetric string model $M_1$ --- directly corresponds to the set of possible 
self-consistent 
$\IZ_2$ orbifolds of that $D$-dimensional string model, or equivalently the set
of $D$-dimensional string models $M_2$ which are related to $M_1$ through a simple 
$\IZ_2$ orbifold.

Given these observations, 
any such $(D-1)$-dimensional interpolating model will have a partition function of the form~\cite{Rohm,finitetemp,AtickWitten,wasKounnasRostand,Kounnas:1989dk}
\beqn
        Z_{\rm string}(\tau,R)  &=&
           Z^{(1)}(\tau) ~ \calE_0(\tau,R) ~+~
           Z^{(2)}(\tau) ~ \calE_{1/2}(\tau,R) \nonumber\\
           && ~~+~  Z^{(3)}(\tau) ~ \calO_{0}(\tau,R) ~+~
           Z^{(4)}(\tau) ~ \calO_{1/2}(\tau,R) ~
\label{EOmix}
\eeqn
where $Z^{(1)} + Z^{(2)}$ reproduces the partition function of the ten-dimensional model $M_1$ and
where $Z^{(1)} + Z^{(3)}$ reproduces the partition function of the ten-dimensional model $M_2$.
Of course, the entire partition function in Eq.~(\ref{EOmix}) must be modular invariant.
Thus, at the partition-function level, the choice of twist involved in compactifying $M_1$ translates into the 
choice of how we separate the partition function of $M_1$ into the separate contributions $Z^{(1)}$ (to be multiplied
by $\calE_0$) and $Z^{(2)}$ (to be multiplied by $\calE_{1/2}$).
Once $Z^{(1,2)}$ are separately chosen, the remaining terms $Z^{(3,4)}$ follow directly from modular invariance:
\beqn
   Z^{(3)} &=& \half\left[
        Z^{(1)} (-1/\tau) -Z^{(2)} (-1/\tau)
            + Z^{(1)} (-1/\tau+1) - Z^{(2)} (-1/\tau+1)\right] \nonumber\\
   Z^{(4)} &=& \half\left[
        Z^{(1)} (-1/\tau) -Z^{(2)} (-1/\tau)
            - Z^{(1)} (-1/\tau+1) + Z^{(2)} (-1/\tau+1)\right] ~.
\label{Z3Z4}
\eeqn

The fact that the $D$-dimensional models~$M_1$ and $M_2$ are directly related to each other through a $\IZ_2$ orbifold
twist implies that these individual partition-function pieces $Z^{(i)}$  
can be written in terms of a fundamental partition function $Z^+_+$,
its projection sector $Z^-_+$, and their corresponding twisted versions $Z^\pm_-$:
\beqn
        Z^{(1)} &=& \half\left( Z^+_+ + Z^-_+ \right)\nonumber\\
        Z^{(2)} &=& \half\left( Z^+_- + Z^-_- \right)\nonumber\\
        Z^{(3)} &=& \half\left( Z^+_+ - Z^-_+ \right)\nonumber\\
        Z^{(4)} &=& \half\left( Z^+_- - Z^-_- \right)~.
\label{Zifactors}
\eeqn
Indeed, any other relation between the different $Z^{(i)}$ factors
will render Eq.~(\ref{EOmix}) inconsistent from an underlying geometric
perspective.
From Eq.~(\ref{Zifactors}) and the limiting behaviors given below Eq.~(\ref{Smixing})
it then follows that
\beqn
   R\to \infty:~~~~~~  && Z_{\rm model} ~=~ Z^{(1)}+Z^{(2)} ~=~ \half( Z^+_+ + Z^-_+ + Z^+_- + Z^-_-)
             \nonumber\\
   R\to 0:~~~~~~  && Z'_{\rm model} ~=~ Z^{(1)}+Z^{(3)} ~=~ Z^+_+ ~.
\label{endpoints}
\eeqn
We thus identify $Z^+_+$ as the partition function of Model~$M_2$,
whereupon it follows that 
 $\half( Z^+_+ + Z^-_+ + Z^+_- + Z^-_-)$ is the partition function of the orbifold of Model~$M_2$ ---
\ie, the partition function of Model~$M_1$.
Note that these results also implicitly allow us to identify our direct and T-dual volumes
of compactification within Eqs.~(\ref{identifyZs}):  we find that
$V\equiv 2\pi R$, as expected, while $\tilde V \equiv 2\pi \tilde R$ with $\tilde R\equiv \alpha'/(2R)$ 
(or $\tilde a = 2/a$).

From this perspective, then, we see that the construction of our $(D-1)$-dimensional interpolating model 
is relatively straightforward.
We begin with the $D$-dimensional string model $M_1$, and choose another $D$-dimensional string model $M_2$
to which it is directly related in $D$ dimensions through the action of an particular $\IZ_2$ orbifold twist $Q$.
We then construct our $(D-1)$-dimensional interpolating model by compactifying $M_2$ on a circle of radius $R$,
and orbifold the resulting $(D-1)$-dimensional theory by the twist $\calT Q$ where $Q$ acts 
on the internal part of the string, as above, while
$\calT$ acts on the compactified circle.
In particular, $\calT$ corresponds to the
$\IZ_2$ shift $y\to y+\pi R$ where $y$ is the (T-dual) coordinate
along the compactified dimension, so that
states with even values of $n$ (such as
those within $\calE_{0,1/2}$) are invariant under
$\calT$, while those with odd values of 
$n$ (such as those within $\calO_{0,1/2}$) 
pick up a minus sign.  
Together, the resulting orbifold procedure yields a partition 
function of the form 
in Eq.~(\ref{EOmix}), with 
the $Z^{(i)}$ factors identified as in Eq.~(\ref{Zifactors}).
The resulting $(D-1)$-dimensional model then interpolates
between $M_1$ at $R\to\infty$ and $M_2$ as $R\to 0$.

If $M_1$ has spacetime supersymmetry but $M_2$ does not, 
the relevant $\IZ_2$ orbifold twist 
$Q$ at least must include $(-1)^F$ (where $F$ is spacetime fermion number).
However, $Q$ may (and indeed often {\it must}\/) include
additional twist factors which act on the purely internal gauge quantum numbers.
These different choices for $Q$ 
then correspond to the possible choices of how we might effect our 
breaking of supersymmetry, which is of course our main interest.  
Thus, specifying the separation of the $M_1$ partition function
into $Z^{(1)}$ and $Z^{(2)}$ is tantamount to 
specifying these additional
gauge twists,
and thereby specifying a choice for how 
the supersymmetry of $M_1$ is ultimately broken.

If Model~$M_1$ is supersymmetric
but $M_2$ is tachyonic  
(which, being supersymmetric, it may well be),
then the $R\to \infty$ limit of the interpolating model
will be tachyon-free.  However, as the radius shrinks towards zero, certain
states which were previously massive for radii $R$ exceeding some critical
radius $R^\ast$ will become massless at $R=R^\ast$ and tachyonic for $R<R^\ast$.
The interpolating model will then be tachyonic
for all radii $R<R^\ast$.
Indeed, this is nothing but the temperature/radius ``dual'' of the 
Hagedorn phenomenon wherein a string winding mode becomes massless and then ultimately tachyonic
as the temperature formally reaches and then increases beyond the critical Hagedorn temperature.
However, it is not required that the model $M_2$ be tachyonic;  indeed, all that is required
is that it be non-supersymmetric.
In that case, our $(D-1)$-dimensional interpolating 
model can be tachyon-free over the entire range of interpolation, \ie, for all radii $0\leq R\leq \infty$.

Finally, we observe that while all of the states within $Z^{(1,2,3)}$ have world-sheet energies $(m,n)$ satisfying
$m-n\in \IZ$, those within $Z^{(4)}$ instead have $m-n\in \IZ+1/2$.  This does not violate modular invariance since
$\calO_{1/2}$ is odd under $T:\tau\to \tau+1$, while $\calE_{0,1/2}$ and $\calO_0$ are even.

\subsection{Two examples}

We now provide two concrete examples of such interpolating models.
Having concrete examples will allow us to examine their spectra
in detail, and determine the manner in which supersymmetry is broken in these
models.  For ease of analysis, our examples 
will literally be the simplest heterotic interpolating models
that can be constructed:  nine-dimensional string models which 
interpolate between different ten-dimensional endpoints.
We shall begin by discussing our possible ten-dimensional endpoints.
We shall then discuss the nine-dimensional models that interpolate between them.


\subsubsection{Ten-dimensional endpoint models}

Our starting point will be the heterotic string theories in $D=10$. Of these, only three shall concern
us here:  the supersymmetric $SO(32)$ string,
the supersymmetric $E_8\times E_8$ string,
and the non-supersymmetric tachyon-free $SO(16)\times SO(16)$ string.

In order to describe the partition functions of these three theories, we begin with the standard 
Dedekind $\eta$-function and Jacobi $\vartheta_i$-theta functions. 
These are defined in Appendix~\ref{notation}.
However, given these ``primitive'' building blocks, we then wish to express 
our string partition functions in terms of combinations 
of these functions which have direct physical interpretations.
As we shall see, this will directly assist us in analyzing the phenomenology of these models.

Given the Dedekind $\eta$-function, 
the partition function of $D$ free coordinate bosons is given by
\beq
         Z^{(D)}_{\rm boson} ~\equiv ~ {\tau_2}^{-D/2}\, (\overline{\eta}\eta)^{-D}~.
\label{bosons}
\eeq
Likewise, 
we can combine the $\eta$- and $\vartheta_i$-functions
in order to construct the
characters of the level-one $SO(2n)$ affine Lie algebras.
Recall that at affine level $k=1$, the $SO(2n)$ algebra for each $n\in\IZ$ has
four distinct representations:  the identity ($I$),
the vector ($V$), the spinor ($S$), and the conjugate spinor ($C$).
In general,
the conformal dimensions of these representations are given by $\lbrace h_I,h_V,h_S,h_C\rbrace=
\lbrace 0,1/2,n/8,n/8\rbrace$, and likewise their corresponding characters are given by
\beqn
 \chi_I &=&  \half\,(\tthree^n + \tfour^n)/\eta^n ~=~ q^{h_I-c/24} \,(1 + n(2n-1)\,q + ...)\nonumber\\
 \chi_V &=&  \half\,(\tthree^n - \tfour^n)/\eta^n ~=~ q^{h_V-c/24} \,(2n + ...)\nonumber\\
 \chi_S &=&  \half\,(\ttwo^n + {\vartheta_1}^n)/\eta^n ~=~ q^{h_S-c/24} \,(2^{n-1} + ...)\nonumber\\
 \chi_C &=&  \half\,(\ttwo^n - {\vartheta_1}^n)/\eta^n ~=~ q^{h_C-c/24} \,(2^{n-1} + ...)~
\label{chis}
\eeqn
where the central charge is $c=n$ at affine level $k=1$.
The vanishing of $\vartheta_1$ implies that
$\chi_S$ and $\chi_C$ have identical $q$-expansions;
this is a reflection of the conjugation symmetry between the spinor
and conjugate spinor representations.
Indeed, when $SO(2n)$ represents a transverse spacetime Lorentz group,
the distinction between $S$ and $C$ can be interpreted as being
equivalent to relative spacetime chirality;  the choice of which spacetime chirality is to
be associated with $S$ or $C$ is then a matter of convention.
Note that the special case $SO(8)$
has a further so-called ``triality'' symmetry under which
the vector and spinor representations are
indistinguishable.  Thus, for $SO(8)$, we find that $\chi_V=\chi_S$.
Indeed, this is nothing but the 
identity already given
below Eq.~(\ref{etathetadefs})
in terms of $\vartheta_i$-functions.

Given these $SO(2n)$ characters,
we can now write down the partition functions of our three relevant 
heterotic string theories in $D=10$.
We adopt the convention that right-moving degrees of freedom
(associated with the supersymmetric side of the heterotic string)
are collected in anti-holomorphic characters $\chi_V$ of the 
transverse $SO(8)$ Lorentz group, while the left-moving degrees
of freedom are collected in the holomorphic characters $\chi_i \chi_j$
of $SO(16)\times SO(16)$, with $i,j\in\lbrace I, V, S, C\rbrace$.

Let us  begin with the 
supersymmetric $SO(32)$ heterotic string, which has the partition function
\beq
         Z_{\rm model} ~=~ Z^{(8)}_{\rm boson}~ (\overline{\chi}_V-\overline{\chi}_S)
         \, \left( \chi_I^2 +  \chi_V^2 + \chi_S^2 + \chi_C^2 \right)~.
\label{start}
\eeq
The spacetime supersymmetry follows from the factor $(\overline{\chi}_V-\overline{\chi}_S)$
coming from the right-movers:   in terms of actual $SO(8)$ Lorentz representations,
this means that any left-moving $SO(32)$ state from the left-movers simultaneously comes
not only as a spacetime Lorentz vector 
but also as a spacetime Lorentz spinor.
As a useful exercise in reading partition functions, 
let us explicitly read off the massless states in this theory.
The heterotic string has vacuum energies $(E_R,E_L)=(-1/2,-1)$, so we are looking for states
which have $(h_R,h_L)=(1/2,1)$ in order to make massless states.
For the right-movers, both the $SO(8)$ vector (V) and spinor (S) have $h=1/2$, which is why
their ground states (the vector and spinor representations) describe the spacetime Lorentz symmetries of
the massless fields.
Let us  now look at the left-movers.  Here each $\chi$ is an $SO(16)$ character, so $\chi_I$ has $h=0$
while $\chi_V$ has $h=1/2$ and $\chi_{S,C}$ each have $h=1$.  Therefore massless states can only come from
the ground state of $\chi_I\chi_I$ (along with left-moving coordinate excitations to produce $h=1$), 
from the {\it first descendants}\/ of $\chi_I\chi_I$ (with no external coordinate excitations), or the ground state of 
$\chi_V\chi_V$ (again with no external coordinate excitations).
The first group of states produces a left-moving vector which, when tensored with the right-moving tensor and/or spinor, produces the supergravity multiplet.
The second group of states is a bit more complicated.
As indicated below Eq.~(\ref{chis}), the first descendant of the identity sector is always the adjoint.
Thus the first descendants of $\chi_I\chi_I$ transform in the ${\bf (adj,1)\oplus (1,adj)}$ representation of
$SO(16)\times SO(16)$, and tensored with the right-movers these are either spacetime vectors or spacetime spinors.
Likewise,
the ground state of $\chi_V\chi_V$ transforms as ${\bf (vec,vec)}$.
Thus the third group of states are spacetime vectors or spinors transforming as 
${\bf (vec,vec)}$.
Together, these latter two groups of states fill out the adjoint of $SO(32)$.
Thus, we learn that the massless states in this theory consist of simply the supergravity multiplet as well
as the gauge bosons (and gauginos) of $SO(32)$.

The $E_8\times E_8$ heterotic string is similar.  Its partition function is
given by
\beq
     Z^{(8)}_{\rm boson} \,
      (\chibar_V -\chibar_S) \,(\chi_I + \chi_S)^2~.
\label{e8pf}
\eeq
Again spacetime supersymmetry is clear, as is the supergravity multiplet coming from $\chi_I\chi_I$.
The gauge bosons (and gauginos) of $SO(16)\times SO(16)$ come from the first descendants within $\chi_I\chi_I$
while the ground states associated with the cross terms $\chi_I\chi_S$ and $\chi_S\chi_I$ 
produce spacetime vector and spinor states transforming as
${\bf (spinor,1)\oplus (1,spinor)}$
of $SO(16)\times SO(16)$.
This enhances the gauge group to $E_8\times E_8$, and we see that there are no other massless states
in this theory.
In passing, we further note that the $SO(16)$ characters satisfy an identity
\beq
          \chi_I \chi_S + \chi_S \chi_I ~=~ \chi_V^2 + \chi_C^2~
\label{identy}
\eeq
which holds at the level of their $q$-expansions.
This implies that 
Eqs.~(\ref{start}) and (\ref{e8pf})
have identical $q$-expansions.
This in turn implies that the ten-dimensional supersymmetric $SO(32)$ and $E_8\times E_8$ 
heterotic strings have the same bosonic and fermionic state degeneracies at each mass level.

Finally, let us consider the non-supersymmetric but tachyon-free $SO(16)\times SO(16)$ heterotic string model~\cite{SOsixteen}.
This string has partition function
\beq
 Z^{(8)}_{\rm boson} \,\biggl\lbrace
         \chibar_I \,(\chi_V\chi_C+\chi_C\chi_V) 
    ~+~\chibar_V \,(\chi_I^2 + \chi_S^2)  
 ~-~  \chibar_S \,(\chi_V^2+\chi_C^2) 
      ~-~ \chibar_C \,(\chi_I\chi_S+\chi_S\chi_I)\biggr\rbrace~.
\label{so16}
\eeq
First, we observe that this partition function does not factorize;  spacetime SUSY is clearly broken.
Despite this, we observe that spacetime bosons (whose contributions
are proportional to $\chibar_I$ or $\chibar_V$) come with a plus sign,
as required,
while the spacetime fermions come with a minus sign.
Because $\chi_V\chi_C$ has $h=3/2$, we see that the first term
(proportional to $\chibar_I$) cannot give rise to physical tachyons or massless states.
Since this is the only term which could have produced physical tachyons, 
this model is tachyon-free.
However, massless states can emerge from 
$\chibar_V \chi_I^2$, $\chibar_S\chi_V^2$, and $\chibar_C (\chi_I\chi_S+\chi_S\chi_I)$.
Let us therefore consider each of these in turn.
The term $\chibar_V\chi_I^2$ contains the gravity multiplet as well
as gauge bosons transforming in the ${\bf (adj,1)\oplus (1,adj)}$ representation
of $SO(16)\times SO(16)$.  The term $\chibar_S \chi_V^2$ contains spacetime
fermions with a certain chirality transforming in the ${\rm (vec,vec)}$ representation,
while the final term contains 
spacetime fermions with the opposite chirality transforming in the ${\rm (spinor,1)\oplus (1,spinor)}$
representation.    Note that this configuration of massless states guarantees the cancellation
of the irreducible gravitational anomaly (even without supersymmetry), as required for a consistent
string model.  Finally, we note that the ground state of the $\chibar_V\chi_I^2$ term   
is an off-shell Lorentz $SO(8)$ vector of off-shell tachyons with world-sheet energies
$(m,n)=(0,-1)$.  These are nothing but our {\it proto-graviton}\/ states, discussed above.   

There are six other perturbative heterotic strings in $D=10$, all 
of which are non-supersymmetric and tachyonic~\cite{KLTclassification}.  However, we shall not need them for this discussion.


\subsubsection{Nine-dimensional interpolations}

Given the above partition functions for our ten-dimensional theories,
we can now present the partition functions corresponding to their nine-dimensional interpolations.
In the following discussion 
we shall restrict our attention to the cases
in which $M_1$ is either the supersymmetric $SO(32)$ 
or $E_8\times E_8$ string model, and in which $M_2$ is the $SO(16)\times SO(16)$ model.
Thus we obtain two distinct nine-dimensional interpolating models, both of which are tachyon-free
over the entire range of their interpolations.

The partition functions for these two cases are respectively given as
\beqn
    Z_{SO(32)}  ~=~  Z^{(8)}_{\rm boson} \,\times \,\bigl\lbrace
    &\phantom{+}&
    \lbrack \chibar_V \,(\chi_I^2 + \chi_S^2)  ~-~  \chibar_S \,(\chi_V^2+\chi_C^2) \rbrack ~\calE_0
           \nonumber\\
   &+&
     \lbrack \chibar_V \,(\chi_V^2 + \chi_C^2)  ~-~  \chibar_S \,(\chi_I^2+\chi_S^2) \rbrack~ \calE_{1/2}
           \nonumber\\
   &+&
    \lbrack  \chibar_I \,(\chi_V\chi_C+\chi_C\chi_V) ~-~ \chibar_C \,(\chi_I\chi_S+\chi_S\chi_I) \rbrack ~\calO_0
           \nonumber\\
   &+&
     \lbrack \chibar_I \,(\chi_I\chi_S+\chi_S\chi_I) ~-~ \chibar_C\,(\chi_V\chi_C+\chi_C\chi_V) \rbrack~ \calO_{1/2}
       ~~\bigr\rbrace~
\label{firstso16}
\eeqn
and 
\beqn
    Z_{E_8\times E_8} ~=~  Z^{(8)}_{\rm boson} \,\times \,\bigl\lbrace
    &\phantom{+}&
           \lbrack
          \chibar_V \,(\chi_I^2 + \chi_S^2)  ~-~  \chibar_S \,(\chi_I\chi_S + \chi_S\chi_I) \rbrack~ \calE_0
             \nonumber\\
   &+&
         \lbrack
          \chibar_V \,(\chi_I\chi_S+\chi_S\chi_I)  ~-~  \chibar_S \,(\chi_I^2+\chi_S^2) \rbrack ~\calE_{1/2}
             \nonumber\\
   &+&
           \lbrack
           \chibar_I \,(\chi_V\chi_C+\chi_C\chi_V) ~-~ \chibar_C \,(\chi_V^2 +\chi_C^2) \rbrack ~\calO_0
             \nonumber\\
   &+&
             \lbrack
           \chibar_I \,(\chi_V^2+\chi_C^2) ~-~ \chibar_C \,(\chi_V\chi_C+\chi_C\chi_V) \rbrack ~\calO_{1/2}
      ~~\bigr\rbrace~.
\label{secondso16}
\eeqn
Once again, as a result of Eq.~(\ref{identy}), these two partition functions have identical $q$-expansions
and thus behave identically in any numerical sense.
Note, in particular, that the partition function in Eq.~(\ref{secondso16}) is nothing but the ``Twist~II'' 
interpolating model of Ref.~\cite{Itoyama:1986ei}, only written 
in terms of $SO(2n)$ characters.
We shall henceforth restrict our attention to the partition function
in Eq.~(\ref{secondso16})
and its corresponding one-loop cosmological constant.
We note, however, that nine-dimensional interpolations of this sort generally exist 
between our two supersymmetric ten-dimensional models and 
{\it all}\/ of the non-supersymmetric ten-dimensional models,
including those that are supersymmetric.
These models are described and classified in Refs.~\cite{julie1,julie2,Dienes:2012dc}. 

Analyzing Eq.~(\ref{secondso16}),
we can easily understand the physical effects of the SUSY breaking that occurs in this model.
As $R\to\infty$, we reproduce the ten-dimensional $E_8\times E_8$ string model whose partition function is given in
Eq.~(\ref{e8pf}).  However, for any finite $R$, spacetime supersymmetry is broken.
Let us consider large but finite $R$.
In such cases, we can read off exactly what happens to each of the massless states of the $E_8\times E_8$ model.
For small $a$ (\ie, large $R$), 
any state with a non-zero winding mode becomes extremely heavy.
Since all of the states in the ${\cal O}_{0,1/2}$ sectors have non-zero winding modes,
we can disregard these sectors and concentrate on the $n=0$ states within the ${\cal E}_{0,1/2}$ sectors.
The lightest excitations within the ${\cal E}_0$ sector have $m=0$, while the
lightest excitations within the ${\cal E}_{1/2}$ sector have $m=\pm 1/2$.
Thus, we see that the massless states within $Z^{(1)}$ [\ie, the term multiplying ${\cal E}_0$] 
remain massless after SUSY breaking:
these are the gravity multiplet, the gauge bosons of $SO(16)\times SO(16)$, and Lorentz {\it spinors}\/
transforming in the ${\bf (spinor,1)\oplus (1,spinor)}$ representation of 
$SO(16)\times SO(16)$.
By contrast, the massless states within $Z^{(2)}$ 
[\ie, the term multiplying ${\cal E}_{1/2}$] 
now accrue small masses which 
are proportional to $a\sim 1/R$:
these are the superpartners of the gravity multiplet, 
the $SO(16)\times SO(16)$ gauginos,
and
Lorentz {\it vectors}\/
transforming in the ${\bf (spinor,1)\oplus (1,spinor)}$ representation of 
$SO(16)\times SO(16)$.
Thus, in this model, we see that the $\IZ_2$ 
twisting has two separate effects on the lightest states: 
the supersymmetry is broken, with $1/R$ setting the effective scale for 
mass splittings within the former SUSY multiplets,
and the gauge symmetry is broken from $E_8\times E_8$ to $SO(16)\times SO(16)$.
Indeed, as $R\to\infty$, the supersymmetry is restored as each state and its (former) superpartner come back into alignment and become degenerate.  Likewise, the states transforming
in the ${\bf (spinor,1)\oplus (1,spinor)}$ representation of
$SO(16)\times SO(16)$ now combine to enhance the gauge symmetry back to $E_8\times E_8$.
The higher values of $m$ within the $\calE_{0,1/2}$ functions then correspond to the KK excitations
of these different states.

The existence of the proto-graviton states in this model
can also be seen from the partition function in Eq.~(\ref{secondso16}).
Indeed, the proto-graviton state is nothing but the ground state of the $\chibar_V \chi_I^2$ sector 
within $Z^{(1)}$ --- it is a gauge singlet which transforms as a spacetime Lorentz vector,
with eight components, and indeed we find that the partition function for this interpolating
model begins with $Z^{(1)}\sim 8/q + ...$ regardless of the radius.
Moreover, as indicated above, it is in fact a general theorem that {\it any}\/ interpolating function 
will have such a sector within $Z^{(1)}$, since this is the sector from which the gravity multiplet must arise.

\begin{figure*}[h!]
\begin{center}
  \epsfxsize 2.5 truein \epsfbox {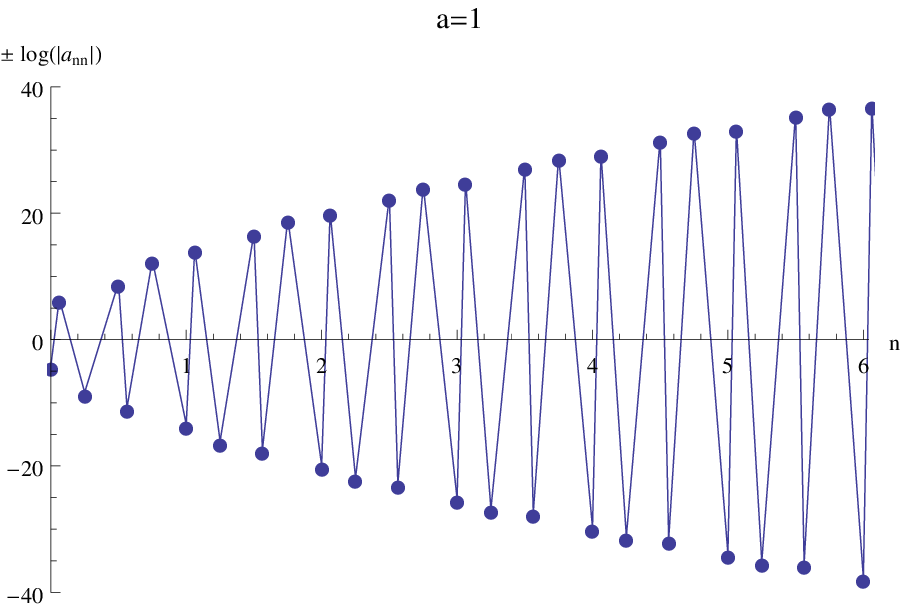}    
\hskip 0.2 truein
  \epsfxsize 2.5 truein \epsfbox {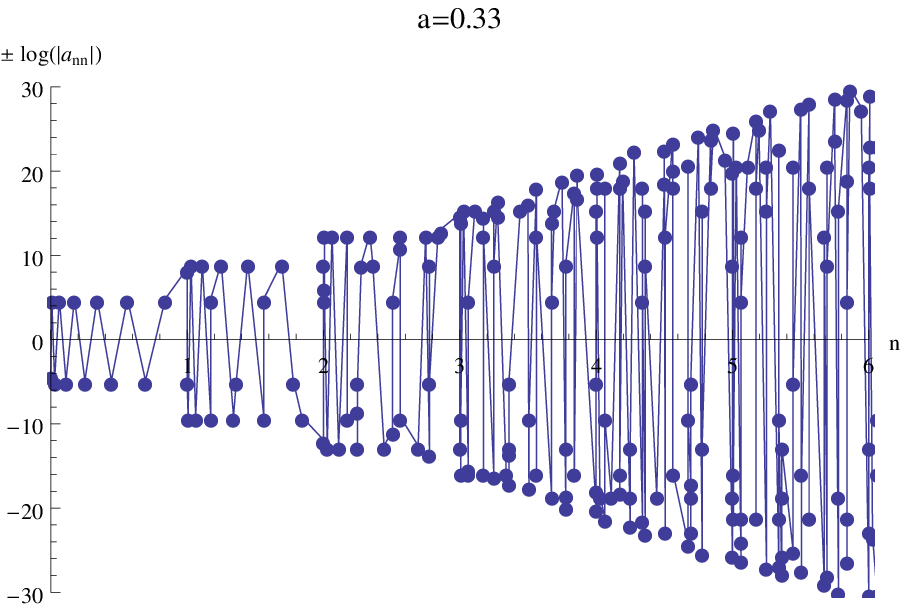}
\end{center}
\begin{center}
  \epsfxsize 2.5 truein \epsfbox {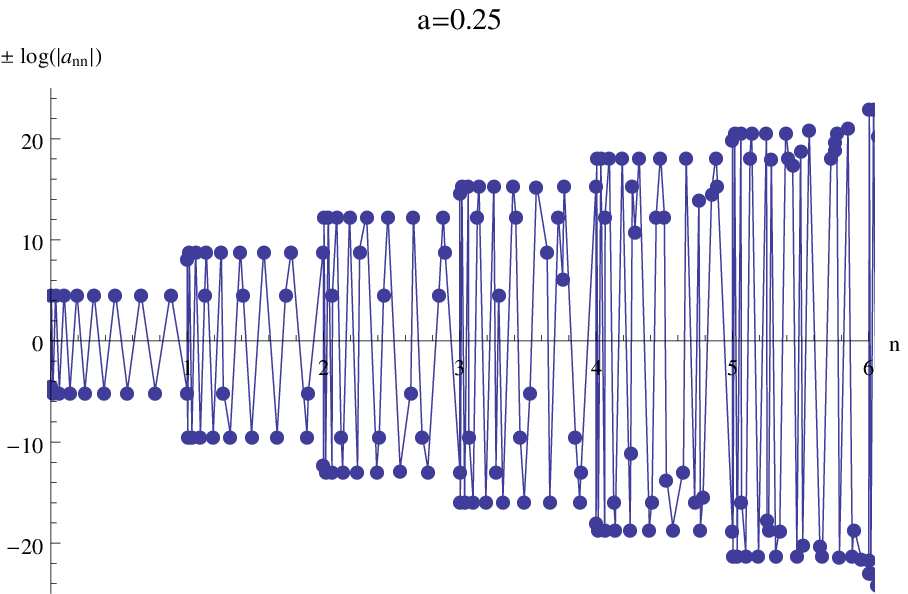}
\hskip 0.2 truein
  \epsfxsize 2.5 truein \epsfbox {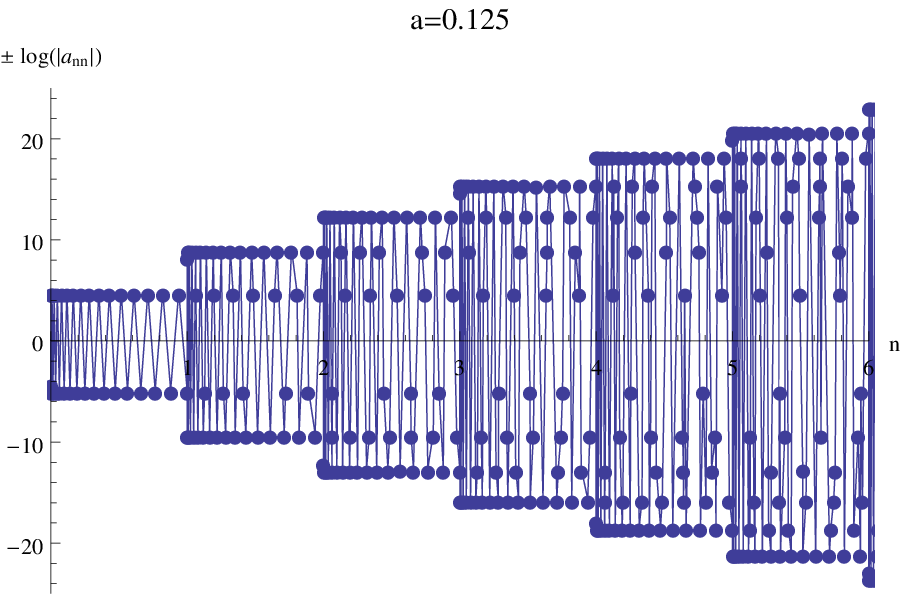}
\end{center}
\caption{Degeneracies of physical states for the interpolating model in Eq.~(\ref{secondso16}) with
$a=1$ (upper left), $a=0.33$ (upper right), $a=0.25$ (lower left), $a=0.125$ (lower right).  
Within each plot, points are connected in order of increasing world-sheet energy $n$.  In all
cases we see that surpluses of bosonic states alternate with surpluses of fermionic states
as we proceed upwards in $n$;  this behavior is the signal of an underlying ``misaligned  
supersymmetry'' which exists within all modular-invariant non-supersymmetric tachyon-free 
string theories and which is ultimately responsible for the finiteness of closed strings  --- even 
in the absence of spacetime supersymmetry.  For $R=\sqrt{\alpha'}$ (or $a=1$), we see that this
oscillation between bosonic and fermionic surpluses occurs within 
the exponentially growing envelope function $|a_{nn}|\sim e^{c\sqrt{n}}$ associated
with a Hagedorn transition.
However, as the compactification radius increases (or
equivalently as $a\to 0$), we see that a hierarchy begins to emerge between the oscillator 
states and their KK excitations;
the oscillator states continue to experience densities of states which are exponentially growing as 
functions of $n$,  but 
their corresponding KK excitations are densely packed within 
each interval $(n,n+1)$ and, as expected, exhibit constant state degeneracies.}
\label{fig:interplots}
\end{figure*}

We can also understand the spectrum of such interpolating models more globally at all energies
by calculating
their net degeneracies of physical bosonic minus fermionic states as functions of the
world-sheet energy level $E_L=E_R=n$.
Specifically, for a given interpolating model, we expand the total partition function  
in the form $Z= \tau_2^{1-D/2}\sum_{mn} a_{mn} \qbar^m q^n$
and plot $\pm \log(|a_{nn}|)$ versus $n$, where we choose the overall sign
to match the sign of $a_{nn}$.
Our results are shown in Fig.~\ref{fig:interplots} for 
the model in Eq.~(\ref{secondso16}),
evaluated at several different values of $a$.

Several features are immediately evident from Fig.~\ref{fig:interplots}.
First, we observe that
in all cases, surpluses of bosonic states alternate with surpluses of fermionic states
as we proceed upwards in $n$.   As discussed in Ref.~\cite{missusy}, this behavior is 
the signal of an underlying ``misaligned
supersymmetry'' which exists within all modular-invariant non-supersymmetric tachyon-free
string theories and which is ultimately responsible for the finiteness of closed strings  --- even
in the absence of spacetime supersymmetry. 

But perhaps even more interestingly, we see that  
the densities of states begin to exhibit
an interesting behavior as the radius increases to infinity (or equivalently as $a\to 0$).
In general, string models in which all compactification radii are at or near 
the string scale
exhibit densities of states which resemble that in the upper left panel of Fig.~\ref{fig:interplots}:
surpluses of bosonic and fermionic states oscillate within an exponentially rising envelope
function $|a_{nn}|\sim e^{c\sqrt{n}}$.  Indeed, this exponential rise in the total state
degeneracies is triggered by the exponential rise in the number of string {\it oscillator}\/ states
(as opposed to KK or winding states),  
and ultimately leads to a Hagedorn transition.
However, as $a\to 0$, we see that a hierarchy
begins to emerge between the oscillator 
states and their corresponding KK excitations.
The oscillator states 
have world-sheet energies which are quantized in units of $n$
and continue to exhibit exponentially growing state degeneracies
(even though the rate of growth becomes somewhat smaller as $a$ decreases). 
By contrast,
the Kaluza-Klein (KK) excitations of these oscillator states
have world-sheet energies which are quantized in units of $a\ll n$.
Thus, as $a\to 0$, we see that the overall  
spectrum of these models becomes increasingly 
dense, with each interval $(n,n+1)$ 
populated by the different KK excitations of the
oscillator states at level $n$.
Moreover, {\it within}\/ these 
intervals, we see that the degeneracies of states continue to oscillate
between bosonic and fermionic surpluses (as required by misaligned supersymmetry),
but do so only within an envelope function of constant amplitude.
This is exactly as expected for KK states.

Ultimately, as $a$ reaches zero, the KK states become infinitely dense.  The cancellations
between bosonic and fermionic surpluses fall into alignment, and spacetime supersymmetry is restored.
The cosmological constant then vanishes exactly. 

As discussed in Sect.~II, the spectra of non-supersymmetric 
strings generally do not have bosonic and fermionic states which can be
identified as belonging to the same supermultiplet.  However, interpolating models
of this sort are somewhat special in that they contain a tunable parameter (namely the compactification
radius) which allows them to connect smoothly back to a supersymmetric limit.
Despite this fact, these models continue to exhibit all of the properties
of misaligned supersymmetry discussed in Sect.~II, 
including supertrace relations of the form in Eqs.~(\ref{supertraces1}) and
(\ref{supertraces2}) which ultimately depend on $\Lambda$ rather than $M_c$
and which rely on the subtle interplay amongst the contributions from the KK modes,
the winding modes, and
the oscillator modes.
Indeed, the contributions from {\it all}\/ of these modes are inextricably tied together 
through modular invariance, and none of these can be altered in isolation. 
Thus, as we shall see, situations in which $\Lambda$ is suppressed 
give rise to supertraces whose overall magnitudes are smaller
(as a result of conspiracies across {\it all}\/ string energy levels)
than they would have been if they had been evaluated supermultiplet by supermultiplet.

\section{Suppression of the cosmological constant}
\setcounter{footnote}{0}

Having set the stage, we now seek to understand 
the behavior of the cosmological constant $\Lambda(R)$ 
associated with such interpolating models for large $R$.
Clearly, as $R\to\infty$, we know that $\Lambda\to 0$, reflecting the restoration of SUSY in this limit.
The question is to derive the leading correction to this result that emerges when $R$ is large but finite
(or equivalently, for $a\ll 1$). 
We shall keep the following discussion as general as possible, referring to the model in Eq.~(\ref{secondso16}) only
as an example where appropriate.

\subsection{Leading terms}

First, 
since we are assuming that SUSY is restored in the $R\to\infty$ limit, we know that $Z^{(2)}= -Z^{(1)}$ at the level of their $q$-expansions.
Since our main interest here is in the numerical behavior of $\Lambda$, we are only concerned with the $q$-expansions
that these functions have, and consequently we shall take $Z^{(2)}= -Z^{(1)}$ without further comment.
As a result, our general partition in Eq.~(\ref{EOmix}) takes the form
\beq
        Z_{\rm string}(R) ~=~ 
           Z^{(1)}\, [\calE_0(R) - \calE_{1/2}(R)]~+~
           Z^{(3)}\, \calO_0(R) ~+~ 
           Z^{(4)}\, \calO_{1/2}(R)~. 
\label{EminusE}
\eeq

Next, we observe that for large $R$ (or small $a$),
all states within 
the ${\cal O}_0$ and ${\cal O}_{1/2}$ sectors are extremely heavy as a result of non-vanishing winding modes $n\not=0$.
In general, the contributions from heavy states 
to the cosmological constant are exponentially suppressed.
As a result, contributions from such sectors will not generally yield the leading behavior for 
$\Lambda$, and we will need not consider such sectors further.
This then leaves the contributions from the ${\cal E}_{0,1/2}$ sectors:
\beq
        Z_{\rm string}(R) ~=~ 
           Z^{(1)}\, [\calE_0(R) - \calE_{1/2}(R)]~+~...
\label{Eonly}
\eeq
As a result, we see that the leading behavior generally depends on
the $q$-expansion of $Z^{(1)}$ alone, and does not depend on $Z^{(3)}$ or $Z^{(4)}$.

Let us assume that 
massless states make the dominant contributions to $\Lambda$ in theories that are devoid of physical tachyons.  
(This assertion shall be demonstrated explicitly below.)  
We shall therefore restrict our attention to the leading contributions to $\Lambda$ which come from the massless states within $Z^{(1)}$ in Eq.~(\ref{Eonly}).
Writing
\beq
         Z^{(1)} =   \tau_2^{-4}\,  \biggl\lbrack (N_b- N_f) (\qbar q)^0 + {\rm others...}\biggr\rbrack ~,
\label{preexpr}
\eeq
we thus wish to evaluate
\beq
   \Lambda ~=~ (N_f-N_b) \, \int_{\cal F} {d^2 \tau\over {\tau_2}^2}   \,\tau_2^{-4} \, (\calE_0 - \calE_{1/2}) ~+~ ...
\label{expr}
\eeq
where we are writing our expressions for $\Lambda$ in units of $(\half {\calM}^9)$.
Note that with this notation, $N_b$ and $N_f$ represent the numbers of states
which remain massless in our theory {\it after SUSY breaking has already occurred}\/.
Despite this fact, the expression in Eq.~(\ref{expr}) actually represents the contributions
from all of those states which {\it would have been massless in the absence of SUSY breaking}\/:
the contributions from those that are massless even after SUSY breaking are encapsulated within
${\cal E}_0$, while the contributions from those that are no longer massless after SUSY breaking are encapsulated
within ${\cal E}_{1/2}$.  

Since the only $R$-dependence comes from the ${\cal E}$-functions,
we must first evaluate $\calE_0 - \calE_{1/2}$ for large $R$ (or small $a$).
Clearly, as 
$R\to\infty$, we know that $\calE_0 - \calE_{1/2} \to 0$:  the difference between integer-moded and half-integer-moded KK momentum modes becomes immaterial for a truly infinite radius.
Our goal, however, is to evaluate the leading correction to this result for large but finite $R$.

 For small $a$, the winding modes with $n\not=0$ all produce contributions which are exponentially suppressed compared to those
with $n=0$.  We therefore restrict our attention to the $n=0$ contributions.  We then have
\beqn
  {\cal E}_0-{\cal E}_{1/2}
       &=& \sqrt{\tau_2} \,
           \sum_{m\in\IZ} \left[ (\qbar q)^{m^2 a^2/4} - (\qbar q)^{(m+1/2)^2 a^2/4}\right] + ... \nonumber\\
       &=& \sqrt{\tau_2} \,
           \sum_{m\in\IZ} \left[ e^{-\pi \tau_2 m^2 a^2} - e^{-\pi \tau_2 (m+1/2)^2 a^2} \right] + ... \nonumber\\
       &=& \sqrt{\tau_2} \, \left[ \vartheta_3( i\tau_2 a^2) - \vartheta_2(i\tau_2 a^2) \right] + ...\nonumber\\
       &=& {1\over a} \, \left[ \vartheta_3\left( {i\over \tau_2 a^2}\right) - 
                                \vartheta_4\left( {i\over \tau_2 a^2} \right) \right] + ...\nonumber\\
       &=& {1\over a} \, \sum_{m\in\IZ} 
           \left[ e^{-\pi m^2/\tau_2 a^2} - (-1)^m  e^{-\pi m^2 /\tau_2 a^2 }\right] + ...\nonumber\\
       &=& {2\over a} \, \sum_{m={\rm odd}} e^{-\pi m^2/\tau_2 a^2}  + ...
\label{step}
\eeqn
where we have used the modular transformations $\vartheta_{3,2}(\tau) = \vartheta_{3,4}(-1/\tau)/\sqrt{-i \tau}$ in
passing from the third to the fourth line. 
These transformations effectively resum the infinite series into a form from which the leading small-$a$ behavior can be reliably extracted.  (The method of steepest descents could also be used in order to extract the leading behavior.)
Indeed, short of omitting the contributions from winding modes at the top line, we have made no further approximations.
However, we now recognize that as $a\to 0$, the leading behavior is given by the $m=\pm 1$ terms.
We thus find that 
\beq
   {\cal E}_0-{\cal E}_{1/2} ~\sim ~ {4\over a}\, e^{-\pi/\tau_2 a^2}~~~~~~~{\rm as}~~a\to 0~,
\label{step2}
\eeq
whereupon we have
\beqn
   \Lambda &=&  {4 \over a} \,(N_f-N_b) \,  
             \int_{\cal F} {d^2 \tau\over {\tau_2}^6}  \,e^{-\pi/\tau_2 a^2} ~+~... \nonumber\\
          &\approx&  {96\over \pi^5}\, (N_f-N_b) \, a^9~+~...
\label{preleading}
\eeqn
In passing to the final line above, we have restricted the range of integration to the upper $(\tau_2\geq 1)$ portion of the fundamental domain of the modular group, performed the resulting integral analytically,
 and disregarded subleading terms of order ${\cal O}(e^{-\pi/a^2})$ which result from the integration.
It may easily be verified that the contribution from the lower portion of the fundamental domain is significantly smaller than that of the upper portion for all $a\ll 1$.  
Thus, we see that $\Lambda\sim 1/R^9$.

For $D=4$, the calculation goes through as above, resulting in the conclusion that
\beqn
   \Lambda &=&  {4 \over a} \,(N_f-N_b) \,  
             \int_{\cal F} {d^2 \tau\over {\tau_2}^3}  \,e^{-\pi/\tau_2 a^2} ~+~... \nonumber\\
          &\approx&  {4\over \pi^2}\, (N_f-N_b) \, a^3~+~...
\label{preleadingD}
\eeqn
Indeed, for all spacetime dimensions $D$ we find that 
\beq
       \Lambda~=~  {4(D/2-1)!\over \pi^{D/2}} \,(N_f-N_b)\, a^{D-1}~ +~...~, 
\label{leadingD}
\eeq
with the
understanding (relevant for odd $D$) that $(1/2)!=\Gamma(3/2)=\sqrt{\pi}/2$, {\it etc}\/.

This scaling behavior is not a complete surprise, and is essentially dictated by dimensional analysis.  
Moreover, this result is also familiar in another context.  
Recall that in this setup, we are considering
a ten-dimensional string theory 
compactified on a $\IZ_2$-orbifolded circle of radius $R$.  
However, the associated mathematics is identical to that which we would encounter if
we were instead considering our original ten-dimensional string theory at finite temperature:
we would simply identify $T=(2\pi R)^{-1}$ and interpret the vacuum energy $\Lambda$ as the  
thermal (finite-temperature) vacuum amplitude ${\cal V}(T)$.  
Indeed, the corresponding free energy $F(T)$ would then be given by   $F(T)\equiv T {\cal V}(T)$.
However, it is well known that for $T\ll M_{\rm string}$, any string theory in $D$ spacetime dimensions
which is originally supersymmetric at $T=0$ will have a free energy
which scales as $F(T)\sim T^{D}$.
(Note that this accords, as expected, with
our expectations for the {\it high}\/-temperature limit of field theory.)
This then implies that ${\cal V}(T)\sim T^{D-1}$, which in our case implies that we should 
expect $\Lambda\sim 1/R^9$.

For the model in Eq.~(\ref{secondso16}), we have $N_f-N_b = 64$.
Inserting this into Eq.~(\ref{preleading}),
we thus find the expected leading behavior~\cite{Itoyama:1986ei} 
\beq
                \Lambda ~\sim~ {6144\over \pi^5} \, a^9 ~\sim~ (20.08) \, a^9~~~~~~~{\rm as}~~a\to 0~.
\label{leading}
\eeq

In order to verify this expectation, we now explicitly evaluate 
$\Lambda$ associated with the partition function in Eq.~(\ref{secondso16}) 
as a function of $a$.
We do this numerically, keeping all terms in Eq.~(\ref{secondso16}) and
performing full numerical integrations over the entire fundamental domain
of the modular group.

\begin{figure*}[h!]
\begin{center}
  \epsfxsize 3.5 truein \epsfbox {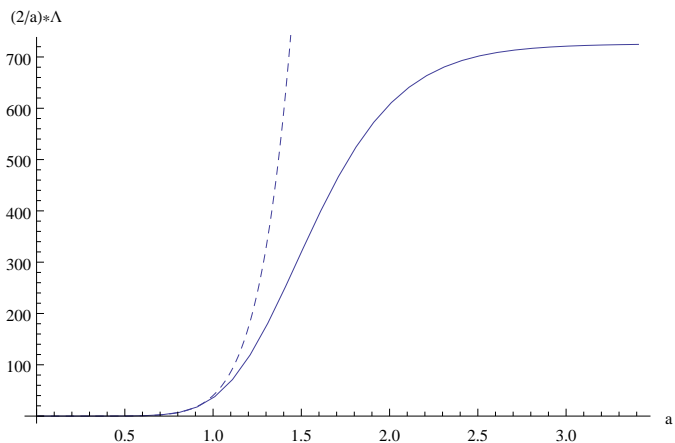}
\end{center}
\caption{The rescaled cosmological constant 
   $(2/a)\Lambda$ 
associated with the model in Eq.~(\ref{secondso16}), plotted versus $a\equiv \sqrt{\alpha'}/R$ (solid line).
    We see that $(2/a)\Lambda$ indeed interpolates
        between $\Lambda=0$ at $a=0$ and $\Lambda\approx 725$ as $a\to\infty$, where $\Lambda\approx 725$
        is the cosmological constant associated with the ten-dimensional $SO(16)\times SO(16)$ heterotic string.
        Note that the factor of $(a/2)$ is the effective (T-dual) 
           ``volume'' of compactification in the $a\to\infty$ limit;
        dividing by this factor allows our interpolating nine-dimensional cosmological constant to 
        asymptote to a finite ten-dimensional limit as $a\to \infty$.
      The dashed line shows the small-$a$ behavior indicated in Eq.~(\ref{leading}).}
\label{fig:lambda}
\end{figure*}

Our results are shown in Fig.~\ref{fig:lambda}, where we plot
    $(2/a)\Lambda$ 
versus $a\equiv \sqrt{\alpha'}/ R$ (solid line).
        Note that the factor of $(a/2)$ is the effective (T-dual) 
           ``volume'' of compactification in the $a\to\infty$ limit;
        dividing by this factor allows our interpolating nine-dimensional cosmological constant to 
        asymptote to a finite ten-dimensional limit as $a\to \infty$.
Indeed, we see from Fig.~\ref{fig:lambda} that
this model successfully interpolates
        between $\Lambda=0$ at $a=0$ and $\Lambda\approx 725$ at $a\to\infty$, where $\Lambda\approx 725$
        is the cosmological constant associated with the ten-dimensional $SO(16)\times SO(16)$ heterotic string.
(Note that a similar plot appears in Ref.~\cite{Itoyama:1986ei}.)
Moreover, we see from Fig.~\ref{fig:lambda}
that Eq.~(\ref{leading})
indeed provides 
an excellent approximation to the cosmological constant for $a\ll 1$,
holding to several significant digits throughout the relevant range.
This verifies not only the overall radius-dependence (scaling power-law behavior) in Eq.~(\ref{leading}) 
but also the numerical coefficient which precedes it.

\subsection{Subleading terms}

Along the way from Eq.~(\ref{EminusE}) to 
Eq.~(\ref{preleading}), six independent approximations were made.
Our result in Eq.~(\ref{preleading})
therefore comes with a variety of corresponding subleading correction terms.

Our first approximation comes from disregarding the contributions from the $\calO_{0,1/2}$ sectors
in passing from Eq.~(\ref{EminusE}) to Eq.~(\ref{Eonly}).
Our second approximation comes from disregarding the contributions
from states which are not massless within $Z^{(1)}$ 
in passing from Eq.~(\ref{Eonly}) to Eq.~(\ref{preexpr}).
Our third approximation comes from disregarding contributions
from the winding modes within $\calE_{0,1/2}$ when writing the first
line of Eq.~(\ref{step}), 
and our fourth approximation comes from disregarding the
higher $|m|>1$ modes in passing from the last line of Eq.~(\ref{step})
to Eq.~(\ref{step2}).
Our fifth approximation comes from disregarding the contribution from the lower ($\tau_2<1$) 
portion of the fundamental domain $\calF$ in evaluating the integral in Eq.~(\ref{preleading}), 
and our sixth approximation comes from disregarding the subleading terms which come from
the integration over the upper ($\tau_2\geq 1$) portion of $\calF$.

As a question of mathematical accuracy, to each of these approximations there corresponds
a subleading correction term which in principle should be included in our  main result.
However, not all of these correction terms are of equal interest to us.
In particular, the third through sixth approximations listed above are particularly uninteresting
because their contributions  cannot ever be isolated from the dominant contribution we have already
computed.  
Indeed, these approximations all fed into our evaluation of $\calE_0-\calE_{1/2}$,
and thus the corresponding corrections can never be relevant in isolation  
from the leading value that we already found for this quantity 
in Eq.~(\ref{step2}).
By contrast, the first two approximations are of a different nature, as there  
may exist string models for which the leading terms fail to appear for
various reasons.  
In particular, both of these approximations 
assume that $Z^{(1)}$ gives rise to a non-zero value for $N_f-N_b$, and yet
there may be models for which $N_b=N_f$ within $Z^{(1)}$.
Note that this does not necessarily mean that the massless spectrum is supersymmetric,
or even that all such massless states are observable;
indeed our observable massless spectrum could consist
of states with $N_b\not=N_f$, provided a possible hidden sector
contains the remaining massless states needed to fill out the matching condition $N_b=N_f$. 
In such cases, the subleading corrections can potentially 
be of critical importance.
We shall therefore consider each of these two corrections in turn.
 
\subsubsection{Contributions from massive and off-shell states within $Z^{(1)}$}

We begin by discussing the contributions from massive and off-shell states within
$Z^{(1)}$.  As we pointed out in Sect.~IIB, the spectrum of every non-supersymmetric string model contains
off-shell proto-graviton states 
whose contributions to the partition function are completely uncancelled.
Indeed, there will generally be a plethora of other similar light (or even off-shell tachyonic) states 
in the spectra of such strings, and we have already seen in Sect.~III that such states often make
the largest contributions
to the one-loop cosmological constant.
Given these considerations, one might suspect that the approximation
we made 
in passing from Eq.~(\ref{Eonly}) to Eq.~(\ref{preexpr}) --- \ie, the approximation 
in which we focused on only the contributions from the massless states within $Z^{(1)}$ ---
will lead to a particularly huge correction term.

\begin{figure*}[h!]
\begin{center}
  \epsfxsize 3.5 truein \epsfbox {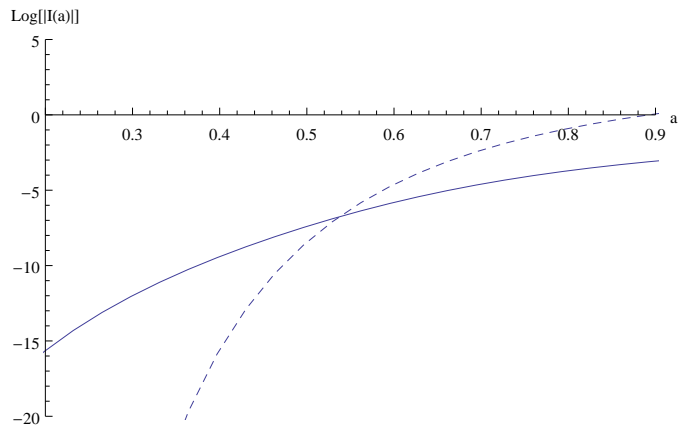}
\end{center}
\caption{Vacuum-energy contributions from massless states (solid line) versus proto-gravitons (dashed line).  
   In this figure, we have plotted $\log(|\tilde I_{0,0}^{(10)}(a)|)$ 
   and  $\log(|\tilde I_{0,-1}^{(10)}(a)|)$ versus $a$, respectively. 
   We see that the former exceeds the latter in the $a\to 0$ limit, 
   and indeed exceeds the latter for all $a\lsim 0.54$.}
\label{lambdacompare}
\end{figure*}

However, this is ultimately not the case because the integral that describes
the contributions to the cosmological constant from the individual states within the $Z^{(1)}$ sector
of such interpolating models is not given by Eq.~(\ref{integral}), but rather by 
\beq
        \tilde I_{m,n}^{(D)}(a) ~\equiv~ {4\over a} \, 
    \int_{\cal F} {d^2\tau\over \tau_2^2} \, \tau_2^{1-D/2} \, \qbar^m q^n\,
        e^{-\pi /\tau_2 a^2}  ~.
\label{integral2}
\eeq
Comparing the results for massless states and proto-graviton states, we find the 
results shown in Fig.~\ref{lambdacompare}.
We see that the contributions from the massless states exceed the contributions from the proto-graviton
states in the $a\to 0$ limit, and indeed exceed these latter contributions for all $a\lsim 0.54$.
  
This demonstrates that the massless physical states indeed provide the dominant contributions as $a\to 0$.
However, in general, the largest subdominant contributions 
actually come from light but massive  
physical states within $Z^{(1)}$ --- \ie,  states with $m=n$ for small $m$.
In general, we find that 
\beq
        \tilde I_{m,m}^{(D)}(a) ~\approx~ 
            4 \, \left( 2\sqrt{m} a\right)^{(D-1)/2}\, e^{-4\pi \sqrt{m} /a} ~~~~~~ {\rm as}~~a\to 0~.
\label{res}
\eeq
Thus, the leading subdominant contributions come from those physical
string states with the smallest non-zero values of $m$.
In general, this value of $m$ is determined by examining the individual terms 
within $Z^{(1)}$;  the result is generally model-dependent since it is sensitive
to the nature of the orbifold twists involved in the construction
of the model at hand.
We thus find these states produce an overall correction
term given by
\beq
        {\rm correction~term}~\approx~ 
            4 \, \left[N_f(m)-N_b(m)\right ]\, 
        \left( 2\sqrt{m} a\right)^{(D-1)/2}\, e^{-4\pi \sqrt{m} /a} ~~~~~~ {\rm as}~~a\to 0~,
\label{correction1}
\eeq
where $N_f(m)-N_b(m)$ indicates the net degeneracy of such states with $m=n$.
As an example, for the partition function in Eq.~(\ref{secondso16}), 
we find that $m=1$, with $N_f(1)-N_b(1)= -4608$.
Note, in particular, that the overall exponential 
suppression for such contributions scales as $e^{-1/a}$ rather than $e^{-1/a^2}$. 

Although the nature of the above correction is model-dependent, 
there is one model-independent subleading correction of this sort
which is always guaranteed to exist.
This is, as discussed, the contribution from the
proto-graviton states with $(m,n)=(0,-1)$.
Although these contributions are smaller than those from the $(m,m)$ states discussed
above, there may be situations where the latter states  
fail to exist for sufficiently small $m$.  Even in such cases, however, we are nevertheless certain 
to accrue contributions from the proto-graviton states.
It turns out that as $a\to 0$, these contributions 
scale approximately as
\beq
                \tilde I^{(D)}_{0,-1}(a) ~\approx~ - {4\sqrt{2}\over \pi} \, e^{2\pi}  \, a^2\, e^{-\pi/a^2} ~~~~~~ {\rm as}~~a\to 0~.
\label{scalingbeh}
\eeq
Indeed, in the $a\to 0$ limit, the result in Eq.~(\ref{scalingbeh}) is 
independent of the spacetime dimension $D$, and scales as $e^{-1/a^2}$ rather than $e^{-1/a}$.  
Thus, the subleading correction term to the cosmological constant
from proto-graviton states scales approximately as
\beq
                {\rm correction~term}~\approx~
            {4\sqrt{2}\over \pi} \, e^{2\pi}  \, N_{\rm proto}\, a^2\, e^{-\pi/a^2} ~+~... ~~~~~~ {\rm as}~~a\to 0~,
\label{correction2}
\eeq
where $N_{\rm proto}$ is the number of proto-graviton states in the model.  For example, for the model
in Eq.~(\ref{secondso16}), we have $N_{\rm proto}=8$.
In general, since proto-graviton states always emerge from a sector of the form $\chibar_V\chi_I$ where $\chi_I$ denotes the ground state of the left-moving sector, our proto-graviton states generally transform as a vector of the transverse Lorentz group $SO(D-2)$.  We thus expect that $N_{\rm proto}= D-2$.

Of all unphysical states (\ie, states for which $m\not=n$), it turns out that the contributions from the 
proto-gravitons are the largest.
In fact, the result in Eq.~(\ref{scalingbeh})
is only a special case of the more general result
\beq
                \tilde I^{(D)}_{m,n}(a) ~\approx~ - {4\sqrt{2}\over \pi} \, e^{-2\pi (m+n)} \, a^2\, e^{-\pi/a^2} ~~~~~~ {\rm as}~~a\to 0~.
\label{scalingbeh2}
\eeq
Quite remarkably, this result holds for all $D$ as well as for all $(m,n)$, as long as $m\not=n$;
this result is even independent of $|m-n|$. 
Indeed it is only in the $a\to 0$ limit that our modular integrals take such simple forms.
We thus see from Eq.~(\ref{scalingbeh}) that the largest contributions from unphysical states
come from those with the smallest $m+n$.  These are indeed the proto-graviton states, as discussed above.

To summarize, we see that when $N_b=N_f$, the largest contributions to the cosmological
constant in the $a\to 0$ limit
come from the lightest 
physical states within $Z^{(1)}$.
This contribution is given in Eq.~(\ref{correction1}), and scales as $e^{-1/a}$.
Failing this, however, there will always be a correction from the proto-graviton states.
This contribution is given in Eq.~(\ref{correction2}), and scales as $e^{-1/a^2}$.

\subsubsection{Contributions from $\calO_{0,1/2}$ sectors}

Finally, we now seek to evaluate the possible subleading contributions
from the $\calO_0$ and $\calO_{1/2}$ sectors. 
In models with $N_b=N_f$, it is possible that these subleading contributions 
might surpass those described above.

Although $\calO_{0}$ and $\calO_{1/2}$ each vanish as $a\to 0$, they do produce non-zero contributions
for any non-zero $a$.
As a result, we must first
evaluate these functions in the limit of small but non-zero $a$.
To do this, we follow an algebraic procedure which parallels our analysis of
$\calE_{0}-\calE_{1/2}$ in Eq.~(\ref{step}).  Indeed, the most significant change is that the $\calO_{0,1/2}$
functions in the $a\to 0$ limit are now related to the {\it generalized}\/ Jacobi $\vartheta$-functions:
\beqn
           \calO_0     &=&  2\, \sqrt{\tau_2}\, e^{-\pi \tau_2/a^2}\, \vartheta_3 (\tau_1, i\tau_2 a^2)~+~...\nonumber\\
           \calO_{1/2} &=&  2\, \sqrt{\tau_2}\, e^{-\pi \tau_2/a^2}\, \vartheta_2 (\tau_1, i\tau_2 a^2)~+~...~,
\eeqn
where the generalized Jacobi $\vartheta_i(z,\tau)$ functions are 
defined in Eqs.~(\ref{eq:jacobi}) and (\ref{shorthand}).
Using the modular transformations in Eq.~(\ref{modulartransfs})
and keeping only the leading terms, we then find
\beq
           \calO_{0,1/2} ~\approx ~   {2\over a}\, e^{-\pi |\tau|^2 / \tau_2 a^2}~~~~~~~~~ {\rm as}~a\to 0~.
\eeq
Thus, a given state with world-sheet energies $(m,n)$ within the $\calO_{0,1/2}$ sectors 
makes a contribution to the cosmological constant given by 
\beq
        \hat I_{m,n}^{(D)}(a) ~\equiv~ {2\over a} \, 
    \int_{\cal F} {d^2\tau\over \tau_2^2} \, \tau_2^{1-D/2} \, \qbar^m q^n\,
        e^{-\pi |\tau|^2 /\tau_2 a^2}  ~
\label{integral3}
\eeq
in the $a\to 0$ limit.

Given the result in Eq.~(\ref{integral3}),
our final task is to survey the possible states within $Z^{(3,4)}$ which might provide the dominant subleading
contributions to $\Lambda$.  Note that this task is independent of the survey of the states within 
$Z^{(1)}$, given the fact that the contribution in Eq.~(\ref{integral3}) differs significantly in structure
from that in Eq.~(\ref{integral2}). 
We also recall that the states within $Z^{(4)}$ 
have world-sheet energies $(m,n)$ with $m-n\in\IZ+1/2$, whereas those within $Z^{(3)}$ have $m-n\in\IZ$.
We find, however, that
\beq
      \hat I^{(D)}_{m,n}(a) ~\approx~   {2\sqrt{2}\over \pi} \, e^{-2\pi (m+n)} \, a^2\, e^{-\pi/a^2} ~~~~~~ {\rm as}~~a\to 0~.
\label{scalingbeh3}
\eeq
Indeed, this result holds for all spacetime dimensions $D$ as well as all
energy configurations $(m,n)$, regardless of whether $m=n$ or not.

This result is quite remarkable, since it exactly duplicates
the result we found in Eq.~(\ref{scalingbeh2}) for the $\calE_{0,1/2}$ sectors up to an overall sign
and a factor of two!
Indeed, this duplication occurs {\it even though}\/ the form of the integral in Eq.~(\ref{integral2}) is
quite different from the form of the integral in Eq.~(\ref{integral3}), and {\it even though}\/
the result in Eq.~(\ref{scalingbeh2}) is subject to a restriction (namely $m\not=n$) which
does not apply to the result in Eq.~(\ref{scalingbeh3}).
Clearly, the mathematical elegance of these features
resides in the power of the asymptotic $a\to 0$ limit.
But there are also potentially important phenomenological implications.
For example, as a result of this observation, we now see that the contribution from a state
in the ${\cal E}_{0,1/2}$ sectors
~with energy configuration $(m,n)$ (with $m\not=n$)
can be completely cancelled
by the contribution of a state in the ${\cal O}_{0}$ or ${\cal O}_{1/2}$ sectors
with a completely different energy configuration $(m',n')$, so long as $m+n= m'+n'$.
Even more remarkably, focusing on the special case with $m'=n'= \half(m+n)$,
we see that the contribution of an {\it unphysical}\/ state
in the $\calE_{0,1/2}$-sectors can be cancelled by the contribution
of a {\it physical}\/ state in the $O_0$-sector!
These results clearly imply a myriad of potential new ways of further suppressing
the contributions to the cosmological constant.

\subsubsection{Contributions to cosmological constant:~  Summary}

Pulling together the different results above, we thus find
that the contributions to the cosmological constant in the
$a\to 0$ limit from a given state with world-sheet energies $(m,n)$
in the different $\calE/\calO$-sectors are given as
\beq
\begin{tabular}{||c|c|| c ||}
\hline
\hline
~~ {\rm sector}~~ & ~~{\rm state}~~& ~~contribution~to~$\Lambda$~~\\
\hline
\hline
~~$\calE_0-\calE_{1/2}$~~ & ~~ $m=n=0$~~ & ~~$-\displaystyle{{[4 (D/2-1)!/ \pi^{D/2}}] \, a^{D-1} }$ \\[5 pt]
\hline
~~$\calE_0-\calE_{1/2}$~~ & ~~ $m=n\not= 0$~~ & ~~$\displaystyle{4 (2\sqrt{m} a)^{(D-1)/2} e^{-4\pi \sqrt{m}/a}}$~~\\
\hline
~~$\calE_0-\calE_{1/2}$~~ & ~~ $m\not =n$~~ & ~~$\displaystyle{-[{4\sqrt{2} / \pi]} e^{-2\pi (m+n)} a^2 e^{-\pi/a^2}}$~~\\
\hline
~~$\calO_{0,1/2}$~~ & ~~ any~$(m,n)$~~ & ~~ $\displaystyle{{[2\sqrt{2} / \pi]} e^{-2\pi (m+n)} a^2 e^{-\pi/a^2}}$~~\\
\hline
\hline
\end{tabular}
\label{tableofcontributions}
\eeq
Note that in this table, $D$ represents the dimensionality of the theory in question {\it prior}\/ 
to the compactification on the twisted circle.
Thus, if we expand each of the $Z^{(i)}$ in Eq.~(\ref{EOmix}) in the form 
$Z^{(i)}=\tau_2^{1-D/2} \sum_{mn} a_{mn}^{(i)}\qbar^m q^n$,
we see that the leading contribution takes the form
\beq
      {\rm leading}:~~~~~  -{4\, (D/2-1)! \, \alpha'^{(D-1)/2} \over \pi^{D/2}} \, a_{00}^{(1)} \, {1\over R^{D-1}}~,
\eeq
while the dominant subleading contribution takes the form
\beq
      {\rm dominant~subleading}:~~~~~  -4 \sum_{m>0} a_{mm}^{(1)}\, (2\sqrt{m}a)^{(D-1)/2}\, 
             e^{-4\pi \sqrt{m} R /\sqrt{\alpha'} }~. 
\label{subsizes}
\eeq
Indeed, the values of $a_{mm}^{(1)}$ for the model in Eq.~(\ref{secondso16}) 
are plotted exactly in Fig.~\ref{fig:missusy} (thereby allowing that figure to do double duty).
We thus see that as a result of the misaligned supersymmetry inherent in the oscillations
in the values of the $a_{nn}^{(1)}$ coefficients, the magnitudes of the subleading terms
in Eq.~(\ref{subsizes}) for any small but non-zero $a$ are significantly smaller than they otherwise
would have been.

\subsection{Interpolating models on a twisted two-torus}

Finally, let us briefly comment on the leading behavior of the cosmological constant in situations
in which more than a single dimension is compactified on a twisted circle.  Clearly, the resulting
sensitivity of $\Lambda$ will depend on the particular compactification geometry.  For example, 
two-dimensional compactifications could occur on twisted two-tori or on spheres.
The resulting phenomenologies would then be markedly different.
Despite this fact,   
we expect the general behavior in which the leading contributions to $\Lambda$ 
are inverse to the compactification volume to remain intact.

As an example, let us consider the case of a 
supersymmetric theory in $D$ dimensions compactified $S^1\times S^1$, where each circle $S^1$
has its own radius $R_i$ and is subject to its own supersymmetry-breaking $\IZ_2$ orbifold twist.
The partition function of such a theory then takes the general form
\beq
        Z ~=~ \sum_{a,b=1}^4   Z^{(a,b)} \, \calC_a \tilde \calC_b~,
\label{twocircles}
\eeq
where $C_a=\lbrace \calE_0,\calE_{1/2},\calO_0,\calO_{1/2}\rbrace$ and where quantities with (without)
the tilde refer to the first (second) circle with radius $R_1$ ($R_2$).
In complete analogy with Eq.~(\ref{EOmix}), $\sum_{a,b=1}^2 Z^{(a,b)}$ 
is the partition function of our original uncompactified $D$-dimensional theory,
while the quantities $Z^{(1,1)} + Z^{(1,2)}$,
$Z^{(1,1)} + Z^{(2,1)}$,
and $Z^{(1,1)}$ 
respectively represent the traces over those string states which are invariant under
the first twist,
the second twist,
and both twists simultaneously.
In the large-$R_1$ and large-$R_2$ limits,
the contributions from the $\calE_{0,1/2}$ sectors will dominate;  consequently
we can restrict our attention to these sectors, and reshuffle the $\calE$-related 
terms within Eq.~(\ref{twocircles}) 
into the form
\beqn
    Z &=& \phantom{~+~} 
              \textstyle{1\over 4} \left[ Z^{(1,1)} + Z^{(1,2)} + 
                            Z^{(2,1)} + Z^{(2,2)} \right]
               \, (\calE_0+\calE_{1/2}) \, (\tilde\calE_0+\tilde\calE_{1/2}) \nonumber\\
      && ~+~   
              \textstyle{1\over 4} \left[ Z^{(1,1)} + Z^{(1,2)} - 
                            Z^{(2,1)} - Z^{(2,2)} \right]
               \, (\calE_0-\calE_{1/2}) \, (\tilde\calE_0+\tilde\calE_{1/2}) \nonumber\\
      && ~+~   
              \textstyle{1\over 4} \left[ Z^{(1,1)} - Z^{(1,2)} + 
                            Z^{(2,1)} - Z^{(2,2)} \right]
               \, (\calE_0+\calE_{1/2}) \, (\tilde\calE_0-\tilde\calE_{1/2}) \nonumber\\
      && ~+~   
              \textstyle{1\over 4} \left[ Z^{(1,1)} - Z^{(1,2)} - 
                            Z^{(2,1)} + Z^{(2,2)} \right]
               \, (\calE_0-\calE_{1/2}) \, (\tilde\calE_0-\tilde\calE_{1/2}) ~+~...
\label{Zexpanded}
\eeqn
Note that the coefficient in the top line of
Eq.~(\ref{Zexpanded})
is nothing but the partition function of our original $D$-dimensional
string model prior to compactification.
However, this model was assumed to be supersymmetric 
by construction.  Thus, at the level of the corresponding $q$-expansions,
this coefficient vanishes identically and the remaining terms simplify to
\beqn
    Z &=& \phantom{~+~} 
              \textstyle{1\over 2} \left[ Z^{(1,1)} + Z^{(1,2)} \right]
               \, (\calE_0-\calE_{1/2}) \, (\tilde\calE_0+\tilde\calE_{1/2}) \nonumber\\
      && ~+~   
              \textstyle{1\over 2} \left[ Z^{(1,1)} + Z^{(2,1)} \right]
               \, (\calE_0+\calE_{1/2}) \, (\tilde\calE_0-\tilde\calE_{1/2}) \nonumber\\
      && ~+~   
              \textstyle{1\over 2} \left[ Z^{(1,1)} + Z^{(2,2)} \right]
               \, (\calE_0-\calE_{1/2}) \, (\tilde\calE_0-\tilde\calE_{1/2}) ~+~...
\label{Zexpanded2}
\eeqn
Recall that in the $R\to\infty$ (or $a\to 0$) limit,
  $\calE_0+\calE_{1/2} \to 2/a$ while
  $\calE_0-\calE_{1/2} \to (4/a) e^{-\pi/\tau_2 a^2}$.  
Thus, {\it a priori}\/, the leading terms come from the top two lines of 
Eq.~(\ref{Zexpanded2}), whereupon we have
\beq
    Z ~\sim~ {4\over a_1 a_2} \left\lbrace
               \left[ Z^{(1,1)} + Z^{(1,2)} \right] e^{-\pi/\tau_2 a_1^2}        
       ~+~   
      \left[ Z^{(1,1)} + Z^{(2,1)} \right] e^{-\pi/\tau_2 a_2^2}        
         \right\rbrace~+~...  ~~~~~~~~ {\rm as}~~ a_1,a_2\to 0~.
\eeq
However, as indicated above, the coefficients of these two terms are nothing but the
traces over those states which are invariant under the first and second twists,
respectively.
Recognizing, as before, that physical massless states will make 
the largest contributions to $\Lambda$, we 
therefore find that
\beq
       \Lambda ~\sim~  {4 (D/2-1)!\over\pi^{D/2}} {1\over a_1 a_2}\,
       \left[  (N_f-N_b) \,a_1^D + (\tilde N_f-\tilde N_b)   \,a_2^D\right]~+~...  ~~~~~~~~~~ {\rm as}~~a_1,a_2\to 0~,
\label{twistedtwo}
\eeq
where $N_f-N_b$ and $\tilde N_f-\tilde N_b$ denote
the net numbers of physical massless states 
(fermionic minus bosonic)
which are invariant under the first and second twists, respectively.
Note that the leading factor $\sim {1/a_1 a_2}$ is entirely expected, since this is nothing but the volume factor
for our two-torus.
Moreover, taking $a_1\to 0$ or $a_2\to 0$ in Eq.~(\ref{twistedtwo}) 
reproduces the result in Eq.~(\ref{leadingD})
for a single twisted circle. 
It is also possible to show that the subleading corrections to Eq.~(\ref{twistedtwo}) are
exponentially suppressed, as before.

\section{Constructing string models with $N_b=N_f$:  ~Our ``roadmap'' and 6D starting point}
\setcounter{footnote}{0}

We now turn to the task of constructing phenomenologically viable models 
that incorporate all the features described in the previous sections. 
Ultimately, our goal is to find a non-supersymmetric Standard-Model-like 
theory that has $N_b=N_f$ for the massless states, and hence an exponentially suppressed cosmological constant.
In this context, we hasten to add that demanding $N_b=N_f$ 
for the massless states does {\it not}\/ imply that we
are demanding any form of supersymmetry.
Indeed, it is only the counting of bosonic degrees of freedom which needs to match the counting
of fermionic degrees of freedom.
Moreover, it is not even required that all of these states be observable.  Many string models (including the models
we shall eventually construct) have not only an observable sector but also a hidden sector.
An exponentially suppressed cosmological constant will be assured  so long as the {\it total}\/ numbers of bosonic  
and fermionic degrees of freedom match, even if they do not match within the observable or hidden sectors separately.
Thus, there is no reason why our observed low-energy world (for which there are apparently unequal numbers
of bosonic and fermionic states) could not nevertheless be
among the class of string theories exhibiting $N_b=N_f$ --- all without any additional visible states required.

\subsection{Roadmap:~  Our approach towards realizing models with $N_b=N_f$}

Models with $N_b=N_f$ are not easy to construct;  an even more difficult
task is to ensure that they also simultaneously exhibit low-lying spectra resembling either the Standard Model
or one of its numerous extensions.
As a result, the construction of our models follows a rather deliberate, step-by-step
approach which involves starting in six dimensions and 
then performing a so-called ``coordinate-dependent
compactification'' (CDC) down to four dimensions.
CDC's are generalizations of ordinary Scherk-Schwarz compactifications~\cite{scherkschwarz}
which were introduced and developed in Refs.~\cite{Ferrara:1987es,Ferrara:1987qp,Ferrara:1988jx,Kounnas:1989dk, Antoniadis:1992fh}.

\begin{figure*}[h!]
\begin{center}
  \epsfxsize 7.0 truein \epsfbox{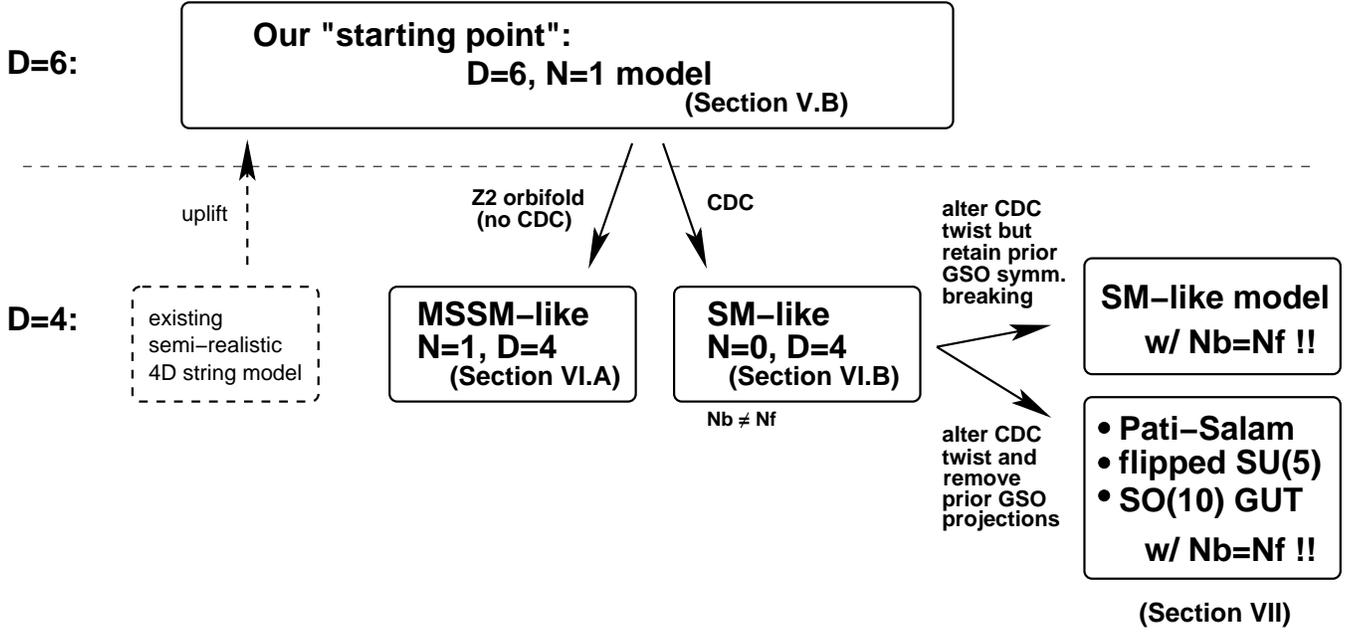}
\end{center}
\caption{Roadmap illustrating our procedure for constructing 
semi-realistic non-supersymmetric string models with $N_b=N_f$, as discussed in the text.}
\label{fig:map}
\end{figure*}

In Fig.~\ref{fig:map} we illustrate 
the specific model-construction route we shall be taking 
in this paper.
As discussed in Sect.~III, one aspect of the  class of model we ultimately seek to construct is
that it should  have one or more adjustable radii $R_i$;   moreover, we require the
$R_i\to\infty$ limits to be supersymmetric, with vanishing cosmological constant.
To accomplish this, 
we therefore start with a supersymmetric semi-realistic model in higher dimensions,
and then compactify on some sort of twisted manifold.
This procedure thereby introduces the needed radii and ensures that 
the corresponding cosmological constant vanishes as they are taken to infinity. 

For technical reasons it actually proves advantageous to begin in {\it six}\/ dimensions rather
than five.
We therefore begin our discussion by presenting a six-dimensional string model
with $\calN=1$ supersymmetry.
This model serves as the starting point for our analysis, and is ultimately
derived by lifting into six dimensions 
several semi-realistic four-dimensional 
$\calN=1$ string models~\cite{Antoniadis:1989zy, Antoniadis:1990hb, 
  Faraggi:1991be, Faraggi:1991jr, Faraggi:1992fa, Faraggi:1994eu, Dienes:1995bx}
which are already on the market.
This six-dimensional model therefore 
already inherits many of the desirable phenomenological features of its
four-dimensional parents, and will be discussed in Sect.~V.B.

Once we have this model, our next step is to compactify back down to four dimensions.
This will be discussed in Sect.~VI.~
For pedagogical purposes,
 it will prove useful to compare the four-dimensional $\calN=1$ model that results from
compactifying back to four dimensions on a $\IZ_2$ orbifold 
with the four-dimensional $\calN=0$ model that results from a coordinate-dependent compactification on the 
same orbifold. 
This will allow us to see directly how the breaking of supersymmetry manifests itself in the partition function,
following the lines discussed in Sect.~II, and
to see how this in turn affects the leading contribution to the cosmological constant,
as outlined in Sect.~IV.  

Our final step is to take the $\calN=0$ model produced in Sect.~VI and introduce modifications
which render $N_b$ and $N_f$ equal, thereby assuring an exponentially suppressed cosmological constant.
This will be discussed in Sect.~VII.~
As we demonstrate, there are several different ways in which this can be done.
One way is to alter the final CDC twist but retain the prior GSO symmetry breaking:  as we shall see, this produces 
an SM-like model with $N_b=N_f$.
By contrast, altering the final CDC twist and also removing prior GSO projections
can lead to a variety of additional models:  a  Pati-Salam-like model, a flipped-$SU(5)$ ``unified'' model,
and an $SO(10)$ ``unified'' model,  each also with $N_b=N_f$.
Undoubtedly these models are only several within an entire new terrain 
which deserves exploration.

Throughout this paper we shall work in the so-called free-fermionic formalism 
of Refs.~\cite{Kawai:1986ah, Antoniadis:1986rn, Kawai:1987ew}.  Within this formalism,
the cancellation of the conformal anomaly is achieved through the introduction of world-sheet fermions. 
The notation we use is that of 
Ref.~\cite{Kawai:1987ew}, which is summarized in Appendix~\ref{ffs}.
Moreover, the orbifold compactification from six to four dimensions
can be treated using the ``unified'' formalism of Ref.~\cite{Chamseddine:1989mz}, which is a straightforward 
generalization.  In particular, since only the untwisted sectors feel the CDC, the presence of the orbifold does not change the physics of supersymmetry breaking.

\subsection{Our starting point:~  An $\mathcal{N}=1$, 6D model}

We begin by presenting our six-dimensional theory. 
Following Appendix~\ref{ffs}, we assign a set of boundary conditions to the two-dimensional world-sheet
fermions so as to preserve modular invariance and space-time supersymmetry. 
The spin structure of the model is summarized by the set of basis vectors in Table~\ref{table:A}.
Along with these vectors is a matrix $k_{ij}$ which specifies the phases involved
in the corresponding GSO projections:
\beq
k_{ij}~=~
\begin{pmatrix}
 0 & 0 & 0 & 0 & \frac{1}{2} & \frac{1}{4}\\ \vspace{3pt}
 0 & 0 & \frac{1}{2} & 0 & 0 & 0 \\ \vspace{3pt}
 0 & 0 & 0 & \frac{1}{2} & 0 & 0 \\ \vspace{3pt}
 0 & 0 & 0 & 0 & 0 & \frac{1}{4} \\ \vspace{3pt}
 \frac{1}{2} & 0 & 0 & 0 & 0 & \frac{3}{4} \\ \vspace{3pt}
 0 & 0 & 0 & 0 & 0 & \frac{3}{4} \\
\end{pmatrix} 
\eeq
Note that different choices of $k_{ij}$ may break supersymmetry by discrete torsion. This will be discussed
further in Appendix~C. 

\begin{table}[t]
\centering
\begin{tabular}{|c|*{1}c|{l}|}
\hline
~Sector~& {\small $\xx \psi^{34}\yyr \psi^{56}\yyr \chi^{34}\:y^{34}\yyr\omega^{34}\yyr\chi^{56}y^{56}\yyr\omega^{56}$} &{\small $\overline{y}^{34}\yyr\overline{\omega}^{34}\yyr\overline{y}^{56}\yyr\overline{\omega}^{56}\yyr\overline{\psi}^{1}\yy\overline{\psi}^{2}\yy\overline{\psi}^{3}\yy\overline{\psi}^{4}\yy\overline{\psi}^{5}\yy\overline{\eta}^{1}\yy\overline{\eta}^{2}\yy\overline{\eta}^{3}\yy\overline{\phi}^{1} \yy\overline{\phi}^{2}\yy\overline{\phi}^{3}\yy\overline{\phi}^{4}\yy\overline{\phi}^{5}\yy\overline{\phi}^{6}\yy\overline{\phi}^{7}\yy\overline{\phi}^{8}$}
\\ \hline
$V_0$&1\xx 1\xx 1\xx 1\xx 1\xx 1\xx 1\xx 1\xx &1\xxl  1\xxl  1\xxl 1\xxl  1\xxl 1\xxl 1\xxl  1\xxl 1\xxl 1\xxl 1\xxl 1\xxl 1\xxl 1\xxl 1\xxl 1\xxl 1\xxl 1\xxl 1\xxl 1 \\
$V_1$&1\xx 1\xx 1\xx 0\xx 0\xx 1\xx 0\xx 0\xx&0\xxl  0\xxl 0\xxl 0\xxl 0\xxl  0\xxl 0\xxl  0\xxl 0\xxl 0\xxl  0\xxl 0\xxl 0\xxl 0\xxl 0\xxl 0\xxl 0\xxl 0\xxl 0\xxl 0 \\
$V_2$&1\xx 1\xx 0\xx 1\xx 0\xx 0\xx 1\xx 0\xx&1\xxl  0\xxl  1\xxl  0\xxl  1\xxl  1\xxl 1\xxl  1\xxl 1\xxl 1\xxl  0\xxl 0\xxl 0\xxl 0\xxl 0\xxl 0\xxl 0\xxl 0\xxl 0\xxl 0 \\
$V_5$&0\xx 0\xx 0\xx 0\xx 0\xx 0\xx 1\xx 1\xx&0\xxl  1\xxl  0\xxl  0\xxl  1\xxl  1\xxl 1\xxl  0\xxl 0\xxl 0\xxl  0\xxl 0\xxl 1\xxl 1\xxl 1\xxl 1\xxl 0\xxl 0\xxl 1\xxl 1 \\
$V_6$&0\xx 0\xx 0\xx 0\xx 0\xx 0\xx 0\xx 0\xx&1\xxl  1\xxl  0\xxl  1\xxl  1\xxl  1\xxl 1\xxl  0\xxl 0\xxl 0\xxl  0\xxl 0\xxl 0\xxl 1\xxl 1\xxl 1\xxl 1\xxl 1\xxl 1\xxl 0 \\
$V_7$&0\xx 0\xx 0\xx 1\xx 1\xx 0\xx 0\xx 0\xx&1\xxl  0\xxl  1\xxl  0\xxl  $\frac{1}{2}$\xxh  $\frac{1}{2}$\xxh $\frac{1}{2}$\xxh  $\frac{1}{2}$\xxh $\frac{1}{2}$\xxh $\frac{1}{2}$\xxh $\frac{1}{2}$\xxh $\frac{1}{2}$\xxh 0\xxl 1\xxl $\frac{1}{2}$\xxh $\frac{1}{2}$\xxh $\frac{1}{2}$\xxh 1\xxl 1\xxl $\frac{1}{2}$ \vspace{0.8pt} \\
\hline
\end{tabular}
\caption{~Spin structure of the $\mathcal{N}=1$ $6D$ model, where all entries are understood to be multiplied by a 
factor of $-1/2$.  Thus the `1' entries denote Ramond ground states, while '0' entries are Neveu-Schwarz
and `$1/2$' entries denote phases of $-1/4$ for the corresponding complexified fermions.
These conventions will apply to all subsequent tables in which explicit spin structures are listed.
As has become standard practice in string theory,
the space-time states listed on the left above are the right-moving 
fermions while (for consistent confusion) those on the right are left-moving.}
\label{table:A}
\end{table}

The vectors $\{V_0, V_1, V_2\}$ correspond to the so-called NAHE~\cite{Antoniadis:1989zy} 
vectors $\{\mathbf{1}, \mathbf{S}, \mathbf{b}_1\}$, lifted to six dimensions. 
By contrast, the additional vectors $V_5$, $V_6$, and $V_7$ are inspired by the 4D MSSM-like models 
listed in the Appendix of Ref.~\cite{Dienes:1995bx}, which in turn are based on the earlier models of
Refs.~\cite{Antoniadis:1990hb, Faraggi:1991be, Faraggi:1991jr, Faraggi:1992fa, Faraggi:1994eu}.
The vector ${V_1}$ is the supersymmetry generator of the model (with the superpartners of states in sector $\overline{\alpha {V}}$ residing in ${\overline{V_1+ \alpha V}}$). The internal right-moving $\chi^{34}, \chi^{56}$ fields
carry the supersymmetric charges. Of course $V_1$ also projects out the tachyonic states from the spectrum thanks to the generalized GSO projections in Appendix~\ref{ffs}. By themselves, the vectors
${V_0,V_1}$ generate an ${\cal N}=4$ theory which is broken to ${\cal N}=2$ by $V_2$. 
The Neveu-Schwarz sector of the theory ($\overline{\alpha V} = {0}$)
gives rise to
not only the gravity multiplet but 
also the massless scalar states required to build ${\cal N}=2$ gauge multiplets, as well as hypermultiplets, 
while the sector $V_2$ produces the sets of fermions in the spinorial of the parent $SO(16)\supset SO(10)$ `visible sector' gauge group defined by  the internal left-moving complex fermions $\overline{\psi}^1\dots\overline{\psi}^5$. (Superpartners arise in $V_1$ and $\overline{V_1+V_2}$ accordingly.) 
Introducing $V_{5, 6, 7}$ breaks the gauge group to  $[SU(3)]^2\otimes[SU(2)]^2\otimes[SO(4)]\otimes[U(1)]^8$. The additional vectors do not overlap with $V_1$, and therefore space-time supersymmetry is not broken further at this stage. However, these vectors are needed to break the horizontal symmetries embedded in the gauge group. 
As we shall see later, these horizontal symmetries arise from the fermions that are not complexified with a phase of $\frac{1}{2}$ in the $V_7$ sector. Since the horizontal symmetries are generation-dependent, their breaking reduces the number of generations of matter fields obtained from the $V_2$ sector.

\section{Two different ways of compactifying to 4D}
\setcounter{footnote}{0}

We now take the next step in our model-construction
procedure, namely the compactification of the model in Sect.~V.B to
four dimensions.
There are two different methods that we shall consider.
The first is to compactify on an ordinary $\IZ_2$ orbifold.
As we shall discuss in Sect.~VI.A, 
under certain choices for GSO projections
this can produce an MSSM-like model with $\calN=1$ SUSY.
The second method, by contrast, is to spontaneously break supersymmetry, by performing a
so-called ``coordinate-dependent compactification'' (CDC). This can 
be considered as a generalization of Scherk-Schwarz SUSY breaking.
We shall review this method in Sect.~VI.B;  
as we shall see, in practice this corresponds to 
compactifying on the same 
$\IZ_2$ orbifold as before,
but with an additional CDC-induced shift in the masses that breaks supersymmetry.
Indeed, we shall find that with certain choices it can lead 
to a non-supersymmetric
model whose particle content resembles that of the Standard-Model (SM).
In Sects.~VI.C and VI.D we will respectively examine
the partition functions and cosmological constants  
associated with these models.

\subsection{Compactifying on a $\mathbb{Z}_2$ orbifold:~ An $\mathcal{N}=1$, 4D MSSM-like model}  

First, we perform a normal compactification of the theory on a  $\mathbb{Z}_2$  orbifold (without CDC) to arrive at an unbroken ${\cal N}=1$ 4D model.  Note that in this process we must introduce at least a single twist in order to obtain $\calN=1$ SUSY (rather than $\calN=2$ SUSY) in four dimensions.  In the present case, we
we actually introduce two twists, corresponding to the vectors $b_3$ and $b_4$, describing the possible actions of the orbifold on the world-sheet degrees of freedom.  The resulting model then has  the spin structure\footnote{
             In writing our four-dimensional model in terms of a six-dimensional spin structure, we
             are implicitly recognizing that the remaining two dimensions are not fermionized;  
             they are instead retained as bosons compactified on a twisted
             two-torus with arbitrary radii, as discussed below.  Their treatment is thus 
             outside the free-fermionic formalism.
             However, as discussed in the Appendix of Ref.~\cite{julie2}, 
             an alternative way of specifying the resulting four-dimensional  
             string model 
             is to pass to the free-fermionic radii $R_i= \sqrt{2\alpha'}$.
             We can then describe the resulting model in terms of a {\it four}\/-dimensional 
             free-fermionic spin structure, and subsequently extend the resulting model back to arbitrary radii
             following the procedure outlined in Ref.~\cite{julie2}.}   
 as summarized in Table~\ref{table:D}. 

\begin{table}[H]
\centering
\begin{tabular}{|c|*{1}c|{l}|}
\hline
~Sector~& {\small\xx $\psi^{34}\yyr \psi^{56}\yyr \chi^{34}\:y^{34}\yyr\omega^{34}\yyr\chi^{56}y^{56}\yyr\omega^{56}$} &{\small $\overline{y}^{34}\yyr\overline{\omega}^{34}\yyr\overline{y}^{56}\yyr\overline{\omega}^{56}\yyr\overline{\psi}^{1}\yy\overline{\psi}^{2}\yy\overline{\psi}^{3}\yy\overline{\psi}^{4}\yy\overline{\psi}^{5}\yy\overline{\eta}^{1}\yy\overline{\eta}^{2}\yy\overline{\eta}^{3}\yy\overline{\phi}^{1} \yy\overline{\phi}^{2}\yy\overline{\phi}^{3}\yy\overline{\phi}^{4}\yy\overline{\phi}^{5}\yy\overline{\phi}^{6}\yy\overline{\phi}^{7}\yy\overline{\phi}^{8}$}
\\ \hline
$V_0$&1\xx 1\xx 1\xx 1\xx 1\xx 1\xx 1\xx 1\xx &1\xxl  1\xxl  1\xxl 1\xxl  1\xxl 1\xxl 1\xxl  1\xxl 1\xxl 1\xxl 1\xxl 1\xxl 1\xxl 1\xxl 1\xxl 1\xxl 1\xxl 1\xxl 1\xxl 1 \\
$V_1$&1\xx 1\xx 1\xx 0\xx 0\xx 1\xx 0\xx 0\xx&0\xxl  0\xxl 0\xxl 0\xxl 0\xxl  0\xxl 0\xxl  0\xxl 0\xxl 0\xxl  0\xxl 0\xxl 0\xxl 0\xxl 0\xxl 0\xxl 0\xxl 0\xxl 0\xxl 0 \\
$V_2$&1\xx 1\xx 0\xx 1\xx 0\xx 0\xx 1\xx 0\xx&1\xxl  0\xxl  1\xxl  0\xxl  1\xxl  1\xxl 1\xxl  1\xxl 1\xxl 1\xxl  0\xxl 0\xxl 0\xxl 0\xxl 0\xxl 0\xxl 0\xxl 0\xxl 0\xxl 0 \\
$b_3$&1\xx 0\xx 1\xx 0\xx 0\xx 0\xx 0\xx 1\xx&0\xxl  0\xxl  0\xxl  1\xxl  1\xxl  1\xxl 1\xxl  1\xxl 1\xxl 0\xxl  1\xxl 0\xxl 0\xxl 0\xxl 0\xxl 0\xxl 0\xxl 0\xxl 0\xxl 0 \\
$b_4$&1\xx 0\xx 0\xx 0\xx 1\xx 1\xx 0\xx 0\xx&0\xxl  1\xxl  0\xxl  0\xxl  1\xxl  1\xxl 1\xxl  1\xxl 1\xxl 0\xxl  0\xxl 1\xxl 0\xxl 0\xxl 0\xxl 0\xxl 0\xxl 0\xxl 0\xxl 0 \\
$V_5$&0\xx 0\xx 0\xx 0\xx 0\xx 0\xx 1\xx 1\xx&0\xxl  1\xxl  0\xxl  0\xxl  1\xxl  1\xxl 1\xxl  0\xxl 0\xxl 0\xxl  0\xxl 0\xxl 1\xxl 1\xxl 1\xxl 1\xxl 0\xxl 0\xxl 1\xxl 1 \\
$V_6$&0\xx 0\xx 0\xx 0\xx 0\xx 0\xx 0\xx 0\xx&1\xxl  1\xxl  0\xxl  1\xxl  1\xxl  1\xxl 1\xxl  0\xxl 0\xxl 0\xxl  0\xxl 0\xxl 0\xxl 1\xxl 1\xxl 1\xxl 1\xxl 1\xxl 1\xxl 0 \\
$V_7$&0\xx 0\xx 0\xx 1\xx 1\xx 0\xx 0\xx 0\xx&1\xxl  0\xxl  1\xxl  0\xxl  $\frac{1}{2}$\xxh  $\frac{1}{2}$\xxh $\frac{1}{2}$\xxh  $\frac{1}{2}$\xxh $\frac{1}{2}$\xxh $\frac{1}{2}$\xxh $\frac{1}{2}$\xxh $\frac{1}{2}$\xxh 0\xxl 1\xxl $\frac{1}{2}$\xxh $\frac{1}{2}$\xxh $\frac{1}{2}$\xxh 1\xxl 1\xxl $\frac{1}{2}$ \vspace{0.8pt}\\
\hline
\end{tabular}
\caption{~Spin structure of the world-sheet fermions of the $\mathcal{N}=1$, $D=4$ model before applying the CDC. This spin structure is accompanied by two bosonic degrees of freedom compactified on a $\IZ_2$ orbifold with twist action corresponding to the vectors $b_{3,4}$.}
\label{table:D}
\end{table}
In the remainder of our discussion of this model,
the terms ``twisted'' and ``untwisted'' will be used to refer to the two dimensions 
we compactify to get from six to four dimensions. 
The $\mathbb{Z}_2$ projection is 
\begin{equation}
\label{eqn: 15}
\mathtt{\widehat{g}}\phi~=~
\begin{cases}
      \phantom{-} \phi\widehat{{\mathtt g}} &  ~~{\rm for}\, \phi\notin b_3\, \mbox{ or } \, b_4 \nonumber\\
               -  \phi\widehat{{\mathtt g}} &  ~~{\rm for}\, \phi \in   b_3\, \mbox{ or } \, b_4 
\end{cases}
\end{equation}
where $\mathtt{\widehat{g}}$ is a generator of the $\mathbb{Z}_2$ orbifold. 
Furthermore, this particular choice of $b_3$ and $b_4$ is consistent with global invariance of the world-sheet supercurrent as defined in Eq.~\eqref{eqn: 4}:
\begin{equation}
\label{eqn: 10}
\mathtt{\widehat{g}}\, T_F(z)~=~-T_F(z)\,\mathtt{\widehat{g}}~.
\end{equation}
The $b_3$ and $b_4$ \textit{right-moving} entries are assigned so as to break the extended supersymmetries after compactification, leaving only ${\cal N}=1$. 
As the vector 
\begin{eqnarray}
V_4 &=& \overline{b_3+b_4}
\nonumber \\
&~=~& -{\scriptstyle{\frac{1}{2}}} [ ~00~101~101~|~ 0101~00000~011~00000000~]\, 
\end{eqnarray}
itself provides an additional untwisted sector, an entirely equivalent route is to start in 6D with an ${\cal N}=1$ fermionic theory that has $V_4$ as an additional vector, and from there to compactify on the $\mathbb{Z}_2$  orbifold with a single twist action  ($b_3$, for example). 
In the $\lbrace V_0,V_1,V_2,b_3,V_4,V_5,V_6,V_7\rbrace$ basis, the structure constants we choose to define our projection phases are as follows: 
\beq
k_{ij}~=~
\begin{pmatrix}
 0 & 0 & 0 & 0 & 0 &0 & \frac{1}{2} & \frac{1}{4}\\ \vspace{4pt}
 0 & 0 & 0 & 0 & 0 & 0 & 0 & 0 \\ \vspace{4pt}
 0 & \frac{1}{2} & 0 & \frac{1}{2} & 0 & \frac{1}{2} & 0 & 0 \\ \vspace{4pt}
 0 & \frac{1}{2} & \frac{1}{2} & 0 & 0 & 0 & 0 & \frac{3}{4} \\ \vspace{4pt}
 0 & \frac{1}{2} & 0 & 0 & 0 &0 & 0 & 0 \\ \vspace{4pt} 
 0 & 0 & 0 & \frac{1}{2} & 0 & 0 & 0 & \frac{1}{4}\\\vspace{4pt}
 \frac{1}{2} & 0 & 0 & 0 & 0 & 0 & 0 & \frac{3}{4}\\ \vspace{4pt}
 0 & 0 & 0 & 0 & 0 & 0 & 0 & \frac{3}{4} 
\end{pmatrix}~ .
\eeq
Considerable care has to be taken assigning these constants. As we shall see later and in Appendix~C, an incorrect  
choice can break supersymmetry explicitly or, conversely, disallow spontaneous supersymmetry 
breaking in our discussion of coordinate-dependent compactification.

The \textit{left-moving} entries are assigned so as to initially break the $SO(16)$ gauge group to $SO(10)\otimes SO(6)$, ensuring that the matter fields carry the correct Standard-Model charges once 
$V_5$,  $V_6$, and $V_7$ are added.  The final gauge group of the theory is then found to be 
\beq
 SU(3)\otimes SU(2)\otimes U(1)_Y \otimes G'_{\rm hidden}
\eeq
where
\beq
-\frac{1}{2}U(1)_Y ~\equiv ~ \frac{1}{3}\left[U(1)_{\overline{\psi}^1}+U(1)_{\overline{\psi}^2}+U(1)_{\overline{\psi}^3}\right]+\frac{1}{2}\left[U(1)_{\overline{\psi}^4}+U(1)_{\overline{\psi}^5}\right]~.
\eeq

Let us now consider the massless spectrum of 
this $\mathcal{N}=1$, $4D$ model.
To do this, 
we apply the rules of Appendix~\ref{ffs} but also impose the additional effects of the orbifolding. 
These can be found in Ref.~\cite{Chamseddine:1989mz}, but they can also be deduced from the form of the partition function presented in Sect~\ref{sec:part}.~  One first evaluates the would-be projections on the states (\ie, one evaluates the projection coefficients $C^\alpha_\beta$ 
with $b_{3}$ and $b_{4}$, following the same rules as for the other basis vectors).
However one must also take into account the additional phase shifts coming from the 
fact that oscillators in a given state may 
be odd under the orbifolding. Indeed a generic state with total winding number (in untwisted sectors) $n=n_1+n_2$ and 
$m=m_1+m_2$ transforms under the orbifold (with action ${\beta\cdot V} \equiv {b_3,b_4}$) as 
\begin{equation}
\prod_{i=1}
X^{(a_i)}_{-n_{i}}
\prod_{k=1}
\Psi^{(b_k)}_{-m_{k}}
\prod_{j=1}
\tilde{X}^{(a_j)}_{-n_{j}}
\prod_{\ell=1}
\tilde{\Psi}^{(b_\ell)}_{-m_{\ell}}
|m,n\rangle 
                ~\longrightarrow~
    ({-}1)^A
\prod_{i=1}
X^{(a_i)}_{-n_{i}}
\prod_{k=1}
\Psi^{(b_k)}_{-m_{k}}
\prod_{j=1}
\tilde{X}^{(a_j)}_{-n_{j}}
\prod_{\ell=1}
\tilde{\Psi}^{(b_\ell)}_{-m_{\ell}}
|{-}m,{-}n\rangle 
\label{eqn: 33}
\end{equation}
where the overall phase $A$ is calculated by 
introducing an overall minus sign
for each excitation in the $X_5$ or $X_6$
direction and then 
maintaining the other GSO phases
exactly as they were. 
For the untwisted states, invariance under the orbifold action is then simply equivalent to shifting the GSO projections of $b_3,b_4$ in Appendix~\ref{ffs} by the {\it additional}\/ phase coming from the compact bosons, $\frac{1}{2}\sum_{a_i,a_j \in 5,6}$ (since $\frac{1}{2}\sum_{b_k,b_\ell \in b_{3,4}}$ is already included). For states with non-zero $n$ or $m$ one then has opposite projections for the 
even/odd wave-functions, so the remaining invariant combination is $\frac{1}{\sqrt{2}}\left(|n,m\rangle + ({-}1)^A|{-}n,{-}m\rangle\right)$, while any {\it zero-modes} that are odd under the orbifolding are projected out in the usual way.  

The resulting observable sector content is summarized in Tables~\ref{table:E1}-\ref{table:F}. We shall now discuss the contributions from the untwisted and the ${b_3}$ and ${b_4}$ twisted sectors in turn. \\

\noindent \textbf{$\bullet$ Untwisted sectors not involving $V_4=\overline{b_{3}+b_4}$}: 
Here none of the projections are altered by the orbifold action on the bosonic oscillators except for that of the radion, and the vacuum energies obey 
\begin{equation}
\label{eqn: 11}
E_{L,R}~=~\frac{1}{2}\sum_l \left[(\overline{\alpha V^l})^2-\frac{1}{12}\right]-\frac{(D-2)}{24}-\frac{1}{12}~. 
\end{equation}
Here $D=4$, the sum is over complex fermions,  and the factor of $-1/12$ 
accounts for the two  {real compactified}\/ bosons. 
The Neveu-Schwarz sector, ${\overline{\alpha V}}=\mathbf{0}$, yields the massless bosons for the gauge and gravity sector [including the complex radion for dimensions (5,6)], as well as 
three pairs of complex Higgs scalars and three pairs of singlet scalar states. 
The ${V_1}$ sector generates their superpartners. 

\begin{table}[H]
\centering
\begin{tabular}{||c|l|c|l||}
\hline
~Sector~ & ~~States remaining after CDC~~ & ~Spin~ & ~Particles \\
\hline &&& \\
\multirow{8}{*}{${\bf{0}} $}  & \multirow{3}{*}{~$\psi^{34}_{-\frac{1}{2}}|0\rangle_R \otimes X^{34}_{-1}|0\rangle_L$} & \multirow{2}{*}{2} & ~Graviton $g_{\mu\nu}$, \\
& & & ~Antisymmetric tensor $B_{[\mu\nu]}$, \\
& & 0 & ~Dilaton $\phi$  \\ [0.5em]
 & ~$\psi^{56}_{-\frac{1}{2}}|0\rangle_R \otimes X^{56}_{-1}|0\rangle_L$ & 0 & ~Complex radion $\Phi$   \\[0.5em]
 & ~$\psi^{34}_{-\frac{1}{2}}|0\rangle_R \otimes \Psi^i_{-\frac{1}{2}}\Psi^j_{-\frac{1}{2}}|0\rangle_L$ & 1 & ~Gauge bosons $V_{\mu}$   \\[0.5em]
& $~\psi^{56}_{-\frac{1}{2}}|0\rangle_R \otimes \Psi^i_{-\frac{1}{2}}\Psi^j_{-\frac{1}{2}}|0\rangle_L$ & 0 & ~Complex scalars $H_{U_1}$, ${H}_{D_1}$, $\Xi_1$, $\Xi'_1$  ~\\ [0.5em]
\hline &&& \\
\multirow{3}{*}{${{V}_1} $} & ~$|\alpha\rangle_R \otimes \Psi^i_{-\frac{1}{2}}\Psi^j_{-\frac{1}{2}}|0\rangle_L$ & $\frac{1}{2}$ & ~Weyl spinors $\tilde{H}_{U_2}$, $\tilde{H}_{D_2}$, $\tilde{\Xi}_2$, $\tilde{\Xi}'_2$  \\ [0.5em]
& ~$|\alpha\rangle_R \otimes \Psi^i_{-\frac{1}{2}}\Psi^j_{-\frac{1}{2}}|0\rangle_L$ & $\frac{1}{2}$ & ~Weyl spinors $\tilde{H}_{U_3}$, $\tilde{H}_{D_3}$, $\tilde{\Xi}_3$, $\tilde{\Xi}'_3$ \\  [0.5em]
\hline
\end{tabular}
\caption{The  ($\mathbb{Z}_2$-untwisted) visible-sector states of the $\mathcal{N}=1$,  $D=4$ model.
As we shall see in Sect.~VI.B, all of these states will remain massless after the CDC is imposed.
The $\Psi_i$ refer to generic left-moving degrees of freedom, with indices $i, j =1 \ldots  20$. 
 \label{table:E1}}
\end{table}

\begin{table}[H]
\centering
\setlength{\fboxsep}{0pt}
\begin{tabular}{||c|l|c|l||}
\hline
~Sector~ & ~~States removed by CDC~~ & ~Spin~ & ~Particles \\
\hline &&&\\
\multirow{7}{*}{${V_1}$} &  \multirow{2}{*}{~$|\alpha\rangle_R \otimes X^{34}_{-1}|0\rangle_L$}  & \multirow{1}{*}{$\frac{3}{2}$} & ~Gravitino $\psi_{\mu}$, \\
& &  $\frac{1}{2}$ & ~Dilatino $\widetilde{\phi}$  \\ 
& ~$|\alpha\rangle_R \otimes X^{56}_{-1}|0\rangle_L$ &  $\frac{1}{2}$ & ~Radino $\widetilde{\Phi}$  \\[0.5em]
& ~ $|\alpha\rangle_R \otimes \Psi^i_{-\frac{1}{2}}\Psi^j_{-\frac{1}{2}}|0\rangle_L$  & $\frac{1}{2}$ & ~Gauginos $\lambda_{\mu}$   \\[0.5em]
&  ~$|\alpha\rangle_R \otimes \Psi^i_{-\frac{1}{2}}\Psi^j_{-\frac{1}{2}}|0\rangle_L$ & $\frac{1}{2}$ &  ~Weyl spinors $\tilde{H}_{U_1}$, $\tilde{H}_{D_1}$, $\tilde{\Xi}_1$, $\tilde{\Xi}'_1$  \\ [0.5em]
\hline &&& \\
\multirow{3}{*}{${\bf {0}} $} 
& ~$\chi^{34}_{-\frac{1}{2}}|0\rangle_R \otimes \Psi^i_{-\frac{1}{2}}\Psi^j_{-\frac{1}{2}}|0\rangle_L$ & 0 & ~Complex scalars ${H}_{U_2}$, ${H}_{D_2}$, $\Xi_2$, $\Xi'_2$  \\ [0.5em]
& ~$\chi^{56}_{-\frac{1}{2}}|0\rangle_R \otimes \Psi^i_{-\frac{1}{2}}\Psi^j_{-\frac{1}{2}}|0\rangle_L$ & 0 & ~Complex scalars ${H}_{U_3}$, ${H}_{D_3}$, $\Xi_3$, $\Xi'_3$ \\ [0.5em]
\hline
\end{tabular}
\caption{
Additional ($\mathbb{Z}_2$-untwisted) visible-sector states of the $\mathcal{N}=1$,  $D=4$ model.
As we shall see in Sect.~VI.B, these states 
will no longer remain in the massless spectrum after the CDC is imposed,
and will instead accrue masses $\frac{1}{2}\sqrt{ R_1^{-2}+R_2^{-2} }$.
\label{table:E}}
\end{table}

An attractive feature of this model is that there are no massless Higgs triplets, and the only massless visible sector scalars transform in the \textbf{$\left({\bf 2}, \pm \frac{1}{2}\right)$} representation of $SU(2)\otimes U(1)_Y$. Specifically the three pairs of Higgs doublets $H_{U_i}$ and $H_{D_i}$ that survive the GSO and orbifold projection are 
\begin{align}
\label{eqn: 17.1}
\left[H_{U_1}\right]_{1,0,0,0,0,0,0}, ~\left[H_{D_1}\right]_{-1,0,0,0,0,0,0}~=~ \psi^{56}_{-\frac{1}{2}}|0\rangle_R&\otimes \overline{\psi}^{4,5}_{-\frac{1}{2}} \:\overline{\eta}^{1}_{-\frac{1}{2}}|0\rangle_L  \nonumber\\
\left[H_{U_2}\right]_{0,1,0,0,0,0,0}, ~\left[H_{D_2}\right]_{0,-1,0,0,0,0,0}~=~\chi^{34}_{-\frac{1}{2}}|0\rangle_R&\otimes \overline{\psi}^{4,5}_{-\frac{1}{2}}\:\overline{\eta}^{2}_{-\frac{1}{2}}|0\rangle_L  \nonumber\\
\left[H_{U_3}\right]_{0,0,1,0,0,0,0}, ~\left[H_{D_3}\right]_{0,0,-1,0,0,0,0}~=~\chi^{56}_{-\frac{1}{2}}|0\rangle_R&\otimes \overline{\psi}^{4,5}_{-\frac{1}{2}}\:\overline{\eta}^{3}_{-\frac{1}{2}}|0\rangle_L  \, .
\end{align}

In addition to these, the three pairs of singlet scalar states $\Xi_i$ and $\Xi'_i$ that survive the GSO and orbifold projection are given by
\begin{align}
\label{eqn: 17.4}
\left[\Xi_{1}\right]_{0,1,-1,0,0,0,0}, ~\left[\Xi'_{1}\right]_{0,-1,1,0,0,0,0}~=~\psi^{56}_{-\frac{1}{2}}|0\rangle_R&\otimes \overline{\eta}^{2}_{-\frac{1}{2}} \: \overline{\eta}^{3}_{-\frac{1}{2}}|0\rangle_L  \nonumber \\
\left[\Xi_{2}\right]_{1,0,-1,0,0,0,0}, ~\left[\Xi'_{2}\right]_{-1,0,1,0,0,0,0}~=~\chi^{34}_{-\frac{1}{2}}|0\rangle_R&\otimes \overline{\eta}^{1}_{-\frac{1}{2}} \: \overline{\eta}^{3}_{-\frac{1}{2}}|0\rangle_L  \nonumber\\
\left[\Xi_{3}\right]_{1,-1,0,0,0,0,0}, ~\left[\Xi'_{3}\right]_{-1,1,0,0,0,0,0}~=~ \chi^{56}_{-\frac{1}{2}}|0\rangle_R&\otimes \overline{\eta}^{1}_{-\frac{1}{2}} \: \overline{\eta}^{2}_{-\frac{1}{2}}|0\rangle_L   \, .
\end{align}

The generalized GSO projections pick out the relevant components (\ie, ${b}^{4,5}_{-{1}/{2}}$ or 
${d}^{4,5}_{-{1}/{2}}$ in the notation of Ref.~\cite{Kawai:1987ew}) of the $\overline{\psi}^{4,5}_{-{1}/{2}}$ operators, for the electroweak doublets, and of the $\overline{\eta}^{1,2,3}$ for the singlet states. The presence of these scalar doublets (and their superpartners) in the massless spectrum is correlated with the existence of $U(1)$ horizontal symmetries embedded in the larger broken gauge group.  The subscripts on the states above are the charges under the horizontal $U(1)$ symmetries embedded in the larger gauge group; more details on these will be given in the upcoming sections.

At the level of four-dimensional theories, the ${V_2}$ sector gives rise to 16 generations of massless fermionic states which transform in the $\mathbf{16}$ of the $SO(10)$ with scalar superpartners in the $\overline{{V_2+V_1}}$ sector. However, half of the generations are projected out by the $\mathbb{Z}_2$ twists, while $V_5$, $V_6$, and $V_7$ overlap non-trivially with $V_2$, $b_3$, and $b_4$, constraining the total number of generations to two from the $V_2$ sector and just one from each of the $b_3$ and $b_4$ sectors. The states are listed in the Table~\ref{table:F1} and their superpartners in Table~\ref{table:F} --- they are the usual decomposition of a $\bf 16$ representation of $SO(10)$ under $SU(3)\times SU(2)\times U(1)_Y$. At this point it should be noted that it is possible to use another choice of $V_5$, $V_6$, and $V_7$ vectors, along with real fermions instead of complexified ones, in order to break further the horizontal symmetries and get one generation of matter fields from each of the $V_2$, $b_3$, and $b_4$ sectors. However the number of generations within this model is not critical for our immediate purposes since our primary goal is to construct an SM-like model with vanishing cosmological constant.

\begin{table}[H]
\centering
\begin{tabular}{|c|l|c|l|}
\hline
~Sector~ & ~~States remaining after CDC~~ & ~Spin~ & ~Particles~ \\
\hline &&&\\
\multirow{9}{*}{$V_2$} & ~$|\alpha\rangle_R \otimes |\hat{\alpha}\rangle_L$  & $\frac{1}{2}$ & ~$e_R$ \\ [0.5em]
& ~$|\alpha\rangle_R \otimes {\overline{\psi}}^4_0{\overline{\psi}}^5_0|\hat{\alpha}\rangle_L$  & $\frac{1}{2}$ &  ~$\nu_R$ \\ [0.5em]
& ~$|\alpha\rangle_R \otimes {\overline{\psi}}^i_0{\overline{\psi}}^a_0|\hat{\alpha}\rangle_L$  & $\frac{1}{2}$ &  ~$Q_L$ \\ [0.5em]
& ~$|\alpha\rangle_R \otimes {\overline{\psi}}^i_0{\overline{\psi}}^j_0|\hat{\alpha}\rangle_L$ & $\frac{1}{2}$ &  ~$d_R$ \\  [0.5em]
& ~$|\alpha\rangle_R \otimes {\overline{\psi}}^i_0{\overline{\psi}}^j_0{\overline{\psi}}^4_0{\overline{\psi}}^5_0|\hat{\alpha}\rangle_L$ & $\frac{1}{2}$ & ~$u_R$ \\  [0.5em]
& ~$|\alpha\rangle_R \otimes {\overline{\psi}}^1_0{\overline{\psi}}^2_0{\overline{\psi}}^3_0{\overline{\psi}}^a_0|\hat{\alpha}\rangle_L$ & $\frac{1}{2}$ & ~$L_L$ \\ [0.5em]
\hline
\end{tabular}
\caption{
($\mathbb{Z}_2$-untwisted) chiral multiplets of the $\mathcal{N}=1$, $D=4$ model, where $i,j\in SU(3)$ 
   and $a\in SU(2)$.  Just as with the states in Table~\ref{table:E1},
we shall see in Sect.~VI.B that all of these states will remain massless after the CDC is imposed.
 The $|\alpha\rangle_R$ represent right-moving Ramond ground states (\ie, space-time spinors), while $|\hat{\alpha}\rangle_L$ represents the left-moving Ramond excitations that do not overlap with the SM gauge group. The multiplets are essentially the decomposition of the $\bf 16$ of $SO(10)$. 
The same decomposition applies for the two generations of  $b_3$ and $b_4$ twisted-sector matter fields, but we shall find that those sectors are unaffected by the CDC and therefore will remain (globally) supersymmetric.}
\label{table:F1}
\end{table}

\begin{table}[H]
\centering
\setlength{\fboxsep}{0pt}
\begin{tabular}{|c|l|c|l|}
\hline
~Sector~ & ~~States removed by CDC~~ & ~Spin~ & ~Particles~ \\
\hline& & & \\
\multirow{9}{*}{$\overline{V_1\hspace{-0.1cm}+\hspace{-0.1cm}V_2}$} & ~$|\alpha\rangle'_R \otimes |\hat{\alpha}\rangle_L$  & 0 & ~$\widetilde{e}_R$ \\ [0.5em]
& ~$|\alpha\rangle'_R \otimes {\overline{\psi}}^4_0{\overline{\psi}}^5_0|\hat{\alpha}\rangle_L$  & 0 & $~\widetilde{\nu}_R$ \\ [0.5em]
& ~$|\alpha\rangle'_R \otimes {\overline{\psi}}^i_0{\overline{\psi}}^a_0|\hat{\alpha}\rangle_L$ & 0 & $~\widetilde{Q}_L$ \\ [0.5em]
& ~$|\alpha\rangle'_R \otimes {\overline{\psi}}^i_0{\overline{\psi}}^j_0|\hat{\alpha}\rangle_L$  & 0 & ~$\widetilde{d}_R$ \\ [0.5em]
& ~$|\alpha\rangle'_R \otimes {\overline{\psi}}^i_0{\overline{\psi}}^j_0{\overline{\psi}}^4_0{\overline{\psi}}^5_0|\hat{\alpha}\rangle_L$ & 0 & ~$\widetilde{u}_R$ \\ [0.5em]
& ~$|\alpha\rangle'_R \otimes {\overline{\psi}}^1_0{\overline{\psi}}^2_0{\overline{\psi}}^3_0{\overline{\psi}}^a_0|\hat{\alpha}\rangle_L$  & 0 &  ~$\widetilde{L}_L$ \\ [0.5em]
\hline
\end{tabular}
\caption{
($\mathbb{Z}_2$-untwisted) chiral multiplets of the $\mathcal{N}=1$, $D=4$ model 
where $i,j\in SU(3)$ and $a\in SU(2)$. 
Just as with the states in Table~\ref{table:E},
we shall see in Sect.~VI.B that these states will 
accrue masses $\frac{1}{2}\sqrt{ R_1^{-2}+R_2^{-2} }$ by the CDC. The $|\alpha\rangle'_R$ represent right-moving Ramond ground states that are not space-time spinors.}
\label{table:F}
\end{table}

\noindent \textbf{$\bullet$ Untwisted sectors involving $V_4=\overline{b_3+ b_4}$}: As mentioned above, this combination of the two orbifold twist actions is effectively just another untwisted sector, and we give it its own name, $V_4$, for later convenience.  Checking the entirety of sectors containing the combination $b_3+ b_4$, we find that the only massless states are either singlets or additional Higgs-like doublets --- albeit with charges that, as we shall see, prohibit their direct coupling to the matter fields in Yukawa couplings. \\
 
\noindent \textbf{$\bullet$ Twisted sectors}: In the twisted sectors, in addition to the orbifold itself acting on the compact bosonic oscillators, the vacuum energies  are given by
\begin{equation}
\label{eqn: 13}
E_{L,R}~=~\frac{1}{2}\sum_l\left[(\overline{\alpha V^l})^2-\frac{1}{12}\right]-\frac{(D-2)}{24}+\frac{1}{24}
\end{equation}
where now $1/24$ accounts for the twisted complex boson.
Similar to the untwisted $V_2$ sector, the twisted ${b_3}$ and ${b_4}$ sectors each give rise to another set of 16 generations of massless chiral massless fields, each of which are projected down to a single generation. 
In total, then, 48 generations (including those from $V_2$) are projected down to four.  Of course, there are many other linear combinations of twisted sectors ${\overline{b_3+\alpha V}}$ and ${\overline{b_4+\alpha V}}$ that contribute extra hidden states and singlets to the massless spectrum of the theory. However,  these twisted sectors states are ultimately of minor phenomenological importance. The other phenomenological properties such as Yukawa couplings will be discussed later. \\


\subsection{Coordinate-dependent compactification (CDC):~ An $\mathcal{N}=0$, 4D SM-like model}  
\label{sec:cdc}

We shall now consider the effects that arise when this model is compactified with a CDC.~
It turns out that the CDC does not change the spectrum of the twisted sectors, which remain (globally) supersymmetric.
Therefore our focus will be on what happens to the spectrum of the untwisted sectors. 
Relative to the regular compactification of Sect.~VI.A, compactification with a CDC can in some sense be viewed as following a similar path
but with an alternative final step.
Indeed, the CDC model we shall consider will be based on the 
${\cal N}=1$, 4D model we have just constructed, supplemented with a final CDC-induced set of projections.  

In order to develop a general approach,  
we shall find it convenient to begin with a
toroidally compactified $\calN=2$ model consisting of only the untwisted vectors
$V_{0,1,2,4,5,6,7}$. 
This model, which serves only as a stepping stone towards our final result,
will be discussed in Sect.~VI.B.1.
The final stage of achieving chirality, \ie,  the orbifolding 
with action $b_3$ and $b_4$, will then require a slight adjustment to $V_{5,7}$. This
will be treated in Sect.~VI.B.2.
The end result will then be a chiral, non-supersymmetric, SM-like model in four dimensions. 

\subsubsection{Non-chiral $\calN=2\rightarrow \calN=0$ model}

As already mentioned, we may view the CDC as a generalized Scherk-Schwarz SUSY breaking mechanism.
In particular, the CDC will lift the mass of some of the states (including the gravitino) 
and split the spectrum at a scale $\sim 1/R$ where $R$ is a generic radius of the compact (5,6) dimensions. 

The resulting SUSY breaking occurs in an interesting way. Without the CDC the theory is of course ${\cal N}=2$ supersymmetric.  With the CDC, by contrast, the theory is non-supersymmetric.  However we will see that  the theory {\it with}\/ the CDC but {\it without}\/ $V_4$ is {also}\/ ${\cal N}=2$ supersymmetric --- but in this case it is a  {\it different}\/ supersymmetry which is preserved. The conflict between these two supersymmetries is ultimately a matter of the choice of structure constant $k_{ij}$, and thus an alternative choice would 
also leave the theory supersymmetric. 

This breaking of supersymmetry by choice of structure constant is 
equivalent to ``discrete torsion'', and it is interesting to conjecture that the interpolation to $R\sim 1$ of such CDC
theories, like the one we are about to present, yields a 4D model with supersymmetry broken by discrete torsion.
(As indicated in the footnote in the previous subsection, this is certainly true for the theory {\it prior}\/ 
to the CDC.)
In Appendix~C we briefly demonstrate one aspect of SUSY breaking by discrete torsion that  is particularly 
suggestive, namely that it admits a large class of non-supersymmetric string theories that are
tachyon-free.

In general, the CDC is a deformation that is able to incorporate the super-Higgs phenomena of the
Scherk-Schwarz procedure but in more general string configurations~\cite{Kounnas:1989dk}.
From the space-time perspective, the super-Higgs mechanism involves the auxiliary field of some supermultiplet acquiring a non-vanishing VEV and the gravitino becoming massive by absorbing a goldstino. From the world-sheet perspective, however, the occurrence of a super-Higgs phenomenon implies that the world-sheet Lagrangian $\mathfrak{L}_w$ is deformed by a non-BRST invariant operator that must preserve only a \textit{discrete} subgroup of a $U(1)$ symmetry~\cite{Kounnas:1989dk}. The super-Higgs effect manifests itself when this discrete symmetry is spontaneously broken and auxiliary fields on the world-sheet develop a VEV. 

In practice this means deforming an already existing model through the addition of a local generator $\mathbf{Q}$ of the  parent $U(1)$ world-sheet symmetry which at least partly involves the $R$-symmetry (in order that gravitini and graviton have different charges). For the breaking to be spontaneous, the world-sheet supercurrent defined in Eq.~(\ref{eqn: 4}) has to be invariant under the discrete symmetry, but it does not commute with the local generator $\mathbf{Q}$:
\begin{equation}
\label{eqn: 20}
[T_F(z), \mathbf{Q}(z)]~\neq~ 0\, .
\end{equation}

In order to apply a CDC to our case, 
we begin with the $\calN=2$ theory with the
complexification of the internal right-moving fermions given by
\begin{align}
\label{eqn: 21}
\chi_c^{(1)}~\equiv~ \chi_{34}&~=~\frac{1}{\sqrt{2}}(\chi^3+i\chi^4) && \chi_c^{(2)}~\equiv~ \chi_{56} ~=~\frac{1}{\sqrt{2}}(\chi^5+i\chi^6)
\nonumber \\
\omega_c^{(1)}~\equiv~ \omega_{34}&~=~\frac{1}{\sqrt{2}}(\omega^3+i\omega^4) &&\omega_c^{(2)}~\equiv~ \omega_{56}~=~\frac{1}{\sqrt{2}}(\omega^5+i
\omega^6) ~.
\end{align} 
(The $y$ fields will be largely irrelevant for this discussion.)  This complexification admits two \textit{discrete} world-sheet symmetries,
$\mathfrak{J}_1$ and $\mathfrak{J}_2$, 
which are subgroups of the internal $SO(4)$ group of the compactification from ten to six dimensions.  Each of these
symmetries is defined by the eight transformations
\begin{align}
\label{eqn: 22}
\chi^3&\rightarrow~ - \chi^3 && \omega^3~\rightarrow~ -\omega^3 
\nonumber \\
\chi^4&\rightarrow~ -\chi^4 && \omega^4~\rightarrow~ -\omega^4 
\nonumber \\
\chi^5&\rightarrow~ -\epsilon\chi^5 && \omega^5~\rightarrow~ -\epsilon\omega^5 
\nonumber \\
\chi^6&\rightarrow~ -\epsilon\chi^6 && \omega^6~\rightarrow~ -\epsilon\omega^6 \, ,
\end{align}
where the $y$ fields remain invariant under each symmetry and 
where $\epsilon=+1$ for  $\mathfrak{J}_1$  and $\epsilon=-1$ for  $\mathfrak{J}_2$. 
Since the $y$ fields do not acquire a phase, 
the CDC can be expressed in terms of the $U(1)$ charges of the complex 
states $f_c\equiv \lbrace \chi_c^{(i)},\omega_c^{(i)}\rbrace$:
\begin{equation}
\label{eqn: 23a}
 \mathfrak{J}_{1,2}: ~~~~~f_{c}~\rightarrow~ e^{2\pi i \mathbf{e}^{(1,2)}}  {f}_{c}
\end{equation}
where $\mathbf{e}_i^{(1)}$ and $\mathbf{e}_i^{(2)}$ take the values
\begin{equation}
\label{eqn: 23b}
\mathbf{e}^{(1)}_i, ~\mathbf{e}^{(2)}_i~=~\left\{\begin{array}{ll}
                                          \frac{1}{2} & \mbox{ for } \:\chi_c^{(1)}, \omega_c^{(1)} \\
                                          \frac{1}{2}\epsilon & \mbox{ for }\: \chi^{(2)}_c, \omega^{(2)}_c \\
                                           0 & \mbox{ otherwise}\, .
                                          \end{array}\right.
\end{equation}
We shall henceforth make the choice $\epsilon=+1$. The advantage of this complexification is of course that the operator associated with the $\mathfrak{J}_{1}$ symmetry can easily be written in the basis of the original ${\cal N}=1$ model of 
Table~\ref{table:D}: 
\begin{equation}
\label{eqn: 25}
       \widehat{J}_1~=~e^{2\pi i \mathbf{e\cdot Q}}\, ,
\end{equation}
where 
\begin{eqnarray}
\mathbf{e\cdot Q} &=& \frac{1}{2} \frac{1}{2\pi i} \int dz \left( \overline{\chi}_c^{(i)}(z)\chi_c^{(i)}(z) + \overline{\omega}_c^{(i)}(z)\omega_c^{(i)}(z) \right) \nonumber \\
&=& \frac{1}{2} \left( Q_{\chi^{34}}+Q_{\chi^{56}} + Q_{\omega^{34}}+Q_{\omega^{56}}\right) \, .
\end{eqnarray}
It is then convenient to work with the $\mathbf{e}$ charges defined in this basis:
\begin{equation}
\label{eqn: 71}
{\bf e} ~=~ {\scriptstyle\frac{1}{2}}[~00~101~101~ | ~0000~00000~000~00000000~]\, .
\end{equation}
For later reference we note that  $|\mathbf{e}^2|=1$. 

In order to understand the effects of the CDC on the resulting spectrum, it is convenient to study 
the CDC-deformed one-loop partition function in the ``charge-lattice'' formalism. This is given by 
\begin{equation}
\label{eqn: 53}
    Z(\tau)~=~
   {\rm Tr}\, \sum_{m_{1,2}, n_{1,2}}   
   \mathtt{g}  \, q^{[L'_0]}\overline{q}^{[\overline{L'}_0]}
\end{equation}
where the primes indicate that these expressions are CDC deformations of the traditional
supersymmetric expressions and where $\mathtt{g}$ is the generalized GSO fermion-number projection operator.
The latter are independent of the values of the charges $\mathbf{e}$. 
Following the conventions of Eq.~(\ref{Zcircdef}), we can write
the Virasoro operators for the left- and right-moving sectors of the 
tachyon-free, non-supersymmetric model as
\beq
    {[L'_0]}=\alpha' p_L^2/2 + {\rm osc.}~, ~~~~~
    {[\overline{L'}_0]}=\alpha' p_R^2/2 + {\rm osc.}
\eeq
We can then follow the procedure in Refs.~\cite{Ferrara:1987es, Ferrara:1987qp,Ferrara:1988jx}, but with two additional bosonic coordinates $(X^5, X^6)$ 
compactified with radii $R_1=r_1/\sqrt{\alpha'}$ and $R_2=r_2 /\sqrt{\alpha'}$. 
(Thus the dimensionless inverse radii $a_i$ of Sects.~III and IV are given by $a_i=r_i^{-1}$.)
Defining the respective winding and Kaluza-Klein numbers to be $n_{1,2}$ and $m_{1,2}$, 
we then find that the {general} forms of the Virasoro operators are given by
\begin{align}
\label{eqn: 55b}
L'_0&=~\frac{1}{2}\left[\mathbf{Q}_L-\mathbf{e}_L(n_1+n_2)\right]^2+\frac{1}{4}\left[\frac{m_1+\mathbf{e\cdot Q}-\frac{1}{2}(n_1+n_2)\mathbf{e}^2}{r_1}+n_1r_1\right]^2 \nonumber \\
&~~~~~~~~+\frac{1}{4}\left[\frac{m_2+\mathbf{e\cdot Q}-\frac{1}{2}(n_1+n_2)\mathbf{e}^2}{r_2}+n_2r_2\right]^2-1+ \mbox{ other oscillator contributions}~, \nonumber \\
\nonumber\\
\overline{L'}_0&=~\frac{1}{2}\left[\mathbf{Q}_R-\mathbf{e}_R(n_1+n_2)\right]^2+\frac{1}{4}\left[\frac{m_1+\mathbf{e\cdot Q}-\frac{1}{2}(n_1+n_2)\mathbf{e}^2}{r_1}-n_1r_1\right]^2 \nonumber \\
&~~~~~~~~+\frac{1}{4}\left[\frac{m_2+\mathbf{e\cdot Q}-\frac{1}{2}(n_1+n_2)\mathbf{e}^2}{r_2}-n_2r_2\right]^2 -\frac{1}{2}+ \mbox{ other oscillator contributions}\, ,~
\end{align}
where $L_0$ and $\overline{L}_0$ are the Virasoro operators of the original supersymmetric model in four dimensions (\ie, the Virasoro operators with ${\bf e}={\bf 0}$). It follows that
\begin{align}
\label{eqn: 56}
L'_0+\overline{L'}_0&=~L_0+\overline{L}_0+\frac{1}{2}\left[\mathbf{e\cdot Q}-\frac{(n_1+n_2)}{2}\mathbf{e}^2\right]^2\left(\frac{1}{{r_1}^2}+\frac{1}{{r_2}^2}\right) -{(n_1+n_2)}\left(\mathbf{e}_L\cdot \mathbf{Q}_L+\mathbf{e}_R\cdot \mathbf{Q}_R\right)\nonumber \\
&~~~~~~~~+\frac{1}{2}(n_1+n_2)^2\left(\mathbf{e}_L^2+\mathbf{e}_R^2\right)+\left(\frac{m_1}{r^2_1}+\frac{m_2}{r^2_2}\right)\left[\mathbf{e\cdot Q}-\frac{(n_1+n_2)}{2}\mathbf{e}^2\right] ~,\nonumber \\
L'_0-\overline{L'}_0&=~L_0-\overline{L}_0\, ,
\end{align}
where ${\bf e}_{R,L}$ and ${\bf Q}_{R,L}$ refer to just the right- or left-moving elements of these vectors
and where ${\bf e}\cdot {\bf Q}$ denotes a Lorentzian dot product. 

From these expressions we may easily read off the effect of the CDC 
on the particle spectrum shown in Tables~\ref{table:E1} through \ref{table:F}.
The end result is that the states in Tables~\ref{table:E1} and \ref{table:F1} remain in the massless spectrum,
while the states in Tables~\ref{table:E} and \ref{table:F} gain masses and are eliminated.
(Of course we emphasize that this is merely a subset of the spectrum 
at this point, since the toroidally compactified theory before CDC has $\calN=2$ supersymmetry.)  

It is straightforward to understand this result.
First, in the NS-NS sector, it is clear that no massless states receive masses since all the charges overlapping $\mathbf{e}$ are zero. Likewise, all winding and KK masses are also unshifted.  However, in the ${V_1}$ sector, there are charges overlapping $\mathbf{e}$, and these can be $\pm \frac{1}{2}$, depending on the  chirality. In order to see which states remain massless, let us consider the generalized GSO projections on the gravitinos.  These projections
include 
\begin{eqnarray}
V_{0}\cdot N+\frac{1}{4}\left(1-\gamma_{\psi_{34}}\gamma_{\psi_{56}}\gamma_{\chi_{34}}\gamma_{\chi_{56}}\right) & = & k_{01}+\frac{1}{2}-V_{0}\cdot V_{1}\nonumber \\
\frac{1}{4}\left(1-\gamma_{\psi_{34}}\gamma_{\psi_{56}}\gamma_{\chi_{34}}\gamma_{\chi_{56}}\right) & = & k_{11}+\frac{1}{2}-V_{1}\cdot V_{1}\nonumber \\
V_{4}\cdot N+\frac{1}{4}\left(1-\gamma_{\chi_{34}}\gamma_{\chi_{56}}\right) & = & k_{41}-V_{4}\cdot V_{1} \,\,\,\,\,\,\: ~~\mbox{mod} \,(1)~,
\end{eqnarray}
where $N$ corresponds to the number operator associated with the non-Ramond degrees of freedom.  We also have 
the general constraints
\begin{eqnarray}
\label{eqn: 35}
&& V_{4}\cdot V_{1}~ =~ k_{14}+k_{41}~=~\frac{1}{2}\,\,\:\nonumber \\
&& V_0\cdot V_1 ~=~ k_{01}+k_{10}~=~ 0 \:\:\:~~~\mbox{mod}\,(1)~.
\end{eqnarray}
We thus see that the $V_4$ projection removes those gravitinos of the ${\cal N}=4$ theory 
for which
\beq
\label{compare}
\frac{1}{4} \left(1-\gamma_{\chi_{34}}\gamma_{\chi_{56}}\right)~=~k_{14} ~~~~\mbox{mod}\,(1)~,
\eeq
leaving behind an ${\cal N}=2$ theory. 
At the same time, given the form of the Virasoro operators in Eq.~(\ref{eqn: 55b}),  
we see that the CDC shifts the masses of those states with non-zero charges overlapping ${\bf e}$: 
\begin{equation}
\label{eqn: 58}
\alpha'\mathtt{\mathbf{m}}^2~=~|\mathbf{e\cdot Q}|^2\left(\frac{1}{{r_1}^2}+\frac{1}{{r_2}^2}\right)~.
\end{equation}
We thus find that states with chiralities such that $\mathbf{e\cdot Q} = \pm 1/2$ acquire masses of $\frac{1}{2}\sqrt{ R_1^{-2}+R_2^{-2} }$, while states with $\mathbf{e\cdot Q}=0$ mod\,(1) remain 
massless.  This clearly breaks the degenerate multiplet structure.

Thus, we see that the $\gamma_{\chi_{34}}\gamma_{\chi_{56}}=1$ states are made heavy by the CDC while 
the $\gamma_{\chi_{34}}\gamma_{\chi_{56}}=-1$ states are unaffected.  Comparison with Eq.~(\ref{compare}) shows that if $k_{14}=0$, there are no massless gravitinos remaining in the spectrum, whereupon  the theory is non-supersymmetric. 
Conversely, the choice $k_{14}=1/2$ leaves the ${\cal N}=2$ symmetry of the original theory intact. 
(Indeed, this discussion is similar to that in Appendix~C concerning SUSY breaking by discrete torsion.) 
Clearly the same splitting applies to all fermions in the $V_1$ sector, as per Table~\ref{table:E}. The mechanism whereby fermions gain masses while scalars (a.k.a.\ Higgses) remain massless is essentially the same as the one described in Ref.~\cite{Antoniadis:1992fh} for the Higgs/Higgsino, namely that the fermion masses are supersymmetric ``$\mu$-terms'' while the scalars have soft terms that precisely cancel this contribution to their squared masses.

\subsubsection{The chiral $\calN=1\rightarrow \calN=0$ model}
\label{chiraln=1}

Given the previous model, 
we now re-introduce the orbifolding required in order to produce a chiral theory.
As noted in Ref.~\cite{Ferrara:1987es}, an orbifold 
in the $X_{56}$ dimensions reverses the sign of the KK and winding modes. 
As a result, we see from the discussion surrounding Eq.~(\ref{eqn: 33}) 
as well as from additional terms in the exponents of Eq.~(\ref{eqn: 56}) that 
the orbifolding does not form a sensible projection on states with degenerate masses 
unless an odd element of the orbifolding acts on the charges as 
${\mathbf {e\cdot Q}}\rightarrow -{\mathbf {e\cdot Q}}$. 
More succinctly, 
a sufficient condition that the theory correspond 
to a four-dimensional $\calN=1$ theory with all symmetries spontaneously broken by ${\mathbf e}$
is that our operator $\widehat{J}_1$ obey the condition~\cite{Ferrara:1987es}
\begin{equation}
\label{eqn: 26a}
   \{\mathfrak{L}, \mathtt{\widehat{g}}\}~=~0~ ,
\end{equation}
where $\mathfrak{L}=\mathbf{e\cdot Q}$ 
and where $\mathtt{\widehat{g}}$ corresponds to the possible odd actions $b_{3,4}$ 
under the $\mathbb{Z}_2$ orbifold in Eq.~(\ref{eqn: 15}). 

Orbifold actions that obey Eq.~(\ref{eqn: 26a}) then 
act on the fields appearing in ${\mathbf e}$ as a generalized conjugation, $\widehat{\tt g} \chi_c^{(i)}=\pm \overline{\chi}_c^{(i)}$ and $\widehat{\tt g} \omega_c^{(i)}=\pm \overline{\omega}_c^{(i)}$. This can be achieved by taking the original $b_3$ and $b_4$ but applying them in a rotated complexification where they overlap with the CDC vector ${\bf e}$, with 
\begin{eqnarray}
&& b_3:~~~~(\chi_{35},\omega_{46})\rightarrow (-\chi_{35},-\omega_{46})~ \nonumber \\ 
&& b_4:~~~~(\chi_{46},\omega_{35})\rightarrow (-\chi_{46},-\omega_{35})~ ,
\end{eqnarray}
so that for example $b_3:\chi_{34} \rightarrow -\overline{\chi}_{34}$. This may of course be written using real-fermion boundary conditions 
(about which much more later) but in order to clarify the connection with the original $\calN=1$ model it is convenient to use the complex basis in which the ${\bf e}$ vectors are diagonal and introduce the notation that a boundary condition $\chi\rightarrow \overline{\chi}$ is represented with `$\overline{0}$' while $\chi\rightarrow -\overline{\chi}$ is represented with `$\overline{1}$'.  Thus, we have
   ${0}\equiv (00)_r$ 
   and ${1}\equiv (11)_r$ as before,
 while 
   $\overline{0}\equiv (01)_r$ 
   and
   $\overline{1}\equiv (10)_r$.

The CDC'd model is then given by the boundary conditions in Table~\ref{table:Dp} with the same set of structure constants. There are a number of additional adjustments that have been made which immediately follow from the rotation of the orbifold basis, namely that certain entries within the vectors $V_{5,7}$ have also acquired bars in order to keep them aligned with the orbifold actions. We shall discuss these and other aspects of handling models with real fermions in more detail below. This distinction is irrelevant for the other vectors because it is explicit that the barred and unbarred vectors are the same. Moreover, $V_4=\overline{b_3+b_4}$ is also unchanged (since the bars cancel), and therefore the action of the ${\bf e}$ shift on the untwisted massless spectrum is precisely as described in the $\calN=2$ theory with the appropriate change of basis. 

\newcommand{\bo}{\mbox{$\overline{1}$}} 
\newcommand{\bz}{\mbox{$\overline{0}$}} 
\begin{table}[H]
\centering
\begin{tabular}{|c|*{1}c|{l}|}
\hline
~Sector~& {\small\xx $\psi^{34}\yyr \psi^{56}\yyr \chi^{34}\:y^{34}\yyr\omega^{34}\yyr\chi^{56}y^{56}\yyr\omega^{56}$} &{\small $\overline{y}^{34}\yyr\overline{\omega}^{34}\yyr\overline{y}^{56}\yyr\overline{\omega}^{56}\yyr\overline{\psi}^{1}\yy\overline{\psi}^{2}\yy\overline{\psi}^{3}\yy\overline{\psi}^{4}\yy\overline{\psi}^{5}\yy\overline{\eta}^{1}\yy\overline{\eta}^{2}\yy\overline{\eta}^{3}\yy\overline{\phi}^{1} \yy\overline{\phi}^{2}\yy\overline{\phi}^{3}\yy\overline{\phi}^{4}\yy\overline{\phi}^{5}\yy\overline{\phi}^{6}\yy\overline{\phi}^{7}\yy\overline{\phi}^{8}$}
\\ \hline
$V_0$&1\xx 1\xx 1\xx 1\xx 1\xx 1\xx 1\xx 1\xx & 1\xxl  1\xxl  1\xxl 1\xxl  1\xxl 1\xxl 1\xxl  1\xxl 1\xxl 1\xxl 1\xxl 1\xxl 1\xxl 1\xxl 1\xxl 1\xxl 1\xxl 1\xxl 1\xxl 1 \\
$V_1$&1\xx 1\xx 1\xx 0\xx 0\xx 1\xx 0\xx 0\xx & 0\xxl  0\xxl 0\xxl 0\xxl 0\xxl  0\xxl 0\xxl  0\xxl 0\xxl 0\xxl  0\xxl 0\xxl 0\xxl 0\xxl 0\xxl 0\xxl 0\xxl 0\xxl 0\xxl 0 \\
$V_2$&1\xx 1\xx 0\xx 1\xx 0\xx 0\xx 1\xx 0\xx & 1\xxl  0\xxl  1\xxl  0\xxl  1\xxl  1\xxl 1\xxl  1\xxl 1\xxl 1\xxl  0\xxl 0\xxl 0\xxl 0\xxl 0\xxl 0\xxl 0\xxl 0\xxl 0\xxl 0 \\
$b_3$&1\xx 0\xx \bo\xx 0\xx \bz\xx \bz\xx 0\xx \bo\xx & 0\xxl  0\xxl  0\xxl  1\xxl  1\xxl  1\xxl 1\xxl  1\xxl 1\xxl 0\xxl  1\xxl 0\xxl 0\xxl 0\xxl 0\xxl 0\xxl 0\xxl 0\xxl 0\xxl 0 \\
$b_4$&1\xx 0\xx \bz\xx 0\xx \bo\xx \bo\xx 0\xx \bz\xx & 0\xxl  1\xxl  0\xxl  0\xxl  1\xxl  1\xxl 1\xxl  1\xxl 1\xxl 0\xxl  0\xxl 1\xxl 0\xxl 0\xxl 0\xxl 0\xxl 0\xxl 0\xxl 0\xxl 0 \\
$V_5$&0\xx 0\xx 0\xx \bz\xx \bz\xx 0\xx \bo\xx \bo\xx & 0\xxl  1\xxl  0\xxl  0\xxl  1\xxl  1\xxl 1\xxl  0\xxl 0\xxl 0\xxl  0\xxl 0\xxl 1\xxl 1\xxl 1\xxl 1\xxl 0\xxl 0\xxl 1\xxl 1 \\
$V_6$&0\xx 0\xx 0\xx 0\xx 0\xx 0\xx 0\xx 0\xx & 1\xxl  1\xxl  0\xxl  1\xxl  1\xxl  1\xxl 1\xxl  0\xxl 0\xxl 0\xxl  0\xxl 0\xxl 0\xxl 1\xxl 1\xxl 1\xxl 1\xxl 1\xxl 1\xxl 0 \\
$V_7$&0\xx 0\xx 0\xx \bo\xx \bo\xx 0\xx \bz\xx \bz\xx&1\xxl  0\xxl  1\xxl  0\xxl  $\frac{1}{2}$\xxh  $\frac{1}{2}$\xxh $\frac{1}{2}$\xxh  $\frac{1}{2}$\xxh $\frac{1}{2}$\xxh $\frac{1}{2}$\xxh $\frac{1}{2}$\xxh $\frac{1}{2}$\xxh 0\xxl 1\xxl $\frac{1}{2}$\xxh $\frac{1}{2}$\xxh $\frac{1}{2}$\xxh 1\xxl 1\xxl $\frac{1}{2}$ \vspace{0.8pt}\\
\hline
\end{tabular}
\caption{Spin structure of the world-sheet fermions of the $\mathcal{N}=0$, $D=4$ model after applying the CDC,
  with overlined entries in these vectors as defined in the text. 
This spin structure is accompanied by two bosonic degrees of freedom compactified on 
a $\IZ_2$ orbifold with twist action corresponding to the vectors $b_{3,4}$. 
As always, every entry in this table is understood to be multiplied by $-\frac{1}{2}$.}
\label{table:Dp}
\end{table}

Since the orbifolding and the CDC are effectively operating in different complexifications, it is worth elucidating how this works for a particular state. The theory neglecting the CDC is clearly identical to the original theory, but with a different complexification, \eg, $\chi^{36}=\chi^3+i \chi^6$ and $\chi^{45}=\chi^4+i \chi^5$. 
Following the notation of Ref.~\cite{Kawai:1987ew}, 
we shall let $b_n$ and $d_n$ denote the positive and negative coefficients 
in a normal-mode expansion
of these fields;
with boundary condition $v$ 
we see that the $b_3$ projection then generally 
takes the form
\beq
 {b_3 \cdot \mathbf {Q}} ~=~ \frac{1}{2} \sum_{n} \left(b^\dagger_{n+v-\frac{1}{2}}b_{n+v-\frac{1}{2}} - d^\dagger_{n+\frac{1}{2}-v}b_{n+\frac{1}{2}-v}\right) +\ldots ~~~~~\mbox{ mod\,(1)}\, .
\eeq
Suppose there is a massless state $\chi^{36}_{-\frac{1}{2}} |0\rangle_R \equiv (b^\dagger_{\chi^{36},\frac{1}{2} }\oplus d^\dagger_{\chi^{36},\frac{1}{2} })|0\rangle_R $ allowed by the $b_3$ projection. 
We see that such a projection cannot distinguish $b^\dagger$ from $d^\dagger$. 
Meanwhile ${\mathbf {e\cdot Q}}$ is shifting the spectrum of states. We may write this shift in terms of the coefficients of the real fermions, \eg, $b_{\chi^{34},n} = \frac{1}{\sqrt{2}}(\chi^3_n+i\chi^4_n)$  and $d_{\chi^{34},n} = \frac{1}{\sqrt{2}}(\chi^3_n-i\chi^4_n)$, with all states remaining unshifted if they satisfy
\begin{eqnarray}
{\mathbf {e\cdot Q}} &=&
 \frac{1}{2} \sum_{\ell=34,56} \left( b^\dagger_{\chi^{\ell}{,\frac{1}{2}}} b_{\chi^{\ell},{\frac{1}{2}}} +
b^\dagger_{\omega^{\ell},{\frac{1}{2}}} b_{\omega^\ell,{\frac{1}{2}}} - 
d^\dagger_{\chi^{\ell},{\frac{1}{2}}} d_{\chi^{\ell},{\frac{1}{2}}} - 
d^\dagger_{\omega^{\ell},{\frac{1}{2}}} d_{\omega^{\ell},{\frac{1}{2}}} 
\right) +\ldots = 0 ~~~~\mbox{ mod\,(1)}\,\nonumber \\
&=&  \frac{i}{2} \sum_{\ell=3,5}
\left( {\chi^{\ell\dagger}_{\frac{1}{2}}} {\chi^{\ell+1}_{\frac{1}{2}}} -
 {\chi^{\ell+1\dagger}_{\frac{1}{2}}} {\chi^{\ell}_{\frac{1}{2}}} 
\,+\,
 {\omega^{\ell\dagger}_{\frac{1}{2}}} {\omega^{\ell+1}_{\frac{1}{2}}} -
 {\omega^{\ell+1\dagger}_{\frac{1}{2}}} {\omega^{\ell}_{\frac{1}{2}}}
\right) = 0 ~~~~\mbox{ mod\,(1)}\,. 
\end{eqnarray}
In other words, in this case the state would become massive (or more precisely the KK tower could be shifted by integers, still leaving a massless state). By contrast, states such as $b^\dagger_{\frac{1}{2} } d^\dagger_{\frac{1}{2} }|0\rangle_R $ (which in this example already has a string scale mass) would remain unshifted. Conversely, one may consider the same states written in the original complex basis as, \eg,
$\chi_{-\frac{1}{2}}^{34} |0\rangle_R \equiv (b^\dagger_{\chi^{34},\frac{1}{2} }\oplus d^\dagger_{\chi^{34},\frac{1}{2} })|0\rangle_R $. 
Such states would gain degenerate masses from the CDC of order $\sim 1/2R$. However the $b_3$ projection leaves only the conjugation-invariant linear combination 
$\frac{1}{\sqrt{2} }(b^\dagger_{\chi^{34},\frac{1}{2} }+ d^\dagger_{\chi^{34},\frac{1}{2} })|0\rangle_R = \chi^{3}_{-\frac{1}{2}} |0\rangle_R $. In a 
similar fashion $\chi^{56}_{-\frac{1}{2}}|0\rangle_R$ leaves behind only $\chi^{6}_{-\frac{1}{2}} |0\rangle_R$. Thus, either way, we consistently find the states $\chi^{36}_{-\frac{1}{2}} |0\rangle_R$ remaining in the spectrum with mass $\sim 1/2R$.
Not surprisingly, since the orbifold action  acts as a conjugation in the original basis, it is blind to states such as $b^\dagger_{\chi^{34},\frac{1}{2} } d^\dagger_{\chi^{34},\frac{1}{2} }|0\rangle_R$
that are neutral under the corresponding CDC charges.

In this discussion we have employed real fermions only as they appear in the 
complexification of the shifted charge lattice.
However, as suggested above, the entire formalism for this class of models can actually be recast in a more straightforward manner using real fermions from the start.  
As we shall see,
this makes makes the previous discussion self-evident and also clarifies 
how the CDC interacts with the $V_{5,7}$ vectors. 

Recall that, because the relevant phase in the GSO projection is either $0$ or $-1/2$ and therefore 
blind to relative signs, in order to write a model in terms of real fermions it is convenient to use the trick of 
Ref.~\cite{Kawai:1987ew} whereby we reverse the sign of the $d^\dagger d$ entries in the charge operators. 
In this construction,
we can no longer define the theory 
in terms of a charge lattice, and indeed
 the charge operator appearing in the GSO projections is replaced by 
the sum of number operators associated with the real fermions in addition to a vacuum ``charge'':
\begin{eqnarray}
{\mathbf {e\cdot Q}}_r &\equiv&
 \frac{1}{2} \sum_{\ell=34,56} \left( b^\dagger_{\chi^{\ell}{,\frac{1}{2}}} b_{\chi^{\ell},{\frac{1}{2}}} +
b^\dagger_{\omega^{\ell},{\frac{1}{2}}} b_{\omega^\ell,{\frac{1}{2}}} +
d^\dagger_{\chi^{\ell},{\frac{1}{2}}} d_{\chi^{\ell},{\frac{1}{2}}} +
d^\dagger_{\omega^{\ell},{\frac{1}{2}}} d_{\omega^{\ell},{\frac{1}{2}}} 
\right) +\ldots = 0 ~~~~\mbox{mod\,(1)}\,\nonumber \\
&=&  \frac{1}{2} \sum_{\ell=3,6,4,5}\left( {\chi^{\ell\dagger}_{\frac{1}{2}}} {\chi^{\ell}_{\frac{1}{2}}} +
{\omega^{\ell\dagger}_{\frac{1}{2}}} {\omega^{\ell}_{\frac{1}{2}}}\right)+\ldots = 0 ~~~~~\mbox{mod\,(1)~.}\, 
\end{eqnarray}
We must therefore revisit the action of the CDC on the Virasoro operators. 
Defining the real number operator  at level $n^\ell$ as $N_{n^\ell+\overline{\alpha V}^\ell-\frac{1}{2}} = 
b^\dagger_{n^\ell+\overline{\alpha V}-\frac{1}{2}} b_{n^\ell+\overline{\alpha V}-\frac{1}{2}}$, 
we see that the Virasoro operators before the CDC take the form
\begin{equation} 
L_0 ~=~ \sum_\ell \sum_{n^\ell=1}(n^\ell+\overline{\alpha V}^\ell-\frac{1}{2})N_{n^\ell+\overline{\alpha V}^\ell-\frac{1}{2}} + \frac{1}{2} (\overline{\alpha V}^\ell)^2 + \ldots
\, ,\end{equation}
and similar for ${\overline{L}}_0$. 
Here the quadratic piece is the relevant vacuum-energy contribution, 
and the dots indicate terms that do not depend on the fermion boundary condition. 
In order to maintain modular invariance, all that is required is for the CDC to induce the 
correct shift in Eq.~(\ref{eqn: 56}). 
Shifting the vacuum ``charge'' 
as $\overline{\alpha V}\rightarrow \overline{\alpha V} - {\mathbf e} (n_1+n_2)$ indeed generates the shift 
\begin{equation} 
L_0\rightarrow L'_0 ~=~ L_0  - (n_1+n_2)\, {\mathbf e} \cdot   (N_{n^\ell+\overline{\alpha V}^\ell-\frac{1}{2}} +\overline{\alpha V})  + \frac{1}{2} {\mathbf e}_L\cdot {\mathbf e}_L (n_1+n_2)^2 
\, ,\end{equation}
and similar for ${\overline L}_0\rightarrow {\overline L}'_0$. 
In the real-fermion formalism therefore we should simply replace ${\mathbf Q}$ in the CDC with 
\beq
   {\mathbf Q}_r~\equiv~ N_{n^\ell+\overline{\alpha V}^\ell-\frac{1}{2}} +\overline{\alpha V}^\ell
\eeq
 and perform real Lorentz products accordingly. 
It is in this notation that we may most easily analyze the spectra of theories 
which include vectors such as $V_{5,7}$
with barred components (and thus implicitly real fermions).

As a self-consistency check, we can ask what in the real formalism 
becomes of the requirement in Eq.~(\ref{eqn: 26a}) that the orbifold action 
should act as a conjugation on the charges overlapping the CDC vector. 
Let us return to the example above of a state $\chi^{3\dagger}_{\frac{1}{2}} |0\rangle_R $ 
allowed by the orbifold action and examine this state in the real-fermion notation. 
The mass of a single real state with CDC shift ${e} = 1/2$ would clearly be shifted by $1/2R$ 
in the real formalism as in the complex.  However, the massive KK modes of a real excitation fall 
into pairs, $\chi^{\dagger}_{\frac{1}{2}} |\pm m\rangle_R$.  
After the CDC, these states would have different masses, $(m\pm \frac{1}{2})/R$ respectively, 
and an orbifold projection cannot be defined for them individually. 
Consistency then requires that there initially be a {\it second}\/ set of real states $\chi^{\prime\dagger}_{\frac{1}{2}} |\pm m\rangle_R$ with the {\it opposite} shift $e=-1/2$
 which obtain masses $(m\mp \frac{1}{2})/R$ respectively
and which map as $\chi'\leftrightarrow \chi$ under the orbifold. 
Masses then naturally arise by forming 
invariant linear combinations of $\chi$ and  $\chi'$ 
excitations with the same mass. 
We then obtain two towers of orbifold eigenstates,
\beq
     \frac{1}{\sqrt{2}} \left( \chi^{\dagger}_{\frac{1}{2}}|m\rangle_R \pm 
          \chi^{\prime\dagger}_{\frac{1}{2}} |-m\rangle_R \right)~,
\eeq
one with positive orbifold eigenvalue and the other negative.
Those states with positive eigenvalue are nothing but the 
KK modes of $\chi^3|0 \rangle_R$, while those with 
negative eigenvalue are the KK modes of 
$\chi^4|0\rangle_R$ and are projected out entirely. (The 
winding numbers come along for the ride.) Note that the negative sign required for the CDC shift 
of $\chi'$ is a natural consequence of the fact that we reversed the sign of $d^\dagger d$ in 
the ``charge'' operator.
We thus conclude that in the real formalism it is not possible to have a positive eigenstate of 
the orbifold overlapping the CDC vector without also having a negative one. 
This is the real-formalism equivalent of the conjugation requirement.

We also see that there is no restriction on the action of the {\it non}\/-orbifold boundary conditions on 
the pair of states $\chi_3$ and $\chi_4$, and in particular they do not always need to be the same. 
Indeed one could for example have a non-orbifold sector in which 
$\chi_3\rightarrow \chi_3$ and $\chi_4\rightarrow \pm\chi_4$. In the original basis $\chi_4\rightarrow \chi_4$ would correspond to $\chi \rightarrow \chi $ and $ \chi' \rightarrow  \chi'$ while $\chi_4\rightarrow - \chi_4$ 
would correspond to the same permutation action as for the orbifold, \ie, $\chi \leftrightarrow \chi' $, 
and would thus clearly commute with it.
Conversely, this action, combined with the orbifolding, would produce a sector in which neither $\chi_3$ nor $\chi_4$ excitations were projected out. However, due to the modular-invariance conditions involving real fermions, some other excitations inevitably would be. 
It is of course not possible to have a complex phase (such as the ones in $V_7$) overlapping
the above operations on real fermions as they do not commute. Note that in practice 
we may determine the massless spectrum remaining after CDC by simply adding
the general constraint ${\mathbf {e\cdot Q}}= 0 \mbox{ mod\,(1)}$ (with real-fermion contributions incorporated as described above) to the GSO projections.

Putting all the pieces together in the context of the model at hand, we see that
the states ${\mathbf {e\cdot Q}} \neq 0$~mod$\,$(1) are lifted. In this particular example, 
two pairs of Higgs fields and two of the singlet scalar states accrue masses while the scalars 
\begin{align}
H_{U_1}, H_{D_1}~=~\psi^{56}_{-\frac{1}{2}}|0\rangle_R&\otimes \overline{\psi}^{4,5}_{-\frac{1}{2}} \: \overline{\eta}^{1}_{-\frac{1}{2}}|0\rangle_L \nonumber \\
\Xi_1,\, \Xi'_1 ~=~\psi^{56}_{-\frac{1}{2}}|0\rangle_R&\otimes \overline{\eta}^{2}_{-\frac{1}{2}} \: \overline{\eta}^{3}_{-\frac{1}{2}}|0\rangle_L\,
\end{align}
remain massless. Likewise, there remain two pairs of massless Higgsinos and Weyl spinors in the spectrum: 
\begin{align}
\tilde{H}_{U_2}, \tilde{H}_{D_2}~=~\{ \chi^{34}_{0}\} |0\rangle_R&\otimes \overline{\psi}^{4,5}_{-\frac{1}{2}}\:\overline{\eta}^{2}_{-\frac{1}{2}}|0\rangle_L  \nonumber\\
\tilde{H}_{U_3}, \tilde{H}_{D_3}~=~\{ \chi^{56}_{0}\}|0\rangle_R&\otimes \overline{\psi}^{4,5}_{-\frac{1}{2}}\:\overline{\eta}^{3}_{-\frac{1}{2}}|0\rangle_L \nonumber\\
\tilde{\Xi}_2,\, \tilde{\Xi}'_2~=~\{ \chi^{34}_{0}\} |0\rangle_R&\otimes \overline{\eta}^{1}_{-\frac{1}{2}}\:\overline{\eta}^{3}_{-\frac{1}{2}}|0\rangle_L  \nonumber\\
\tilde{\Xi}_3,\, \tilde{\Xi}'_3~=~\{ \chi^{56}_{0}\} |0\rangle_R&\otimes \overline{\eta}^{1}_{-\frac{1}{2}}\:\overline{\eta}^{2}_{-\frac{1}{2}}|0\rangle_L  \,. 
\end{align}
As we shall see, the Higgsinos 
can be coupled to the singlet scalars through Yukawa couplings of the form
\beq
\tilde{H}_{D_2}\tilde{H}_{U_3}\Xi_1+ \tilde{H}_{U_2} \tilde{H}_{D_3}\Xi'_1~.
\eeq
We will discuss these and the other Yukawa couplings in Sect.~IX.B.~ 
Thus, if the singlet scalars accrue a VEV, the Higgsinos will effectively become massive.

The fermionic matter fields in the $V_2$ sector are somewhat similar to those in the NS-NS sector in the sense that they have no charges overlapping with $\bf e$.  Thus, their masses are unshifted. However their superpartners in the  ${\overline{V_1+V_2}}$ sector behave precisely as for the $V_1$ sector, and all receive the same masses as the gravitinos if $k_{14}=0$, as per Table~\ref{table:F}. 

From a phenomenological perspective, this splitting is therefore very appealing:   one pair of Higgs scalars and the chiral matter fields (in other words, the field content of the Standard Model) all remain massless while their superpartners gain large masses. It is for this reason that we refer to this as an ``SM-like'' model.  Note, however, that the twisted sectors are unaffected, so only one generation is truly split while two generations remain (quasi-) supersymmetric.

\subsection{Partition functions of the MSSM- and SM-like models}

\label{sec:part}
Thus far, we have presented two models of interest:
an MSSM-like model in Sect.~VI.A and an SM-like model in Sect.~VI.B.2.~
In order to further elucidate the structure of these models,
we now turn to a comparison of their partition functions.
This will also enable us to consider quantum effects, such as the cosmological constant or contributions to soft terms. 
Moreover, we noted below Eq.~(\ref{eqn: 58}) that the alternative phase choice $k_{14}=1/2$ produces
a different CDC model in which the gravitino remains massless.
An analysis of the partition function of this variant model will also enable us to demonstrate that 
it really is supersymmetric at all mass levels.

\subsubsection{For the ${\cal N}=1$ MSSM-like model}

In order to construct the partition function for the supersymmetric theory presented in Sect.~VI.A, 
we will use results and definitions from Appendix~\ref{notation}. 
In the untwisted sector, the modular-invariant partition function for the two compact 
bosonic degrees of freedom in terms of the 
winding numbers ${\vec{\ell}}=\{ \ell_1 , \ell_2 \}$, $\vec n=\{ n_1 , n_2 \}$, and the radii $r_1$, $r_2$ is
given by 
\beq
\label{eqn: 29}
Z_{\substack{\mathbf{B}}}
{\tiny\begin{bmatrix}
									0\\
									0
									\end{bmatrix}}(\tau) = ~\sum_{\vec{\ell},\vec n} Z_{\vec{\ell},\vec{n}}
\eeq
where
\beq
Z_{\vec{\ell},\vec{n}} ~=~\frac{{\cal M}^2 r_1r_2}{\tau_2\overline{\eta}^2 \eta^2}\sum_{{\vec{\ell},\vec{n}}}\exp \Big\{-\frac{\pi}{\tau_2}\left[r_1^2|\ell_1-n_1\tau|^2+r_2^2|\ell_2-n_2\tau|^2\right]\Big\}~.
\eeq
In Eq.~(\ref{eqn: 29}), the notation ${\tiny\begin{bmatrix}
				0\\
				0
				\end{bmatrix}}$ indicates untwisted boundary conditions in both the spacelike and timelike toroidal directions.  By contrast, the contributions from the twisted sectors are given by 
\beq
\label{eqn: 30}
   Z_{\substack{\mathbf{B}}}{\tiny\begin{bmatrix} 
                 \overline{(\alpha_{3}+\alpha_{4})/2}\\
                 \overline{(\beta_{3}+\beta_{4})/2}
              \end{bmatrix}}(\tau)   
         ~=~   
     2\left|\frac{\eta}{\vartheta{\tiny\begin{bmatrix}
              1/2-\overline{(\alpha_{3}+\alpha_{4})/2}\\
              1/2-\overline{(\beta_{3}+\beta_{4})/2}
                    \end{bmatrix}}}\right|^{2}
\eeq
where $\overline{\alpha_{3}+\alpha_{4}}$ and $\overline{\beta_{3}+\beta_{4}}$ indicate
the $\mathbb{Z}_{2}$ twists on the complex boson (see, \eg,  Ref.~\cite{Chamseddine:1989mz}). 
Of course, it is assumed in Eq.~(\ref{eqn: 30})
that either
$\overline{\alpha_{3}+\alpha_{4}}$
or
$\overline{\beta_{3}+\beta_{4}}$ 
is odd.
These functions have the following modular transformations:
\beqn
&  Z_{\mathbf B}{\tiny\begin{bmatrix}
1/2\\
1/2
\end{bmatrix}}(\tau+1) = Z_{\mathbf B}{\tiny\begin{bmatrix}
1/2\\
0
\end{bmatrix}}(\tau)
~, ~~~~~~~& 
Z_{\mathbf B}{\tiny\begin{bmatrix}
1/2\\
1/2
\end{bmatrix}}(-1/\tau) =  Z_{ \mathbf B}{\tiny\begin{bmatrix}
1/2\\
1/2
\end{bmatrix}}(\tau)
 \nonumber \\
&  Z_{\mathbf B}{\tiny\begin{bmatrix}
1/2\\
0
\end{bmatrix}}(\tau+1) =  Z_{\mathbf B}{\tiny\begin{bmatrix}
1/2\\
1/2
\end{bmatrix}}(\tau)
~, ~~~~~~~& 
Z_{\mathbf B}{\tiny\begin{bmatrix}
1/2\\
0
\end{bmatrix}}(-1/\tau) =  Z_{\mathbf B}{\tiny\begin{bmatrix}
0\\
1/2
\end{bmatrix}}(\tau)
 \nonumber \\
&  Z_{\mathbf B}{\tiny\begin{bmatrix}
0\\
1/2
\end{bmatrix}}(\tau+1) =  Z_{\mathbf B}{\tiny\begin{bmatrix}
0\\
1/2
\end{bmatrix}}(\tau)
~, ~~~~~~~& 
Z_{\mathbf B}{\tiny\begin{bmatrix}
0\\
1/2
\end{bmatrix}}(-1/\tau) =  Z_{\mathbf B}{\tiny\begin{bmatrix}
1/2\\
0
\end{bmatrix}}(\tau)~.
\eeqn
Note that the modular transformations of the $Z_{\substack{\mathbf{B}}}(\tau)$ bosonic factors do not introduce any additional phases.   This is why (as we remarked when we were considering the spectra of these models) 
the physical projections induced by the orbifolding are virtually the same as they would have been in the non-compact (6D) theory in which one treats $b_{3}$ and $b_{4}$ as regular additional boundary-condition vectors. 
In addition, the untwisted $Z_{\mathbf B}{\tiny\begin{bmatrix}
0\\
0
\end{bmatrix}}$ partition function is modular invariant by itself;  in general its divergent contribution to the total
partition function is cancelled by vanishing contributions from world-sheet fermions 
in order to yield finite results.
The complete one-loop partition function $Z(\tau)$  
for the ${\cal N}=1$, 4D model is therefore  
\begin{align}
\label{eqn: 31b}
Z(\tau)~=~\frac{{\cal M}^{2}}{{\tau_{2}}|\eta|^{4}}\frac{1}{[\eta(\tau)]^{8}[\overline{\eta}(\overline{\tau})]^{20}}\sum_{\left\{ \alpha,\beta\right\} }C_{\mathbf{\beta}}^{\mathbf{\alpha}}Z_{\substack{\mathbf{B}}}{\tiny\begin{bmatrix}
\overline{(\alpha_{3}+\alpha_{4})/2}\\
\overline{(\beta_{3}+\beta_{4})/2}
\end{bmatrix}} (\tau) \prod_{\substack{i_R}}\vartheta{\tiny\begin{bmatrix}
\overline{\alpha\mathbf{V}_i}\\
-\beta\mathbf{V}_i \\
\end{bmatrix}}(\tau)\prod_{i_L}\overline{\vartheta}{\tiny\begin{bmatrix}
\overline{\alpha\mathbf{V}_i}\\
-\beta\mathbf{V}_i \\
\end{bmatrix}}(\overline{\tau})
\end{align}
where the generalized GSO-projection coefficients are given by
\begin{equation}
\label{eqn: 32}
C_{\mathbf{\beta}}^{\mathbf{\alpha}}~=~\exp\left[2\pi i
     \left(\alpha s+\beta s+\beta_{i}k_{ij}\alpha_{j}\right)\right]\, .
\end{equation}
Note that in writing the projection coefficients in this form we have
 followed the conventions in Ref.~\cite{Kawai:1987ew}
and absorbed the factor of 
$e^{2\pi i \beta V \cdot \overline{\alpha V}}$ 
from the partition function in Eq.~(\ref{trace-formula}) into our definition $C_\beta^\alpha $.

Since the CDC-twisted sectors of the theory remain supersymmetric,
we now focus on the explicit form of the contributions to the partition function from the untwisted sectors.
Indeed, this is where  all the action takes place once the CDC is introduced.

In general, as with any $\IZ_2$-orbifolded theory,
the contributions to the total partition function are of the form 
\beq
  Z~=~\frac{1}{2} \left( Z{\tiny\begin{bmatrix} 0 \\ 0 \end{bmatrix}} + Z{\tiny\begin{bmatrix} g \\ 0 \end{bmatrix}} + Z{\tiny\begin{bmatrix} 0 \\ g \end{bmatrix}} + Z{\tiny\begin{bmatrix} g \\ g \end{bmatrix}}\right) 
\eeq
where `0' and `$g$' indicate the sum over all untwisted and twisted sectors, respectively. 
As mentioned above and as discussed in Ref.~\cite{Ferrara:1987qp}, contributions with a twist on either cycle are independent of the vector $\mathbf e$.  This is obvious when there is a twist on the $a$-cycle, but less so
for the term $Z{\tiny\begin{bmatrix} 0 \\ g \end{bmatrix}} $. However, the reason the latter also does not depend on  
$\mathbf e$ is that the orbifold reverses  charges, windings and/or KK modes, as we have seen,
 and therefore precisely half of these states are projected out, leaving the invariant combination 
$(|n,m,Q\rangle + |-n,-m,-Q \rangle)$. Since there is an overall factor of $1/2$ in the 
projection, all states with non-zero $n,m$ or with a $Q$ that conjugates under the orbifolding are already counted by the untwisted $\frac{1}{2} Z{\tiny\begin{bmatrix} 0 \\ 0 \end{bmatrix}}$ contribution. 
However, these are the only states that have ${\mathbf e}$-dependence in their Hamiltonian, and therefore $Z{\tiny\begin{bmatrix} 0 \\ g \end{bmatrix}} $ simply provides extra contributions from 
the orbifold projections on the rest of the spectrum.  
Consequently, for the purposes of determining the cosmological constant $\Lambda$, the CDC'd partition function can be written
\begin{equation}
\label{eq:triv}
Z({\mathbf e}) ~=~Z(0)+ \frac{1}{2}\left( Z{\tiny\begin{bmatrix} 0 \\ 0 \end{bmatrix}}({\mathbf e})  - Z{\tiny\begin{bmatrix} 0 \\ 0 \end{bmatrix}}(0) \right) ~=~ \frac{1}{2}Z{\tiny\begin{bmatrix} 0 \\ 0 \end{bmatrix}} ({\mathbf e}) \, , 
\end{equation}
since the supersymmetric partition functions provide cancelling contributions to $\Lambda$. 
In order to see the effect on $\Lambda$, one can therefore work entirely within the toroidally compactified theory as long as all the untwisted-sector contributions are included, including 
those containing the combination $V_4=\overline{b_3+b_4}$. 

Let us in addition separate out the action of $V_{1}$ from the partition function, as this is what governs the supersymmetric cancellations. In addition to $V_{5,6,7}$, a convenient basis corresponds to the following 
linear combination of the vectors defined in Table~\ref{table:D}: 
\begin{eqnarray}
\label{eqn: 34}
V_{0}'~=~\overline{V_{0}+V_{1}}&=&-{\scriptstyle\frac{1}{2}}[~00~011~011~|~1...1]\nonumber \\
V_{1} & = & -{\scriptstyle\frac{1}{2}}[~11~100~100~|~...~]\nonumber \\
V_{2}'~=~\overline{V_{2}+V_{0}+b_{3}+b_{4}}&=&-{\scriptstyle\frac{1}{2}}[~00~000~000~|~...~]\nonumber \\
V_{4} ~=~\overline{b_{4}+b_{3}}&=&-{\scriptstyle\frac{1}{2}}[~00~101~101~|~...~]\, .
\end{eqnarray}
In this basis, only $V_{4}$ overlaps with $V_{1}$;  consequently terms that cancel due to supersymmetry will 
largely factor out.

The vectors $V_{i}$ can be divided into two sets:  $\lbrace V_{1},V_{4}\rbrace$ and $\lbrace V_{a}\rbrace$ where 
$a\not\in \lbrace1,4\rbrace$. 
Without loss of generality we can choose $k_{1a}=0$, and write the partition function as 
\begin{align}
\label{eqn: 36}
Z(\tau)~=~ \frac{{\cal M}^{2}}{\tau_{2} \eta^{10} \overline{\eta}^{22}} \sum_{\substack{\alpha_{1,4}\\\beta_{1,4}}}Z_{\substack{\mathbf{B}}}{\tiny\begin{bmatrix}
0\\
0
\end{bmatrix}}(\tau)\:{C}^\alpha_\beta\prod_{\substack{i_R \\\in\{1,2,3,6\}}}\vartheta{\tiny\begin{bmatrix}
\overline{\alpha\mathbf{V}_i}\\
-\beta\mathbf{V}_i \\
\end{bmatrix}}(\tau)\sum_{\substack{\alpha_{0',2',5,6,7} \\ \beta_{0',2',5,6,7}}} \prod_{\substack{i_R\\ \notin\{1,2,3,6\}}}\vartheta{\tiny\begin{bmatrix}
\overline{\alpha\mathbf{V}_i}\\
-\beta\mathbf{V}_i \\
\end{bmatrix}}(\tau)\prod_{i_L}\overline{\vartheta}{\tiny\begin{bmatrix}
\overline{\alpha\mathbf{V}_i}\\
-\beta\mathbf{V}_i \\
\end{bmatrix}}(\overline{\tau})~.
\end{align}
The factors $C_{\beta}^{\alpha}$ can also be split. Letting $a\equiv 0',2',5,6,7$, we have 
\begin{eqnarray}
\label{eqn: 37}
C_{\mathbf{\beta}}^{\mathbf{\alpha}} & = & \exp\left[2\pi i\left(\alpha s+\beta s+\beta_{i}k_{ij}\alpha_{j}\right)\right]\nonumber \\
 & = & e^{\pi i(\alpha_{1}+\beta_{1})}\exp\left[2\pi i\left(\beta_{a}k_{ab}\alpha_{b}+\beta_{4}k_{4b}\alpha_{b}+\beta_{a}k_{a4}\alpha_{4}+\beta_{1}k_{14}\alpha_{4}+\beta_{4}k_{41}\alpha_{1}\right)\right]\nonumber \\
& = & e^{\pi i(\alpha_{1}+\beta_{1})}\exp\left[2\pi i\left(\beta_{4}k_{4b}\alpha_{b}+\beta_{a}k_{a4}\alpha_{4}+\beta_{1}k_{14}\alpha_{4}+\beta_{4}k_{41}\alpha_{1}\right)\right]\hat{C}_{\mathbf{\beta}}^{\mathbf{\alpha}}~.
\end{eqnarray}
One can then identify contributions involving different $V_{4}$ contributions to the spin structure: 
\begin{equation}
\label{eqn: 38}
Z(\tau)~=~\frac{{\cal M}^{2}}{\tau_{2} \eta^{10} \overline{\eta}^{22}}Z_{\substack{\mathbf{B}}}{\tiny\begin{bmatrix}
0\\
0
\end{bmatrix}}\sum_{\alpha_{4},\beta_{4}}\Omega{\tiny\begin{bmatrix}
\alpha_{4}\\
\beta_{4}
\end{bmatrix}}
\end{equation}
where, using the double-index shorthand for the $\vartheta$-functions (see Appendix~\ref{notation}),
we have
\begin{align}
\label{eqn: 40}
\Omega{\tiny\begin{bmatrix}
0\\
0
\end{bmatrix}} & = ~\left[\vartheta_{00}^{4}-\vartheta_{01}^{4}-\vartheta_{10}^{4}+\vartheta_{11}^{4}\right]\times 
\sum_{\substack{\alpha_{0',2',5,6,7} \\ \beta_{0',2',5,6,7}}}\hat{C}^\alpha_\beta  
\prod_{\substack{i_R \\ \notin\{1,2,3,6\}}}\vartheta{\tiny\begin{bmatrix}
\overline{\alpha\mathbf{V}_i}\\
-\beta\mathbf{V}_i \\
\end{bmatrix}}
\prod_{j_L}\overline{\vartheta}{\tiny\begin{bmatrix}
\overline{\alpha\mathbf{V}_j}\\
-\beta\mathbf{V}_j \\
\end{bmatrix}}
~,\nonumber\\
\nonumber\\
\Omega{\tiny\begin{bmatrix}
1\\
0
\end{bmatrix}} & = ~  \left[\vartheta_{00}^{2}\vartheta_{10}^{2}-(-1)^{2k_{14}}\vartheta_{01}^{2}\vartheta_{11}^{2}-\vartheta_{10}^{2}\vartheta_{00}^{2}+(-1)^{2k_{14}}\vartheta_{11}^{2}\vartheta_{01}^{2}\right] \nonumber \\
&\qquad\qquad~~~\times~ \sum_{\substack{\alpha_{0',2',5,6,7} \\ \beta_{0',2',5,6,7}}}\hat{C}^\alpha_\beta  
e^{2\pi i \beta_ak_{a4}} \prod_{\substack{i_R \\ \notin\{1,2,3,6\}}}\vartheta{\tiny\begin{bmatrix}
\overline{\alpha\mathbf{V}_i}\\
-\beta\mathbf{V}_i 
\end{bmatrix}}
\prod_{j_L}\overline{\vartheta}{\tiny\begin{bmatrix}
\overline{\alpha\mathbf{V}_j}\\
-\beta\mathbf{V}_j
\end{bmatrix}} 
~,\nonumber\\
\nonumber\\
\Omega{\tiny\begin{bmatrix}
0\\
1
\end{bmatrix}} & = ~ \left[\vartheta_{00}^{2}\vartheta_{01}^{2}-\vartheta_{01}^{2}\vartheta_{00}^{2}-(-1)^{2k_{41}}\vartheta_{10}^{2}\vartheta_{11}^{2}+(-1)^{2k_{41}}\vartheta_{11}^{2}\vartheta_{10}^{2}\right]\nonumber \\
&\qquad\qquad~~~\times~ \sum_{\substack{\alpha_{0',2',5,6,7} \\ \beta_{0',2',5,6,7}}}\hat{C}^\alpha_\beta  e^{2\pi i k_{4a}\alpha_a}\prod_{\substack{i_R\\ \notin\{1,2,3,6\}}}\vartheta{\tiny\begin{bmatrix}
\overline{\alpha\mathbf{V}_i}\\
-\beta\mathbf{V}_i 
\end{bmatrix}}
\prod_{j_L}\overline{\vartheta}{\tiny\begin{bmatrix}
\overline{\alpha\mathbf{V}_j}\\
-\beta\mathbf{V}_j
\end{bmatrix}} 
~,\nonumber\\
\nonumber\\
\Omega{\tiny\begin{bmatrix}
1\\
1
\end{bmatrix}} & = ~\left[\vartheta_{00}^{2}\vartheta_{11}^{2}-(-1)^{2k_{14}}\vartheta_{01}^{2}\vartheta_{10}^{2}-(-1)^{2k_{41}}\vartheta_{10}^{2}\vartheta_{01}^{2}+(-1)^{2(k_{14}+k_{41})}\vartheta_{11}^{2}\vartheta_{00}^{2}\right]\nonumber \\
&\qquad\qquad~~~\times~\sum_{\substack{\alpha_{0',2',5,6,7} \\ \beta_{0',2',5,6,7}}}\hat{C}^\alpha_\beta e^{\left[2\pi i(k_{4a}\alpha_a+\beta_ak_{a4}+k_{44})\right]}\prod_{\substack{i_R \\ \notin\{1,2,3,6\}}}\vartheta{\tiny\begin{bmatrix}
\overline{\alpha\mathbf{V}_i}\\
-\beta\mathbf{V}_i 
\end{bmatrix}}
\prod_{j_L}\overline{\vartheta}{\tiny\begin{bmatrix}
\overline{\alpha\mathbf{V}_j}\\
-\beta\mathbf{V}_j
\end{bmatrix}} \,\, .
\end{align}
As the theory is supersymmetric, these contributions all cancel, as they should: 
the first term vanishes by the so-called ``abstruse'' identity [\ie,
the first identity listed below Eq.~(\ref{etathetadefs})],
while thanks to Eq.~(\ref{eqn: 35}) the three other terms vanish by inspection.

\subsubsection{For the ${\cal N}=0$ SM-like model}

We now move on to the partition function of the non-supersymmetric model, after CDC, 
following the same steps as in Refs.~\cite{Ferrara:1987es, Ferrara:1987qp, Ferrara:1988jx}. 
It is convenient to define $n=(n_{1}+n_{2})$~mod$\,$(1) and $\ell=(\ell_{1}+\ell_{2})$~mod$\,$(1). 
We also define $\overline{n}=1-n$ and $\overline{\ell}=1-\ell$. 
We can then write 
\begin{equation}
\label{eqn: 38p}
Z'(\tau)~=~\frac{{\cal M}^{2}}{\tau_{2} \eta^{10} \overline{\eta}^{22}}\sum_{\vec{\ell},\vec{n}}Z_{\bf \vec{\ell},\vec{n}}\sum_{\alpha_{4},\beta_{4}} \Omega_{\ell,n}{\tiny\begin{bmatrix}
\alpha_{4}\\
\beta_{4}
\end{bmatrix}}
\end{equation}
where
\begin{equation}
\label{eqn: 47}
\Omega_{\ell,n}
{\tiny\begin{bmatrix}
\alpha_4\\
\beta_4\\
\end{bmatrix}}~=~\prod_{\substack{i_R\\\in\{1,2,3,6\}}}\vartheta{\tiny\begin{bmatrix}
\overline{\alpha\mathbf{V}_i}-n\mathbf{e}_i\\
-\beta\mathbf{V}_i+\ell\mathbf{e}_i \\
\end{bmatrix}}  \sum_{\substack{\alpha_{0',2',5,6,7} \\ \beta_{0',2',5,6,7}}} {\tilde{C}}^\alpha_\beta \prod_{\substack{i_R \\ \notin\{1,2,3,6\}}}\vartheta{\tiny\begin{bmatrix}
\overline{\alpha\mathbf{V}_i}-n\mathbf{e}_i\\
-\beta\mathbf{V}_i +\ell\mathbf{e}_i\\
\end{bmatrix}}
\prod_{j_L}\overline{\vartheta}{\tiny\begin{bmatrix}
\overline{\alpha\mathbf{V}_j}-n\mathbf{e}_j\\
-\beta\mathbf{V}_j +\ell\mathbf{e}_j
\end{bmatrix}}\, .
\end{equation}
In this expression, the
coefficients of the partition function are given by
\begin{equation}
\label{eqn: 45}
{\tilde{C}}^\alpha_\beta~=~\exp\left\lbrack
        -2\pi i \left(n\mathbf{e}\cdot\beta{V}-\frac{1}{2}n\ell\mathbf{e}^2\right) \right\rbrack \,C^\alpha_\beta\, ,
\end{equation}
where $C^\alpha_\beta$ are the coefficients of the untwisted partition function before CDC, as given 
in Eq.~(\ref{eqn: 37}). 

The new partition function is the same as the old one except that the CDC vector $\mathbf{e}$ defined in Eq.~(\ref{eqn: 71}) shifts the $i=3, 6, 5, 8$ arguments by a half-unit when $n$ or $\ell$ is odd, and there is a phase $e^{-2\pi in {\bf e}\cdot\beta V}$.  This phase is only sensitive to $V_{1}$ and $V_{4}$ as these are the only \textit{untwisted}\/ vectors that overlap with $\mathbf{e}$. This phase is trivial when $\beta_{1}+\beta_{4}$ is even, and 
gives a factor $(-1)^{n}$ when $\beta_{1}+\beta_{4}$ is odd. 
In total, then, we find 
\begin{align}
\label{eqn: 49}
\Omega_{\ell,n}{\tiny{\tiny\begin{bmatrix}
0\\
0
\end{bmatrix}}}&=~ (-1)^{n\ell}\left[\vartheta_{00}^{2}\vartheta_{n\ell}^{2}-(-1)^{n}\vartheta_{01}^{2}\vartheta_{n\overline{\ell}}^{2}-\vartheta_{10}^{2}\vartheta_{\overline{n}\ell}^{2}+(-1)^{n}\vartheta_{11}^{2}\vartheta_{\overline{n}\overline{\ell}}^{2}\right] \nonumber \\
&\qquad\qquad~~\times~\sum_{\substack{\alpha_{0',2',5,6,7} \\ \beta_{0',2',5,6,7}}}\hat{C}^\alpha_\beta\prod_{\substack{i_R\in \{4,7\}}}\vartheta{\tiny {\tiny{\tiny\begin{bmatrix}
\overline{\alpha\mathbf{V}_i}\\
-\beta\mathbf{V}_i \\
\end{bmatrix}}}}\prod_{\substack{i_R\in\{5,8\}}}\vartheta{\tiny{\tiny\begin{bmatrix}
\overline{\alpha\mathbf{V}_i}-n\mathbf{e}_i\\
-\beta\mathbf{V}_i+\ell\mathbf{e}_i \\
\end{bmatrix}}}
\prod_{j_L}\overline{\vartheta}{\tiny{\tiny\begin{bmatrix}
\overline{\alpha\mathbf{V}_j}\\
-\beta\mathbf{V}_j \\
\end{bmatrix}}} 
~,\nonumber\\
\nonumber\\
\Omega_{\ell,n}{\tiny{\tiny\begin{bmatrix}
1\\
0
\end{bmatrix}}}&=~(-1)^{n\ell}\left[\vartheta_{00}^{2}\vartheta_{\overline{n}\ell}^{2}-(-1)^{2k_{14}+n}\vartheta_{01}^{2}\vartheta_{\overline{n}\overline{\ell}}^{2}-\vartheta_{10}^{2}\vartheta_{n\ell}^{2}+(-1)^{2k_{14}+n}\vartheta_{11}^{2}\vartheta_{n\overline{\ell}}^{2}\right] \nonumber \\
&\qquad\qquad~~\times~\sum_{\substack{\alpha_{0',2',5,6,7} \\ \beta_{0',2',5,6,7}}}\hat{C}^\alpha_\beta e^{2\pi i \beta_ak_{a4}}\prod_{\substack{i_R\in \{4,7\}}}\vartheta{\tiny{\tiny\begin{bmatrix}
\overline{\alpha\mathbf{V}_i}\\
-\beta\mathbf{V}_i \\
\end{bmatrix}}}\prod_{\substack{i_R\in\{5,8\}}}\vartheta{\tiny{\tiny\begin{bmatrix}
\overline{\alpha\mathbf{V}_i}-n\mathbf{e}_i\\
-\beta\mathbf{V}_i+\ell\mathbf{e}_i \\
\end{bmatrix}}}
\prod_{j_L}\overline{\vartheta}{\tiny{\tiny\begin{bmatrix}
\overline{\alpha\mathbf{V}_j}\\
-\beta\mathbf{V}_j \\
\end{bmatrix}}}
~,\nonumber\\
\nonumber\\
\Omega_{\ell,n}{\tiny{\tiny\begin{bmatrix}
0\\
1
\end{bmatrix}}} &=~(-1)^{n\ell}\left[(-1)^{n}\vartheta_{00}^{2}\vartheta_{n\overline{\ell}}^{2}-\vartheta_{01}^{2}\vartheta_{n\ell}^{2}-(-1)^{2k_{41}+n}\vartheta_{10}^{2}\vartheta_{\overline{n}\overline{\ell}}^{2}+(-1)^{2k_{41}}\vartheta_{11}^{2}\vartheta_{\overline{n}\ell}^{2}\right] \nonumber \\
&\qquad\qquad~~\times~\sum_{\substack{\alpha_{0',2',5,6,7} \\ \beta_{0',2',5,6,7}}}\hat{C}^\alpha_\beta e^{2\pi i k_{4a}\alpha_a}\prod_{\substack{i_R\in \{4,7\}}}\vartheta{\tiny{\tiny\begin{bmatrix}
\overline{\alpha\mathbf{V}_i}\\
-\beta\mathbf{V}_i \\
\end{bmatrix}}}\prod_{\substack{i_R\in\{5,8\}}}\vartheta{\tiny{\tiny\begin{bmatrix}
\overline{\alpha\mathbf{V}_i}-n\mathbf{e}_i\\
-\beta\mathbf{V}_i+\ell\mathbf{e}_i \\
\end{bmatrix}}}
\prod_{j_L}\overline{\vartheta}{\tiny{\tiny\begin{bmatrix}
\overline{\alpha\mathbf{V}_j}\\
-\beta\mathbf{V}_j \\
\end{bmatrix}}} 
~,\nonumber\\
\nonumber\\
\Omega_{\ell,n}{\tiny{\tiny\begin{bmatrix}
1\\
1
\end{bmatrix}}}&=~(-1)^{n\ell}\left[(-1)^{n}\vartheta_{00}^{2}\vartheta_{\overline{n}\overline{\ell}}^{2}-(-1)^{2k_{14}}\vartheta_{01}^{2}\vartheta_{\overline{n}\ell}^{2}-(-1)^{2k_{41}+n}\vartheta_{10}^{2}\vartheta_{n\overline{\ell}}^{2}+(-1)^{2(k_{14}+k_{41})}\vartheta_{11}^{2}\vartheta_{n\ell}^{2}\right]\nonumber \\
&\qquad\qquad~~\times~\sum_{\substack{\alpha_{0',2',5,6,7} \\ \beta_{0',2',5,6,7}}}\hat{C}^\alpha_\beta e^{2\pi i(k_{4a}\alpha_a+\beta_ak_{a4}+k_{44})}
\prod_{\substack{i_R\in \{4,7\}}}\vartheta{\tiny{\tiny\begin{bmatrix}
\overline{\alpha\mathbf{V}_i}\\
-\beta\mathbf{V}_i \\
\end{bmatrix}}}\prod_{\substack{i_R\in\{5,8\}}}\vartheta{\tiny{\tiny\begin{bmatrix}
\overline{\alpha\mathbf{V}_i}-n\mathbf{e}_i\\
-\beta\mathbf{V}_i+\ell\mathbf{e}_i \\
\end{bmatrix}}}
\prod_{j_L}\overline{\vartheta}{\tiny{\tiny\begin{bmatrix}
\overline{\alpha\mathbf{V}_j}\\
-\beta\mathbf{V}_j \\
\end{bmatrix}}}\, .
\end{align}

We may now test our previous expectation (given the appearance of a massless 
gravitino in the spectrum) that 
the theory still has ${\cal N}=1$ supersymmetry
when  $k_{14}=\frac{1}{2}$. 
It is straightforward to check by inspection and through the ``abstruse'' identity 
that the prefactors in the above expressions all cancel for any $n$ and $\ell$ when $k_{14}=\frac{1}{2}$, and that they do not when $k_{14}=0$. 
A corollary is that the theory without $V_{4}$ is inevitably still supersymmetric despite the CDC:  
although there is a shift in the spectrum due to the CDC, this shift is simply tantamount to shifting 
the R-charges of the states on the KK tower. This can be deduced from the fact that $\Omega_{\ell,n}{\tiny\begin{bmatrix}
0\\
0
\end{bmatrix}}=0$ regardless of $\ell,n$ and $k_{14}$.

\subsection{Cosmological constant of the SM-like model}

We now turn to an analytical evaluation of the one-loop cosmological constant $\Lambda$ for these models. 
Because the MSSM-like model of Sect.~VI.A has an unbroken spacetime supersymmetry,
its cosmological constant is identically zero.
Thus we shall focus on the  cosmological constant of our SM-like model. 
We shall first evaluate $\Lambda$ for this theory, and find the same leading and subleading terms anticipated in Sect.~IV;
indeed a direct comparison with the circle-compactification case discussed in Sect.~IV
can be obtained by taking the $r_2\gg r_1$ limit. 
Moreover, as anticipated in Sect.~IV, we shall also see why a model with 
equal numbers of massless bosons and fermions has
an exponentially small cosmological constant. 

As discussed above, the CDC affects only the untwisted sectors of the theory. 
Hence all the twisted sectors of the theory are still supersymmetric and make no net contribution 
to $\Lambda$. Moreover, as we saw, the model without $V_4$ is supersymmetric even in the presence of the CDC 
and therefore $\Omega_{\ell,n}{\tiny {\tiny\begin{bmatrix}
0\\
0
\end{bmatrix}}}$ does not contribute.  
As a result,
in order to have a non-supersymmetric model,
we shall henceforth take $k_{14}=0$. 

While the $\ell$ indices correspond to resummed KK modes,
the sectors with $n\neq 0$ correspond to winding-mode contributions 
and not surprisingly by Eq.~(\ref{eqn: 29}) their contributions are extremely suppressed.  
Indeed, as anticipated in Sect.~IV, this is a suppression by a factor of at least 
$\exp(-\pi r^2)$ for generic radius $r$. Therefore we will again neglect such terms. 
(Of course in the $r_{1,2}\rightarrow 0$ limit the reverse would be true, and we would have to resum both $n$ and $\ell$ in order to pick out the winding-mode contributions instead and neglect the KK contributions.)

We shall therefore focus on the terms with $n=0$ and arbitrary $\ell$ in Eq.~(\ref{eqn: 49}).
For odd $\ell$,
the only {non-vanishing} contributions to the partition function arise 
when the winding number $n$ is zero (or even):
\begin{align}
\label{eqn: 71b}
\ell=\mbox{odd}: \qquad & \Omega_{\ell,0}{\tiny\begin{bmatrix}
1\\
0
\end{bmatrix}} ~=~ -\left[\vartheta_{00}^{2}\vartheta_{11}^{2}-\vartheta_{01}^{2}\vartheta_{10}^{2}-\vartheta_{10}^{2}\vartheta_{01}^{2}+\vartheta_{11}^{2}\vartheta_{00}^{2}\right] + \ldots 
~=~ 2\vartheta_{10}^2\vartheta_{01}^2 + \ldots \nonumber \\
\ell=\mbox{odd}: \qquad  & \Omega_{\ell,0}{\tiny\begin{bmatrix}
0\\
1
\end{bmatrix}}~=~ -\left[\vartheta_{00}^{4}-\vartheta_{01}^{4}+\vartheta_{10}^{4}\right] + \dots \nonumber \\
\ell=\mbox{odd}: \qquad  & \Omega_{\ell,0}{\tiny\begin{bmatrix}
1\\
1
\end{bmatrix}}~=~-[\vartheta_{00}^{2}\vartheta_{10}^{2}-\vartheta_{01}^{2}\vartheta_{11}^{2}+\vartheta_{10}^{2}\vartheta_{00}^{2}-\vartheta_{11}^2\vartheta_{01}^2] + \dots ~=~ -2\vartheta_{00}^2\vartheta_{10}^2+ \dots 
\end{align} 
Although it is possible to evaluate the entire integral methodically, in order to examine the effect
of the CDC for the $n=0$ contributions it is simpler to go back to the original expression for the partition function in Eq.~(\ref{eqn: 38p}).  When $n=0$, there are no shifts on the $a$-cycle and the only change is on the $b$-cycle due to the phase $2\pi {\bf e\cdot Q}$.  In particular, from Eq.~(\ref{eqn: 45}), we see that $C_\beta^\alpha$ is not shifted, so the GSO projections are the same as for the supersymmetric theory. 
Defining the vectors
\beq
\label{eqn: 77}
\vec{\ell}~\equiv~ (r_1\ell_1, r_2\ell_2)~,~~~~~ \vec{n}~\equiv~ (r_1n_1, r_2n_2)~,
\eeq
we can then use the
trace formula in Eq.~(\ref{trace-formula}) in order to write the
non-vanishing contributions as  
\begin{align}
\label{eqn: 38pp}
Z'_{n=0}(\tau)&=~\frac{{\cal M}^{2}}{\tau_{2} \eta^{10} \overline{\eta}^{22}}\sum_{\vec{\ell}={\rm odd}} Z_{\bf \vec{\ell},0}
\sum_{\alpha_{4},\beta_{4}} e^{2\pi i {\bf e\cdot Q}} 
\, \Omega{\tiny\begin{bmatrix}
\alpha_{4}\\
\beta_{4}
\end{bmatrix}}\nonumber \\
&=~\frac{{\cal M}^{4}r_1r_2}{\tau^2_{2} \eta^{12} \overline{\eta}^{24}}\sum_{\vec{\ell}={\rm odd}}e^{-\frac{\pi}{\tau_2}|\vec{\ell}|^2 } 
e^{2\pi i {\bf e\cdot Q}} \sum_{\alpha_{4},\beta_{4}} 
\Omega{\tiny\begin{bmatrix}
\alpha_{4}\\
\beta_{4}
\end{bmatrix}}
\end{align}
where the $\Omega$'s are the expressions for the supersymmetric non-CDC theory. 
Note that $e^{2\pi i {\bf e\cdot Q}} $ is an operator that does not depend on $\alpha,\beta$ so in the sum 
it becomes a simple overall factor.  
As a result, 
the net effect of the CDC on the Poisson-resummed partition function can be 
summarized as one in which we simply reverse the
signs of the 
contributions with ${\bf e\cdot Q}=1/2$
in the supersymmetric theory.
This is especially straightforward in the large-$\tau_2$ region which dominates the one-loop integral, where 
the $\tau_1$-integral projects onto the physical spectrum and we can simply count the physical states. 
Every fermion that is lifted by the CDC counts $+2$ and every boson $-2$. Conversely the 
partition-function contribution is proportional to the states remaining unshifted in the spectrum after CDC, namely $2(N^i_b-N^i_f)$, at some degenerate mass level $i$. 
 
After the $\tau_1$-integral has fixed the level-matching condition, we then find from Eq.~(\ref{cosconstdef})
that
\begin{eqnarray}
\label{eqn: 79}
\Lambda&=& {r_1r_2}{\cal M}^4 \int_{\frac{1}{\mu^2}\approx 1}^\infty\frac{d\tau_2}{\tau_2^4}\sum_{\substack{\vec{\ell} = {\rm odd}\/ \\ {\rm level} \: i}}(N^i_f-N^i_b) e^{-\frac{\pi}{\tau_2}|\vec{\ell}|^2}e^{-\pi\tau_2 {\alpha'm^2_{i}}} 
\end{eqnarray}
 where ${m^2_i}$ is the physical mass of the state $i$, and where we reiterate that $N_b^i$ and $N_f^i$ counts the number of {\it unshifted} bosons and fermions at the $i$'th mass level. The lower limit $\mu^{-2}$ reflects the fact that we are neglecting the $\tau_1$ dependence of the UV end of the fundamental domain. Note that this expression is completely general for any CDC. 
 
Writing $\Lambda = \sum_{{\rm level}~i}\Lambda_i$, we see that there are then two types of contribution, depending on whether the states are massless or massive. Assuming that $r_1$ is the smaller radius, we find that the 
contribution from massless states is given by
\begin{align}
\Lambda_0&=~ \frac{{2r_1r_2}{\cal M}^4}{\pi^3}(N^0_f-N^0_b) \sum_{\vec{\ell}={\rm odd}} {|\vec{\ell} |^{-6}} \left[1- {\cal O} (e^{-\pi|\vec{\ell}|^2 {\mu^2}}) \right] \nonumber \\
& =~ \frac{4r_1r_2{{\cal M}^4}}{\pi^3}(N^0_f-N^0_b)  \, (2r_1)^{-6}\, \zeta\left(6,\frac{1}{2}\right) + ...\nonumber \\
&=~ {r_1r_2}{{\cal M}^4}\, (N^0_f-N^0_b)\, \frac{\pi^3}{240 r_1^6} +... \, ,
\end{align}
where $\zeta(a,b)$ is the Hurwitz zeta-function and where 
a factor of two arises from $\ell=\pm 1$. Decompactifying to five dimensions by taking the $r_2\rightarrow \infty $ limit and factoring out the infinite volume $r_2$ reproduces the single compactified-dimension result  of
Eq.~(\ref{preleadingD}) for $D=6$. 

 For the contributions from massive states, we can use a saddle-point approximation with the saddle at $\tau_2=|\vec{\ell}|/\sqrt{\alpha'} m_i$ (which is valid for $\sqrt{\alpha '} m_i \ll 1$) to obtain
\beq
\Lambda_{i\neq 0} ~=~ r_1r_2{\cal M}^4 (N^i_f-N^i_b) \sum_{\vec{\ell}={\rm odd}} |\vec{\ell}|^{-7/2}(\sqrt{\alpha'} m_i)^{5/2} e^{-2\pi \sqrt{\alpha'}m_i |\vec{\ell}|} \left[1- {\cal O} \left( \frac{1}{2\pi |{\vec{\ell}}| \sqrt{\alpha'} m_i }\right)\right]\, . 
\eeq
Again the $r_2\rightarrow \infty$ decompactification limit yields the 5D correction 
already anticipated in Eq.~\eqref{correction1}.
Note that the subleading terms which we neglect in the saddle-point 
approximation are larger than the $n\neq 0$ contributions, 
so it would not be appropriate to consider the latter at this order of approximation.

\section{String models with exponentially suppressed cosmological constants}
\setcounter{footnote}{0}

Given the important preparatory model-building work in Sects.~V and VI,
we are now finally ready for the centerpiece of this paper:
the construction of non-supersymmetric string models with exponentially suppressed 
cosmological constants.  As discussed in Sect.~II, only such models 
will have suppressed one-loop dilaton tadpoles 
of the sort that would be required for stability, even without
supersymmetry.  In particular,
the exponential suppression of their cosmological constants suggests that we might have a hope
of stabilizing such models within field theory, with the stability issues having the same status and degree of severity
as in a supergravity theory with supersymmetry softly broken by non-perturbative 
field-theory effects. Indeed, it is entirely 
possible that field-theory effects can dominate over the exponentially suppressed dilaton-tadpole term, 
or that there is some interesting interplay between the two effects. 
Moreover, even if these models are not fully stabilized, the strong suppression
of their cosmological constants
suggests that any ``rolling'' which might arise is likely to occur sufficiently slowly 
as to be tolerable within the cosmological history of the universe.

As anticipated in Fig.~\ref{fig:map}, these models will have 
a number of semi-realistic characteristics that will enable us to
identify them as being either SM-like, Pati-Salam-like, or resembling a unified extension
thereof [such as flipped $SU(5)$ or $SO(10)$].
Despite these features,
we nevertheless emphasize 
that these models are not fully realistic and contain a number of
direct phenomenological flaws.
However, our main purpose in this paper is not to construct a single candidate
model describing the universe, but rather to provide an existence proof
illustrating that
non-supersymmetric models with suppressed cosmological constants can
indeed be constructed --- some even with semi-realistic features.
Moreover, our construction techniques suggest that such models are only the tip of the iceberg
in terms of what is logically possible.
This can then hopefully pave the way for more refined model-building and future
enhanced phenomenological studies.

In this section, we shall present our models.
As discussed in Sect.~IV, one way to guarantee exponential suppression
of the cosmological constant
is to construct string models exhibiting equal numbers of massless 
bosonic and fermionic degrees of freedom --- \ie, models for which $N_b^0=N_f^0$ ---
and that is the route we shall follow here.
A summary of the models we shall present also appears in Appendix~D.~
The subsequent three sections will then analyze some of their formal and 
phenomenological features.

\subsection{Warm-up:  ``Unified'' $SO(10)$ and flipped $SU(5)$ theories}

We begin with the warm-up exercise of constructing $N^0_b=N^0_f$ unified theories. 
By ``unified'', we simply mean that the gauge group is (semi-)simple and
contains the Standard-Model gauge group as a subgroup.
In particular, at this level of analysis we shall not be concerned
with whether the Higgses required for breaking the GUT symmetry are present.

In order to build such unified models, we can strip away the $V_{5,6,7}$ vectors from our previous constructions.
However, in order to construct a model with the desired leading-order cancellation of the cosmological constant, 
we shall make use of two additional facts.  The first is that the CDC vector ${\bf e}$ can act both on the right-moving (space-time) side, where it adjusts the structure of the supersymmetry breaking, and simultaneously on the left-moving (gauge) side, where it adjusts the structure of the gauge group. As discussed above, the only constraint on the form of this vector is that one should retain ${\bf e\cdot e}=1$ in order to have modular invariance.  Indeed, with the vectors $V_{1,...,4}$ as above, we find that the CDC 
vector
\begin{equation}
{\bf e} ~=~ {\scriptstyle\frac{1}{2}}[~00~101~101~ | ~1011~00000~000~00011111~]
\end{equation}
generates $N^0_b=N^0_f$ in the toroidally compactified $\calN=2\rightarrow \calN=0$ theory.
(Note that for convenience --- and also so that we can more easily identify bosonic and fermionic sectors --- the basis we are using here is related to that in Table~\ref{table:D} as $V_1\rightarrow \overline{V_1+V_0}$ and $V_2\rightarrow \overline{V_2+V_0}$.)

The second simplifying feature arises from the fact that, as per Eq.~(\ref{eq:triv}), 
only the untwisted sectors can contribute to the non-vanishing partition function 
of the spontaneously broken theory. Therefore we can start by finding an 
$\calN=2\rightarrow \calN=0$ theory with $N^0_b=N^0_f$, such as the one above. 
Any orbifold twisting we later add in order to generate a chiral theory is then guaranteed 
(with a suitable adjustment of structure constants) to preserve the 
cancellation of $N^0_b - N^0_f$ because it halves the number of degrees of freedom of bosons and fermions that contribute with an ${\mathbf e}$ dependence. (Indeed often --- but not always --- this orbifold twisting simply halves the 
massless degrees of freedom.) The only constraint on the orbifold action is that in at least one sector it should overlap with precisely one-half of the CDC elements when written in the real formalism (with the possibility that more general twisted sectors may arise through the introduction of additional untwisted boundary-condition vectors). 
This additional constraint on the orbifolding means that it is somewhat more difficult to find twisted matter. 

We should stress that the choice of structure constants $k_{ij}$ throughout this process is crucial: 
just as for our pedagogical example with $N^0_f\neq N^0_b$, a poor choice of $k_{ij}$ can break supersymmetry by discrete torsion before the CDC is even applied,  or it can leave the model with ${\cal N}=1$ supersymmetry 
after the CDC.  In each case the presence or absence of spacetime supersymmetry can also be seen directly
at the level of the partition function.    One should therefore generally check 
that removing the CDC restores supersymmetry, as does reversing the choice of $k_{14}$. 
Overall, as summarized in the Appendix, this means the models are rather constrained.

In Appendix~D, we present an $SO(10)$ example of a unified model with $N_b^0=N_f^0$.  This model has 
fundamental ${\bf 10}$'s as well as eight complex ${\bf 16}$'s in the untwisted sector --- all quasi-supersymmetric ---
 with 328 complex degrees of freedom in total. 
Moreover, by re-introducing a $V_7$ vector, we can construct a flipped-$SU(5)$ model with $N_b^0=N_f^0$;  this too 
appears in Appendix~D.  This model has twisted massless matter as well as untwisted matter.
As usual, the $\overline{\alpha V} = 0$ sector gives rise to the gravity multiplet and adjoint gauge bosons that are not removed by this particular choice of $\mathbf{e}$. Furthermore, there is no massless gravitino or dilatino, and likewise there are no corresponding massless gauginos in the $\overline{V_0+V_1}$ sector. 
We have confirmed that these models
each have $N_b^0=N_f^0$ at the level of the
spectrum, and in the original and large-radius
Poisson-resummed partition functions.
This can also be seen in the untwisted partition function, which contains no constant term.

\subsection{An SM-like model}

In a similar way, it is also possible
to construct an SM-like model with $N_b^0=N_f^0$.
Recall that our pedagogical model in 
Sect.~VI.B.2
had a  single complex Higgs pair which remains massless as well as four generations of 
matter fields in total, two from the untwisted $V_2$ sector and two from the twisted sectors. 
Given this, it is possible to construct similar models
with $N_b^0=N_f^0$;  one such model is presented in 
Appendix~D.  In this example, there are $N_b^0=N_f^0=136$ complex massless bosons and fermions. 

The full gauge group of this model is given in Eq.~(\ref{SMgg}), and we retain
the same identification of $U(1)_Y$ as  in the previous model.
The resulting mass spectrum from the $\overline{\alpha V} = 0$ and $\overline{V_0+V_1}$ sectors 
is the same as that
summarized in the Tables~\ref{table:E1} and~\ref{table:E} above, including the ``Higgses'' 
of the $V_0$ sector. Note that the orbifold condition is satisfied because $b_3$ overlaps half of the entries in ${\mathbf e}$.  

The model has two entire supersymmetric chiral generations arising in the $V_0+V_2$ sector with the spectrum shown in Table~\ref{table:F1}. There also appears to be a third untwisted generation in the $V_0+V_1+V_4+V_7$ and $V_0+V_1+V_4+3V_7$ sectors.  In this model the twisted ($b_3$ and $b_3+V_4$) sectors provide mainly singlets with additional Higgs/Higgsinos.  
In our view this is a remarkable model. It has chiral generations of SM-like matter and is clearly non-supersymmetric, but it nevertheless also has 
equal numbers of massless bosons and fermions and hence an exponentially small one-loop cosmological constant. 

\subsection{A Pati-Salam-like model}

An alternative route to achieving $N_b^0=N_f^0$ is to remove the final breaking to unitary gauge groups which is driven by $V_7$. As we mentioned earlier, this vector makes the task of building a consistent modular-invariant model significantly more difficult.  As a result, constructions without $V_7$ are significantly less constrained than those with $V_7$.
The enlarged theory is in this case Pati-Salam-like.  One can then achieve $N^0_b=N^0_f$ by choosing parameters as presented in Appendix~D.  Note that this is essentially the SM-like model presented above but without the additional $V_6$ and $V_7$ vectors.   There are now $N_b^0=N_f^0=208$ complex bosons and fermions. 

The full gauge group for this model is given in Eq.~(\ref{PSgg}).
The spectrum for $SO(2N)$ representations can be decomposed under the corresponding $U(N)$ in a complex basis of world-sheet fermions, so that for example the adjoint of $SO(4)\sim SU(2)_L\otimes SU(2)_R$ is a ${\bf 1\oplus 4\oplus \overline{1}}$ of $U(2)$ [which is related as $U(2)\supset SU(2)\equiv SU(2)_L$]. 
The NS-NS sector produces the usual gravity multiplet as well as the adjoint gauge bosons, including 
the ${\bf 15}$ and ${\bf 6}$ adjoint gauge bosons of the visible sector. Again there is no massless gravitino or dilatino and there are no corresponding massless gauginos in the $\overline{V_0+V_1}$ sector. 

In addition, as expected from the SM-like model from which this theory derives, 
the removal of both $V_6$ and $V_7$ naturally enables more pairs of light Higgs scalars and singlets to survive the GSO and orbifold projections in the NS-NS sector. Indeed, we find more than one pair of Higgses and singlets that remain massless after CDC. Likewise more Higgsino-like states are left massless;  these are states which 
appear in the $\overline{V_0+V_1}$ sector, where we refer to the basis used in the Appendix. 
The resulting mass spectrum from the $\overline{\alpha V} = 0$ and the $\overline{V_0+V_1}$ sectors 
is summarized in Tables~\ref{table:G} and~\ref{table:H}.       

Specifically, the complex scalar electroweak doublets $\mathbb{H}$ that survive the GSO and the orbifold projections 
are
\beqn
\label{eqn: 95}
\mathbb{H}_1~\equiv~\{H_{U_1}, H_{D_1}\}&=&\psi^{56}_{-\frac{1}{2}}|0\rangle_R\otimes \overline{\psi}^{4,5}_{-\frac{1}{2}}\overline{\eta}^1_{-\frac{1}{2}}|0\rangle_L\nonumber\\
\mathbb{H}_2~\equiv~\{H_{U_2}, H_{D_2}\}&=&\psi^{56}_{-\frac{1}{2}}|0\rangle_R\otimes \overline{\psi}^{4,5}_{-\frac{1}{2}}\overline{y}^{36,45}_{-\frac{1}{2}}|0\rangle_L\nonumber\\
\mathbb{H}_3~\equiv~\{H_{U_3}, H_{D_3}\}&=&\chi^{36}_{-\frac{1}{2}}|0\rangle_R\otimes \overline{\psi}^{4,5}_{-\frac{1}{2}}\overline{\eta}^3_{-\frac{1}{2}}|0\rangle_L\nonumber\\
\mathbb{H}_4~\equiv~\{H_{U_4}, H_{D_4}\}&=&\chi^{36}_{-\frac{1}{2}}|0\rangle_R\otimes \overline{\psi}^{4,5}_{-\frac{1}{2}}\overline{\omega}^{45}_{-\frac{1}{2}}|0\rangle_L\nonumber\\
\mathbb{H}_5~\equiv~\{H_{U_5}, H_{D_5}\}&=&\chi^{45}_{-\frac{1}{2}}|0\rangle_R\otimes \overline{\psi}^{4,5}_{-\frac{1}{2}}\overline{\eta}^2_{-\frac{1}{2}}|0\rangle_L\, .
\eeqn
Note that in these expressions, the labels of the left-moving internal complex fermions are 
different from those of the pedagogical SM-like theory discussed above. 
The purpose of this labelling is to ensure that the horizontal 
symmetries in the PS-model are completely aligned.
Similarly, the singlets $\mathbb{X}$ and exotic states $\mathbb{E}$ that survive the projections are
\beqn
\label{eqn: 96}
\mathbb{X}_1~\equiv~\{X_1, X'_1\}&=&\psi^{56}_{-\frac{1}{2}}|0\rangle_R\otimes \overline{\eta}^2_{-\frac{1}{2}}\overline{\eta}^3_{-\frac{1}{2}}|0\rangle_L\nonumber\\
\mathbb{X}_2~\equiv~\{X_2, X'_2\}&=&\psi^{56}_{-\frac{1}{2}}|0\rangle_R\otimes \overline{\omega}^{45}_{-\frac{1}{2}}\overline{\eta}^2_{-\frac{1}{2}}|0\rangle_L\nonumber\\
\mathbb{X}_3~\equiv~\{X_3, X'_3\}&=&\chi^{36}_{-\frac{1}{2}}|0\rangle_R\otimes \overline{\eta}^1_{-\frac{1}{2}}\overline{\eta}^2_{-\frac{1}{2}}|0\rangle_L\nonumber\\
\mathbb{X}_4~\equiv~\{X_4, X'_4\}&=&\chi^{36}_{-\frac{1}{2}}|0\rangle_R\otimes \overline{y}^{36,45}_{-\frac{1}{2}}\overline{\eta}^2_{-\frac{1}{2}}|0\rangle_L\nonumber\\
\mathbb{X}_5~\equiv~\{X_5, X'_5\}&=&\chi^{45}_{-\frac{1}{2}}|0\rangle_R\otimes \overline{\eta}^1_{-\frac{1}{2}}\overline{\eta}^3_{-\frac{1}{2}}|0\rangle_L\nonumber\\
\mathbb{X}_6~\equiv~\{X_6, X'_6\}&=&\chi^{45}_{-\frac{1}{2}}|0\rangle_R\otimes \overline{y}^{36,45}_{-\frac{1}{2}}\overline{\omega}^{45}_{-\frac{1}{2}}|0\rangle_L\nonumber\\
\mathbb{X}_7~\equiv~\{X_7, X'_7\}&=&\chi^{45}_{-\frac{1}{2}}|0\rangle_R\otimes \overline{\omega}^{45}_{-\frac{1}{2}}\overline{\eta}^1_{-\frac{1}{2}}|0\rangle_L\nonumber\\
\mathbb{X}_8~\equiv~\{X_8, X'_8\}&=&\chi^{45}_{-\frac{1}{2}}|0\rangle_R\otimes \overline{y}^{36,45}_{-\frac{1}{2}}\overline{\eta}^3_{-\frac{1}{2}}|0\rangle_L\nonumber\\
\mathbb{E}&=& \chi^{45}_{-\frac{1}{2}}|0\rangle_R\otimes \overline{\psi}^{1,2,3}_{-\frac{1}{2}}\overline{\omega}^{36}_{-\frac{1}{2}}|0\rangle_L\, .
\eeqn
Of course, there are additional hidden-sector states that also survive the GSO and orbifold projections. 
However, beyond their contributions to enforcing $N_0^b=N_0^f$, 
these states will not be relevant for this discussion.

\begin{table}[H]
\centering
\begin{tabular}{|c|l|c|l|l|}
\hline
~Sector~ & ~~States remaining after CDC~~ & ~Spin~ & ~$SU(4)\otimes SU(2)_L\otimes SU(2)_R$~ & ~Particle \\
\hline
\multirow{17}{*}{${\bf{0}} $}  & \multirow{3}{*}{$~\psi^{34}_{-\frac{1}{2}}|0\rangle_R \otimes X^{34}_{-1}|0\rangle_L$} & \multirow{2}{*}{2} & \multirow{2}{*}{~(1, 1, 1)}& ~$g_{\mu\nu}$, $B_{[\mu\nu]}$ \\[0.5em]
& & 0 & &~Dilaton $\phi$  \\[0.5em]
& $~\psi^{56}_{-\frac{1}{2}}|0\rangle_R \otimes X^{56}_{-1}|0\rangle_L$ & 0 & ~(1, 1, 1) & ~Complex radion $\Phi$   \\[0.5em]
& \multirow{4}{*}{$~\psi^{34}_{-\frac{1}{2}}|0\rangle_R \otimes \Psi^i_{-\frac{1}{2}}\Psi^j_{-\frac{1}{2}}|0\rangle_L$} & \multirow{4}{*}{1} &$~({\rm Adj}\/, 1, 1)+$ & \multirow{4}{*}{~Gauge bosons $\mathbb{V}_{\mu}$}   \\
& & &$~(1, {\rm Adj}\/, 1) +$&\\
& & &$~(1, 1, {\rm Adj}\/) +$&\\
& & &$~{\rm Adj}\: G_{\rm hidden}$ &\\[0.5em]
& \multirow{3}{*}{$~\psi^{56}_{-\frac{1}{2}}|0\rangle_R \otimes \Psi^i_{-\frac{1}{2}}\Psi^j_{-\frac{1}{2}}|0\rangle_L$} & \multirow{2}{*}{0} & ~(1, \bf{2}, \bf{2}) & ~Complex scalar $\mathbb{H}_1$ \\[0.5em]
& & &~(1, 1, 1) & ~Complex scalar $\mathbb{X}_1$ \\[0.5em]
& \multirow{2}{*}{$~\chi^{36}_{-\frac{1}{2}}|0\rangle_R \otimes \Psi^i_{-\frac{1}{2}}\Psi^j_{-\frac{1}{2}}|0\rangle_L$} & \multirow{2}{*}{0} & ~(1, \bf{2}, \bf{2}) & ~Complex scalar $\mathbb{H}_4$ \\[0.5em]
& & &~(1, 1, 1) & ~Complex scalar $\mathbb{X}_4$ \\[0.5em]
& \multirow{2}{*}{$~\chi^{45}_{-\frac{1}{2}}|0\rangle_R \otimes \Psi^i_{-\frac{1}{2}}\Psi^j_{-\frac{1}{2}}|0\rangle_L$} & \multirow{2}{*}{0} & ~(1, 1, 1) & ~Complex scalar $\mathbb{X}_7$ \\[0.5em]
& & &~(1, 1, 1) & ~Complex scalar $\mathbb{X}_8$~ \\[0.5em]
\hline
\multirow{12}{*}{~$\overline{V_0+V_1}$~} & \multirow{2}{*}{$~\psi^{56}_0|\alpha\rangle_R  \otimes \Psi^i_{-\frac{1}{2}}\Psi^j_{-\frac{1}{2}}|0\rangle_L$} & \multirow{2}{*}{$\frac{1}{2}$} & ~(1, \bf{2}, \bf{2}) & ~Weyl spinor $\tilde{\mathbb{H}}_2$ \\[0.5em]
& & & ~(1, 1, 1) & ~Weyl spinor $\tilde{\mathbb{X}}_2$ \\[0.5em]
&\multirow{2}{*}{$~\chi^{36}_0|\alpha\rangle_R  \otimes \Psi^i_{-\frac{1}{2}}\Psi^j_{-\frac{1}{2}}|0\rangle_L$} & \multirow{2}{*}{$\frac{1}{2}$} & ~(1, \bf{2}, \bf{2}) & ~Weyl spinor $\tilde{\mathbb{H}}_3$ \\[0.5em]
& & & ~(1, 1, 1)& ~Weyl spinor $\tilde{\mathbb{X}}_3$ \\[0.5em]
&\multirow{4}{*}{$~\chi^{45}_0|\alpha\rangle_R  \otimes \Psi^i_{-\frac{1}{2}}\Psi^j_{-\frac{1}{2}}|0\rangle_L$} & \multirow{3}{*}{$\frac{1}{2}$} & ~(1, \bf{2}, \bf{2}) & ~Weyl spinor $\tilde{\mathbb{H}}_5$ \\[0.5em]
& & & ~(1, 1, 1)& ~Weyl spinor $\tilde{\mathbb{X}}_5$ \\[0.5em]
& & & ~(1, 1, 1)& ~Weyl spinor $\tilde{\mathbb{X}}_6$ \\[0.5em]
& & & ~({\bf 4}, 1, 1)& ~Exotic spinor $\tilde{\mathbb{E}}$ \\[0.5em]
\hline
\end{tabular}
\caption{The $\mathbb{Z}_2$-untwisted visible-sector states of the $\mathcal{N}=1$ $D=4$ 
Pati-Salam model which remain massless after the CDC.  The $\Psi_i$ refer to generic left-moving degrees of freedom, with indices $i, j =1 \ldots  20$.  Here $|\alpha\rangle_R$ refers to  the remaining unspecified Ramond ground states.}
\label{table:G}
\end{table}

Fermionic matter arises in the untwisted $\overline{V_0+V_2}$ sector, transforming  in
the $({\bf 4, 2, 1 }) $ and $( {\bf \bar{4}, 1, 2 )} $  representations
of $SU(4)\times SU(2)_L\times SU(2)_R$, where the SM matter fields are embedded as 
\beqn
\mathbb{F}_L&\equiv&\{Q_L, L_L\} \nonumber \\
\mathbb{F}_R&\equiv&\{e_R, \nu_R, u_R, d_R\}\, .
\label{eqn: 97b}
\eeqn
[Note that the identification of $SU(2)_L\times SU(2)_R \sim SO(4)$ implies that the ${\bf \overline{2}}$ spinor of $SO(4) $ is in the fundamental of $SU(2)_L$ while the ${\bf {2}}$ spinor  is in the fundamental of $SU(2)_R$. Meanwhile the electroweak Higgses $\mathbb{H}$ are in the fundamental of $SO(4)$ which corresponds to the ${\bf ( 2,  2)}$.] The visible matter in this particular example is still quasi-supersymmetric.  We emphasize that in this theory there are additional, extraneous generations of visible matter, since there are no $V_6$ and $V_7$ vectors to break the horizontal symmetries embedded in the gauge group. In particular, there is an unbroken $SO(4)$ horizontal symmetry, arising from the $\overline{y}^{36}, \overline{y}^{45}$ fermions. This allows four chiral generations from the $\overline{V_0+V_2}$ sector rather than the two of the SM-like  theory.  However, there is no massless twisted-sector matter.

\begin{table}[t!]
\centering
\setlength{\fboxsep}{0pt}
\begin{tabular}{|c|l|c|l|l|}
\hline
~Sector~ & ~~States removed by CDC~~ & ~Spin~ & ~$SU(4)\otimes SU(2)_L\otimes SU(2)_R$~ & ~Particle \\
\hline
& & & &\\
\multirow{17}{*}{$~\overline{V_0+V_1}~$}  &\multirow{2}{*}{$~|\alpha\rangle_R \otimes X^{34}_{-1}|0\rangle_L$} & \multirow{1}{*}{$\frac{3}{2}$} & \multirow{2}{*}{~(1, 1, 1)} &  ~Gravitino $\psi_{\mu}$ \\[0.5em]
& & $\frac{1}{2}$ & & ~Dilatino $\widetilde{\phi}$  \\ [0.5em]
& $~|\alpha\rangle_R \otimes X^{56}_{-1}|0\rangle_L$ & $\frac{1}{2}$ & ~(1, 1, 1) & ~Radino $\widetilde{\Phi}$   \\
& \multirow{4}{*}{$~|\alpha\rangle_R \otimes \Psi^i_{-\frac{1}{2}}\Psi^j_{-\frac{1}{2}}|0\rangle_L$ } & \multirow{4}{*}{$\frac{1}{2}$}& $~({\rm Adj}, 1, 1)+$ & \multirow{4}{*}{~Gauginos $\mathbf{\lambda}_{\mu}$}   \\
& & &$~(1, {\rm Adj}, 1) +$&\\
& & &$~(1, 1, {\rm Adj}) +$&\\
& & &$~{\rm Adj}\: G_{\rm hidden}$ &\\[0.5em]
& \multirow{2}{*}{$~\psi^{56}_0~|\alpha\rangle_R \otimes \Psi^i_{-\frac{1}{2}}\Psi^j_{-\frac{1}{2}}|0\rangle_L$} & \multirow{2}{*}{$\frac{1}{2}$} & ~(1, \bf{2}, \bf{2}) &  ~Weyl spinor $\tilde{\mathbb{H}}_1$ \\[0.5em]
& & &~(1, 1, 1) & ~Weyl spinor $\tilde{\mathbb{X}}_1$ \\[0.5em]
& \multirow{2}{*}{$~\chi^{36}_0~|\alpha\rangle_R \otimes \Psi^i_{-\frac{1}{2}}\Psi^j_{-\frac{1}{2}}|0\rangle_L$} &  \multirow{2}{*}{$\frac{1}{2}$} & ~(1, \bf{2}, \bf{2}) & ~Weyl spinor $\tilde{\mathbb{H}}_4$ \\[0.5em]
& & &~(1, 1, 1) & ~Weyl spinor $\tilde{\mathbb{X}}_4$ \\[0.5em]
&  \multirow{2}{*}{$~\chi^{45}_0~|\alpha\rangle_R \otimes \Psi^i_{-\frac{1}{2}}\Psi^j_{-\frac{1}{2}}|0\rangle_L$} & \multirow{2}{*}{$\frac{1}{2}$} & ~(1, 1, 1) & ~Weyl spinor $\tilde{\mathbb{X}}_7$ \\[0.5em]
& & &~(1, 1, 1) & ~Weyl spinor $\tilde{\mathbb{X}}_8$ \\[0.5em]
\hline
\multirow{12}{*}{${\bf{0}}$} &\multirow{2}{*}{$~\psi^{56}_{-\frac{1}{2}}|0\rangle_R\otimes \Psi^i_{-\frac{1}{2}}\Psi^j_{-\frac{1}{2}}|0\rangle_L$} & \multirow{2}{*}{$\frac{1}{2}$} & ~(1, \bf{2}, \bf{2}) & ~Complex scalar $\mathbb{H}_2$ \\[0.5em]
& & & ~(1, 1, 1) & ~Complex scalar $\mathbb{X}_2$ \\[0.5em]
&\multirow{2}{*}{$~\chi^{36}_{-\frac{1}{2}}|0\rangle_R\otimes \Psi^i_{-\frac{1}{2}}\Psi^j_{-\frac{1}{2}}|0\rangle_L$} & \multirow{2}{*}{$\frac{1}{2}$} & ~(1, \bf{2}, \bf{2}) & ~Complex scalar $\mathbb{H}_3$ \\[0.5em]
& & & ~(1, 1, 1) & ~Complex scalar $\mathbb{X}_3$ \\[0.5em]
&\multirow{4}{*}{$~\chi^{45}_{-\frac{1}{2}}|0\rangle_R\otimes \Psi^i_{-\frac{1}{2}}\Psi^j_{-\frac{1}{2}}|0\rangle_L$} & \multirow{2}{*}{$\frac{1}{2}$} & ~(1, \bf{2}, \bf{2}) & ~Complex scalar $\mathbb{H}_5$~~ \\[0.5em]
& & & ~(1, 1, 1) & ~Complex scalar $\mathbb{X}_5$ \\[0.5em]
& & & ~(1, 1, 1) & ~Complex scalar $\mathbb{X}_6$ \\[0.5em]
& & & ~({\bf 4}\/, 1, 1) & ~Exotic boson $\mathbb{E}$ \\[0.5em]
\hline
\end{tabular}
\caption{The $\mathbb{Z}_2$-untwisted visible-sector states of the $\mathcal{N}=1$  $D=4$ Pati-Salam model that are given masses $\frac{1}{2}\sqrt{ R_1^{-2}+R_2^{-2} }$  by the CDC. The $\Psi_i$ refer to generic left-moving degrees of freedom, with indices $i, j =1 \ldots  20$. } 
\label{table:H}
\end{table}

Despite the globally supersymmetric matter spectra, the scalar partners would be expected to pick up masses from RG running in the usual way.  As a result,  the theory is somewhat ``no-scale'' from the point of view of the visible sector, with gauginos dominating the contributions.  (We will discuss this calculation in more detail below.)
However, we also note that the $({\bf \bar{4}, 1,2})$ scalars can play the role of the Higgs field $\mathbb{K}$ for breaking the Pati-Salam gauge symmetry down to the Standard-Model gauge symmetry. The mass spectrum for the generations of matter fields in the theory is summarized in Tables~\ref{table:I} and~\ref{table:J}.
Of course, in presenting a Pati-Salam-like model, there is an implicit assumption that the final stage of symmetry breaking can be consigned to the effective field theory without destabilizing the original theory. 
As evident above, our Pati-Salam model indeed has the necessary Higgses for 
this additional stage of breaking.  

In summary, then, we have presented four different models in this section, all with $N_b=N_f$ and therefore all with exponentially suppressed one-loop cosmological constants and dilaton tadpoles.   While none of these models is completely realistic from all points of view, each can potentially represent a different starting point for a subsequent, more refined model-construction effort.  Moreover, we have seen that our different models exhibit varying levels of success for different phenomenological features. 

In the following sections, we shall primarily be concerned with those aspects of the phenomenology which relate directly to hierarchy and stability issues.  These include scalar masses as well  as Yukawa couplings.  We shall tend to concentrate on our Pati-Salam model as a benchmark in what follows, basing our subsequent analysis on the explicit light spectrum presented here.  We shall also, as needed, refer back to the SM-like model we presented in Sect.~VI.B.2, to which our Pati-Salam model is closely related.

\begin{table}[H]
\centering
\begin{tabular}{|c|l|c|l|l|}
\hline
~Sector~ & ~~States remaining after CDC~~ & ~Spin~ & ~$SU(4)\otimes SU(2)_L\otimes SU(2)_R$~ & ~Particle~ \\
\hline
& & & &\\
\multirow{8}{*}{${V_0+V_2}$}  & $~|\alpha\rangle_R \otimes {\overline{\psi}}^i_0{\overline{\psi}}^a_0|\hat{\alpha}\rangle_L$  & \multirow{2}{*}{$\frac{1}{2}$} & \multirow{2}{*}{~$(\bf{4}, \bf{2}, 1)$} & \multirow{2}{*}{~$\mathbb{F}_L$} \\[0.5em]
&$~|\alpha\rangle_R \otimes {\overline{\psi}}^1_0{\overline{\psi}}^2_0{\overline{\psi}}^3_0{\overline{\psi}}^a_0|\hat{\alpha}\rangle_L$ &  &  &    \\[0.5em]
& $~|\alpha\rangle_R \otimes |\hat{\alpha}\rangle_L$  & \multirow{4}{*}{$\frac{1}{2}$} &\multirow{4}{*}{~$(\bf{4}, 1, \bf{2})$} & \multirow{4}{*}{~$\mathbb{F}_R$} \\[0.5em]
& $~|\alpha\rangle_R \otimes {\overline{\psi}}^4_0{\overline{\psi}}^5_0|\hat{\alpha}\rangle_L$ & & &\\[0.5em]
&$~|\alpha\rangle_R \otimes {\overline{\psi}}^i_0{\overline{\psi}}^j_0|\hat{\alpha}\rangle_L$ & & & \\[0.5em]
&$~|\alpha\rangle_R \otimes {\overline{\psi}}^i_0{\overline{\psi}}^j_0{\overline{\psi}}^4_0{\overline{\psi}}^5_0|\hat{\alpha}\rangle_L$ & & & \\[0.5em]
\hline
\multirow{2}{*}{$~\overline{V_1+V_2}~$} & $~|\alpha\rangle_R \otimes |\beta\rangle_L$ & $0$ & ~$(\bf{4}, \bf{2}, 1)$ & ~Exotic spinor $\mathbb{E}$ \\[0.5em]
&  $~|\alpha\rangle_R \otimes |\beta\rangle_L$ & $0$ & ~$(\bf{4}, 1, \bf{2})$ & ~Complex scalar $\mathbb{K}$~ \\[0.5em]
\hline
\end{tabular}
\caption{Chiral ($\mathbb{Z}_2$-untwisted) multiplets of the $\mathcal{N}=1$, $D=4$ Pati-Salam model 
that remain massless after the CDC. 
Here $i,j\in SU(4)$ and $a\in SU(2)_L\otimes SU(2)_R$. 
The $|\alpha\rangle_R$ represent right-moving Ramond ground states (space-time spinors), 
while $|\hat{\alpha}\rangle_L$ (respectively $|\beta\rangle_L$) 
represent the left-moving Ramond excitations that do 
not (respectively do) overlap with the Pati-Salam gauge group.
Again the multiplets are essentially the decomposition of the $\bf 16$ of $SO(10)$. 
The same decomposition applies for the two massless generations of the $b_3$- and $b_4$- 
 twisted-sector matter fields.
 \label{table:I}}
\end{table}

\begin{table}[H]
\bigskip
\centering
\setlength{\fboxsep}{0pt}
\begin{tabular}{|c|l|c|l|l|}
\hline
~Sector~ & ~~States removed by CDC~~ & ~Spin~ & ~$SU(4)\otimes SU(2)_L\otimes SU(2)_R$~ & ~Particle \\
\hline
& & & &\\
\multirow{2}{*}{${V_1+V_2}$}  & $~|\alpha\rangle'_R \otimes |\beta\rangle_L$ & $\frac{1}{2}$ & ~$(\bf{4}, \bf{2}, 1)$ & ~Spinor  $\tilde{\mathbb{E}}$ \\[0.5em]
& $~|\alpha\rangle'_R \otimes |\beta\rangle_L$ & $\frac{1}{2}$ & $~(\bf{4}, 1, \bf{2})$ & ~Spinor $\tilde{\mathbb{K}}$~~\\[0.5em]
\hline
\multirow{8}{*}{~$V_0+V_2$}~ & $~|\alpha\rangle'_R \otimes {\overline{\psi}}^i_0{\overline{\psi}}^a_0|\hat{\alpha}\rangle_L$  & \multirow{2}{*}{$0$} & \multirow{2}{*}{$~(\bf{4}, \bf{2}, 1)$} & \multirow{2}{*}{~$\tilde{\mathbb{F}}_L$} \\[0.5em]
&$~|\alpha\rangle'_R \otimes {\overline{\psi}}^1_0{\overline{\psi}}^2_0{\overline{\psi}}^3_0{\overline{\psi}}^a_0|\hat{\alpha}\rangle_L$ &  &  &    \\[0.5em]
&$~|\alpha\rangle'_R \otimes |\hat{\alpha}\rangle_L$  & \multirow{4}{*}{$0$} &\multirow{4}{*}{$~(\bf{4}, 1, \bf{2})$} & \multirow{4}{*}{~$\tilde{\mathbb{F}}_R$} \\[0.5em]
&$~|\alpha\rangle'_R \otimes {\overline{\psi}}^4_0{\overline{\psi}}^5_0|\hat{\alpha}\rangle_L$ & & &\\[0.5em]
&$~|\alpha\rangle'_R \otimes {\overline{\psi}}^i_0{\overline{\psi}}^j_0|\hat{\alpha}\rangle_L$ & & & \\[0.5em]
&$~|\alpha\rangle'_R \otimes {\overline{\psi}}^i_0{\overline{\psi}}^j_0{\overline{\psi}}^4_0{\overline{\psi}}^5_0|\hat{\alpha}\rangle_L$ & & & \\[0.5em]
\hline
\end{tabular}
\caption{Chiral ($\mathbb{Z}_2$-untwisted) multiplets 
of the $\mathcal{N}=1$, $D=4$ Pati-Salam model 
which are given masses $\frac{1}{2}\sqrt{ R_1^{-2}+R_2^{-2} }$ by the CDC. 
Here $i,j\in SU(4)$ while  $a\in SU(2)_L\otimes SU(2)_R$.  
The $|\alpha\rangle'_R$ represent right-moving Ramond ground states that are not space-time spinors.}
\label{table:J}  
\end{table}

\section{Interpolation properties of models with suppressed cosmological constants} 
\setcounter{footnote}{0}

Given the exponential suppression of their cosmological constants, it is interesting to investigate the behavior of models with equal numbers of massless bosons and fermions under interpolation. There are several issues of importance. First there is the question of what the exponential suppression means for the spectrum at large radius, and if one expects the small radius limit to exhibit the same form. Connected with this is the interpolation of the cosmological constant $\Lambda(r)$ from  large to small $r$. Two other important issues are whether tachyons can appear at some small critical radius $r\sim 1$, signalling a Hagedorn-like instability, and whether there is restoration of gauge symmetry. 

\begin{figure*}
\begin{center}
  \epsfxsize 7.0 truein \epsfbox {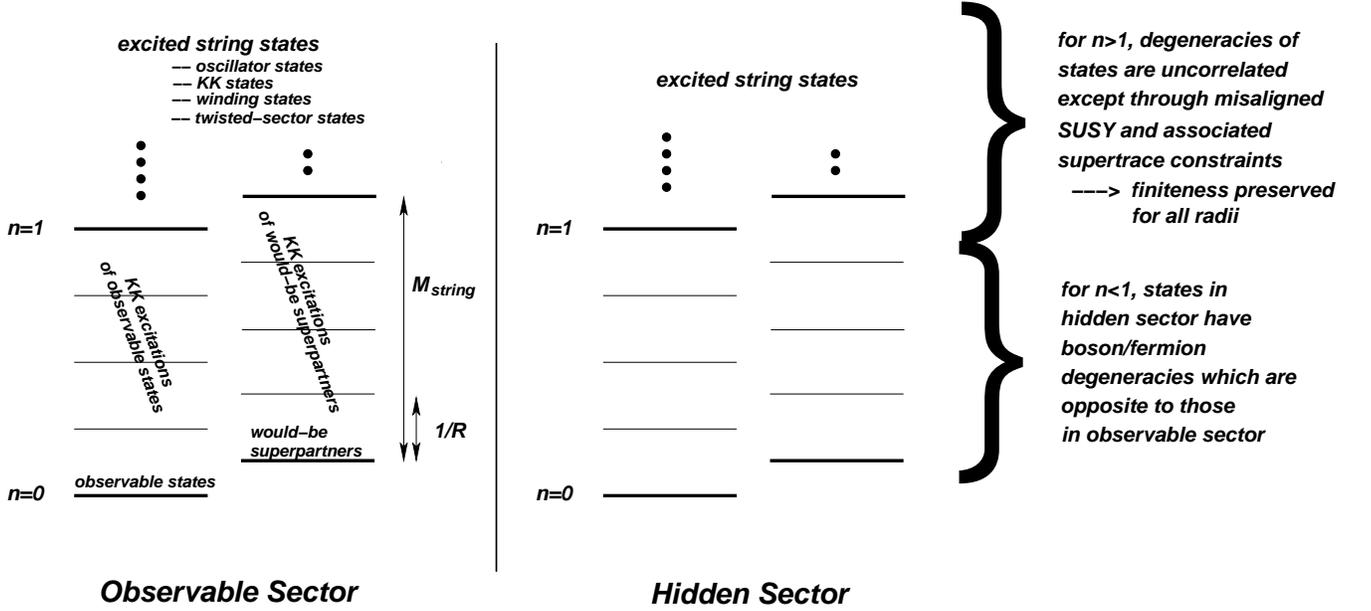}
\end{center}
\caption{The structure of the spectrum of a generic interpolating model with suppressed
cosmological constant in the limit of large interpolating radius.
States with masses below $M_{\rm string}$ (or below $n=1$)  consist of massless observable states,
massless hidden-sector states,
their would-be superpartners, and their lightest KK excitations.
For these lightest states,
the net (bosonic minus fermionic) numbers of degrees of freedom 
from the hidden sector are exactly equal and opposite
to those from the observable sector for all large radii. 
Note that this cancellation of net physical-state degeneracies between the observable and hidden sectors
bears no connection with any supersymmetry, either exact or approximate, in the string spectrum.
Nevertheless, it is this conspiracy between the observable and hidden sectors which suppresses
the overall cosmological constant and enhances the stability of these strings.
For the heavier states, by contrast, the observable and hidden sectors need no longer 
supply equal and opposite numbers of degrees of freedom.  The properties of these
sectors are nevertheless governed by misaligned-supersymmetry constraints, and the
entire string spectrum continues to satisfy the supertrace relations in Eqs.~(\ref{supertraces1}) and
(\ref{supertraces2}).
These relations thereby continue to maintain the finiteness of the overall string theory,
even without spacetime supersymmetry.} 
\label{fig:cdcspectra}
\end{figure*}

We begin by discussing the low-lying spectra of models with suppressed cosmological constants in the limit of large
interpolating radius (\ie, the limit $a\equiv 1/r \to 0$).  
In general, these spectra have the general structure schematically illustrated in Fig.~\ref{fig:cdcspectra}.  
First, there is a natural division 
into a visible sector and a ``hidden'' sector, where by ``visible'' we refer to
the states associated with the Standard Model (or one of its unified extensions)
and where by ``hidden'' we refer to states which do not carry Standard-Model gauge quantum numbers.
Each sector contains not only these states but also their would-be superpartners whose masses are 
generally shifted by an amount $\sim 1/(2R)$;  thus neither sector is supersymmetric 
except in the infinite-radius limit.
The lightest states in each sector consist of the $n=0$ string states, along with their KK excitations.
However, at higher mass levels $M > M_{\rm string}$,
these sectors generically
also contain string oscillator states, winding states, and states coming from twisted sectors. 
This much is general for all semi-realistic interpolating models.
However, for interpolating models with suppressed cosmological constants, 
the observable and hidden sectors have one additional feature:  they contain
exactly equal and opposite net numbers of
bosons and fermionic states with masses $M<M_{\rm string}$.
Indeed, this property holds for all sufficiently small radii;  in particular,
the strict $a\to 0$ limit is not required.

It is easy to understand these properties in terms of the general structure 
of the $\IZ_2$ interpolating models presented in Sects.~III and IV.
(Similar arguments can also be given for other kinds of interpolating models as well, including
the CDC models which have been our primary focus in this paper.)
Recall that the $\IZ_2$ models in Sects.~III and IV generally contain four sectors whose states
contribute to the partition traces $Z^{(i)}$ in Eq.~(\ref{EOmix})
with $i=1,2,3,4$, respectively.
As discussed above Eq.~(\ref{EminusE}), 
the assumption that the $a\to 0$ limit is supersymmetric implies that
$Z^{(2)}= -Z^{(1)}$ at the level of their $q$-expansions.  Indeed, $Z^{(2)}$ contains the contributions
from the would-be superpartners of the states in $Z^{(1)}$, and for low-lying mass levels
the functions $\calE_0$ and $\calE_{1/2}$
serve to tally the KK excitations of these states, which are shifted relative to each other by masses $\sim 1/(2R)$.
However, we also know that $Z^{(1)}$ contains the contributions from our observable massless states, and by assumption
(or through experimental observation) we know that these states do not, by themselves, contain equal numbers of bosonic
and fermionic degrees of freedom.  As a result, the only way to achieve $N_b^0=N_f^0$ (where the superscript `$0$'
signifies the massless level) is to recognize that in such cases $Z^{(1)}$ must itself
contain contributions from both an observable sector and a separate hidden sector.
This hidden sector need not bear any relation to the observable sector except that it must have an equal
and opposite net (bosonic minus fermionic) number of massless states in order to 
satisfy the condition $N_b^0=N_f^0$ for $Z^{(1)}$ as a whole.
Since $Z^{(2)}= -Z^{(1)}$, the same must also be true of the would-be superpartners, whereupon
their respective multiplications by  $\calE_0$ and $\calE_{1/2}$ guarantee that the same will also be true
for their low-lying KK states.

Above $M_{\rm string}$, by contrast, $Z^{(1)}$ may generally have a non-zero net number of physical states.
As a result, the contributions from the observable and hidden sectors need no longer cancel directly.
Nevertheless, the constraints from misaligned supersymmetry (and in particular, the associated supertraces)
will continue to hold across the entire string spectrum.  This includes
all kinds of string states, both observable and hidden, at all mass levels.
In this way, through the contributions of these heavier states,
the string maintains the extraordinary finiteness properties we discussed in Sects.~II and III.

\begin{figure*}
\begin{center}
  \epsfxsize 3.0 truein \epsfbox {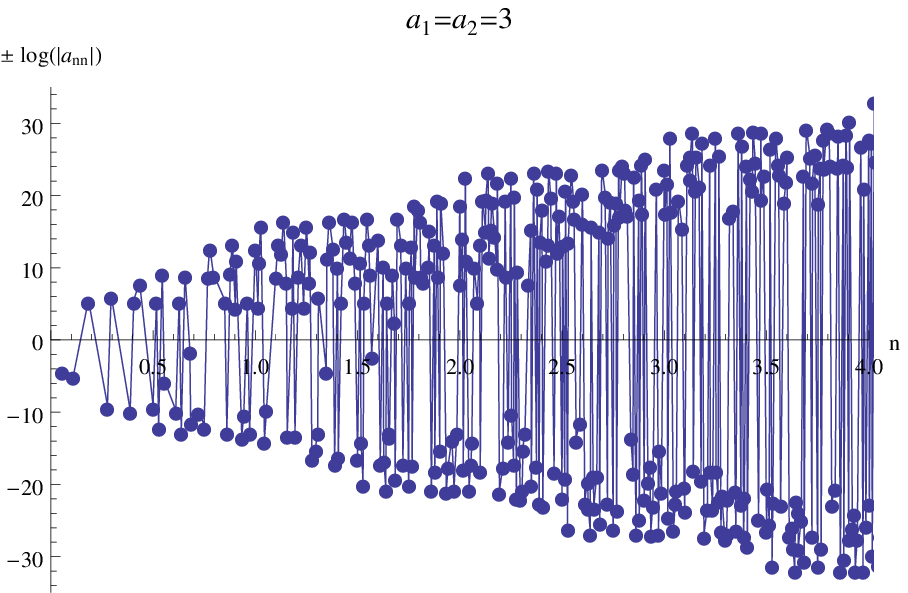}    
\hskip 0.2 truein
  \epsfxsize 3.0 truein \epsfbox {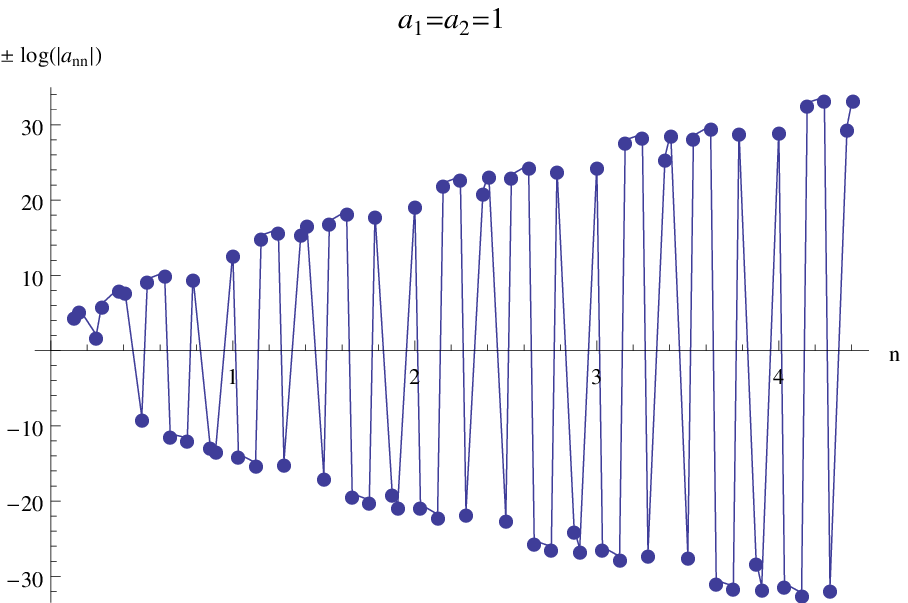}
\end{center}
\begin{center}
  \epsfxsize 3.0 truein \epsfbox {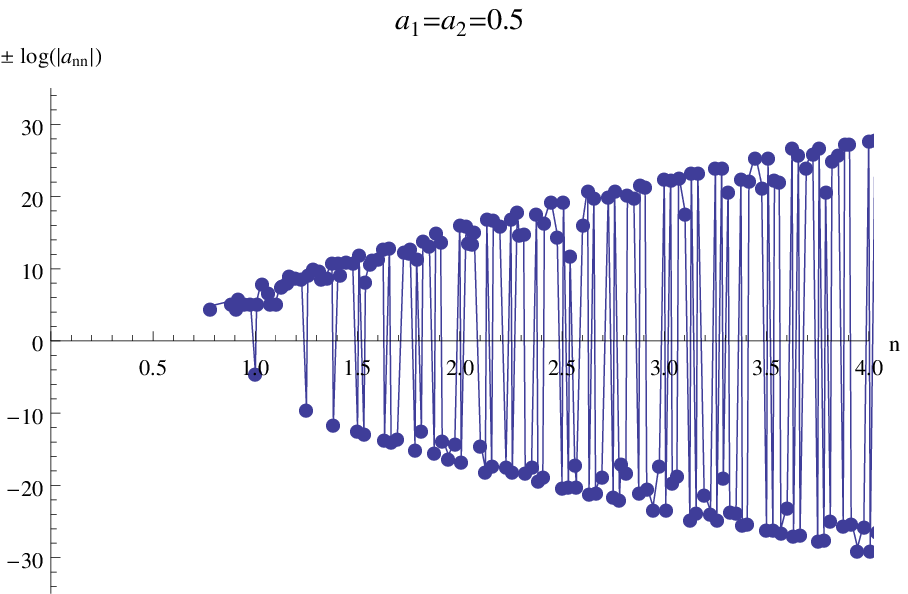}
\hskip 0.2 truein
  \epsfxsize 3.0 truein \epsfbox {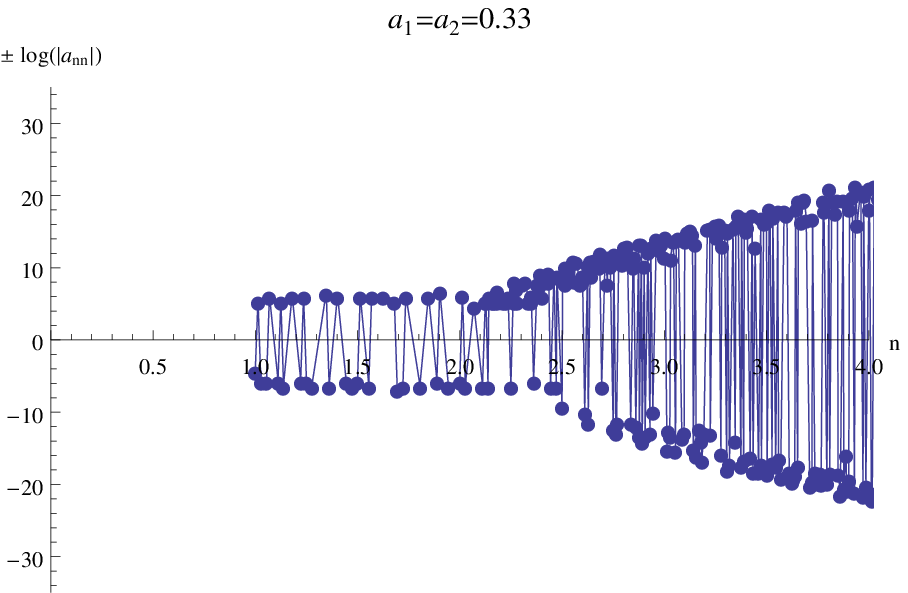}
\end{center}
\begin{center}
  \epsfxsize 3.0 truein \epsfbox {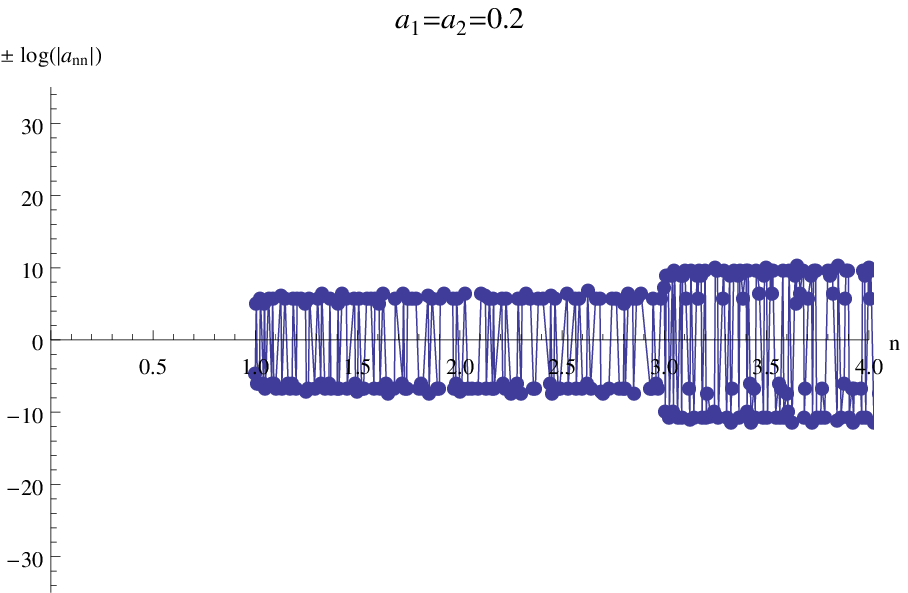}
\hskip 0.2 truein
  \epsfxsize 3.0 truein \epsfbox {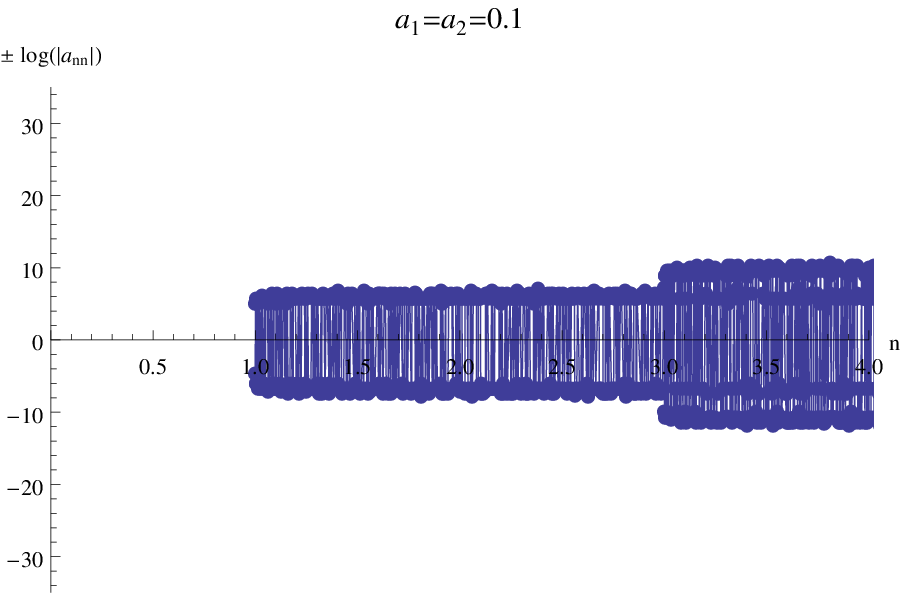}
\end{center}
\caption{Degeneracies of physical states for our Pati-Salam model 
with exponentially suppressed cosmological constant.
In this figure the inverse radius $a=1/r$ varies from $a=3$ (upper left)
to $a=0.1$ (lower right).
Comparing with Fig.~\ref{fig:interplots}, we see that all of the general features associated with general interpolating models in Fig.~\ref{fig:interplots} 
survive here as well, including 
a smoothly growing exponential envelope function for $a\sim {\cal O}(1)$ which slowly deforms into a discretely step-wise growing exponential function as $a\to 0$.   As discussed in Sect.~III, 
this reflects the emerging hierarchy between KK states and oscillator states.
However, we also observe a critical new feature which reflects the fact that this model has an exponentially    
suppressed cosmological constant:  the removal or ``evacuation'' of all 
non-zero net state degeneracies $a_{nn}$ for $n\leq 1$ for
sufficiently small $a$.  Thus, for sufficiently large radius, the spectrum of such models develops an exact 
boson/fermion degeneracy  
for all relevant mass levels $n<1$, even though there is no supersymmetry anywhere in the spectrum.
Indeed, as illustrated in Fig.~\ref{fig:cdcspectra}, this degeneracy does {\it not}\/ occur through a pairing of states 
with their would-be superpartners, but rather as the result of the balancing of non-zero net degeneracies
associated with a non-supersymmetric  {\it observable}\/ sector 
against the degeneracies associated with a non-supersymmetric {\it hidden}\/ sector.
This is why the evacuation of non-zero state degeneracies for $n<1$ is exact for a sufficiently
small (non-zero) inverse radius $a$, even though supersymmetry is only restored in the strict $a\to 0$ limit.}
\label{fig:cdcmissusyplots}
\end{figure*}

Given this structure, we can now examine how the actual physical-state degeneracies
$a_{nn}$ behave as a function of $n$ for such interpolating models with suppressed
cosmological constants.
Our results for the Pati-Salam model in Sect.~VII 
are shown in Fig.~\ref{fig:cdcmissusyplots} 
which may be compared with Fig.~\ref{fig:interplots}.
In all cases, bosonic/fermionic oscillations 
are readily apparent, as required by misaligned supersymmetry.
Moreover, for $a\sim {\cal O}(1)$, our physical-state degeneracies oscillate within smoothly
growing exponential functions;  these envelope functions are determined by the oscillator states,
yet for $a\sim {\cal O}(1)$ the KK and winding states have masses similar to those of
the oscillator states and thus their contributions are not readily distinguishable from those
of the oscillator states.
However, as $a\to 0$, we see that the KK states begin to separate out from the 
oscillator states, which leads to the discretized, step-wise growth in the envelope function
evident in Fig.~\ref{fig:cdcmissusyplots}.
Moreover, we see from 
Fig.~\ref{fig:cdcmissusyplots} that for sufficiently small $a$, the physical-state degeneracies
$a_{nn}$ for all $n<1$ vanish exactly;  this reflects the cancellation of bosonic states
and fermionic states between the observable and hidden sectors respectively, as illustrated
in Fig.~\ref{fig:cdcspectra}.
This ``evacuation'' of net physical-state degeneracies below $n=1$ as $a\to 0$ is thus the hallmark
of interpolating string models 
with suppressed cosmological constants.

The form of these physical-state degeneracies can also be understood from the expressions in Eq.~(\ref{eqn: 55b}). 
At large $r$ (or small $a$) we see that we may neglect terms with non-zero $n_{1,2}$ (just as in the evaluation of the cosmological constant).   Thus the low-mass spectrum is identical to the massless spectrum except for the shift in ``effective'' Kaluza-Klein number induced by ${\bf e\cdot Q}$. However ${\bf{e\cdot Q}}$ shifts an equal number of bosons and fermions in the massless sector,  as well as all of their Kaluza-Klein modes.  Thus the spectrum actually exhibits  
$N_b=N_f$ {\it for all states up to the first-excited string oscillator mass level}\/,
even though there is no supersymmetry in the string spectrum.
Indeed, as remarked above, we have what might be called a ``fake supersymmetry''
which is actually the result of a conspiracy between the observable and hidden sectors of such models.
However, this ``fake supersymmetry'' holds 
merely at the level of counting physical states. 
In particular, we cannot associate these cancelling
bosonic and fermionic states as being superpartners in any way.

The above discussion describes the large-radius (small-$a$) limiting behavior of these models.
We now turn to the opposite limit, namely that with small radius (large $a$). 
The situation here is quite different from that at small $a$. 
Here it is the winding modes $n_1$, $n_2$ that are becoming close packed,
and states with non-zero ``net KK number'' $m_{1,2}+{\bf e\cdot Q}-(n_1+n_2)/2$ are very heavy. 
Let us consider first the winding modes of the massless sector with ${\bf e\cdot Q}=0$ with $N_b^0=N_f^0$. 
Denoting by ${\bf Q}^0$
the charges in this sector,
we see that  
at small $r$ different winding modes require a shift, ${\bf Q} = {\bf Q}^0+{\bf e}(n_1+n_2)$, in order to remain light, with a corresponding shift in KK number, $m_i=-\frac{1}{2}(n_1+n_2)$, to cancel the net KK contribution. Hence at generic but small $r$ only the even winding modes of the ${\bf e\cdot Q}=0$ states are light (where we use `even' to refer to $n_1+n_2$). 

At this point, one might erroneously conclude that these low-lying states correspond to simply taking the physical states with ${\bf e\cdot Q}^0=0$ and mapping them to a set of even winding modes with charge and net KK number given respectively by ${\bf Q}={\bf Q}^0+{\bf e}(n_1+n_2)$ and $m_i=-\frac{1}{2}(n_1+n_2)$, mimicking what happens for the KK modes at large radius.  However this would not be correct because the shift in ${\bf Q}$ also affects the GSO projection, \ie, 
it affects the factor ${\bf g}$ in Eq.~(\ref{eqn: 53}) which includes a phase $2\pi i \beta V\cdot{\bf Q}$. This phase is shifted by a factor $2\pi i (n_1+n_2) \beta V\cdot{\bf e}$ with respect to the non-winding sector, and some of the overlaps $V_i\cdot{\bf e}$ generate $1/4$-integer values. Thus while one particular subset of the winding modes --- namely,
 those with $n_1+n_2=0\,\, $mod(4) ---  still exhibit the $N^0_b=N^0_f$ cancellation of the massless sector, 
the remainder --- those with $n_1+n_2=4 k+2, \,\, k\in {\mathbb{Z}}$ --- 
have different projections and generally do not exhibit this cancellation. 
This is evident within the large-$a$ (small-$r$) plot within Fig.~\ref{fig:cdcmissusyplots},
where we observe that the low-mass states no longer exhibit such cancellations. 
We therefore do not expect the exponential suppression of the cosmological constant 
to be a feature of these models when interpolated to $r\rightarrow 0$ or $a\to\infty$. 

Meanwhile, the winding modes of the states that were given masses by ${\bf e\cdot Q}\neq 0$ at large $r$ have different behavior. Because of the shift in KK number, at small $r$ these states can have low-lying {\it odd}\/ winding modes, with $m_{1,2}$ again compensating to make the net KK contribution vanish. Denoting
by ${\bf Q}^1$ the charges of the original non-winding states, 
we see that  the low-lying winding states have charges shifted as ${\bf Q}={\bf Q}^1+{\bf e}(n_1+n_2)$, with 
\beq
       m_{1,2}+{\bf e\cdot Q}-\frac{1}{2}(n_1+n_2){\bf e}^2~=~0~,
\eeq
so there are indeed odd-winding/KK states with no net KK number. 

Of particular importance now is the fact that the $V_1$ projection is reversed for these low-lying odd-winding modes, but the $V_0$ projection is unaffected, again due to the $2\pi i (n_1+n_2) \beta V\cdot{\bf e}$ shift in the GSO phase. 
Thus the projection in Eq.~(\ref{eqn: 27}) removes the odd winding modes of the gravitini, and moreover odd winding modes of NS-NS tachyons are {\it in principle}\/ allowed. Among the low-lying states at small $r$ there could therefore be winding modes of the NS-NS tachyon with squared masses
\begin{equation}
\frac{\alpha' M^2}{4}~=~-\frac{1}{2} +\frac{n^2_1 r^2_1+n^2_2 r^2_2}{4} ~ .
\end{equation}
Note that such states would become tachyonic when $r_1,r_2<\sqrt{2}$.
Such tachyonic states, if they existed, would have ${\bf e\cdot Q}=\frac{1}{2} \,\, \mbox{mod(1)}$, but it is easy to check that in these models they are projected out (in the shifted GSO projections of the odd-winding sector) by the projections associated with the $V_4$ vector (in the basis of Appendix~D).
Thus we do not expect to find a tachyon-induced (Hagedorn-like) instability at $R\sim 1$, and this
can indeed be verified by inspection of the partition function. 

It is tempting to conclude that the absence of 
such tachyon-induced Hagedorn-like instabilities is a special feature of models with exponentially suppressed cosmological constants.  Such a property certainly applies for generic non-supersymmetric string models 
which interpolate between an $r\to \infty$ endpoint which is supersymmetric
and an $r\to 0$ endpoint which is non-supersymmetric but tachyon-free.
The question, however, is whether all interpolating models with $N_b=N_f$ must be in this category.

Further analyzing the generic spectrum as a function of radius,
we also note that the point $T=U=i$ (or $r_1=r_2=1$) is normally a point of 
enhanced gauge symmetry for $T_2$ compactifications 
where the entire theory can be fermionized in order to take the form of a
fermionic string (see, \eg, Ref.~\cite{Nooij:2004cz}).
Accordingly, in the theory without CDC, it is evident from 
Eq.~(\ref{eqn: 55b}) that additional massless states appear with $m_{1,2}=n_{1,2}=\pm1$ when either $r_1=1$ or $r_2=1$. 
However, {\it in the CDC theory}\/, we observe that such symmetry enhancement cannot occur for states with ${\bf e\cdot Q}=0\,\, \mbox{mod(1)}$ because such enhancement implies $n_1+n_2=odd$ --- \ie,  there would inevitably be a 
non-zero net KK number. Moreover the lightest squared masses
would correspond to
\begin{eqnarray}
\frac{\alpha' M_L^2}{4}&=&\frac{1}{4}\left(-\frac{n_1}{2r_1}\right)^2 +\frac{1}{4}\left(-\frac{n_1}{2r_1}-n_1 r_1\right)^2\nonumber \\
\frac{\alpha' M_R^2}{4}&=&\frac{1}{4}\left(-\frac{n_1}{2r_1}\right)^2 +\frac{1}{4}\left(-\frac{n_1}{2r_1}+n_1 r_1\right)^2~,
\end{eqnarray}
which can satisfy level-matching constraints only when there are equal numbers of left-moving and right-moving
excitations.  These states are therefore eliminated from the spectrum through the projection associated with $V_0$. 
By similar arguments, winding modes of states with ${\bf e\cdot Q}^1=\frac{1}{2}\,\, \mbox{mod(1)}$ cannot lead to additional massless states either. Therefore at the traditional enhanced symmetry point there does not seem to be a direct link to the 4D fermionic string although (as mentioned previously) it is interesting to conjecture that the model there corresponds to a conventional 4D model broken to a tachyon-free non-supersymmetric model by discrete torsion. 

\begin{figure*}[t!]
\begin{center}
  \epsfxsize 5.0 truein \epsfbox {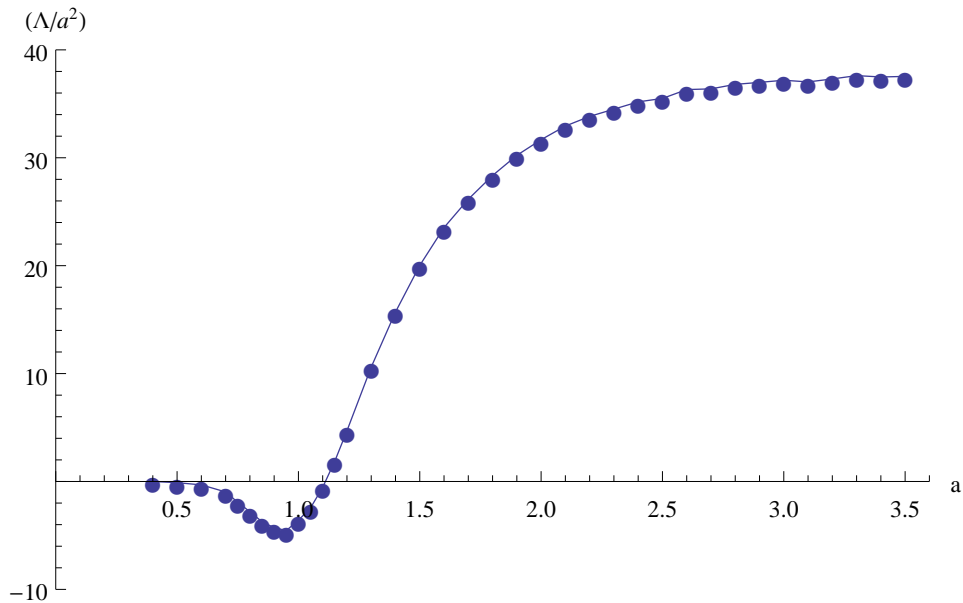}    
\end{center}
\caption{The rescaled cosmological constant $\Lambda/a^2$ for the Pati-Salam model of Sect.~VII, 
        plotted versus $a\equiv \sqrt{\alpha'}/R$.
  For large $a$,  we find that $\Lambda/a^2$ tends to a constant, a feature which is similar to that 
  in Fig.~\ref{fig:lambda} and which indicates that the $a\to \infty$ limit of this model is 
  non-supersymmetric and tachyon-free.
  We also see that the entire curve is finite, which indicates that no tachyons emerge at any intermediate radii.
  Thus this model lacks Hagedorn-like instabilities, as we have already noted in the text.
  However we observe that the small-$a$ behavior of this curve is radically different from that in 
  Fig.~\ref{fig:lambda}.
  First, we see that $\Lambda$ is not power-law suppressed, as in Fig.~\ref{fig:lambda}, but rather
  exponentially suppressed.
  Second, and somewhat surprisingly, 
  we observe that $\Lambda$ has the opposite sign as $a\to 0$ as it does when $a\to\infty$.
  Indeed, we see that the cosmological constant appears to have
  a stable minimum near (but not precisely at) the self-dual radius,
  and moreover that the cosmological constant crosses zero at yet another (slightly higher) radius.
  It is not clear whether there might exist a hidden unbroken supersymmetry at either of these specific radii.}
\label{fig:cdcinterplot}
\end{figure*}

Given these observations, we now numerically evaluate 
the cosmological constant $\Lambda(a)$ of the Pati-Salam model as a function of the inverse radius $a=1/r$.
Our results are shown in Fig.~\ref{fig:cdcinterplot}, and are consistent with the gross
features that one would expect from the above discussion, namely that the cosmological constant is
finite for all radii, exponentially suppressed in the large-radius limit, and radius-independent in the small-radius limit.  This last observation suggests the existence of a zero-radius endpoint model with an entirely non-supersymmetric but tachyon-free spectrum --- one which might well correspond to 
a 6D fermionic string constructed with discrete torsion.
More surprisingly, however,
just above (but not at) the self-dual radius, we find a stable anti-de Sitter minimum. 
This turn-over could indicate a restoration of gauge symmetry and/or supersymmetry, and is similar to the situation encountered in the Type~II models of Ref.~\cite{Angelantonj:2006ut}.

Of course, despite these interesting features which suggest the possibility of stabilizing the {\it radius}\/ modulus
near $a\sim {\cal O}(1)$,
our interest in this paper remains focused on the large-radius limit 
where the one-loop cosmological constant --- and hence the one-loop
dilaton tadpole --- is exponentially suppressed.
Indeed, as discussed in Sects.~II through IV, only under such conditions can we claim to have addressed 
the (more fundamental) {\it dilaton}\/-related stability issues associated with non-supersymmetric strings.
Moreover, in this limit, $\partial \Lambda(r)/dr$ is also suppressed.  Thus we have
an effectively 
flat potential for the radius modulus in this region, similar to what 
exists for softly-broken supersymmetric models of this type. 
Indeed, as already noted at the end of Sect.~II,
any ``rolling'' of the radius modulus which happens sufficiently slowly 
can certainly be tolerated within the cosmological history of the universe.

\section{Phenomenological properties of models with suppressed cosmological constants} 
\setcounter{footnote}{0}

We now turn to some of the phenomenological aspects associated with these models. 
Some of these are quite general and would apply to any non-supersymmetric string model of this type.  However, 
others --- such as natural particle assignments and Yukawa couplings --- are quite model-specific  and are best analyzed within the context of a specific model. But perhaps most interestingly, some features --- in particular, possible exponentially suppressed  radiative contributions to scalar masses --- are closely related to the exponential suppression of the cosmological constant.


\subsection{Natural particle assignments}

We first consider the more  model-independent aspects, such as 
how one might wish to identify the particles of the Standard Model in terms of specific string states. 
Returning to the SM-like compactification presented in Section~VI.B.2 (and neglecting its cosmological constant), 
let us recall that supersymmetry breaking is manifest only for the untwisted generations at leading order. While this supersymmetry breaking will inevitably appear in the twisted generations as well due to RG running 
(for the colored particles the contributions would largely come from the gluino), 
the breaking in the twisted generations will almost certainly be smaller than that 
in the untwisted ones, assuming that untwisted matter exists at all.

Recently it has been argued~\cite{Murayama:2012jh,Dimopoulos:2014aua} that Scherk-Schwarz configurations may enhance ``naturalness'' in the sense advanced in Ref.~\cite{Shadmi:2011hs}. From this point of view, a natural assignment is to take the untwisted generation to be the first generation of the Standard Model and the two twisted generations to be the second and third generations. The large SUSY-splitting within the first generation does indeed then indicate a certain degree of naturalness, and as such it seems like a good starting point 
for having reduced third generation masses 
while evading experimental bounds on, \eg, the gluino, which one would expect to be heavy.
Given that the second generation is also relatively light, there are many flavor-related issues that 
would need to be addressed;  indeed, one would hope that a sufficiently heavy first generation, along with relatively mild partial cancellation of radiative contributions to the third generation, would suffice. 

Within field theory, these would be difficult questions to address as there are threshold contributions from the entire tower of states in the spectrum.  Probably one would have to resort to setting soft terms as boundary conditions at the compactification scale.  However, one of the appealing aspects of the present constructions is that one does not have to rely on field theory;  indeed all the leading effects including thresholds and  RG running can in principle be simply computed from scratch within string perturbation theory.  We shall discuss how such terms would be calculated below, although a comprehensive study would be worthwhile only within a completely realistic SM-like setup. It would of course be interesting to see if the results indeed favored any kind of naturalness in such a model. 


\subsection{Yukawa couplings}

It is useful for pedagogical reasons to begin by considering Yukawa couplings 
within the framework of the SM-like theory of Sect.~VI.B.2 which had $\calN=0$ but $N^0_b\neq N^0_f$. 
In particular, this will inform our eventual discussion below for the Pati-Salam model with realistic Higgs sectors, which 
is the model most closely resembling the pedagogical SM-like theory.

\bigskip
{\underline {\it The SM-like theory with $N_b\not= N_f$}}:~~
For the moment, let us focus on the fact that supersymmetry breaking in the twisted generations is 
generally suppressed, and see how this squares with the couplings.
Along the way, we shall also determine which of the three Higgses in this model is best suited to be {\it the}\/ Higgs. 
The possible Yukawa couplings are determined by the non-SM charges under horizontal $U(1)$ symmetries. The number of these depends on the assignment of the boundary conditions in the defining spin structure. Every model has at least three horizontal $U(1)_{L_k}$ $(k=1,2,3)$ symmetries for the left-moving world-sheet currents and three matching global $U(1)_{R_k}$ $(k=1,2,3)$ symmetries from the right-moving world-sheet currents. With $J_\psi \equiv \frac{1}{2\pi}\int dz \:\psi^\dagger_{-\frac{1}{2}}(z) \psi_{-\frac{1}{2}}(z) $,  these world-sheet currents may be denoted
\begin{align}
\label{eqn: 89}
J_{\overline{\eta}_{i=1..3}}~\rightarrow~U(1)_{L_{i=1..3}} ~; 
~~~~~~~~~ J_{\psi^{56}, \chi^{34}, \chi^{56}}~\rightarrow~U(1)_{R_{1,2,3}}~.
\end{align}
A quick glance at the spectrum reveals that the chiral  matter states arising from the $V_2$, $b_3$, and $b_4$ sectors
sectors and the Higgses, $H_{U_{i=1,2,3}} $ and  $H_{D_{i=1,2,3}} $, carry charges under $U(1)_{L_{i=1,2,3}}$ and $U(1)_{R_{i=1,2,3}}$ respectively. Given that supersymmetry breaking is concentrated in the $V_2$ sector, we adopt the assignment that states from the $V_2$ sector are first generation while those from the $b_3$ and $b_4$ twisted sectors are second and third generations, respectively.

In addition to these symmetries  there are other 
 horizontal $U(1)_{L_{k=4,5, \dots}}$ symmetries from the complexification of pairs of real world-sheet fermions from the subsets $\big\{\overline{y}^{3,4,5,6}\big\}$, $\big\{\overline{\omega}^{3,4}\big\}$ and $\big\{\overline{\omega}^{5,6}\big\}$. Correspondingly the complexified right-moving fermions from the subsets $\big\{{y}^{3,4,5,6}\big\}$, $\big\{{\omega}^{3,4}\big\}$ and $\big\{{\omega}^{5,6}\big\}$ give rise to four $U(1)_{R_{k=4',4,5,6}}$ symmetries assigned as
\begin{align}
\label{eqn: 91}
J_{\overline{y}^{34},\overline{y}^{56}, \overline{\omega}^{56}, \overline{\omega}^{34}}~\rightarrow~U(1)_{L_{4',4,5,6}} 
    ~; ~~~~~~~~~ 
J_{y^{34},y^{56}, \omega^{56}, \omega^{34}}~\rightarrow~U(1)_{R_{4',4,5,6}}~.
\end{align}
The non-vanishing Yukawa couplings for the states from the sectors $V_2$, $b_3$, and $b_4$ then depend on the boundary conditions assigned to the real fermions of the $V_7$ sector.  Indeed, we are ultimately just imposing the requirement
that the charge vectors for the states in a given coupling sum to zero,  ${\bf Q}_1+{\bf Q}_2+{\bf Q}_3={\bf 0}$. 
The states with non-vanishing Yukawa couplings are determined according to the value of~\cite{Faraggi:1992fa}
\begin{equation}
     |V_{7,R_{k+3}}-V_{7,L_{k+3}}|\, ,
\label{eqn: 93}
\end{equation}
where the subscript refers to the element of $V_7 $ corresponding to that particular world-sheet fermion. The value of this parameter, which can be either $0$ or $\frac{1}{2}$, determines which type of coupling is generated, \ie, involving $d_R, e_R$, or $u_R,\nu_R$ respectively. Both couplings cannot be present for the same generation. As mentioned in the previous section, the $V_2$ sector gives rise to two generations which are correlated with the $U(1)_4$ and $U(1)_{4'}$ horizontal symmetries. The $b_3$ and $b_4$ sectors give rise to one generation each, correlated with the existence of $U(1)_5$ and $U(1)_6$, respectively. In total, from all the sectors of the unbroken ${\cal N}=1$ theory, 
we find the couplings
\beqn
     W & \supset &  u_{R_1}H_{U_1}Q_{L_1}+\nu_{R_1}H_{U_1}L_{L_1}+d_{R_{1'}}H_{D_1}Q_{L_{1'}}+e_{R_{1'}}H_{D_1}L_{L_{1'}}+d_{R_2}H_{D_2}Q_{L_2}+e_{R_2}H_{D_2}L_{L_2}\nonumber \\
& &~~~\quad +u_{R_3}H_{U_3}Q_{L_3}+\nu_{R_3}H_{U_3}L_{L_3}+H_{U_1}H_{D_2}\Xi_3+H_{U_2}H_{D_3}\Xi_1+H_{U_1}H_{D_3}\Xi_2+ H_{D_1}H_{U_2}\Xi'_3\nonumber\\
& &~~~\quad +H_{D_2}H_{U_3}\Xi'_1+H_{D_1}H_{U_3}\Xi'_2+\Xi_1\Xi'_2\Xi_3+\Xi'_1\Xi_2\Xi'_3\, ,
\label{eqn: 94}
\eeqn
where indices `$1$' and `$1'$' on the matter fields label the two generations from the $V_2$ sector while indices `$2$' and `$3$' correspond to the two twisted-sector generations. 
All these Yukawa couplings arise with the same degenerate magnitude. 
Note that in Eq.~(\ref{eqn: 94}) we have written our Yukawa couplings 
in the form of superpotential terms although of course the relevant superpartners have been lifted.

For the original $\mathcal{N}=0$ theory defined with the choice of $\mathbf{e}$ in Eq.~(\ref{eqn: 71}), the non-vanishing Yukawa couplings (again written as superpotential terms) are
given by
\beqn
\label{eqn: 98}
W & \supset&  u_{R_1}H_{U_1}Q_{L_1}+\nu_{R_1}H_{U_1}L_{L_1}+d_{R_{1'}}H_{D_1}Q_{L_{1'}}+e_{R_{1'}}H_{D_1}L_{L_{1'}}+H_{U_2}H_{D_3}\Xi_1+H_{D_2}H_{U_3}\Xi'_1\nonumber \\
& &~~~\quad+\Xi_1\Xi'_2\Xi_3+\Xi'_1\Xi_2\Xi'_3\, .
\eeqn
Clearly $H_{U_1}$ is to be identified as the actual Higgs in this theory, which remains the only massless Higgs after CDC, while  $H_{D_2}$ and $H_{U_3}$ become massive. 

It is interesting that this generic situation is quite similar to the phenomenological `one-Higgs-doublet' model~\cite{Davies:2011mp}, which was adopted in the context of Scherk-Schwarz breaking in Ref.~\cite{Dimopoulos:2014aua}. 
Likewise one would look to `wrong-Higgs' couplings coming from the K\"ahler potential to generate masses for the bottom and tau, and to higher-order corrections (as usual) to fill in the missing charm mass. 
Again it should be possible to calculate these from first principles. 
We observe, in this connection, that any superpotential-like terms would violate the usual non-renormalization theorems.
Such terms would therefore  be present but subject to a suppression given by the scale of supersymmetry breaking.

\bigskip
{\underline {\it The Pati-Salam Theory with $N_b=N_f$}}:~~ 
We now turn to the Pati-Salam theory which has $N_b^0=N_f^0$ together with a realistic Higgs sector. 
This model admits a greater number of horizontal $U(1)$ symmetries than those of the above SM-like theory, 
but it contains the same $U(1)_{L_{1,2,3,5,6}}$ and $U(1)_{R_{1,2,3,5,6}}$ symmetries as expected. 
Since $V_7$ is not present, we may simply determine the Yukawa terms by imposing the condition 
that the overall charge $\mathbf{Q}_{\rm total}$ of the coupled states should vanish. 
It turns out that there exist many possible trilinear terms. 
However, just as for the SM-like theory, the Higgs is associated with that 
state, namely $\mathbb{H}_1$,   
 which is never projected out by the CDC. 
Again we can list the interactions as superpotential terms:  
\beqn
\mathbb{W} &\supset & \mathbb{F}_{R_1}\mathbb{H}_1\mathbb{F}_{L_1}+
  \mathbb{H}_3\mathbb{X}_5\mathbb{H}_{1}+\mathbb{H}_5\mathbb{X}_3\mathbb{H}_{1}
  +\mathbb{H}_4\mathbb{X}_7\mathbb{H}_{1} ~ ,
\label{eqn: 99}
\eeqn
where each coupled term involves the relevant component fields, \ie, the fields of the Standard Model-like theory, as defined in Eqs.~(\ref{eqn: 95}), (\ref{eqn: 96}), and~(\ref{eqn: 97b}). For convenience, the label '$1$' here represents {\it all}\/ the generations of matter fields arising from the $V_2$ sector. 

For reasons that will become clear in the following section, we are actually interested in the larger set of all the Yukawa couplings involving the $\mathbb{H}_1$ Higgs and its friends.  These are given by
\beqn
\mathbb{W} & \supset &  \mathbb{F}_{R_1}\mathbb{H}_1\mathbb{F}_{L_1}+ \mathbb{F}_{R_2}\mathbb{H}_4\mathbb{F}_{L_2}+ \mathbb{H}_3\mathbb{X}_5\mathbb{H}_{1}+\mathbb{H}_5\mathbb{X}_3\mathbb{H}_{1}+\mathbb{H}_2\mathbb{X}_6\mathbb{H}_{4}+\mathbb{H}_5\mathbb{X}_2\mathbb{H}_{4}\nonumber\\
& &~~~ \quad+\mathbb{H}_3\mathbb{H}_5\mathbb{X}_{1}+\mathbb{H}_2\mathbb{H}_5\mathbb{X}_{4}+ \mathbb{H}_2\mathbb{H}_3\mathbb{X}_{8}+\mathbb{X}_5\mathbb{X}_3\mathbb{X}_{1}+\mathbb{X}_2\mathbb{X}_6\mathbb{X}_{4}+\mathbb{X}_2\mathbb{X}_3\mathbb{X}_{7} ~.
\label{eqn: 100}
\eeqn
Note that the Higgsinos $\tilde{\mathbb{H}}$ as well as the spinorial singlet fields $\tilde{\mathbb{X}}$ that are not removed by the CDC can be made massive by their Yukawa couplings to the scalar massless fields that remain after CDC, provided that these fields acquire a VEV.


\subsection{Scalar masses}

One interesting aspect of the Scherk-Schwarz mechanism applied to string compactification is that all threshold corrections, scalar masses, {\it etc}\/., are in principle calculable.  Here we shall consider the contributions to scalar 
squared-masses in order to draw some general conclusions. To be concrete, the discussion here will focus on the 
Higgs, but identical treatments can also be applied to the ``soft'' terms associated with all of the scalars.  
We will consider the squared-mass operator for the states 
\begin{equation}
    H_{U_{1},D_{1}}~\equiv ~\psi_{-\frac{1}{2}}^{56}|0\rangle_{R}\times\tilde{\psi}_{-\frac{1}{2}}^{k}\tilde{\psi}_{-\frac{1}{2}}^{l}|0\rangle_{L}~.
\end{equation}
As we have seen, although our Pati-Salam model contains additional light Higgs-like states, this particular state can never be projected out by the CDC.  Thus, in a configuration  in which all extraneous Higgses are lifted by the CDC, this would be the remaining light state. 

In principle,
we can calculate the typical contributions to the scalar two-point functions either in field theory or in string theory.
However, in order to understand the general form of the string results, it is useful first to display the
results in field theory using the Schwinger-time parametrization. 
For example, for a loop of fermions of mass $m$, one has an integral of the form 
\begin{equation}
\Sigma(k^{2})~=~\int\frac{d^{4}q}{(2\pi)^{4}}\mbox{Tr}\left[\frac{i\slashed q-m}{q^{2}-m^{2}}\frac{i(\slashed q+\slashed k)-m}{(q+k)^{2}-m^{2}}\right].
\end{equation}
The $q^{2}$ piece gives a quadratically divergent contribution to the mass of the scalar, 
while the $m^{2}$ piece gives a logarithmic contribution. Using the general formula 
\begin{equation}
\frac{1}{A^{\nu}}~=~\frac{1}{\Gamma(\nu)}\int_{0}^{\infty}dy\, y^{\nu-1}\, \exp(-yA) ~~~~~~{\rm for}~~  {\rm Re}\, A <0~,
\end{equation}
we can express the Euclideanized integrals in the form 
\beqn
I_{n=0,1} &=&   -i\int\frac{d^{4}q}{(2\pi)^{4}}\frac{q^{2n}}{q^{2}+m^{2}}\frac{1}{(q+k)^{2}+m^{2}} \noindent\\
       &=&  -\frac{i}{16\pi^{2}}\int_{0}^{\infty}\int_{0}^{t}dt\, ds\, \exp\left(-m^{2}t+\frac{s^{2}}{t}k^{2}\right)\times\begin{cases}
           t^{-2} & n=0\\
           2/t^3 + k^2 s^2 /t^4 & n=1~.
           \end{cases}
\label{eq:poodrop}
\eeqn
This form of the integrals will have a direct correspondence with the one-loop string result.

Now for the string computation. 
Using standard techniques~\cite{String-Calc}, we begin with the ``natural'' scalar vertex operator
\begin{equation}
\mathcal{V}_{-1}^{kl}~=~e^{-\phi}g_{c}\psi^{56}\tilde{\psi}^{k}\tilde{\psi}^{l}e^{ikX}~,
\end{equation}
where our scalar is one of the $H_{U_1,D_1}$ fields.  These fields are  
bifundamentals on the gauge side; note that $k\neq l$. 
For our calculation we require the $0$-picture vertex operator which 
is found by acting upon the above vertex operator with the world-sheet supercurrent. 
Using the local CFT behavior 
we find that the vertex operator is given by 
\beqn
\mathcal{V}_{0}^{kl} & = & \lim_{\omega\rightarrow z}\, e^{\phi}\, T_{F}(\omega)\, \mathcal{V}_{-1}(k,z)\nonumber\\
   &=& g_{c}\, \lim_{\omega\rightarrow z}\, (\omega-z) \, :\psi^{\mu}\partial X_{\mu}+ \psi^{56}\partial {\bar Z}^{56}+ {\bar{\psi}}^{56}\partial {Z}^{56}+\chi y\omega::\psi^{56}\tilde{\psi}^{k}\tilde{\psi}^{l}e^{ikX}:\nonumber \\
 & = & g_{c}\, \left(\tilde{\psi}^{k}\tilde{\psi}^{l}\right)\, \left(\partial Z^{56}-ik.\psi\psi^{56}\right)e^{ikX}~.
\eeqn
where for convenience we have temporarily set $\alpha'=2$. 
With one vertex operator for this field and one for its complex conjugate, 
the amplitude is then 
\beqn
A(k,-k) & = & \int_{\mathcal{F}}\frac{d^{2}\tau}{\tau_{2}^{2}}\int d^{2}z\, \langle\mathcal{V}_{0}(k,z)\mathcal{V}_{0}(-k,0)\rangle\nonumber \\
 & = & -g_{c}^{2}\int_{\mathcal{F}}\frac{d^{2}\tau}{\tau_{2}^{2}}\int d^{2}z\, \langle\tilde{\psi}^{k}\tilde{\psi}^{k'}\rangle\langle\tilde{\psi}^{l}\tilde{\psi}^{l'}\rangle\left(\langle\partial Z^{56}\partial {\bar Z}^{56}\rangle-k^{\rho}k^{\sigma}\langle\psi_{\rho}\psi_{\sigma}\rangle\langle\psi^{56}{\bar\psi}^{56}\rangle\right)e^{ikX(z)}e^{-ikX(0)}, \nonumber\\
\label{poodrop2}
\eeqn
where the $\tau_{2}$ factor comes from the volume of the integration of the position of the second vertex,
and where we have used 
\begin{equation}
\left(ik.\psi\right)e^{ikX}\left(ik.\psi\right)e^{ikX}~=~-k^{\rho}k^{\sigma}\langle\psi_{\rho}\psi_{\sigma}\rangle e^{k^{\mu}k^{\nu}\langle X_{\mu}(z,\bar{z})X_{\nu}(0,0)\rangle}~.
\end{equation}
The second piece in Eq.~(\ref{poodrop2}) 
would be the only non-vanishing part in a supersymmetric theory because it involves
space-time spinors and is thus spin-dependent,   
yielding the wave-function renormalization contribution. 
By contrast, the first piece in Eq.~(\ref{poodrop2}) is the object of interest here 
as it represents a direct contribution to the squared-mass of the Higgs. 
In a supersymmetric theory this term would be multiplied by precisely the same abstruse identity factors that make the cosmological constant vanish, giving the string version of the non-renormalization theorem. In the present case,
however,  we expect this term to be non-zero and
to depend on the scale of supersymmetry breaking. 

In order to evaluate the various factors, we require the periodic bosonic correlation functions on the torus, given by 
\begin{equation}
\left\langle X(z,\bar{z})X(0)\right\rangle~=~-\log|\vartheta_{1}(z)|^{2}+\frac{2\pi}{\tau_{2}}z_{2}^{2}~,
\end{equation}
as well as the fermion two-point functions, given by 
\begin{equation}
S_{ab}~=~\frac{\vartheta_{ab}(z)}{\vartheta_{ab}(0)}\frac{\vartheta'_{11}(0)}{\vartheta_{11}(z)}~.
\end{equation}
The integral is thus proportional to
\begin{equation}
I~=~\int d^{2}z\,\langle\bar{S}_{a_{k},b_{k}}\bar{S}_{a_{l},b_{l}}\rangle\langle\partial Z^{56}\partial {\bar Z}^{56}\rangle|\vartheta_{1}(z)|^{-2k^{2}}\exp\left[k^{2}\frac{2\pi}{\tau_{2}}z_{2}^{2}\right],
\end{equation}
where a sum over spin structures is understood. 
The factor $\langle\partial Z^{56}(z)\partial {\bar Z}^{56}(0)\rangle$ 
is dominated by the classical contribution corresponding to world-sheet instantons: 
these are given by
\begin{eqnarray}
\langle\partial Z^{56}(z)\partial {\bar Z}^{56}(0)\rangle & = & \sum_{\vec{\ell},\vec{n}}\frac{\pi^{2}}{\tau_{2}^{2}}\left|\vec{\ell}+\vec{n}\bar{\tau}\right|^{2}Z_{\vec{\ell},\vec{n}}+\ldots\nonumber \\
 & \approx & \sum_{\vec{\ell}}\frac{\pi^{2}}{\tau_{2}^{2}}|{\vec{\ell}}|^{2}Z_{\vec{\ell},0}
\end{eqnarray}
where again we will neglect the exponentially suppressed $\vec n\neq0$ terms. 
This multiplies a factor
\begin{eqnarray}
I' & = & \int d^{2}z\, \bar{S}_{a_{k},b_{k}}\bar{S}_{a_{l},b_{l}}\times|\vartheta_{1}(z)|^{-2k^{2}}\exp\left(k^{2}\frac{2\pi}{\tau_{2}}z_{2}^{2}\right)\nonumber \\
 & = & \int d^{2}z\left[\mathcal{\bar{P}}-4\pi i\partial_{\bar{\tau}}\log\sqrt{\vartheta_{a_{k}b_{k}}(0)\vartheta_{a_{l}b_{l}}(0)}/\eta(\bar{\tau})\right]|\vartheta_{1}(z)|^{-2k^{2}}\exp\left(k^{2}\frac{2\pi}{\tau_{2}}z_{2}^{2}\right)\nonumber \\
 & = & \int d^{2}z\left[-\partial_{\bar{z}}^{2}\log\vartheta_{1}(\bar{z})-4\pi i\partial_{\bar{\tau}}\log\sqrt{\vartheta_{a_{k}b_{k}}(0)\vartheta_{a_{l}b_{l}}(0)}\right]|\vartheta_{1}(z)|^{-2k^{2}}\exp\left(k^{2}\frac{2\pi}{\tau_{2}}z_{2}^{2}\right)~,
\end{eqnarray}
where $\mathcal{P}$ is the Weierstrass P-function. 

Note that in the large-$\tau_{2}$ limit, this expression resembles the field-theory results in 
the Schwinger proper-time formalism. 
Indeed, just as for the cosmological-constant computation, 
the partition function in the large-$\tau_{2}$ limit will yield a prefactor 
$e^{-2\pi\tau_{2}m^{2}}$ for every physical state of mass $m$. 
Comparison with the field-theory expression in
Eq.~(\ref{eq:poodrop}) shows that within the two-point function, 
the quantity $2\pi\tau_{2}$ plays the role of the Schwinger proper-time parameter $t$, 
while $2\pi z_{2}$ plays the role of the Schwinger parameter 
      $s$.\footnote{At any order of perturbation theory, 
   the imaginary parts of the $z$'s of the vertex operators and the $\tau$'s 
    of the moduli are related to Schwinger parameters in the ``least coalesced'' diagram. 
    This is in accord with the following power counting. 
      the ``least coalesced'' diagram has the topology of a scalar field theory with a 
    cubic coupling. The number of propagators of a genus $g$ diagram 
    (\ie, the number of Schwinger parameters) in such a theory is 
   \begin{equation}
   \Delta~=~V_{\mbox{ext}}+3(g-1)~,
   \end{equation}
   where $V_{\mbox{ext}}$ is the number of external vertices. 
    On the string-theory side, the Riemann-Roch theorem tells us that in a closed-string 
   theory, the number of real moduli, $2\mu$, minus the number of real conformal Killing vectors, $2\kappa$, is 
   \begin{equation}
            2\mu-2\kappa~=~-3\chi~=~ 6(g-1)~.
   \end{equation}
   The path integral is over the $\mu$ complex moduli and the complex world-sheet coordinates of 
   the $V_{\mbox{ext}}-\kappa$ vertices that are not fixed by the conformal Killing group, 
   giving $\Delta$ complex integrals for any topology, 
   as required.}
   
In an expansion in $k^{2}$ and in the large-$\tau_{2}$ limit, one finds that $\vartheta_{10}\sim q^{1/8}$ but the remaining spin structures are exponentially suppressed.  Thus the sum over spin structures in this limit yields precisely one contribution of $4\pi^{2}\tau_{2}$ from the second term for every pair of physical states that couples to the Higgs. Hence the leading factor becomes 
\begin{eqnarray}
I' & = & \pi-4\pi\tau_{2}i\partial_{\bar{\tau}}\log\sqrt{\vartheta_{a_{k}b_{k}}(0)\vartheta_{a_{l}b_{l}}(0)}\nonumber \\
     & & \rightarrow ~~ 4\pi^{2}\tau_{2}\, {\rm Tr}\, \left(\frac{1}{4\pi\tau_{2}}-\frac{Y^{2}}{g_{\rm YM}^{2}}\right)~,
\end{eqnarray}
where $Y^{2}/g_{\rm YM}^{2}$ includes those states coupling to the Higgs in this sector, and where the trace is understood to be weighted by the partition function. [Alternatively, this expression can be deduced from the fact that two three-point tree-level vertices should coalesce onto the correct four-point vertex at short distance, and also by noting that because this Higgs is really a component of the 6D gauge field, the term is proportional to the current-current propagator $\langle j^{a}j^{b}\rangle$.] 
Note that here the coupling $Y$ includes the gauge couplings. 

Putting everything together, we find that the amplitude can be written as 
\begin{equation}
A(k,-k)~=~-\left(2\pi\right)^{4}\frac{g_{\rm YM}^{2}}{16\pi^{2}}\int_{\mathcal{F}}\frac{d^{2}\tau}{4\tau_{2}}\sum_{\alpha,\beta,\mathbf{\ell}}\left(\frac{Y^{2}}{g_{\rm YM}^{2}}-\frac{1}{4\pi\tau_{2}}\right)\frac{|\vec{\ell}|^{2}}{\tau_2^{2}}Z_{\mathbf{\ell,0}}Z\left[\begin{array}{c}
\alpha\\
\beta
\end{array}\right].
\end{equation}

Let us first consider the contributions from massless states in the loop. 
First recall that when calculating the cosmological constant we placed 
an upper limit $\tau_{2}\rightarrow\infty$ on the integral. Strictly speaking, at the infrared limit of the integrals, $\tau_{2}\gg R^2$, the initial Poisson resummation is misleading as all KK and winding modes are exponentially suppressed in the partition function. The integral gives a logarithmic divergence depending on the 
infrared cutoff $\mu_{\rm IR}$ proportional to $\left(N_{bH}^{0}-N^0_{fH}\right)\log\left(\mu_{\rm IR}R\right)$, where in this case $N_{bH}$ and $N_{fH}$ count those bosonic and fermionic states that couple to the Higgs, respectively. This is the expected contribution to the logarithmic renormalization-group running of the Higgs mass below the KK scale. Indeed, in a UV-finite theory, there is one and only one cutoff required, namely the physical (Wilsonian) infrared one leading to renormalization, while issues such as UV divergences and counterterms are simply spurious artifacts of working within an incomplete theory. Since we are interested in the origin of the Higgs mass, we can take its value in the Poisson-resummed expression to be the value at the KK scale, and simply note that there will be logarithmic RG running between this scale and the physical Higgs mass.

The rest of the computation closely follows that of the cosmological constant. 
We can split the contributions into those from massless sectors and those from massive ones. 
The term $({4\pi\tau_{2}})^{-1}$ will be proportional to the overall cosmological constant and therefore inevitably exponentially suppressed. The contribution from the massless-sector terms to the physical 4D Higgs squared-masses are then 
\begin{eqnarray}
   M_{H_{1}}^{2} & = & \frac{1}{16\pi^{2}}\int_{\frac{1}{\mu^{2}}\approx1}^{\infty}\frac{d\tau_{2}}{4\tau_{2}^{5}}\sum_{\mathbf{\ell}={\rm odd},i}Y^{2}\, (N_{fH}^{i}-N_{bH}^{i})\, |\vec{\ell}|^{2}\, e^{-\frac{\pi}{\tau_{2}}|\vec{\ell}|^{2}}e^{-\pi\tau_{2}\alpha'm_{i}^{2}}\nonumber \\
 & \approx & \frac{2}{\alpha'}\, \frac{Y^{2}}{16\pi^{2}}\, (N_{fH}^{0}-N_{bH}^{0})\, \frac{\pi^{2}}{320r_{1}^{6}}~.
\end{eqnarray}
In order to find the physical result we remove the compactification volume factor $r_1r_2$ upon normalizing the kinetic terms. 
Similarly the contributions from the massive states are
\begin{equation}
M_{H_{1}}^{2}~=~\frac{2}{\alpha'}\frac{Y^{2}}{16\pi^{2}}(N_{fH}^{i}-N_{bH}^{i})\sum_{\mbox{{\bf \ensuremath{\ell}}}={\rm odd}}|\vec{\ell}|^{-5/2}(\sqrt{\alpha'}m_{i})^{7/2}e^{-2\pi\sqrt{\alpha'}m_{i}|\vec{\ell}|}~.
\end{equation}
The first of these expressions does not necessarily vanish even if its analogue does for the cosmological constant, because the Higgs couples differently to the states that are projected out by the CDC.  Thus generically 
$(N_{fH}^{0}-N_{bH}^{0})\neq0$ even if $N_f^0-N_b^0=0$.
 As mentioned above, one should bear in mind that one should also include the gauge fields, with $Y\equiv g_{\rm YM}$. 

We can see this explicitly in the case of the Pati-Salam model. 
Inspecting the Yukawa and gauge couplings, we see that the matter 
fields $\mathbb{F}$ and their scalar superpartners $\mathbb{\tilde{F}}$ both remain in the massless spectrum at leading order.  Thus  they do not contribute to the squared mass of $\mathbb{H}_{1}$. 
However both the gauge fields and singlets are projected out in a non-supersymmetric fashion, 
with $Y^{2}$ involving contraction over the massless pairs $\tilde{\mathbb{H}}_{3}\mathbb{\tilde{X}}_{5}$, 
$\tilde{\mathbb{H}}_{5}\mathbb{\tilde{X}}_{3}$, $\mathbb{H}_{4}\mathbb{X}_{7}$ and $A^{\mu}\mathbb{H}_{1}$. 
The net result is a factor that is essentially the coefficient of the one-loop quadratic divergence 
of the Higgs mass in the effective field theory of the massless degrees of freedom: 
\beq
        Y^{2}(N_{fH}^{i}-N_{bH}^{i}) ~\equiv~ C_{2}(\square)\, g_{SU(2)_L}^{2}+C_{2}(\square)\, g_{SU(2)_R}^{2}-Y^{2}
         ~=~ \frac{g_{\rm YM}^2}{2}~.
\eeq
Note that the other scalars, and in particular the superpartners of the matter multiplets, naturally receive similar contributions.

\section{Questions of Scale}
\setcounter{footnote}{0}

Our main purpose thus far has been to establish a {\it stable}\/ framework for studying non-supersymmetric models.
In this section, however, we shall finally approach the question of what might constitute reasonable energy and mass scales for the phenomenologies of such models.
One issue, in particular, concerns the all-important question of whether  
some scalars ---  and in particular the Higgs discussed in the previous section --- might naturally remain light. 
However, there are also other pressing scale-related phenomenological issues, 
such as the potentially large
contributions to the gauge-coupling beta functions 
due to the preponderance of KK modes that necessarily appear in such models.
Indeed, as discussed in Refs.~\cite{Kiritsis:1996xd,Kiritsis:1998en,recentFaraggi},
the latter is the well-known ``decompactification problem'' which tends to mitigate against 
large compactification radii in the (weakly coupled) heterotic string.

Let us first recall some well-known relations between the various different mass scales. 
Inspection of the effective potential reveals that for closed strings,
our four-dimensional gauge and gravitational couplings are related to the
underlying ten-dimensional string coupling $g_s$ through the volume $V$ of compactification:
\begin{equation}
      g_{\rm YM}^{-2} ~=~  g_s^{-2} \, V \, \ell_s^{-6}~,~~~~~~
       M_P^{2}~=~g_s^{-2} V \ell_s^{-8}\, ,
\end{equation}
where the string length is given by $\ell_s=\sqrt{\alpha'}=1/M_s = 1/2\pi {\cal M}$.
Together these give the tree-level relation  
\begin{equation}
\label{sscale}
             M_P ~=~ g_{\rm YM}^{-1} \ell^{-1} = \frac{2\pi \calM}{g_{\rm YM}}~.
\end{equation}
This relation, which can be recast into the somewhat more familiar form $M_s = g_{\rm YM} M_P$, 
suggests that we interpret 
$g_{\rm YM}$ is the four-dimensional gauge coupling at the string scale (here interpreted as a unification scale).

Let us henceforth assume that we have succeeded in ensuring $N_b^0 =N_f^0$ 
(thereby producing an exponentially suppressed cosmological constant).
For six-dimensional CDC's as we have been considering in this paper (and assuming that $N_b=N_f$ holds only for the 
$n=0$ modes), this implies
\beq
         \Lambda^{1/4} ~\sim ~  {\cal M} \, r^{-3/8} e^{-\pi r/{2}}~ 
\label{cosvalue}
\eeq
where $r\equiv \calM R$.
There are then two cases to consider, depending on whether we have also ensured
$N_{fH}^{0}= N_{bH}^{0}$ (thereby producing an exponential suppression for the scalar masses).

If the scalar masses are unsuppressed, then
we have seen that their leading radius-dependence takes the form
\beq
            M_{\rm scalar} ~\sim~   \frac{   g_{\rm YM} \pi  }{\sqrt{640}}   \, {\cal M}\,  r^{-3}  ~ .
\label{scalarmasses}
\eeq
Of course, the inferred/measured values are 
\beqn
          \Lambda^{1/4} &\sim &  10^{-12}{\mbox{ GeV}} \nonumber \\
          M_{\rm scalar} &\sim &  10^2 {\mbox{ GeV}}\, .
\eeqn
Solving simultaneously, we then find that
$r\approx 25$ and ${\cal M}\sim 2\times 10^7$~GeV.
Together, this implies a KK scale of $1/R\sim 10^6$~GeV. 
It is hardly surprising that this scale emerges because in this scenario SUSY breaking is essentially gauge-mediated to the previously massless Higgses. 
Therefore, even if the Higgs is identified as a state emerging from the quasi-supersymmetric twisted sectors,
we would not expect very different conclusions.

Such a small string scale is clearly incompatible with Eq.~\eqref{sscale} 
unless there is an extremely small $g_{\rm YM}(M_s)\sim 10^{-12}$. 
Now, it is indeed a logical possibility that such a tiny tree-level gauge coupling at the 
string scale could run down to values of order one at the electroweak scale --- 
precisely because of the power-law running induced by the $\calN=2$ KK 
modes~\cite{Dienes:1998vh,Benakli:1998pw,Bachas:1999es}. 
In such a scenario, the hierarchy problem remains but is softened in overall magnitude and
is in a sense dimensionally transmuted:  a fine-tuning of one part in $10^{24}$ is required for the gauge couplings, where a huge threshold would have to be balanced against a huge $4\pi /g^2|_{\rm tree}$.
However, such a scenario may be of interest because it is in a sense
the opposite of the ``brane-world''  scenario  --- indeed here it is the gauge couplings that do 
the work of differentiating themselves from gravity by growing large! 

An alternative way that such scales could be incorporated is instead to have 
$g_{\rm YM}\sim 1$ at tree-level and 
arrange for the  contribution from $\calN=2$ KK modes 
either to be absent \cite{Antoniadis:1990ew} or to be ameliorated by having the $\calN=4\rightarrow \calN=2$ 
breaking also occurring spontaneously, as in Refs.~\cite{Kiritsis:1996xd,recentFaraggi}. 
A drawback is then that one would have to accommodate strong coupling in the string theory, 
which can be done by mapping to weakly coupled dual theories above certain intermediate energy scales. 
This mapping would of course depend on the geometry of the compactification.
(See Ref.~\cite{Antoniadis:2000vd} for a review.) 
Despite the fact that the perturbative computations we have performed here would then 
be somewhat suspect, it nevertheless seems likely that the 
exponential suppression of the cosmological constant would survive.  
Indeed, it may be possible to incorporate either of these mechanisms using the formalism presented here. 

Let us now assume instead that the contribution to the ``Higgs'' masses are also exponentially suppressed
because $N_{fH}^{0}=N_{bH}^{0}$. (This is actually relatively easy to achieve.) 
The resulting one-loop scalar mass then takes the form
\beq
    M_{\rm scalar} ~\sim~ \frac{g_{\rm YM}}{2} \, \calM \, r^{-5/4} \, e^{-\pi r}~,
\eeq
replacing Eq.~(\ref{scalarmasses}).
In such a setup, the gauge hierarchy is clearly eliminated (although 
further cancellation of the cosmological constant would be required).
Assuming that the exponential suppression continues to higher order and that a final cancellation of the (still exponentially suppressed) cosmological constant happens within field theory, 
it turns out that 
a value of $r\approx 10$ suffices to bring the Higgs masses to $\sim 100$~GeV with 
a canonical heterotic value of ${\cal M}\sim 10^{17-18}$~GeV. 
Such a value of $r$ still leads to an energy interval 
marked by power-law running of the gauge couplings,
yet gauge coupling unification
can still occur without unacceptable fine-tuning~\cite{Dienes:1998vh}.
In some sense, this scenario is more in the spirit of Ref.~\cite{supertraces} since, in terms of the exponential prefactor originating from the partition function, it yields $M_{\rm scalar}^2\propto \Lambda$. 
It is also more in the spirit of gravity mediation. Unfortunately, 
because of course all the gauginos feel supersymmetry breaking directly,
it is not possible to obtain the cleanest option of a Higgs in a globally supersymmetric sector that couples to supersymmetry breaking only through gravitational interactions. 

The above scenarios clearly rest on the assumption that the theory continues to mask gauge-mediation of 
SUSY breaking to higher orders, which of course would be interesting to investigate. 
However, even if this assumption turns out to be incorrect,
these constructions nevertheless supply a framework
for stable non-supersymmetric string model-building in which 
${\cal M}$ takes its canonical value a little below the Planck scale, 
while generic dimensionful operators would be two-loop suppressed with respect to that.
This alone is of considerable interest.

\section{Conclusions}
\setcounter{footnote}{0}

One of the most challenging aspects of non-supersymmetric strings
is their lack of stability.
While many perturbative strings have unfixed moduli,
non-supersymmetric strings have an added difficulty in that they generically
give rise to non-zero dilaton tadpoles.
This feature afflicts even those non-supersymmetric strings which are free
of physical tachyons at tree level,
and represents a fundamental obstacle for the use of such 
strings as the basis for a non-supersymmetric string phenomenology.

In this paper, we demonstrated that this problem can be overcome 
within a class of perturbative four-dimensional heterotic strings based on 
coordinate-dependent compactifications (CDC's).
We began by discussing several crucial aspects associated with 
the spectra of non-supersymmetric string models 
--- including the importance of 
off-shell states such as the proto-gravitons  and proto-gravitinos  ---
and studied the leading and subleading contributions from these and other states to
their one-loop cosmological constants.
We also discussed in detail the behavior of generic interpolations 
between completely supersymmetric and non-supersymmetric models,
 and the importance of ``misaligned supersymmetry'' in ensuring the finiteness of these theories. 
We stress that despite the fact that such breakings of supersymmetry are spontaneous,  
from a four-dimensional perspective this SUSY breaking is not ``soft'' in the usual sense of that term.  Indeed,
the discrepancy between the numbers of bosonic and fermionic degrees of freedom at adjacent energy levels
does not fall to zero
asympotically but actually grows exponentially with energy. 
Such models are therefore genuinely non-supersymmetric by construction and at all energy scales. 

We then went on to construct phenomenologically appealing models that have equal numbers of massless bosonic
and fermionic degrees of freedom.  As we demonstrated by explicit calculation, such models 
therefore have exponentially suppressed one-loop 
cosmological constants and exponentially suppressed dilaton tadpoles. 
We presented an SM-like model, a Pati-Salam model, a flipped $SU(5)$ model, and even an 
$SO(10)$ grand-unified model.
While none of these models has all of the desirable phenomenological features one 
would want in order to serve as the
starting point for a detailed phenomenological study,
they all nevertheless exhibited 
equal numbers of massless bosonic and fermionic degrees of freedom and thus suppressed instabilities.
Indeed, these properties were verified in three independent ways:  through their partition functions, through
their Poisson-resummed large-radius expansions, and through explicit construction and examination of
their low-lying spectra.
These models thereby avoid the most serious problems associated with non-supersymmetric strings, and may 
point the way towards new directions in non-supersymmetric string model-building and string phenomenology.

As we demonstrated, such models are most easily built by starting with 
existing self-consistent supersymmetric four-dimensional string theories, and lifting them to ${\cal N}=1$ in six
dimensions. 
The resulting theories are then re-compactified back down to four dimensions on an orbifold using a CDC,
which may be viewed as a generalization of Scherk-Schwarz compactification. 
Our basic starting four-dimensional 
model was formulated in the so-called free-fermionic construction, but there is no reason
why our procedure cannot be duplicated within other formalisms. 
All of our models were derived from the same  ${\cal N}=1$ 4D model, and 
chiral generations of matter coming from both twisted and untwisted sectors are possible. 
Moreover, we found that it is straightforward to construct other models 
by altering the boundary conditions assigned to the vectors of the theory
as well as adjusting the choice of CDC vector $\mathbf{e}$. 

Such models all exhibit what we have called a ``fake supersymmetry'' 
which characterizes their low-lying KK spectra.
However, this is not a supersymmetry in any literal sense, since it only relates 
an aggregate of bosonic states in an observable sector to an
aggregate of fermionic states in a presumably hidden sector,
and vice versa.
Moreover, this feature applies only to the KK excitations of the massless modes.
Indeed, the massive string-oscillator excitations do not exhibit 
any such bosonic/fermionic degeneracies,
and instead it is only through a so-called ``misaligned supersymmetry'' at all mass levels
that the finiteness of such strings is ensured.

The general phenomenology of such models is that the gauge sector 
and the untwisted matter sectors may or may not feel the supersymmetry 
breaking directly, with states in these sectors
gaining masses of order $1/R$ where $R$ is the generic compactification radius from six to four dimensions. 
By contrast, the twisted sectors are initially unaffected, with states gaining masses only radiatively. 
We demonstrated by explicit calculation within our Pati-Salam model that even if the cosmological constant is exponentially suppressed, the contributions to the latter do not generally cancel at leading (one-loop) order. 
One appealing aspect of 
this setup is that all of these radiative terms (including their RG running) 
are completely calculable within the string theory, and finite.
Note that the terms we calculated are simply ``soft'' squared-mass terms in a potential, 
and as such they can be either positive or negative depending on the net sign 
of $N_{fH}^{0}-N_{bH}^{0}$ --- \ie, depending on whether more bosonic or fermionic
massless degrees of freedom couple to the Higgs.
A negative sign would lead to additional symmetry breaking, 
with the potential being stabilized in the usual way by quartic terms that 
may also be present in the potential. 

It is worth emphasizing that the 
questions we have addressed in this paper --- involving the cosmological constant,
vacuum stability, and the mass hierarchy for scalars such as the Higgs ---
are some of the most challenging and unique problems facing non-supersymmetric string models.
However, they are all related within the framework of the models we have studied because 
there is only one source of supersymmetry breaking which is in a sense responsible for 
all of them.
Indeed, this is precisely because these models are all intrinsically non-supersymmetric,
even at the string scale.
The tight self-consistency constraints of our string constructions therefore unavoidably tie the resolutions
of these different problems to each other.

As we discussed, there is an interesting range of mass scales for which such a configuration can be 
consistent and produce interesting sufficiently small cosmological-constant values 
and at the same time reasonable radiative physical masses for scalars. 
However low string scales appear to be necessary, at least for this scenario, 
implying ultimately either some form of strong coupling or large gauge threshold correction from KK modes.
However, alternative ranges of mass scales may be capable of avoiding these conclusions, and may be compatible
with perturbative unification near the canonical heterotic string scale. 

It was suggested a long time ago that non-supersymmetric string compactifications would benefit from 
exponentially suppressed cosmological constants if the massless bosons and fermions obeyed $N^0_b=N^0_f$.
This paper has developed these ideas, thereby laying the groundwork for new classes of non-supersymmetric models built entirely within a heterotic string framework.  Our view of the phenomenological viability 
of the resulting models is that they are morally equivalent to non-stabilized supersymmetric strings, in the sense that 
the latter (once supersymmetry is broken) typically also have runaway dilaton potentials comparable to the potential 
for the compactification radius we find here. 
Presumably many of the string-theory and field-theory techniques that have been brought to bear 
on those problem in the supersymmetric case 
could now be applied in this case as well, 
the only difference being that supersymmetry is broken in the string construction itself. 
More generally, it will be necessary to study the complete moduli spaces of these theories in order 
to verify their full stability:  in principle all the massless singlets will receive squared masses that,
as we have seen, could be of either sign depending on the particles that couple in the loops associated
with their radiative corrections.  Clearly one would like to avoid the $F$-flat directions becoming tachyonic. 

There are many possibilities for future work beyond those we have already mentioned. 
For example, the precise nature of the model under interpolation 
(including the identity of the tachyon-free non-supersymmetric six-dimensional string
model at the $R=0$ endpoint) is of some interest, as is the precise form of the potential and its 
relation to an effective softly-broken supersymmetry theory. 
Interpreting the various features associated with
potentials such as that in Fig.~\ref{fig:cdcinterplot}
is likely to be important, especially insofar as the corresponding dynamics is concerned.
On the more phenomenological side,
it would obviously be desirable to construct a complete three-generation SM-like theory with 
an exponentially small one-loop cosmological constant and a working Higgs sector. 
The advantage of such a model would be that all scalar masses would (as we have seen) 
be calculable and finite. Furthermore, being non-supersymmetric by construction, 
the Yukawa couplings would also receive radiative corrections suppressed by powers of $1/R{\cal M}$. 
It would then be interesting to see if these would have a bearing on the hierarchies of the Standard Model,
and how they relate to their corresponding expressions in softly-broken supersymmetric field theories. 

One question that certainly deserves further study 
is whether the  exponential suppression continues beyond one-loop order. 
Indeed, such higher-order questions inevitably depend
not only on the spectra of these theories but also on their interactions.
Given the ``fake supersymmetry'' in the spectrum, it is reasonable to speculate that such 
suppressions continue to persist to higher orders,
especially since in the effective field theory all loops would be expected to experience the same 
cancellations in such a theory while the couplings exhibit a high degree of degeneracy. 
Given the promise of these models, a careful analysis beyond one-loop certainly seems warranted.

In summary, then, we believe that the models and methods presented here can potentially serve as a 
starting point for the development of a bona-fide non-supersymmetric string phenomenology.
While numerous unresolved issues --- both theoretical and phenomenological --- clearly remain,
the evident existence of a large number of models with suppressed dilaton tadpoles
suggests the existence of a huge landscape of potentially stable models with 
varying theoretical and phenomenological features and prospects.
It therefore remains to explore this landscape with all the tools at hand 
in order to determine the extent to which a truly successful non-supersymmetric 
string phenomenology is possible.


\begin{acknowledgments}
\setcounter{footnote}{0}
SA and KRD are happy to acknowledge the hospitality of the Galileo Galilei Institute (GGI)
in Florence, Italy, during June 2013, where this work began.
We are also happy to thank Costas Kounnas, Viraf Mehta, and Gary Shiu for discussions,
and Ben Aaronson and Richard Stewart for critical comments on the manuscript.
The research of KRD was supported in part by the U.S.\ Department of Energy
under Grant DE-FG02-13ER-41976. EM acknowledges support from the STFC.
The opinions and conclusions expressed herein are those of the authors,
and do not represent any funding agencies.
\end{acknowledgments}

\bigskip
 
\appendix

\section{~Notation and conventions for partition functions}
\label{notation}
\setcounter{footnote}{0}

The basic $\eta$ and $\vartheta$  functions are given by
\beqn
    \eta(\tau)  &\equiv&  q^{1/24}~ \displaystyle\prod_{n=1}^\infty ~(1-q^n)~=~
                \sum_{n=-\infty}^\infty ~(-1)^n\, q^{3(n-1/6)^2/2}\nonumber\\
    \vartheta_1(\tau)&\equiv&
                 {\displaystyle -i \sum_{n=-\infty}^\infty (-1)^n q^{(n+1/2)^2/2} }\nonumber\\
    \vartheta_2(\tau)&\equiv&  2 q^{1/8} \displaystyle\prod_{n=1}^\infty (1+q^n)^2 (1-q^n)~=~
                 \sum_{n= -\infty}^\infty q^{(n+1/2)^2/2} \nonumber\\
    \vartheta_3(\tau)&\equiv&  \displaystyle\prod_{n=1}^\infty (1+q^{n-1/2})^2 (1-q^n) ~=~
                \sum_{n=-\infty }^\infty q^{n^2/2} \nonumber\\
    \vartheta_4(\tau) &\equiv& \displaystyle\prod_{n=1}^\infty (1-q^{n-1/2})^2 (1-q^n) ~=~
                \sum_{n= -\infty}^\infty (-1)^n q^{n^2/2} ~
\label{etathetadefs}
\eeqn
where $q$ is the square of the nome, \ie, $q\equiv \exp(2\pi i\tau)$, with $\tau_{1,2}$ respectively denoting ${\rm Re}\,\tau$ and ${\rm Im}\,\tau$. These functions satisfy the identities $\tthree^4 =\ttwo^4+\tfour^4$ and $\ttwo\tthree\tfour=2\eta^3$.
Note that $\vartheta_1(q)$ has a vanishing $q$-expansion and is modular invariant; its infinite-product representation has a vanishing coefficient and is thus not shown.  We have nevertheless included this function here 
because within string partition functions 
it can often play the role of the indicator of the chirality of fermionic states, as discussed below.

In order to simplify and unify the notation --- and also in order to be able to handle
more complicated systems --- we shall now introduce several generalizations
of these functions.  
First, we shall define the more general theta-function of two arguments:
\begin{eqnarray}
\vartheta(z,\tau) & \equiv & \sum_{n=-\infty}^{\infty} \, \xi^n\,  q^{n^{2}/2}~,\nonumber\\
   & = & q^{-1/24}\, \eta(\tau)\, \prod_{m=1}^{\infty}(1+\xi q^{m-1/2})\,(1+\xi^{-1}q^{m-1/2})\, ~,
\end{eqnarray}
where $\xi\equiv e^{2\pi iz}$. 
Similarly, the $\vartheta$-functions with characteristics are defined as 
\begin{eqnarray}
\vartheta{\tiny\begin{bmatrix}
a\\
b
\end{bmatrix}}(z,\tau) & \equiv & \sum_{n=-\infty}^{\infty} e^{2\pi i(n+a)(z+b)} \, q^{(n+a)^{2}/2}\nonumber\\ 
 & = & 
e^{2\pi iab}\, 
    \xi^{a}\, 
    q^{a^{2}/2}\, \vartheta(z+a\tau+b,\tau)~;
\label{eq:jacobi}
\end{eqnarray}
of course these latter functions have a certain redundancy, 
depending on only $z+b$ rather than $z$ and $b$ separately.

For $a,b\in \lbrace 0,1/2\rbrace$, a common ``shorthand'' for these functions is given by  
\beqn
&\vartheta_{00} ~\equiv~ \vartheta{\tiny\begin{bmatrix} 0\\ 0 \end{bmatrix}} ~=~ \vartheta_{3}~~& ~~~~~~~~~~~
\vartheta_{10} ~\equiv~ \vartheta{\tiny\begin{bmatrix} 1/2\\ 0 \end{bmatrix}}~=~ \vartheta_{2}\nonumber\\
&\vartheta_{01} ~\equiv~ \vartheta{\tiny\begin{bmatrix} 0\\ 1/2 \end{bmatrix}}~=~ \vartheta_{4}& ~~~~~~~~~~~
\vartheta_{11} ~\equiv~ \vartheta{\tiny\begin{bmatrix} 1/2\\ 1/2 \end{bmatrix}}~=~ -\vartheta_{1}~;
\label{shorthand}
\eeqn
note that in Eq.~(\ref{shorthand}) we are making no restrictions on the (suppressed) $(z,\tau)$ arguments, and 
are thus implicitly defining two-argument Jacobi functions $\vartheta_i(z,\tau)$ for $i=1,...,4$.  
In general, the functions in Eq.~(\ref{eq:jacobi}) have modular transformations
\beqn 
\vartheta{\tiny\begin{bmatrix} a\\ b \end{bmatrix}}(z,-1/\tau) &=& 
     \sqrt{-i\tau}\, e^{2\pi i a b}  e^{i \pi \tau z^2} 
 \vartheta{\tiny\begin{bmatrix} -b\\ a \end{bmatrix}}(-z\tau, \tau)~,  \nonumber\\
\vartheta{\tiny\begin{bmatrix} a\\ b \end{bmatrix}}(z,\tau+1) &=& 
e^{-i\pi (a^2 + a)}
\vartheta{\tiny\begin{bmatrix} a \\ a+b+1/2 \end{bmatrix}}(z,\tau)~.   
\label{modulartransfs}
\eeqn 
Moreover, in the $\tau_2\gg 1$ (or $|q|\ll 1$) limit, these functions have the leading behaviors
\beqn
\label{eqn: 74}
\eta(\tau) & \sim & q^{1/24}+\ldots\nonumber \\
\vartheta_{00}(0|\tau) & \sim & 1+2q^{1/2}+\ldots\nonumber \\
\vartheta_{01}(0|\tau) & \sim & 1-2q^{1/2}+\ldots\nonumber \\
\vartheta_{10}(0|\tau) & \sim & 2q^{1/8}+\ldots\nonumber \\
\vartheta_{11}(0|\tau) & = & 0~.
\eeqn

World-sheet  bosons and fermions give rise to partition-function contributions which
can be expressed in terms of these functions.
For those world-sheet bosons which are spactime coordinates (which is always the case
for the string constructions we employ in this paper),
the partition-function contributions  
also depend on the spacetime compactification metric.
In general, a single complex extra dimension has a metric which is
conventionally parametrized as
\beq
G_{ij} ~=~ \frac{T_{2}}{U_{2}}\left(\begin{array}{cc}
1 & U_{1}\\
U_{1} & |U|^{2}
\end{array}\right) ~,~~~~~~
B_{ij} ~=~ \left(\begin{array}{cc}
0 & -T_{1}\\
T_{1} & 0
\end{array}\right)~
\eeq
where $T\equiv T_1+iT_2$ and $U\equiv U_1+iU_2$.
In this paper, however, we shall only consider diagonal compactification metrics --- \ie, metrics with 
$T_1=U_1=0$.
For $U_{1}=0$,  the corresponding Poisson-resummed partition function for the compactified complex boson is 
given by
\begin{equation}
    Z_{\bf B}{\tiny\begin{bmatrix}
    0\\
    0 \end{bmatrix}}(\tau)~=~{\cal M}^2 \frac{T_{2}}{{\tau_{2}}|\eta(\tau)|^{4}}\sum_{\mathbf{n},\mathbf{m}}\exp\left\{ -\frac{\pi}{\tau_{2}}\frac{T_{2}}{U_{2}}|m_{1}+n_{1}\tau|^{2}-\frac{\pi}{\tau_{2}}T_{2}U_{2}|m_{2}+n_{2}\tau|^{2}\right\} .
    \end{equation}
We can then identify $R_{1}=\sqrt{T_{2}/U_{2}}$ and $R_{2}=\sqrt{T_{2}U_{2}}$. Conversely, $T_{2}=R_{1}R_{2}$ is
a volume modulus while $U_{2}=R_{2}/R_{1}$ is a complex-structure modulus.

By contrast, the contribution to the total partition function 
from a single complex fermion with world-sheet boundary conditions
$v\equiv \overline{\alpha V}_i $ and $u\equiv \beta V_i $
is given by
\begin{eqnarray}
\label{trace-formula}
Z_{u}^{v} & = & \mbox{Tr}\left[q^{\hat{H}_{v}}e^{-2\pi iu\hat{N}_{v}}\right]\nonumber \\
& = & q^{\frac{1}{2}(v^{2}-\frac{1}{12})}\prod_{n=1}^{\infty}(1+e^{2\pi i(v\tau-u)}q^{n-\frac{1}{2}})(1+e^{-2\pi i(v\tau-u)}q^{n-\frac{1}{2}})\nonumber \\
& = & e^{2\pi iuv}\,
\vartheta{\tiny\begin{bmatrix}
v\\
-u
\end{bmatrix}}(0,\tau)/ {\eta(\tau)}~.
\end{eqnarray}

\section{~Conventions and spectrum of the fermionic string}
\label{ffs}

In this paper, the free-fermionic construction~\cite{Kawai:1986ah,Antoniadis:1986rn,Kawai:1987ew}
serves as the anchor underpinning our models.
We shall here present the salient features of this construction for the special case
of heterotic strings in six uncompactified spacetime dimensions.
Indeed, as explained in Sect.~V,
such models serve as our starting point prior to implementing subsequent coordinate-dependent compactifications.

In the free-fermionic construction, all world-sheet conformal anomalies are cancelled through the introduction of
free real world-sheet fermionic degrees of freedom.   
In particular, there are 16 right-moving and 40 left-moving real Majorana-Weyl fermions on the world-sheet, 
and it is convenient throughout to pair them into complex fermions: 
\begin{equation}
\label{eqn: 0.1}
f~\equiv~\{f_R; f_L\}~\equiv~\{f_{i_R}; f_{i_L}\}\,  ,
\end{equation}
where $i_R=1,\dots,8$ and $i_L=1,\dots,20$. Models are defined by the phases acquired under parallel transport around 
non-contractible cycles of the one-loop world-sheet, 
\begin{align}
\label{eqn: 0.3}
\mathbf{1}: ~~~f_{i_{R/L}}&\rightarrow~ -e^{-2\pi i v_{i_{R/L}}}f_{i_{R/L}} \nonumber \\
\tau: ~~~f_{i_{R/L}}&\rightarrow~ -e^{-2\pi i u_{i_{R/L}}}f_{i_{R/L}}\, ,
\end{align}
which we collect in vectors written as
\begin{align}
\label{eqn: 0.2}
v\equiv\{v_R; v_L\}&\equiv~\{v_{i_R}; v_{i_L}\} \nonumber\\
u\equiv\{u_R; u_L\}&\equiv~\{u_{i_R}; u_{i_L}\}\, ,
\end{align}
where $v_{i_R}, v_{i_L},u_{i_R}, u_{i_L} \in [-\frac{1}{2},\frac{1}{2})$. 
The spin structure of the model is then given in terms of a set of  basis vectors ${V}_i$~\cite{Kawai:1987ew}. 
Consistent models are constrained by the modular-invariance conditions, 
invariance of the world-sheet  supercurrent, and correct space-time spin-statistics;  
all of these constraints will be satisfied so long as 
\beqn
\label{eqn: 1.1}
m_jk_{ij}&=&0 ~~~~~~~~~~~~~~~ {\rm mod}\,(1) \nonumber\\
k_{ij}+k_{ji}&=&{V}_i\cdot {V}_j ~~~~~~~~~ {\rm mod}\,(1) 
\nonumber\\
k_{ii}+k_{i0}+s_i &=& \frac{1}{2}{V}_i\cdot {V}_i 
~~~~~~~ {\rm mod}\,(1)~, 
\eeqn
where the $k_{ij}$ are otherwise arbitrary structure constants that completely specify the theory, 
where $m_i$ is the lowest common denominator amongst the components of ${V}_i$,
and where $s_i\equiv V_i^1$ is the spin-statistics associated with the vector $V_i$. 
The basis vectors span a finite additive group $G=\sum_k\alpha_k{V}_k$ where 
$\alpha_k\in \lbrace 0,..., m-1\rbrace$, each element of which describes the boundary conditions
associated with a different individual sector
of the theory.  Within each sector $\overline{\alpha V}$,  the physical states are those
which are level-matched and whose fermion-number operators ${N}_{\overline{\alpha V}}$ 
satisfy the generalized GSO projections
\begin{equation}
\label{eqn: 3b}
{\: V}_i\cdot {N}_{\overline{\alpha V}}~=~\sum_jk_{ij}\alpha_j+s_i-{\: V}_i\cdot\overline{\alpha {\: V}} \:\: \mbox{mod}\:(1) ~~~~~ {\rm for~all}~~i~.
\end{equation}
The world-sheet energies associated with such states are given by 
\beq
\label{eqn: 2}
M^2_{L,R} ~=~\sum_{{\ell}}\left\lbrace E_{\overline{\alpha V^{\ell}}} + \sum_{q=1}^\infty\left[(q-\overline{\alpha V^{\ell}})\overline{n}_q^{\ell}+(q+\overline{\alpha V^{\ell}}-1)n_q^{\ell}\right]\right\rbrace 
           -\frac{(D-2)}{24}+\sum_{i=2}^D\sum_{q=1}^\infty qM_q^i 
\eeq
where $\ell$ sums over left- or right world-sheet fermions, 
where $n_q, \overline{n}_q$ are the occupation numbers for complex fermions,
where $M_q$ are the occupation numbers for complex bosons, 
where $D$ is the number of uncompactified spacetime dimensions, 
and where $E_{\overline{\alpha V^{\ell}}}$ is the vacuum-energy contribution 
of the $\ell^{\rm th}$ complex world-sheet fermion:
\begin{equation}
\label{eqn: 9}
E_{\overline{\alpha V^{\ell}}}~=~\frac{1}{2}\left[(\overline{\alpha V^{\ell}})^2-\frac{1}{12}\right]\, .
\end{equation}
Level-matching then simply requires that 
$M^2_{L} = M^2_{R}$.

When we need to refer to them explicitly, we label the fermions in the conventional manner:
\begin{itemize}
\item two complex space-time fermions, denoted by $\psi^{34}$, $\psi^{56}$, which correspond to the transverse modes of the $\psi^\mu$, where $\mu=1,\dots, 6$; 
\item two complex internal fermions, denoted by $\chi^{34}$, $\chi^{56}$, which are present from the original 10D heterotic string model;  and
\item eight real right-moving, internal fermions, denoted by $y^{3,\dots,6}, \omega^{3,\dots,6}$, which are obtained from the fermionization of each compactified bosonic coordinate in the 6D theory.
\end{itemize}
The left-moving world-sheet fermions consist of 20 complex degrees of freedom:
\begin{itemize}
\item 16 complex left-moving fermions, denoted by $\overline{\psi}^{1,\dots,5}, \overline{\eta}^{1,\dots,3}, \overline{\phi}^{1,\dots,8}$, which are present from the 10D heterotic theory;  and
\item 8 real left-moving, internal fermions, denoted by $\overline{y}^{3,\dots,6}, \overline{\omega}^{3,\dots,6}$, corresponding to the internal right-moving fermions obtained from the fermionization procedure. 
\end{itemize}

In terms of the fields listed above, the world-sheet supercurrent is defined as
\begin{equation}
\label{eqn: 4}
T_F(z)~=~\psi^\mu(z)\partial_zX_{\mu}(z)+\sum_{I=3}^{6}\chi^I y^I\omega^I
\end{equation}
Moreover, the vector of $U(1)$ charges for each complex world-sheet fermion is given by
\begin{equation}
\label{eqn: 5}
\mathbf{Q}~=~{N}_{\overline{\alpha V}}+{\overline{\alpha V}}
\end{equation}
where ${\overline{\alpha V}}$ is $0$ for an NS boundary condition and $-\frac{1}{2}$ for a Ramond. 

This has only been a quick summary of the salient features of the free-fermionic construction.
There are, however, numerous subtleties which come into play when dealing with necessarily real world-sheet fermions,
especially if there is to be a subsequent coordinate-dependent compactification.
For this reason, extreme care is required when constructing and analyzing models, and one must adopt
a consistent set of phase conventions pertaining to the GSO projections and real-fermionic modes.
For this paper, however, the conventions we have adopted are exactly those of Ref.~\cite{Kawai:1987ew}.

 
\section{~SUSY breaking by discrete torsion}

In this Appendix, we demonstrate the claim, made in Sect.~VI, that
SUSY breaking by discrete torsion occurs in the fermionic formulation 
when some combination of boundary condition phases {\it not overlapping the gravitini} has the ``wrong'' choice of structure constants $k_{ij}$.

To see this, we first note that in 4D any ${\cal N}=1$ supersymmetric model in the fermionic formulation can be written without loss of generality in terms of the following vectors:
\begin{eqnarray}
\label{eqn: 26b}
V_{0} & = & -\frac{1}{2}\left[1\left(111\right)^{3}|\left(1\right)^{22}\right]\nonumber \\
V_{1} & = & -\frac{1}{2}\left[1\left(100\right)^{3}|\left(0\right)^{22}\right]\nonumber \\
V_{i\geq2} & = & ...
\end{eqnarray}
This basis is always possible because the ${V_0}$ vector must always be present for modular invariance, and because there must be gravitini in the supersymmetric model. The sector in which these appear can be taken to define the ${V_1}$ sector. In addition we may assume that the right-movers have only $0$ and $-1/2$ boundary conditions. Therefore, since the lowest possible vacuum energy on the right-moving side is $-1/2$, a tachyon can appear \textit{only} if there are no right-moving NS excitations. 

Let us now consider the $V_{0,1}$ projections on the gravitini, 
\begin{eqnarray}
\label{eqn: 27}
V_{0}\cdot N+\frac{1}{4}\left(1-\Gamma\right) & = & k_{01}+\frac{1}{2}-V_{0}\cdot V_{1}\nonumber \\
V_{1}\cdot N+\frac{1}{4}\left(1-\Gamma\right) & = & k_{11}+\frac{1}{2}-V_{1}\cdot V_{1}\,\,\,\,\,\,\: \mbox{mod}\,(1)~,
\end{eqnarray}
where $\Gamma=\Gamma_{V_{1}}=\Gamma_{V_{0}}$ are the chirality projections where the vectors overlap with the Ramond states; for the gravitini they are necessarily degenerate.
By inspection the massless gravitini have no excitations.
It then follows that  $V_{0}\cdot N=V_{1}\cdot N=0$ for them. Thus these equations are compatible if and only if $k_{01}+k_{11}=0$ mod(1), which must be true
since 
$k_{10}+k_{01} = k_{11}+k_{10}= 0$ by the relations in Eq.~(\ref{eqn: 1.1}). 
However, an incompatibility for these states can occur if there is an additional vector (or combination of vectors) that does not overlap with $V_1$. This is because if there were an overlap, then any projection would simply fix the definition of chirality of a subset of the spinors. 

Let us call this additional vector $V_X$, so that 
\begin{eqnarray}
V_{X}\cdot N & = & k_{X1} \,\,\,\,\,\,\: \mbox{mod}\, (1)~.
\end{eqnarray}
If $k_{X1}=\frac{1}{2}$ then the gravitini are projected out and supersymmetry is broken. 
This conclusion is general.  Indeed, if it were instead a combination of vectors that did not overlap with $V_1$, then it would be the corresponding linear combination of $k_{i1}$ that would have to sum to $\frac{1}{2}$ in order
to break supersymmetry.  (Note that a $V_X$ completely overlapping with $V_1$ could also be incompatible, but this is again equivalent to a new vector $V_X\rightarrow \overline{V_X+V_0}$ that has no overlap with $V_1$.) 
Such models correspond to supersymmetry breaking by discrete torsion. 

Given this, one can then prove the following: {\it tachyons can be present in the resulting non-supersymmetric
model only if there are sectors including $V_X$ that have negative vacuum energy}.
To see this, let us assume that there is a would-be tachyonic sector ${\overline{\alpha V}}$. 
Normally this would be just the NS-NS sector;  however, it could also involve some $V_{i}$. 
In the supersymmetric theory, since these states are absent they must be projected out by $V_{1}$ because without $V_{1}$ there is no supersymmetry.  Hence tachyons are absent if and only if
\begin{equation}
\label{eqn: 28}
\sum_{i\in V_{1}={\rm Ramond}}\frac{1}{4}\left(1-\Gamma_{_{i}}\right)~\neq~\alpha_{i}k_{1i}+\frac{1}{2}-V_{1}\cdot\overline{\alpha V} ~~~~~ {\rm for~all}~\Gamma  ~~~~~~~\mbox{mod}\, (1)~.
\end{equation}
Note that $k_{X1} \equiv k_{1X}$ appears in this equation only if $\alpha_X \neq 0$. Thus, in sectors without $V_X$, the projection or otherwise of the gravitini is independent of the projection of the tachyons. 
Since the supersymmetric theory is tachyon-free, it follows that the theories with discrete torsion which have the ``wrong'' choice of $k_{X1}$ also have no tachyons in these sectors, and are thus also tachyon-free.
 It remains to consider sectors that {\it do}\/ contain $V_X$. Let us denote such sectors as 
$V_X+\hat{\alpha}V$. The overlap of $V_X$ with $V_1$ is zero, so the left side of Eq.~(\ref{eqn: 28}) is 
the same as it is for the sector $\hat{\alpha}V$.  Likewise the right side of this equation is $k_{1X}+\hat{\alpha}_i k_{1i}+\frac{1}{2}-V_1\cdot\overline{\hat{\alpha}V}$, which differs only by $k_{1X}$ from the version without $V_X$. Therefore, since there are no tachyons in any $\hat{\alpha}V$ sector without $V_X$, the ``wrong'' choice  $k_{X1}=\frac{1}{2}$ may be consistent with tachyons in any sector that {\it does} contain $V_X$, provided there is negative vacuum energy (on both left and right moving sides). This completes our proof. 

It is of course intuitively correct that the appearance of tachyons must involve the vector responsible for projecting out the gravitino. We now also understand why the ${\cal N}=1 $ model presented in the main text could already be broken even before any Scherk-Schwarz effect simply by a choice of $k_{ij}$: the vector $\overline{V_0+V_4}$ has no overlap with $V_1$. Needless to say it is not hard to avoid such breaking.

In general, it is not difficult to exploit these observations in order to generate non-supersymmetric 
string models whose tree-level spectra are tachyon-free.  
A particularly large collection of such models is presented and analyzed in 
Refs.~\cite{Dienes:1990ij,Dienes:2006ut}.

\section{ ~Non-supersymmetric models with $N_b=N_f$}
\setcounter{footnote}{0}

In this Appendix we collect together the definitions of the 
four non-supersymmetric string models presented in Sect.~VII.~
As discussed in the main text, these models are realized through the
free-fermionic construction followed by a CDC.~  Along the way,
a significant number of constraints are applied: 
\begin{itemize}
\item As always, the boundary-condition vectors and $k_{ij}$ structure 
         constants must satisfy modular-invariance constraints.
\item Likewise, the additional modular-invariance constraints for real fermions must also be satisfied.
\item As part of our construction, we demand that removing the CDC restore supersymmetry.
\item Likewise, we demand that there exist an alternative choice 
     of certain $k_{ij}$'s (such as $k_{14}$ or some other combination) which can also 
       restore supersymmetry.
\item  Finally, we also demand that in at least one twisted sector, the boundary conditions --- including the orbifold vector $b_3$ --- 
must overlap with precisely one half of the entries of the CDC vector ${\bf e}$. 
Moreover, we demand that overlaps of complex phases with the CDC vector not be allowed. 
In this way, we ensure that there is a basis in which the orbifold acts as a 
charge conjugation on the CDC charges (plus possible untwisted phases, depending on the sector).
\end{itemize}
All notation used below is explained in the main text. 

\subsection{$SO(10)$ GUTs}
\label{model-SO10}

A simple $SO(10)$ model is defined by the following spin-structure vectors:
\begin{eqnarray}
V_0&=& - {\scriptstyle\frac{1}{2}}[~11~111~111~ | ~1111~11111~111~11111111~]\nonumber\\ 
V_1&=& - {\scriptstyle\frac{1}{2}}[~00~011~011~ | ~1111~11111~111~11111111~]\nonumber\\ 
V_2&=& - {\scriptstyle\frac{1}{2}}[~00~101~101~ | ~0101~00000~011~11111111~]\nonumber\\ 
b_3&=& - {\scriptstyle\frac{1}{2}}[~10~\overline{1}0\overline{0}~\overline{0}0\overline{1}~ | ~0001~11111~010~10011100~]\nonumber\\ 
V_4&=& - {\scriptstyle\frac{1}{2}}[~00~101~101~ | ~0101~00000~011~00000000~]\nonumber\\ 
{\bf e} 
   &=& {\scriptstyle\frac{1}{2}}[~00~101~101~ | ~1011~00000~000~00011111~]
\end{eqnarray} 
Recall that we are using an overbar in order to indicate conjugation of 
the fermion in a complex notation [\ie, 
   ${0}\equiv (00)_r$, $\overline{0}\equiv (01)_r$, ${1}\equiv (11)_r$ and
   $\overline{1}\equiv (10)_r$]. 
The vector dot products and  $k_{ij}$ structure constants for this model are respectively given by
\begin{equation}
V_i\cdot V_j =\left( \begin{array}{c}1\hspace{0.1cm}0\hspace{0.1cm}0\hspace{0.1cm}0\hspace{0.1cm}0\nonumber \\0\hspace{0.1cm}0\hspace{0.1cm}{\scriptstyle\frac{1}{2}}\hspace{0.1cm}{\scriptstyle\frac{1}{2}}\hspace{0.1cm}{\scriptstyle\frac{1}{2}}\nonumber \\0\hspace{0.1cm}{\scriptstyle\frac{1}{2}}\hspace{0.1cm}0\hspace{0.1cm}1\hspace{0.1cm}0\nonumber \\0\hspace{0.1cm}{\scriptstyle\frac{1}{2}}\hspace{0.1cm}1\hspace{0.1cm}0\hspace{0.1cm}0\nonumber \\0\hspace{0.1cm}{\scriptstyle\frac{1}{2}}\hspace{0.1cm}0\hspace{0.1cm}0\hspace{0.1cm}0\end{array}\right)
\:\: \mbox{mod\, (2)}~,~~~~~
k_{ij} =\left( \begin{array}{c}0\hspace{0.1cm}0\hspace{0.1cm}0\hspace{0.1cm}{\scriptstyle\frac{1}{2}}\hspace{0.1cm}0\nonumber \\0\hspace{0.1cm}0\hspace{0.1cm}0\hspace{0.1cm}{\scriptstyle\frac{1}{2}}\hspace{0.1cm}0\nonumber \\0\hspace{0.1cm}{\scriptstyle\frac{1}{2}}\hspace{0.1cm}0\hspace{0.1cm}0\hspace{0.1cm}0\nonumber \\{\scriptstyle\frac{1}{2}}\hspace{0.1cm}0\hspace{0.1cm}0\hspace{0.1cm}0\hspace{0.1cm}0\nonumber \\0\hspace{0.1cm}{\scriptstyle\frac{1}{2}}\hspace{0.1cm}0\hspace{0.1cm}0\hspace{0.1cm}0\end{array}\right)\, .
\end{equation}
We then find that the resulting gauge group is given by
\beq
         G ~=~ SO(4)\otimes SO(4)\otimes U(1)\otimes
  \underbrace{SO(10)}_{\rm contains~SM}
           \otimes U(1)\otimes U(1)\otimes 
          U(1)\otimes SO(4)\otimes SO(6)\otimes SO(4)~, 
\eeq
where the would-be GUT factor (containing the Standard Model) is indicated explicitly.
This factor can be identified by the appearance of appropriate matter multiplets in 
the $V_0+V_2$ sector. 
The untwisted sectors have $N^0_b=N^0_f=656$ massless real degrees of freedom in total. 
Other examples with, \eg, $N^0_b=N^0_f=400$ are also possible.

\subsection{Flipped $SU(5)$ GUTs}
\label{model-SU5}

An  $SU(5)$ model with $N^0_b=N^0_f=304$ massless real degrees of freedom 
in the untwisted sector can be defined by the following vectors:

\begin{eqnarray}
V_0&=& - {\scriptstyle\frac{1}{2}}[~11~111~111~ | ~1111~11111~111~111~11~111~]\nonumber\\ 
V_1&=& - {\scriptstyle\frac{1}{2}}[~00~011~011~ | ~1111~11111~111~111~11~111~]\nonumber\\ 
V_2&=& - {\scriptstyle\frac{1}{2}}[~00~101~101~ | ~0101~00000~011~111~11~111~]\nonumber\\ 
b_3&=& - {\scriptstyle\frac{1}{2}}[~10~\overline{1}0\overline{0}~\overline{0}0\overline{1}~ | ~0001~11111~010~000~01~111~]\nonumber\\ 
V_4&=& - {\scriptstyle\frac{1}{2}}[~00~101~101~ | ~0101~00000~011~000~01~111~]\nonumber\\ 
V_7&=& - {\scriptstyle\frac{1}{4}}[~00~0\overline{0}\overline{0}~0\overline{2}\overline{2}~ | ~2222~11111~111~222~00~200~]\nonumber\\ 
 {\bf e} &=& {\scriptstyle\frac{1}{2}}[~00~101~101~ | ~0101~00000~000~111~01~101~]\end{eqnarray} 
The dot product of vectors and the $k_{ij}$ are respectively given by
\begin{equation}
V_i.V_j =\left( \begin{array}{c}1\hspace{0.1cm}0\hspace{0.1cm}0\hspace{0.1cm}0\hspace{0.1cm}1\hspace{0.1cm}{\scriptstyle\frac{1}{2}}\nonumber \\0\hspace{0.1cm}0\hspace{0.1cm}{\scriptstyle\frac{1}{2}}\hspace{0.1cm}{\scriptstyle\frac{1}{2}}\hspace{0.1cm}{\scriptstyle\frac{3}{2}}\hspace{0.1cm}{\scriptstyle\frac{1}{2}}\nonumber \\0\hspace{0.1cm}{\scriptstyle\frac{1}{2}}\hspace{0.1cm}0\hspace{0.1cm}1\hspace{0.1cm}1\hspace{0.1cm}{\scriptstyle\frac{3}{2}}\nonumber \\0\hspace{0.1cm}{\scriptstyle\frac{1}{2}}\hspace{0.1cm}1\hspace{0.1cm}0\hspace{0.1cm}1\hspace{0.1cm}1\nonumber \\1\hspace{0.1cm}{\scriptstyle\frac{3}{2}}\hspace{0.1cm}1\hspace{0.1cm}1\hspace{0.1cm}1\hspace{0.1cm}{\scriptstyle\frac{3}{4}}\nonumber \\{\scriptstyle\frac{1}{2}}\hspace{0.1cm}{\scriptstyle\frac{1}{2}}\hspace{0.1cm}{\scriptstyle\frac{3}{2}}\hspace{0.1cm}1\hspace{0.1cm}{\scriptstyle\frac{3}{4}}\hspace{0.1cm}0\end{array}\right)
\:\: \mbox{mod\, (2)}~,~~~~~
k_{ij} =\left( \begin{array}{c}0\hspace{0.1cm}0\hspace{0.1cm}0\hspace{0.1cm}{\scriptstyle\frac{1}{2}}\hspace{0.1cm}0\hspace{0.1cm}0\nonumber \\0\hspace{0.1cm}0\hspace{0.1cm}0\hspace{0.1cm}{\scriptstyle\frac{1}{2}}\hspace{0.1cm}0\hspace{0.1cm}0\nonumber \\0\hspace{0.1cm}{\scriptstyle\frac{1}{2}}\hspace{0.1cm}0\hspace{0.1cm}{\scriptstyle\frac{1}{2}}\hspace{0.1cm}0\hspace{0.1cm}{\scriptstyle\frac{1}{2}}\nonumber \\{\scriptstyle\frac{1}{2}}\hspace{0.1cm}0\hspace{0.1cm}{\scriptstyle\frac{1}{2}}\hspace{0.1cm}0\hspace{0.1cm}0\hspace{0.1cm}{\scriptstyle\frac{1}{2}}\nonumber \\0\hspace{0.1cm}{\scriptstyle\frac{1}{2}}\hspace{0.1cm}0\hspace{0.1cm}0\hspace{0.1cm}{\scriptstyle\frac{1}{2}}\hspace{0.1cm}{\scriptstyle\frac{3}{4}}\nonumber \\{\scriptstyle\frac{1}{2}}\hspace{0.1cm}{\scriptstyle\frac{1}{2}}\hspace{0.1cm}0\hspace{0.1cm}{\scriptstyle\frac{1}{2}}\hspace{0.1cm}0\hspace{0.1cm}{\scriptstyle\frac{1}{2}}\end{array}\right)\, .
\end{equation}
The gauge-group structure is
\begin{equation} G ~=~ SO(4)\otimes U(1)\otimes SO(4)\otimes 
            \underbrace{U(5)}_{\rm contains~SM}
    \otimes U(1)\otimes U(1)\otimes U(1)\otimes SO(6)\otimes U(1)\otimes SO(4)\otimes U(1)\, . 
\end{equation}

This model is potentially interesting because 
it contains not only four complete chiral quasi-supersymmetric untwisted generations of matter but
also a massless twisted generation of fermions arising in the $V_4+b_3$ sector, along with 
superpartners in 
the $V_0+V_1+V_4+b_3$ sector. These states are all in the fermionic representation of 
the parent $SO(10)$, so it is natural to associate the above $U(5)$ gauge-group factor 
with the $SU(5)\times U(1)_X$ gauge group
of the flipped $SU(5)$ unification scenario. The model also has 
the vector-like ${\bf 5}+\overline{\bf 5}$ 
Higgs representations required for electroweak symmetry breaking, but no GUT Higgses.

\subsection{SM-like model}

\label{model3}

\noindent This model is defined by the following vectors: 
\begin{eqnarray}V_0&=& - {\scriptstyle\frac{1}{2}}[~11~111~111~ | ~1111~11111~111~111~11~111~]\nonumber\\ 
V_1&=& - {\scriptstyle\frac{1}{2}}[~00~011~011~ | ~1111~11111~111~111~11~111~]\nonumber\\ 
V_2&=& - {\scriptstyle\frac{1}{2}}[~00~101~101~ | ~0101~00000~011~111~11~111~]\nonumber\\ 
b_3&=& - {\scriptstyle\frac{1}{2}}[~10~\overline{1}0\overline{0}~\overline{0}0\overline{1}~ | ~0001~11111~010~001~11~001~]\nonumber\\ 
V_4&=& - {\scriptstyle\frac{1}{2}}[~00~101~101~ | ~0101~00000~011~000~00~000~]\nonumber\\ 
V_5&=& - {\scriptstyle\frac{1}{2}}[~00~0\overline{0}\overline{0}~0\overline{1}\overline{1}~ | ~0101~11100~010~001~00~111~]\nonumber\\ 
V_7&=& - {\scriptstyle\frac{1}{4}}[~00~0\overline{2}\overline{2}~0\overline{0}\overline{0}~ | ~0202~11111~111~002~20~000~]\nonumber\\ 
 {\bf e} &=& {\scriptstyle\frac{1}{2}}[~00~101~101~ | ~0001~00000~000~111~01~111~]\end{eqnarray} 
The vector dot products and the $k_{ij}$ structure constants are respectively given by
\begin{equation}
V_i.V_j =\left( \begin{array}{c}1\hspace{0.1cm}0\hspace{0.1cm}0\hspace{0.1cm}0\hspace{0.1cm}0\hspace{0.1cm}0\hspace{0.1cm}{\scriptstyle\frac{3}{2}}\nonumber \\0\hspace{0.1cm}0\hspace{0.1cm}{\scriptstyle\frac{1}{2}}\hspace{0.1cm}{\scriptstyle\frac{1}{2}}\hspace{0.1cm}{\scriptstyle\frac{1}{2}}\hspace{0.1cm}0\hspace{0.1cm}{\scriptstyle\frac{3}{2}}\nonumber \\0\hspace{0.1cm}{\scriptstyle\frac{1}{2}}\hspace{0.1cm}0\hspace{0.1cm}1\hspace{0.1cm}0\hspace{0.1cm}{\scriptstyle\frac{3}{2}}\hspace{0.1cm}1\nonumber \\0\hspace{0.1cm}{\scriptstyle\frac{1}{2}}\hspace{0.1cm}1\hspace{0.1cm}0\hspace{0.1cm}0\hspace{0.1cm}{\scriptstyle\frac{3}{2}}\hspace{0.1cm}{\scriptstyle\frac{3}{2}}\nonumber \\0\hspace{0.1cm}{\scriptstyle\frac{1}{2}}\hspace{0.1cm}0\hspace{0.1cm}0\hspace{0.1cm}0\hspace{0.1cm}{\scriptstyle\frac{1}{2}}\hspace{0.1cm}{\scriptstyle\frac{1}{2}}\nonumber \\0\hspace{0.1cm}0\hspace{0.1cm}{\scriptstyle\frac{3}{2}}\hspace{0.1cm}{\scriptstyle\frac{3}{2}}\hspace{0.1cm}{\scriptstyle\frac{1}{2}}\hspace{0.1cm}0\hspace{0.1cm}{\scriptstyle\frac{5}{4}}\nonumber \\{\scriptstyle\frac{3}{2}}\hspace{0.1cm}{\scriptstyle\frac{3}{2}}\hspace{0.1cm}1\hspace{0.1cm}{\scriptstyle\frac{3}{2}}\hspace{0.1cm}{\scriptstyle\frac{1}{2}}\hspace{0.1cm}{\scriptstyle\frac{5}{4}}\hspace{0.1cm}1\end{array}\right)
\:\: \mbox{mod\, (2)}~,~~~~~
k_{ij} =\left( \begin{array}{c}0\hspace{0.1cm}0\hspace{0.1cm}0\hspace{0.1cm}0\hspace{0.1cm}0\hspace{0.1cm}0\hspace{0.1cm}0\nonumber \\0\hspace{0.1cm}0\hspace{0.1cm}0\hspace{0.1cm}{\scriptstyle\frac{1}{2}}\hspace{0.1cm}0\hspace{0.1cm}0\hspace{0.1cm}0\nonumber \\0\hspace{0.1cm}{\scriptstyle\frac{1}{2}}\hspace{0.1cm}0\hspace{0.1cm}0\hspace{0.1cm}0\hspace{0.1cm}{\scriptstyle\frac{1}{2}}\hspace{0.1cm}0\nonumber \\0\hspace{0.1cm}0\hspace{0.1cm}0\hspace{0.1cm}{\scriptstyle\frac{1}{2}}\hspace{0.1cm}0\hspace{0.1cm}0\hspace{0.1cm}0\nonumber \\0\hspace{0.1cm}{\scriptstyle\frac{1}{2}}\hspace{0.1cm}0\hspace{0.1cm}0\hspace{0.1cm}0\hspace{0.1cm}{\scriptstyle\frac{1}{2}}\hspace{0.1cm}{\scriptstyle\frac{1}{2}}\nonumber \\0\hspace{0.1cm}0\hspace{0.1cm}0\hspace{0.1cm}{\scriptstyle\frac{1}{2}}\hspace{0.1cm}0\hspace{0.1cm}0\hspace{0.1cm}{\scriptstyle\frac{3}{4}}\nonumber \\{\scriptstyle\frac{1}{2}}\hspace{0.1cm}{\scriptstyle\frac{1}{2}}\hspace{0.1cm}0\hspace{0.1cm}{\scriptstyle\frac{1}{2}}\hspace{0.1cm}0\hspace{0.1cm}{\scriptstyle\frac{1}{2}}\hspace{0.1cm}0\end{array}\right)\, .
\end{equation}
The gauge-group structure is
\begin{equation} 
     G ~=~ SO(4)\otimes U(1)\otimes U(1)\otimes  \underbrace{U(3)\otimes U(2)}_{\rm contains~SM} \otimes U(1)\otimes U(1)\otimes U(1)\otimes SO(4)\otimes U(1)\otimes U(1)\otimes U(1)\otimes SO(4)\otimes U(1)\, .
\label{SMgg}
\end{equation}
This model has $N^0_b=N^0_f=272$ massless real degrees of freedom in the untwisted sector. 

\subsection{Pati-Salam model}
\label{model2} 

\noindent This model is defined by the following vectors: 
\begin{eqnarray}
V_0&=& - {\scriptstyle\frac{1}{2}}[~11~111~111~ | ~1111~11111~111~11111111~]\nonumber\\ 
V_1&=& - {\scriptstyle\frac{1}{2}}[~00~011~011~ | ~1111~11111~111~11111111~]\nonumber\\ 
V_2&=& - {\scriptstyle\frac{1}{2}}[~00~101~101~ | ~0101~00000~011~11111111~]\nonumber\\ 
b_3&=& - {\scriptstyle\frac{1}{2}}[~10~\overline{1}0\overline{0}~\overline{0}0\overline{1}~ | ~0001~11111~001~10000111~]\nonumber\\ 
V_4&=& - {\scriptstyle\frac{1}{2}}[~00~101~101~ | ~0101~00000~011~00000000~]\nonumber\\ 
V_5&=& - {\scriptstyle\frac{1}{2}}[~00~0\overline{0}\overline{0}~0\overline{1}\overline{1}~ | ~0100~11100~000~11100111~]\nonumber\\ 
{\bf e} &=& {\scriptstyle\frac{1}{2}}[~00~101~101~ | ~1011~00000~000~00011111~]\, .
\end{eqnarray} 
The vector dot products and $k_{ij}$ structure constants for this model are given by
\begin{equation}
V_i\cdot V_j =\left( \begin{array}{c}1\hspace{0.1cm}0\hspace{0.1cm}0\hspace{0.1cm}0\hspace{0.1cm}0\hspace{0.1cm}0\nonumber \\0\hspace{0.1cm}0\hspace{0.1cm}{\scriptstyle\frac{1}{2}}\hspace{0.1cm}{\scriptstyle\frac{1}{2}}\hspace{0.1cm}{\scriptstyle\frac{1}{2}}\hspace{0.1cm}0\nonumber \\0\hspace{0.1cm}{\scriptstyle\frac{1}{2}}\hspace{0.1cm}0\hspace{0.1cm}1\hspace{0.1cm}0\hspace{0.1cm}{\scriptstyle\frac{3}{2}}\nonumber \\0\hspace{0.1cm}{\scriptstyle\frac{1}{2}}\hspace{0.1cm}1\hspace{0.1cm}0\hspace{0.1cm}0\hspace{0.1cm}{\scriptstyle\frac{3}{2}}\nonumber \\0\hspace{0.1cm}{\scriptstyle\frac{1}{2}}\hspace{0.1cm}0\hspace{0.1cm}0\hspace{0.1cm}0\hspace{0.1cm}0\nonumber \\0\hspace{0.1cm}0\hspace{0.1cm}{\scriptstyle\frac{3}{2}}\hspace{0.1cm}{\scriptstyle\frac{3}{2}}\hspace{0.1cm}0\hspace{0.1cm}0\end{array}\right)\:\: 
             \mbox{ mod\, (2)}~,~~~~~~ \:\:\:
k_{ij} =\left( \begin{array}{c}0\hspace{0.1cm}0\hspace{0.1cm}0\hspace{0.1cm}{\scriptstyle\frac{1}{2}}\hspace{0.1cm}0\hspace{0.1cm}0\nonumber \\0\hspace{0.1cm}0\hspace{0.1cm}0\hspace{0.1cm}{\scriptstyle\frac{1}{2}}\hspace{0.1cm}0\hspace{0.1cm}0\nonumber \\0\hspace{0.1cm}{\scriptstyle\frac{1}{2}}\hspace{0.1cm}0\hspace{0.1cm}0\hspace{0.1cm}0\hspace{0.1cm}{\scriptstyle\frac{1}{2}}\nonumber \\{\scriptstyle\frac{1}{2}}\hspace{0.1cm}0\hspace{0.1cm}0\hspace{0.1cm}0\hspace{0.1cm}0\hspace{0.1cm}{\scriptstyle\frac{1}{2}}\nonumber \\0\hspace{0.1cm}{\scriptstyle\frac{1}{2}}\hspace{0.1cm}0\hspace{0.1cm}0\hspace{0.1cm}0\hspace{0.1cm}0\nonumber \\0\hspace{0.1cm}0\hspace{0.1cm}0\hspace{0.1cm}0\hspace{0.1cm}0\hspace{0.1cm}0\end{array}\right)\, .
\end{equation}
 The gauge-group structure is
\begin{equation} 
 G ~=~ SO(4)\otimes U(1)\otimes U(1)\otimes 
                 \underbrace{SO(6)\otimes SO(4)}_{\rm contains~SM}
               \otimes U(1)\otimes U(1)\otimes U(1)\otimes U(1)\otimes SO(4)\otimes SO(4)\otimes SO(6)\, , 
\label{PSgg}
\end{equation}
where the Pati-Salam group corresponding to the visible sector is indicated.
This model, which has four quasi-supersymmetric chiral generations of massless 
        untwisted matter but no twisted matter,
 has $N^0_b=N^0_f=416$ massless degrees of freedom in the untwisted sector. 
Many similar examples can be found.

\bigskip

\bibliographystyle{unsrt}

\begin{thebibliography}{99}

\bibitem{SOsixteen}
   L.~Alvarez-Gaume, P.~H.~Ginsparg, G.~W.~Moore and C.~Vafa,
   ``An O(16) X O(16) Heterotic String,''
   Phys.\ Lett.\ B {\bf 171}, 155 (1986);\\
   L.~J.~Dixon and J.~A.~Harvey,
   ``String Theories In Ten-Dimensions Without Space-Time Supersymmetry,''
   Nucl.\ Phys.\ B {\bf 274}, 93 (1986).

\bibitem{Rohm}
  R.~Rohm,
  ``Spontaneous Supersymmetry Breaking in Supersymmetric String Theories,''
  Nucl.\ Phys.\ B {\bf 237}, 553 (1984).

\bibitem{nonSUSYgauge}
  V.~P.~Nair, A.~D.~Shapere, A.~Strominger and F.~Wilczek,
  ``Compactification of the Twisted Heterotic String,''
  Nucl.\ Phys.\  B {\bf 287}, 402 (1987);\\
  P.~H.~Ginsparg and C.~Vafa,
  ``Toroidal Compactification of Nonsupersymmetric Heterotic Strings,''
  Nucl.\ Phys.\  B {\bf 289}, 414 (1987).

\bibitem{Itoyama:1986ei} 
  H.~Itoyama and T.~R.~Taylor,
  ``Supersymmetry Restoration in the Compactified O(16) $\times$ O(16)-prime Heterotic String Theory,''
  Phys.\ Lett.\ B {\bf 186}, 129 (1987).

\bibitem{Itoyama:1987rc} 
  H.~Itoyama and T.~R.~Taylor,
  ``Small Cosmological Constant in String Models,''
  FERMILAB-CONF-87-129-T, C87-06-25.
  
\bibitem{Moore}
   G.~W.~Moore,
   ``Atkin-Lehner Symmetry,''
   Nucl.\ Phys.\ B {\bf 293}, 139 (1987)
   [Erratum-ibid.\ B {\bf 299}, 847 (1988)];\\
  J.~Balog and M.~P.~Tuite,
  ``The Failure Of Atkin-Lehner Symmetry For Lattice Compactified Strings,''
  Nucl.\ Phys.\ B {\bf 319}, 387 (1989);\\
  K.~R.~Dienes,
  ``Generalized Atkin-Lehner Symmetry,''
  Phys.\ Rev.\ D {\bf 42}, 2004 (1990).

\bibitem{Dienes:1990ij} 
  K.~R.~Dienes,
  ``New string partition functions with vanishing cosmological constant,''
  Phys.\ Rev.\ Lett.\  {\bf 65}, 1979 (1990).

\bibitem{KutasovSeiberg}
  D.~Kutasov and N.~Seiberg,
  ``Number of degrees of freedom, density of states and tachyons in string theory and CFT,''
  Nucl.\ Phys.\ B {\bf 358}, 600 (1991).

\bibitem{missusy}
  K.~R.~Dienes,
  ``Modular invariance, finiteness, and misaligned supersymmetry: 
        New constraints on the numbers of physical string states,''
  Nucl.\ Phys.\ B {\bf 429}, 533 (1994)
  [hep-th/9402006];   
  ``How strings make do without supersymmetry: An Introduction to misaligned supersymmetry,''
   In *Syracuse 1994, Proceedings, PASCOS '94* 234-243
  [hep-th/9409114];
  ``Space-time properties of (1,0) string vacua,''
  In *Los Angeles 1995, Future perspectives in string theory* 173-177
  [hep-th/9505194].

\bibitem{supertraces} 
  K.~R.~Dienes, M.~Moshe and R.~C.~Myers,
  ``String theory, misaligned supersymmetry, and the supertrace constraints,''
  Phys.\ Rev.\ Lett.\  {\bf 74}, 4767 (1995)
  [hep-th/9503055];
  ``Supertraces in string theory,''
  In *Los Angeles 1995, Future perspectives in string theory* 178-180
  [hep-th/9506001].

\bibitem{Kachru:1998hd} 
  S.~Kachru, J.~Kumar and E.~Silverstein,
  ``Vacuum energy cancellation in a nonsupersymmetric string,''
  Phys.\ Rev.\ D {\bf 59}, 106004 (1999);
  [hep-th/9807076];\\
  S.~Kachru and E.~Silverstein,
  ``On vanishing two loop cosmological constants in nonsupersymmetric strings,''
  JHEP {\bf 9901}, 004 (1999)
  [hep-th/9810129].

\bibitem{KachSilvothers}
   J.~A.~Harvey,
  ``String duality and nonsupersymmetric strings,''
  Phys.\ Rev.\ D {\bf 59}, 026002 (1999)
  [hep-th/9807213];\\
  S.~Kachru and E.~Silverstein,
  ``Selfdual nonsupersymmetric type II string compactifications,''
  JHEP {\bf 9811}, 001 (1998)
  [hep-th/9808056];\\
   R.~Blumenhagen and L.~Gorlich,
  ``Orientifolds of nonsupersymmetric asymmetric orbifolds,''
  Nucl.\ Phys.\ B {\bf 551}, 601 (1999)
  [hep-th/9812158];\\
  C.~Angelantonj, I.~Antoniadis and K.~Forger,
  ``Nonsupersymmetric type I strings with zero vacuum energy,''
  Nucl.\ Phys.\ B {\bf 555} (1999) 116
  [hep-th/9904092];\\
   M.~R.~Gaberdiel and A.~Sen,
  ``Nonsupersymmetric D-brane configurations with Bose-Fermi degenerate open string spectrum,''
  JHEP {\bf 9911}, 008 (1999)
  [hep-th/9908060].

\bibitem{Shiu:1998he} 
  G.~Shiu and S.~H.~H.~Tye,
  ``Bose-Fermi degeneracy and duality in nonsupersymmetric strings,''
  Nucl.\ Phys.\ B {\bf 542}, 45 (1999)
  [hep-th/9808095].

\bibitem{Iengo:1999sm} 
  R.~Iengo and C.~J.~Zhu,
  ``Evidence for nonvanishing cosmological constant in nonSUSY superstring models,''
  JHEP {\bf 0004}, 028 (2000)
  [hep-th/9912074].
  

\bibitem{DhokerPhong}
   E.~D'Hoker and D.~H.~Phong,
  ``Two loop superstrings 4: The Cosmological constant and modular forms,''
  Nucl.\ Phys.\ B {\bf 639}, 129 (2002)
  [hep-th/0111040];
  ``Lectures on two loop superstrings,''
  Conf.\ Proc.\ C {\bf 0208124}, 85 (2002)
  [hep-th/0211111].
 
\bibitem{Faraggi:2009xy}
  A.~E.~Faraggi and M.~Tsulaia,
  ``Interpolations Among NAHE-based Supersymmetric and Nonsupersymmetric String Vacua,''
  Phys.\ Lett.\ B {\bf 683} (2010) 314
  [arXiv:0911.5125 [hep-th]].

\bibitem{Angelantonj:2010ic} 
  C.~Angelantonj, M.~Cardella, S.~Elitzur and E.~Rabinovici,
  ``Vacuum stability, string density of states and the Riemann zeta function,''
  JHEP {\bf 1102}, 024 (2011)
  [arXiv:1012.5091 [hep-th]].
  
  \bibitem{Bergman:1997rf} 
  O.~Bergman and M.~R.~Gaberdiel,
  ``A Nonsupersymmetric open string theory and S duality,''
  Nucl.\ Phys.\ B {\bf 499}, 183 (1997)
  [hep-th/9701137];
  ``Dualities of type 0 strings,''
  JHEP {\bf 9907}, 022 (1999)
  [hep-th/9906055];\\
  R.~Blumenhagen and A.~Kumar,
  ``A Note on orientifolds and dualities of type 0B string theory,''
  Phys.\ Lett.\ B {\bf 464}, 46 (1999)
  [hep-th/9906234].

\bibitem{julie1} 
  J.~D.~Blum and K.~R.~Dienes,
  ``Duality without supersymmetry: The Case of the SO(16) $\times$ SO(16) string,''
  Phys.\ Lett.\ B {\bf 414}, 260 (1997)
  [hep-th/9707148].

\bibitem{julie2}
  J.~D.~Blum and K.~R.~Dienes,
  ``Strong / weak coupling duality relations for nonsupersymmetric string theories,''
  Nucl.\ Phys.\ B {\bf 516}, 83 (1998)
  [hep-th/9707160].

\bibitem{Faraggi:2007tj} 
  A.~E.~Faraggi and M.~Tsulaia,
  ``On the Low Energy Spectra of the Nonsupersymmetric Heterotic String Theories,''
  Eur.\ Phys.\ J.\ C {\bf 54}, 495 (2008)
  [arXiv:0706.1649 [hep-th]].
  
\bibitem{Dienes:2006ut} 
  K.~R.~Dienes,
  ``Statistics on the heterotic landscape: Gauge groups and cosmological constants of four-dimensional heterotic strings,''
  Phys.\ Rev.\ D {\bf 73}, 106010 (2006)
  [hep-th/0602286].

\bibitem{Dienes:2012dc} 
  K.~R.~Dienes, M.~Lennek and M.~Sharma,
  ``Strings at Finite Temperature: Wilson Lines, Free Energies, and the Thermal Landscape,''
  Phys.\ Rev.\ D {\bf 86}, 066007 (2012)
  [arXiv:1205.5752 [hep-th]].

  \bibitem{finitetemp}
       E.~Alvarez and M.~A.~R.~Osorio,
       ``Cosmological Constant Versus Free Energy For Heterotic Strings,''
       Nucl.\ Phys.\ B {\bf 304}, 327 (1988)
       [Erratum-ibid.\ B {\bf 309}, 220 (1988)];
       ``Duality Is An Exact Symmetry Of String Perturbation Theory,''
       Phys.\ Rev.\ D {\bf 40}, 1150 (1989); \\
        M.~A.~R.~Osorio,
         ``Quantum fields versus strings at finite temperature,''
         Int.\ J.\ Mod.\ Phys.\ A {\bf 7}, 4275 (1992).

\bibitem{AtickWitten}
     J.~J.~Atick and E.~Witten,
     ``The Hagedorn Transition And The Number Of Degrees Of Freedom Of String Theory,''
     Nucl.\ Phys.\ B {\bf 310}, 291 (1988).

\bibitem{wasKounnasRostand}
     M.~McGuigan,
     ``Finite Temperature String Theory And Twisted Tori,''
     Phys.\ Rev.\ D {\bf 38}, 552 (1988); \\
  I.~Antoniadis and C.~Kounnas,
  ``Superstring phase transition at high temperature,''
  Phys.\ Lett.\ B {\bf 261}, 369 (1991);\\
  I.~Antoniadis, J.~P.~Derendinger and C.~Kounnas,
  ``Non-perturbative temperature instabilities in N = 4 strings,''
  Nucl.\ Phys.\ B {\bf 551}, 41 (1999)
  [arXiv:hep-th/9902032].

\bibitem{Kounnas:1989dk}
      C.~Kounnas and B.~Rostand,
      ``Coordinate Dependent Compactifications And Discrete Symmetries,''
      Nucl.\ Phys.\ B {\bf 341}, 641 (1990).

\bibitem{earlystringpapersfiniteT}
  M.~J.~Bowick and L.~C.~R.~Wijewardhana,
  ``Superstrings At High Temperature,''
  Phys.\ Rev.\ Lett.\  {\bf 54}, 2485 (1985);\\
  S.~H.~H.~Tye,
  ``The Limiting Temperature Universe And Superstring,''
  Phys.\ Lett.\ B {\bf 158}, 388 (1985);\\
  B.~Sundborg,
  ``Thermodynamics Of Superstrings At High-Energy Densities,''
  Nucl.\ Phys.\ B {\bf 254}, 583 (1985);\\
  E.~Alvarez,
  ``Strings At Finite Temperature,''
  Nucl.\ Phys.\ B {\bf 269}, 596 (1986);\\
  E.~Alvarez and M.~A.~R.~Osorio,
  ``Superstrings At Finite Temperature,''
  Phys.\ Rev.\ D {\bf 36}, 1175 (1987);
  ``Thermal Heterotic Strings,''
  Physica {\bf A 158}, 449 (1989)
  [Erratum-ibid.\ A {\bf 160}, 119 (1989)];
  ``Thermal Strings In Nontrivial Backgrounds,''
  Phys.\ Lett.\ B {\bf 220}, 121 (1989);\\
  M.~Axenides, S.~D.~Ellis and C.~Kounnas,
  ``Universal Behavior Of D-Dimensional Superstring Models,''
  Phys.\ Rev.\ D {\bf 37}, 2964 (1988);\\
  Y.~Leblanc,
  ``Cosmological Aspects Of The Heterotic String Above The Hagedorn
  Temperature,''
  Phys.\ Rev.\ D {\bf 38}, 3087 (1988);\\
  B.~A.~Campbell, J.~R.~Ellis, S.~Kalara, D.~V.~Nanopoulos and K.~A.~Olive,
  ``Phase Transitions In QCD And String Theory,''
  Phys.\ Lett.\ B {\bf 255}, 420 (1991).

  \bibitem{Ferrara:1987es} 
  S.~Ferrara, C.~Kounnas and M.~Porrati,
  ``Superstring Solutions With Spontaneously Broken Four-dimensional Supersymmetry,''
  Nucl.\ Phys.\ B {\bf 304}, 500 (1988).

  \bibitem{Ferrara:1987qp} 
  S.~Ferrara, C.~Kounnas and M.~Porrati,
  ``$N=1$ Superstrings With Spontaneously Broken Symmetries,''
  Phys.\ Lett.\ B {\bf 206}, 25 (1988).
  
\bibitem{Ferrara:1988jx} 
  S.~Ferrara, C.~Kounnas, M.~Porrati and F.~Zwirner,
  ``Superstrings with Spontaneously Broken Supersymmetry and their Effective Theories,''
  Nucl.\ Phys.\ B {\bf 318}, 75 (1989).

\bibitem{Kiritsis:1997ca} 
  E.~Kiritsis and C.~Kounnas,
  ``Perturbative and nonperturbative partial supersymmetry breaking: N=4 $\to$ N=2 $\to$ N=1,''
  Nucl.\ Phys.\ B {\bf 503}, 117 (1997)
  [hep-th/9703059].
  
\bibitem{Dudas:2000ff} 
  E.~Dudas and J.~Mourad,
  ``Brane solutions in strings with broken supersymmetry and dilaton tadpoles,''
  Phys.\ Lett.\ B {\bf 486}, 172 (2000)
  [hep-th/0004165].

\bibitem{Scrucca:2001ni} 
  C.~A.~Scrucca and M.~Serone,
  ``On string models with Scherk-Schwarz supersymmetry breaking,''
  JHEP {\bf 0110}, 017 (2001)
  [hep-th/0107159].
 
\bibitem{Borunda:2002ra} 
  M.~Borunda, M.~Serone and M.~Trapletti,
  ``On the quantum stability of IIB orbifolds and orientifolds with Scherk-Schwarz SUSY breaking,''
  Nucl.\ Phys.\ B {\bf 653}, 85 (2003)
  [hep-th/0210075].
 
  \bibitem{Angelantonj:2006ut} 
 C.~Angelantonj, M.~Cardella and N.~Irges,
   ``An Alternative for Moduli Stabilisation,''
    Phys.\ Lett.\ B {\bf 641}, 474 (2006)
    [hep-th/0608022].

\bibitem{scherkschwarz}   
  J.~Scherk and J.~H.~Schwarz,
  ``Spontaneous Breaking of Supersymmetry Through Dimensional Reduction,''
  Phys.\ Lett.\ B {\bf 82}, 60 (1979).

\bibitem{Lust:1986kj}
  D.~Lust,
  ``Compactification Of The O(16) X O(16) Heterotic String Theory,''
  Phys.\ Lett.\ B {\bf 178}, 174 (1986).

\bibitem{Lerche:1986ae}
  W.~Lerche, D.~Lust and A.~N.~Schellekens,
  ``Ten-dimensional Heterotic Strings From Niemeier Lattices,''
  Phys.\ Lett.\ B {\bf 181}, 71 (1986).
 
\bibitem{Lerche:1986cx}
   W.~Lerche, D.~Lust and A.~N.~Schellekens,
  ``Chiral Four-Dimensional Heterotic Strings from Selfdual Lattices,''
  Nucl.\ Phys.\ B {\bf 287}, 477 (1987).

\bibitem{Chamseddine:1988ck} 
  A.~H.~Chamseddine, J.~P.~Derendinger and M.~Quiros,
  ``Nonsupersymmetric Four-dimensional Strings,''
  Nucl.\ Phys.\ B {\bf 311}, 140 (1988).
  
\bibitem{Font:2002pq} 
  A.~Font and A.~Hernandez,
  ``Nonsupersymmetric orbifolds,''
  Nucl.\ Phys.\ B {\bf 634}, 51 (2002)
  [hep-th/0202057].
 
\bibitem{Blaszczyk:2014qoa} 
  M.~Blaszczyk, S.~Groot Nibbelink, O.~Loukas and S.~Ramos-Sanchez,
  ``Non-supersymmetric heterotic model building,''
  JHEP {\bf 1410}, 119 (2014)
  [arXiv:1407.6362 [hep-th]].
  
\bibitem{Angelantonj:2014dia}
  C.~Angelantonj, I.~Florakis and M.~Tsulaia,
  ``Universality of Gauge Thresholds in Non-Supersymmetric Heterotic Vacua,''
  Phys.\ Lett.\ B {\bf 736} (2014) 365
  [arXiv:1407.8023 [hep-th]].

\bibitem{othernonsusy}
         C.~Bachas,
         ``A Way to break supersymmetry,''
         hep-th/9503030;\\
        J.~G.~Russo and A.~A.~Tseytlin,
          ``Magnetic flux tube models in superstring theory,''
          Nucl.\ Phys.\ B {\bf 461}, 131 (1996)
          [hep-th/9508068];\\
         A.~A.~Tseytlin,
          ``Closed superstrings in magnetic field: Instabilities and supersymmetry breaking,''
          Nucl.\ Phys.\ Proc.\ Suppl.\  {\bf 49}, 338 (1996)
          [hep-th/9510041];\\
          H.~P.~Nilles and M.~Spalinski,
          ``Generalized string compactifications with spontaneously broken supersymmetry,''
          Phys.\ Lett.\ B {\bf 392}, 67 (1997)
          [hep-th/9606145];\\
         I.~Shah and S.~Thomas,
         ``Finite soft terms in string compactifications with broken supersymmetry,''
         Phys.\ Lett.\ B {\bf 409}, 188 (1997)
         [hep-th/9705182].


\bibitem{Sagnotti:1995ga} 
  A.~Sagnotti,
  ``Some properties of open string theories,''
  In *Palaiseau 1995, Susy 95* 473-484
  [hep-th/9509080].
  
\bibitem{Sagnotti:1996qj} 
  A.~Sagnotti,
  ``Surprises in open string perturbation theory,''
  Nucl.\ Phys.\ Proc.\ Suppl.\  {\bf 56B}, 332 (1997)
  [hep-th/9702093].
  
\bibitem{Angelantonj:1998gj} 
  C.~Angelantonj,
  ``Nontachyonic open descendants of the 0B string theory,''
  Phys.\ Lett.\ B {\bf 444}, 309 (1998)
  [hep-th/9810214].
  
\bibitem{Blumenhagen:1999ns} 
  R.~Blumenhagen, A.~Font and D.~Lust,
  ``Tachyon free orientifolds of type 0B strings in various dimensions,''
  Nucl.\ Phys.\ B {\bf 558}, 159 (1999)
  [hep-th/9904069].
  
\bibitem{Sugimoto:1999tx} 
  S.~Sugimoto,
  ``Anomaly cancellations in type I D-9 - anti-D-9 system and the USp(32) string theory,''
  Prog.\ Theor.\ Phys.\  {\bf 102}, 685 (1999)
  [hep-th/9905159].
  
\bibitem{Aldazabal:1999tw} 
  G.~Aldazabal, L.~E.~Ibanez and F.~Quevedo,
  ``Standard - like models with broken supersymmetry from type I string vacua,''
  JHEP {\bf 0001}, 031 (2000)
  [hep-th/9909172].

\bibitem{Angelantonj:1999xc} 
  C.~Angelantonj,
  ``Nonsupersymmetric open string vacua,''
  PoS trieste {\bf 99}, 015 (1999)
  [hep-th/9907054].
 
\bibitem{Forger:1999ev} 
  K.~Forger,
  ``On nontachyonic Z(N) x Z(M) orientifolds of type 0B string theory,''
  Phys.\ Lett.\ B {\bf 469}, 113 (1999)
  [hep-th/9909010].
  
\bibitem{Moriyama:2001ge} 
  S.~Moriyama,
  ``USp(32) string as spontaneously supersymmetry broken theory,''
  Phys.\ Lett.\ B {\bf 522}, 177 (2001)
  [hep-th/0107203].
  
\bibitem{Angelantonj:2003hr} 
  C.~Angelantonj and I.~Antoniadis,
  ``Suppressing the cosmological constant in nonsupersymmetric type I strings,''
  Nucl.\ Phys.\ B {\bf 676}, 129 (2004)
  [hep-th/0307254].
  
\bibitem{Angelantonj:2004yt} 
  C.~Angelantonj,
  ``Open strings and supersymmetry breaking,''
  AIP Conf.\ Proc.\  {\bf 751}, 3 (2005)
  [hep-th/0411085].

\bibitem{Dudas:2004vi} 
  E.~Dudas and C.~Timirgaziu,
  ``Nontachyonic Scherk-Schwarz compactifications, cosmology and moduli stabilization,''
  JHEP {\bf 0403}, 060 (2004)
  [hep-th/0401201].
  
\bibitem{GatoRivera:2007yi} 
  B.~Gato-Rivera and A.~N.~Schellekens,
  ``Non-supersymmetric Tachyon-free Type-II and Type-I Closed Strings from RCFT,''
  Phys.\ Lett.\ B {\bf 656}, 127 (2007)
  [arXiv:0709.1426 [hep-th]].
  

\bibitem{Antoniadis:1988jn} 
  I.~Antoniadis, C.~Bachas, D.~C.~Lewellen and T.~N.~Tomaras,
  ``On Supersymmetry Breaking in Superstrings,''
  Phys.\ Lett.\ B {\bf 207}, 441 (1988).

 \bibitem{Antoniadis:1990ew} 
  I.~Antoniadis,
 ``A Possible new dimension at a few TeV,''
  Phys.\ Lett.\ B {\bf 246}, 377 (1990).

\bibitem{Antoniadis:1992fh} 
  I.~Antoniadis, C.~Munoz and M.~Quiros,
  ``Dynamical supersymmetry breaking with a large internal dimension,''
  Nucl.\ Phys.\ B {\bf 397}, 515 (1993)
  [hep-ph/9211309].
  
\bibitem{Antoniadis:1996hk} 
  I.~Antoniadis and M.~Quiros,
  ``Large radii and string unification,''
  Phys.\ Lett.\ B {\bf 392}, 61 (1997)
  [hep-th/9609209].

\bibitem{Benakli:1998pw}
  K.~Benakli,
  ``Phenomenology of low quantum gravity scale models,''
  Phys.\ Rev.\ D {\bf 60}, 104002 (1999)
  [hep-ph/9809582].

\bibitem{Bachas:1999es}
  C.~P.~Bachas,
  ``Scales of string theory,''
  Class.\ Quant.\ Grav.\  {\bf 17}, 951 (2000)
  [hep-th/0001093].

\bibitem{Dudas:2000bn} 
  E.~Dudas,
  ``Theory and phenomenology of type I strings and M theory,''
  Class.\ Quant.\ Grav.\  {\bf 17}, R41 (2000)
  [hep-ph/0006190].
  

\bibitem{Fischler:1986ci}
  W.~Fischler and L.~Susskind,
  ``Dilaton Tadpoles, String Condensates and Scale Invariance,''
  Phys.\ Lett.\ B {\bf 171}, 383 (1986).

\bibitem{Fischler:1986tb}
   W.~Fischler and L.~Susskind,
  ``Dilaton Tadpoles, String Condensates and Scale Invariance. 2.,''
  Phys.\ Lett.\ B {\bf 173}, 262 (1986).

  
\bibitem{Hagedorn}
  R.~Hagedorn,
  ``Statistical Thermodynamics Of Strong Interactions At High-Energies,''
  Nuovo Cim.\ Suppl.\  {\bf 3}, 147 (1965).

\bibitem{HR}
  G.~H.~Hardy and S.~Ramanujan,
  Proc.\ London Math.\ Soc.\  {\bf 17}, 75 (1918);\\
  I.~Kani and C.~Vafa,
  ``Asymptotic Mass Degeneracies In Conformal Field Theories,''
  Commun.\ Math.\ Phys.\  {\bf 130}, 529 (1990).


\bibitem{heretic} 
  K.~R.~Dienes,
  ``Solving the hierarchy problem without supersymmetry or extra dimensions: An Alternative approach,''
  Nucl.\ Phys.\ B {\bf 611}, 146 (2001)
  [hep-ph/0104274].

\bibitem{KLTclassification}
  H.~Kawai, D.~C.~Lewellen and S.~H.~H.~Tye,
  ``Classification Of Closed Fermionic String Models,''
  Phys.\ Rev.\ D {\bf 34}, 3794 (1986).



\bibitem{Antoniadis:1989zy} 
  I.~Antoniadis, J.~R.~Ellis, J.~S.~Hagelin and D.~V.~Nanopoulos,
  ``The Flipped SU(5) $\times$ U(1) String Model Revamped,''
  Phys.\ Lett.\ B {\bf 231}, 65 (1989).

\bibitem{Antoniadis:1990hb} 
  I.~Antoniadis, G.~K.~Leontaris and J.~Rizos,
  ``A Three generation SU(4) x O(4) string model,''
  Phys.\ Lett.\ B {\bf 245}, 161 (1990).

\bibitem{Faraggi:1991be} 
  A.~E.~Faraggi,
  ``Hierarchical top - bottom mass relation in a superstring derived standard - like model,''
  Phys.\ Lett.\ B {\bf 274}, 47 (1992).

\bibitem{Faraggi:1991jr} 
  A.~E.~Faraggi,
  ``A New standard - like model in the four-dimensional free fermionic string formulation,''
  Phys.\ Lett.\ B {\bf 278}, 131 (1992).

\bibitem{Faraggi:1992fa} 
  A.~E.~Faraggi,
  ``Construction of realistic standard - like models in the free fermionic superstring formulation,''
  Nucl.\ Phys.\ B {\bf 387}, 239 (1992)
  [hep-th/9208024].

\bibitem{Faraggi:1994eu} 
  A.~E.~Faraggi,
  ``Custodial nonAbelian gauge symmetries in realistic superstring derived models,''
  Phys.\ Lett.\ B {\bf 339}, 223 (1994)
  [hep-ph/9408333].

\bibitem{Dienes:1995bx} 
  K.~R.~Dienes and A.~E.~Faraggi,
  ``Gauge coupling unification in realistic free fermionic string models,''
  Nucl.\ Phys.\ B {\bf 457}, 409 (1995)
  [hep-th/9505046].


\bibitem{Kawai:1986ah} 
  H.~Kawai, D.~C.~Lewellen and S.~H.~H.~Tye,
  ``Construction of Fermionic String Models in Four-Dimensions,''
  Nucl.\ Phys.\ B {\bf 288}, 1 (1987).
 
  \bibitem{Antoniadis:1986rn} 
  I.~Antoniadis, C.~P.~Bachas and C.~Kounnas,
  ``Four-Dimensional Superstrings,''
  Nucl.\ Phys.\ B {\bf 289}, 87 (1987).
 
 \bibitem{Kawai:1987ew} 
  H.~Kawai, D.~C.~Lewellen, J.~A.~Schwartz and S.~H.~H.~Tye,
  ``The Spin Structure Construction of String Models and Multiloop Modular Invariance,''
  Nucl.\ Phys.\ B {\bf 299}, 431 (1988).

\bibitem{Chamseddine:1989mz} 
  A.~H.~Chamseddine, J.~P.~Derendinger and M.~Quiros,
  ``A Unified Formalism for Strings in Four-dimensions,''
  Nucl.\ Phys.\ B {\bf 326}, 497 (1989).

\bibitem{Nooij:2004cz} 
   S.~E.~M.~Nooij,
  ``Classification of the chiral Z(2) x Z(2) heterotic string models,''
    hep-th/0603035.

  
\bibitem{Murayama:2012jh} 
  H.~Murayama, Y.~Nomura, S.~Shirai and K.~Tobioka,
  ``Compact Supersymmetry,''
  Phys.\ Rev.\ D {\bf 86}, 115014 (2012)
  [arXiv:1206.4993 [hep-ph]].

 \bibitem{Dimopoulos:2014aua}
  S.~Dimopoulos, K.~Howe and J.~March-Russell,
  ``Maximally Natural Supersymmetry,''
  Phys.\ Rev.\ Lett.\  {\bf 113}, 111802 (2014)
  [arXiv:1404.7554 [hep-ph]];\\
  I.~G.~Garcia and J.~March-Russell,
  ``Rare Flavor Processes in Maximally Natural Supersymmetry,''
  arXiv:1409.5669 [hep-ph].
 
\bibitem{Shadmi:2011hs} 
  Y.~Shadmi and P.~Z.~Szabo,
  ``Flavored Gauge-Mediation,''
  JHEP {\bf 1206}, 124 (2012)
  [arXiv:1103.0292 [hep-ph]];\\
  N.~Craig, M.~McCullough and J.~Thaler,
  ``Flavor Mediation Delivers Natural SUSY,''
  JHEP {\bf 1206}, 046 (2012)
  [arXiv:1203.1622 [hep-ph]].

\bibitem{Davies:2011mp} 
  R.~Davies, J.~March-Russell and M.~McCullough,
  ``A Supersymmetric One Higgs Doublet Model,''
  JHEP {\bf 1104}, 108 (2011)
  [arXiv:1103.1647 [hep-ph]].


\bibitem{String-Calc}
   Our conventions are as in J.~Polchinski, ``String Theory'', Vols.\ I and II,  1998.\\ 
   Many results may also be found in:\\
    E.~Kiritsis, ``String Theory in a Nutshell'',  2007 Princeton, NJ: 
    Princeton University Press.\\
 Additional results for Weierstrass $P$-function integrals were derived using identities in:\\
   S.~A.~Abel and B.~W.~Schofield,
  ``One-loop Yukawas on intersecting branes,''
    JHEP {\bf 0506}, 072 (2005)
  [hep-th/0412206];\\
   S.~A.~Abel and M.~D.~Goodsell,
  ``Intersecting brane worlds at one loop,''
  JHEP {\bf 0602}, 049 (2006)
  [hep-th/0512072].


\bibitem{Kiritsis:1996xd}
  E.~Kiritsis, C.~Kounnas, P.~M.~Petropoulos and J.~Rizos,
  ``Solving the decompactification problem in string theory,''
  Phys.\ Lett.\ B {\bf 385} (1996) 87
  [hep-th/9606087].

\bibitem{Kiritsis:1998en}
  E.~Kiritsis, C.~Kounnas, P.~M.~Petropoulos and J.~Rizos,
  ``String threshold corrections in models with spontaneously broken supersymmetry,''
  Nucl.\ Phys.\ B {\bf 540}, 87 (1999)
  [hep-th/9807067].
  
  \bibitem{recentFaraggi}
   A.~E.~Faraggi, C.~Kounnas and H.~Partouche,
  ``Large volume susy breaking with a chiral solution to the decompactification problem,''
  arXiv:1410.6147 [hep-th].


  
\bibitem{Dienes:1998vh}
  K.~R.~Dienes, E.~Dudas and T.~Gherghetta,
  ``Extra space-time dimensions and unification,''
  Phys.\ Lett.\ B {\bf 436}, 55 (1998)
  [hep-ph/9803466];
  ``Grand unification at intermediate mass scales through extra dimensions,''
  Nucl.\ Phys.\ B {\bf 537}, 47 (1999)
  [hep-ph/9806292];
  ``TeV scale GUTs,''
  hep-ph/9807522.

\bibitem{Antoniadis:2000vd}
  I.~Antoniadis and K.~Benakli,
  ``Large dimensions and string physics in future colliders,''
  Int.\ J.\ Mod.\ Phys.\ A {\bf 15}, 4237 (2000)
  [hep-ph/0007226].
  

\end{thebibliography}

\end{document}